\documentclass[12pt]{article}
\usepackage{latexsym}
\usepackage{amsmath}
\usepackage{amssymb}
\usepackage{epsfig,graphics}
\usepackage{graphicx}
\usepackage{booktabs}
\usepackage{multirow}
\usepackage{tikz}
\usepackage{bm}
\usepackage{graphicx}
\usepackage{epsf}
\usepackage{epsfig}
\usepackage{color}
\usepackage{dcolumn}     
\usepackage{amsfonts,amssymb}
\usepackage{dsfont}
\usepackage{slashed}
\usepackage{subcaption}
\usepackage{longtable}
\usepackage{float}

 \usepackage[nottoc]{tocbibind}
\usepackage[colorlinks=true,linkcolor=blue,urlcolor=magenta, citecolor=blue]{hyperref}
\usepackage{fullpage}
\usepackage{amssymb,graphicx}
\usepackage{amsmath}
\usepackage[margin=0.5cm]{caption}
\usepackage{hyperref} 
\usepackage{cite}
\usepackage{upgreek}

\def\d{\partial}
\def\l{\left(}
\def\r{\right)}

\newcommand{\be}{\begin{equation}}
\newcommand{\ee}{\end{equation}}
\newcommand{\bea}{\begin{eqnarray}}
\newcommand{\eea}{\end{eqnarray}}
\newcommand{\bg}{\begin{gather}}
\newcommand{\eg}{\end{gather}}
\newcommand{\bseq}{\begin{subequations}}
\newcommand{\eseq}{\end{subequations}}

\renewcommand{\ln}{\mathop{\rm ln}\nolimits}

\newcommand{\Tr}{{\rm Tr}}

\begin{document}

\begin{titlepage}
\begin{center}
{\large\bf Confining Strings and the Worldsheet Axion from the Lattice}\\
\vspace{0.5cm}
{ \large
Andreas Athenodorou$^{a,b}$, Sergei Dubovsky$^{c,d}$, Conghuan Luo$^{c,e}$, \\ \vspace{0.1cm} and Michael Teper$^{f}$ 
}\\
\vspace{.3cm}
{\small\textit{$^a$ Computation-based Science and Technology Research Center,
The Cyprus Institute, Cyprus
}}\\
\vspace{0.3cm}
{\small\textit{$^b$ Dipartimento di Fisica, Universit\'a di Pisa and INFN,
Sezione di Pisa, Largo Pontecorvo 3, 
56127 Pisa, Italy
}}\\
\vspace{0.3cm}
{\small  \textit{  $^c$Center for Cosmology and Particle Physics, Department of Physics,
      New York University 
      New York, NY, 10003, USA}}\\ 
      \vspace{.3cm}  
       {\small  \textit{  $^d$Institute for Advanced Study, Princeton, NJ 08540, USA}}\\ 
       \vspace{.3cm}  
{\small  \textit{  $^e$Department of Physics and Astronomy, University of Southern California,
Los Angeles, CA, 90089, USA}}\\ 
      \vspace{.3cm}  
      {\small \textit {{$^f$Rudolf Peierls Centre for Theoretical
Physics, Clarendon Laboratory, University of Oxford,  Parks Road, Oxford OX1 3PU, UK\\
\centerline{and}
All Souls College, University of Oxford,
High Street, Oxford OX1 4AL, UK}}}

\end{center}
\begin{abstract}
We present a major update on the spectrum of closed flux tubes in $D=3+1$ $SU(N)$ gauge theories.
We measure the excitation spectrum of confining strings wound around a spatial dimension of a size $R$. We do so for the $SU(N)$ Yang-Mills theory with $N=3,5,6$ and for two different values of the lattice spacing. 
We employ the generalized eigenvalue problem in combination with an extended basis of operators; this enables us to project onto all possible irreducible representations characterised by spin $|J_{\rm modulo \ 4}|$, transverse parity $P_{\perp}$, longitudinal parity $P_{\parallel}$ as well as by longitudinal momentum $p_{\parallel}=\frac{2 \pi q}{R}$, and extract accurate results for approximately $35$ lightest states. 
Applying the Thermodynamic Bethe Ansatz (TBA) technique for calculating the finite volume spectrum, we confirm that the observed states are well described by the low energy effective theory of a long string consisting of  two translational Goldstone bosons (``phonons"), along with a massive pseudoscalar (``the worldsheet axion") coupled to phonons through a $\theta$-term. 
Moreover, we find that the leading axion-axion and axion-phonon interactions are well approximated by the $T\bar{T}$ deformation of a free axion. 
\end{abstract}

\end{titlepage}

\setcounter{page}{1}
\newpage
\pagestyle{plain}

\tableofcontents

\section{Introduction}

In a confining gauge theory, such as $SU(N)$ gluodynamics,  the energy of a gauge field sourced by color charges is squeezed into narrow flux tubes---confining strings. Describing the worldsheet dynamics of confining strings is one of the key steps towards 
understanding color confinement. This question is still open in spite of numerous attempts to attack it both from a pure theory side and also informed by lattice data. 

To fully benefit from lattice data one needs a theoretical framework for relating the excitation spectrum of confining flux tubes, which is measured in lattice simulations, to fundamental parameters of the worldsheet theory. One way to achieve this is to start with a candidate worldsheet theory, calculate the corresponding excitation spectrum and compare it to lattice data. Alternatively, one may attempt to extract  dynamics of the worldsheet theory from lattice data in a model-independent way.

The former approach is most natural to apply in the regime when the confining string is long and smooth. In this regime the worldsheet dynamics is universal and described by a low energy theory of translational Goldstone modes (``phonons") governed by the Nambu--Goto action supplemented with subleading higher dimensional operators\footnote{This is true provided no additional gapless modes are present on the worldsheet. These are not expected to be found for the Yang--Mills strings, and lattice data confirms this expectation.}. In this paper we will focus on excitations of torelons---closed confining strings wound around a compact spatial direction of size $R$. A traditional approach for calculating the excitation spectrum of torelons is based on the straightforward perturbative expansion in powers of   $\ell_s^2/R^2$, where 
\[\ell_s^{-2} \equiv \sigma
\]
is the string tension\footnote{We shall use these two notations interchangeably in this paper. }\cite{Luscher:2004ib,Drummond:2006su,Dass:2006ud,Aharony:2009gg,Dass:2009xe,HariDass:2009ub,HariDass:2009ud,Dubovsky:2012sh,Aharony:2013ipa}. We will review this approach in more details in Section~\ref{subsec:Effective_String_Theory}.

A comparison of these universal results with the high precision measurements of the torelon excitations 
both in 3d \cite{Bringoltz:2006zg,Athenodorou:2011rx} and 4d \cite{Athenodorou:2010cs} Yang--Mills theory has led to the following puzzle. In both cases the ground state energy demonstrates an excellent agreement with the universal perturbative predictions. However, the agreement for  excited states is much less accurate, indicating an earlier breakdown of the perturbation theory. On the other hand, for the majority of excited states their energies agree remarkably well  with the result obtained by the ``exact" light cone quantization of a free bosonic string (also known as Goddard--Goldstone--Rebbi--Thorn (GGRT) spectrum \cite{goddard1973quantum,arvis1983exactqq}), even at very large values of the naive perturbative parameter $\ell_s^2/R^2\sim 1$. Note however, that the latter
quantization is not consistent away from a critical number of space-time dimensions, $D=26$.

These results strongly suggest that there should exist a perturbative approach with  better convergence properties. Indeed, it turns out that the better way to systematically compute the excitation spectrum using perturbation theory is via the Thermodynamic Bethe Ansatz (TBA) \cite{zamolodchikov1990thermodynamic,Dorey:1996re} reviewed in Section~\ref{subsec:TBA}.
In this approach the GGRT formula emerges as the leading approximation to the confining string spectrum \cite{Dubovsky:2012wk}, which explains why it works so well. The main idea of this method is to separate the finite volume spectrum calculation into two steps. First one perturbatively calculates the worldsheet scattering amplitudes using the low energy effective theory. At the second step one finds the corresponding finite volume spectrum. Ideally, the second step is to be performed exactly. In practice, it is only possible  when worldsheet scattering does not exhibit particle production. This is indeed the case for the leading order (tree level) Nambu--Goto scattering. At higher orders one needs to rely on approximate methods to be introduced later. 

 A better convergence of this technique is related to the fact that the actual momenta of excitations on the string are softened compared to the naive estimate $p \ll \frac{2\pi}{R}$ as a consequence of interactions. As a result using $p\ell_s$ instead of $\ell_s/R$ as the expansion parameter makes much better predictions for excited states of short strings.
 
 It is worth noting, that this technique is well familiar in lattice community under the name of  L\"uscher equations, relating the scattering data and the finite volume spectrum \cite{Luscher:1990ux}.
A complication present in the present case is that the worldsheet theory is gapless, which implies that the so  called winding corrections are not exponentially suppressed. In the integrable limit, the L\"uscher equations correspond to the Asymptotic Bethe Ansatz (ABA), which neglects the winding corrections. To account for the latter one may either rely on the full TBA in the integrable approximation or on the $T\bar{T}$ dressing technique, which is reviewed later.

A priori, one expects confining flux tubes to exhibit massive worldsheet excitations reflecting the fact that these are not fundamental strings.
However, detailed lattice studies of the low-lying excited states in 2+1 dimensional $SU(N)$ gauge theories revealed no additional massive  string modes associated with the non-trivial structure of the flux tube in the fundamental representation of the gauge group \cite{Bringoltz:2006zg,Athenodorou:2011rx}.

The situation is different for the flux tube excitation spectrum in $D=3+1$ dimensions. A first comprehensive study of the torelon excitation spectrum in  
$D=3+1$ dimensions was presented in \cite{Athenodorou:2010cs}. This work focused on the fundamental representation of the $SU(N)$ gauge group for $N=3,5$.  Accurate results for about $20$  lightest excitations with different quantum numbers were presented. It was demonstrated that the majority of these low-lying states are well described by the spectrum of the GGRT string theory. However, it was also found that there are anomalous pseudoscalar states, that exhibit large deviations from the GGRT spectrum. 

Application of the TBA technique demonstrates that the additional states  arise due to the presence of a massive pseudoscalar resonance coupled to the Goldstone modes (``the worldsheet axion")\cite{Dubovsky:2013gi,Dubovsky:2014fma}. Intriguingly, it is observed \cite{Dubovsky:2015zey} that the numerical value of the leading cubic axion-Goldstone-Goldstone coupling extracted from the lattice data coincides (within error bars) with its value in an integrable model, which is consistent with the non-linearly realized  4d Poincar\'e symmetry. Note, however, that the flux tube 
worldsheet theory is definitely not integrable,  because integrability requires the axion mass to vanish, which is not the case neither at $N=3,5$  \cite{Athenodorou:2010cs} nor in the infinite $N$ limit~\cite{Athenodorou:2017cmw}. 

Interestingly, a single Goldstone boson in $D=2+1$ can also be integrable consistently with the non-linearly realized 3d Poincar\'e symmetry.
However, the excitation spectrum of 3d flux tubes \cite{Bringoltz:2006zg,Athenodorou:2011rx} is again incompatible with the exact integrability \cite{Dubovsky:2014fma,Chen:2018keo}. 
So both at $D=3+1$ and $D=2+1$ one finds that the matter content of the worldsheet theory coincides with that of a possible integrable model. In both cases, however, the integrability is not exact.
These observations lead to a proposal of the Axionic String Ansatz (ASA)
\cite{Dubovsky:2015zey}, reviewed in Section~\ref{subsec:ASA}. In particular, the ASA expectation is that the integrability gets restored at high energies, at least for a suitable subset of inclusive observables.
An indirect confirmation of the ASA comes from a surprisingly successful proposal for the quantum numbers of $D=2+1$ glueballs  which has been put forward based on the ASA \cite{Dubovsky:2016cog,Conkey:2019blu,Dubovsky:2021cor}.

A crucial test of the ASA is to check that no further massive states appear on the worldsheet as one measures higher torelon excitations. 
Motivated by this, in this paper, we  report a major update on the Monte-Carlo simulations of the spectrum of closed flux tube in $D=3+1$  $SU(N)$ gauge theories, which includes more higher excited states with higher precision. To achieve this we improve on the calculation presented in  \cite{Athenodorou:2017cmw} by increasing the basis of operators. Despite the fact that calculations in $D=3+1$ are generally much slower and thus less accurate than in $D=2+1$, we are able to obtain accurate results with statistics comparable to those in calculations for $D=2+1$. 
 
 The rest of the paper  is organized in the following way. After a brief review of the effective string theory and the ASA in Section~\ref{sec:theory}, we proceed 
 with presenting the lattice data. Namely, we first  review our lattice methods and conventions  in Section~\ref{sec:lattice}. In Section~\ref{sec:data} we present the results for $N=3,5,6$, and with two different lattice spacings for the first two cases. Our results cover a broad range of flux-tube circumference from very short strings close to the deconfining transition $R/\ell_s \sim 1.4$ to the very long ones with $R/\ell_s \sim 7$.

We then proceed with a theoretical analysis and interpretation of the presented results. Namely, using the TBA method in Section~\ref{sec:computation} we compute the spectrum of states containing up to two phonon excitations. Here we adopt the counting where the axion is equivalent to two phonons, because it appears as a resonant state in the two phonon scattering. Then we describe the  $T\bar{T}$ technique, which we use to describe three- and four-phonon excitations. Some of these states can also be described as (one- and two-)axion states possibly interacting with additional phonons. By confronting the results with the lattice data we find that all observed states can be described
using Goldstones and the axion. Interestingly, we also find that in most cases the $T\bar{T}$ deformation provides an accurate quantitative description of the interactions between axions and Goldstones.
We conclude in Section~\ref{sec:conclusion}.

\section{A brief overview of past investigations}
\label{sec:historical_review}

Lattice calculations of both open and closed confining flux tubes (as well as Wilson loops)
began in the mid-1980s, with the focus on establishing the approximate linear increase with
length of the ground state energy and of the L\"uscher correction to this energy.
Computations of excited open string states began in the late-1980s,
e.g. \cite{Perantonis:1988uz},
motivated in large part by their role in model calculations of hybrid meson masses, e.g.
\cite{Griffiths:1983ah,Perantonis:1990dy}. The Torino group calculated
Wilson loop expectation values in various gauge and spin models, comparing these to
Nambu-Goto (GGRT) string theory predictions, e.g. \cite{Caselle:1995fh,Caselle:1996ii}.

{ Open flux tube energies obtain contributions from the world lines of the endpoints (boundary terms in the effective string action).  While this complicates the computation, analysis, and interpretation of the spectrum, the resulting spectrum is more directly related to the one measured in actual experiments.}. Physically, the spectrum
is governed by perturbation theory at small separations $R$, and by string-like excitations
at larger $R$. Early calculations of
a limited part of the spectrum, with a comparison to the Nambu-Goto spectrum, include
\cite{Perantonis:1990dy}, where the calculations were limited to the lightest states in the
various symmetry representations, and \cite{Michael:1994ej} where the first excited state
in the totally symmetric representation was found to agree well with Nambu-Goto. A much more
complete and precise comparison with Nambu-Goto was carried out in \cite{Kuti:2005xg},
and this has recently been superseded in \cite{Bicudo:2022lnq,Sharifian:2023idc}
with the calculation of a significant number of excitations in all the symmetry
representations. Many states agree well with Nambu-Goto, while this work also finds evidence
for a massive state equal to the mass found in the spectrum of closed flux tubes
in \cite{Athenodorou:2011rx} and which is studied in detail in the present paper. 

Torelons, closed flux tubes that wind around a spatial torus, avoid any boundary terms
and provide an effective way to investigate the effective string action of confining
flux tubes. For torelon lengths $R$ below a critical length $R_c$, the system is in a
`deconfined' phase so there is no longer a winding flux tube to consider. This transition
is just a relabeling of the usual finite temperature deconfinement so $R_c=1/T_c$ where {$T_c$ is the usual deconfining temperature, below which the center symmetry is spontaneously broken and we are in a ‘deconfining’ phase in which we no longer
have confining flux tubes around the x-torus.
While for $SU(2)$~\cite{Chernodub:2018aix} the transition is second order for $SU(N\geq 3)$ the transition is first order providing a relatively clean
indication of where the string theory should be tested.} (In contrast to open strings
where the separation in $R$ between the perturbative and string like regimes is
more ambiguous.) Hence our earlier focus on torelons both in the $D=2+1$ context \cite{Athenodorou:2011rx},
and in the $D=3+1$ context \cite{Athenodorou:2010cs}. A caveat in all cases is that
highly excited strings can be unstable, e.g. through the emission of glueballs,
although their stability is gradually restored as we increase $N$. 

The early work on the spectrum of torelons in $D=3+1$ gauge theories focused on the absolute ground state
so as to calculate the string tension and to verify the leading (L\"uscher) correction, e.g. \cite{deForcrand:1984wzs}.
These early calculations were consistent with the only massless modes being the Goldstone modes associated
with the transverse fluctuations of the string but were not very precise. Later more precise calculations,
as in \cite{Michael:1994ej}, confirmed this conclusion. Subsequent calculations of excited $SU(3)$ torelon
states for all (lattice) quantum numbers in \cite{Juge:2003vw} had mixed agreement with
the GGRT string spectrum, perhaps due to substantial systematic errors such as glueball
emission. Calculations in \cite{Meyer:2004hv} controlled this potential error with calculations
in $SU(4)$ and $SU(6)$, the latter showing not only that the first excited state was close to the
GGRT prediction but that, remarkably, this persisted to very short flux tube lengths,
comparable to the expected intrinsic width of the flux tube. Highly accurate calculations of
many excited torelon states, for a large range of lengths $R$, and in $SU(5)$ as well as $SU(3)$
were carried out in \cite{Athenodorou:2010cs} showing that most of the low-lying states are
remarkably well-described by the spectrum of GGRT~\eqref{GGRT} not only for larger values of $R$
but also, unexpectedly, for much shorter lengths. This closely mirrored the results obtained 
in 2+1 dimensions \cite{Athenodorou:2011rx}, except that certain states with $J^{P_\perp}=0^-$
quantum numbers deviated significantly from the GGRT spectrum. These provided the first plausible indication
of massive excitations in the spectrum of the confining flux tube. Shortly afterwards 
the work of \cite{Dubovsky:2013gi} provided a novel theoretical framework for understanding this
massive excitation as a pseudoscalar "world-sheet axion", as well as predicting the unexpected approximate
agreement of the flux tube spectrum with the GGRT string spectrum for short flux tubes. 
The discovery that the world sheet theory would become integrable in the planar limit\cite{Dubovsky:2015zey}
if the axion mass were to vanish in that limit prompted a lattice study of its mass at one lattice spacing,
over the range of  $SU(2)$ to $SU(12)$ \cite{Athenodorou:2017cmw}, which showed that it does not vanish even
if its world sheet coupling does appear to satisfy that constraint. The present paper continues this
investigation, with a systematic search through all quantum numbers of the flux tube, with a large
basis of operators, a range of colors $N=3,5,6$ and two lattice spacings so as to provide some
control over the large-$N$ and continuum behaviors.  A small set of the results produced within the
scope of this work have been presented in a number of conferences proceedings\cite{Athenodorou:2021vkw,Athenodorou:2022tsj}.

\section{From effective string theory to 4D Yang-Mills confining string spectrum}
\label{sec:theory}

\subsection{Effective theory of long strings}
\label{subsec:Effective_String_Theory}

Universal properties of low energy excitations of a long string have been studied extensively in the last few decades (review can be found in, e.g.,  \cite{Dubovsky:2012sh,Aharony:2013ipa}). The main
idea is that a long string describes dynamics of translational $(D-2)$ Goldstone bosons, enjoying a  non-linearly realized $ISO(1,D-1)$ Poincar\'e symmetry, which is spontaneously broken to the $ISO(1,1)\times
O(D-2)$ subgroup. The resulting action is most conveniently written in terms of $D$ worldsheet scalar fields $X^{\mu}$ describing the embedding of the string worldsheet into the target spacetime.
These fields transform as a $ISO(1,D-1)$ vector, and the corresponding action also enjoys the worldsheet reparametrization  invariance,
\begin{equation}
    S = -\int d^2 \sigma \sqrt{-\det h_{\alpha\beta}} \left[ \ell_s^{-2}  +\gamma\ell_s^2 {\cal R}^2  + \dots  \right] \,,
    \label{effective_string}
\end{equation}
where $\ell_s^{-2}$ is the string tension,  $\gamma$ is the first  non-universal Wilson coefficient in this effective theory and dots stand for further ${\cal O}(\ell_s^2)$ and higher order terms. 
Here $h_{\alpha\beta}$ is the induced metric of the world-sheet, also known as the first fundamental form,
\begin{equation}
    h_{\alpha\beta} = \partial_{\alpha}X^{\mu} \partial_{\beta} X_{\mu} \,,
    \label{induced_metric}
\end{equation}
and ${\cal R}$ is the scalar curvature constructed out of this metric. In general, higher dimensional operators in this effective action depend also on the extrinsic geometry of the worldsheet as characterized by the 
extrinsic curvature
$K_{\alpha\beta}^{\mu}$, also known as the second fundamental form,
\begin{equation}
    K_{\alpha\beta}^{\mu} = \nabla_{\alpha}\partial_{\beta} X^{\mu} \,.
\end{equation}

The analysis of small perturbations around a long string background is often performed by fixing the static gauge
\begin{equation}
\label{unitary}
    X^{\mu} = (\sigma^{\alpha}, \ell_s Y^i) \,,
\end{equation}
which leaves one with transverse excitations $Y^i$ which correspond to physical Goldstone modes, also called phonons.

An interesting property of the action (\ref{effective_string}) is that the first non-universal term appears only at the next-to-next-to-leading order ${\cal O}(\ell_s^2)$, {\it i.e.} no ${\cal O}(1)$ is present. This follows from the Gauss--Codazzi equation which relates scalar and extrinsic curvatures, and also from the fact that $(K_{\alpha}^{\alpha \mu})^2$ vanishes on-shell, while  $\int d^2 \sigma \sqrt{-\det h_{\alpha\beta}} \mathcal{R}$ is the total derivative. As a result,  the leading Nambu--Goto action provides universal predictions not only  to tree level processes but also at one loop order. Note that the Goldstones $Y^i$ defined by (\ref{unitary}) have a canonically normalized kinetic term, which makes it straightforward to perform power counting for contributions of different terms in the effective string action (\ref{effective_string}) into physical observables.

The torelon sector is obtained by compactifying one of target spatial dimensions on a circle of circumference $R$, $X_1\sim X_1+R$ and by restricting to strings which wound once around this direction. The most straightforward perturbative approach for calculating the torelon spectrum is  the ``naive" $\ell_s/R$ expansion. The simplest way to implement it is to work in the static gauge with periodic $\sigma^1$ coordinate, $\sigma_1\sim \sigma_1+R$. Then the theory turns into a quantum mechanics of an infinite set of Kaluza--Klein (KK) modes, and the $\ell_s/R$ expansion corresponds to the conventional time-independent perturbation theory. This approach is highly predictive for very long strings, where the leading non-universal contribution to excitation energies associated with the Wilson coefficient $\gamma$ in the effective action (\ref{effective_string}) is of the form $\ell_s^6/R^7$ as follows from the fact that
\[
\ell_s^2 {\cal R}^2 \sim \ell_s^6 (\d^2Y)^4+\dots\;,
\]
where we suppressed all tensor indices.
However, in practice, flux tubes measured in the current lattice simulations are not long enough to be within the radius of convergence of this expansion. The torelon ground state is the only exception, being the only state which is well approximated by the sum of universal contributions in the  $\ell_s/R$ expansion.

On the other hand it was observed that the majority of the torelon excited states are well approximated by the so-called Goddar--Goldstone--Rebbi--Thorn (GGRT) spectrum~\cite{goddard1973quantum,arvis1983exactqq} (sometimes also 
referred to as the Arvis formula and also the Nambu--Goto spectrum),
\begin{equation}
    E_{\text{GGRT}}(N_L, N_R)=\sqrt{\frac{4 \pi^2(N_L - N_R)^2}{R^2}+\frac{R^2}{\ell_s^4}+\frac{4 \pi}{\ell_s^2}\left(N_L + N_R-\frac{D-2}{12}\right)} \,,
    \label{GGRT}
\end{equation}
where $N_L$ and $N_R$ are levels of the excited string states, counting the left- and right- moving KK momenta along the strings. This spectrum is obtained by performing a light cone quantization\footnote{More precisely, one needs to use a version of the light cone quantization adapted to the winding sector.} of a bosonic string. Famously, this quantization suffers from the Virasoro anomaly at $D=4$, so (\ref{GGRT}) is not a spectrum of a consistent relativistic flux tube theory.
In particular, when expanded  at large $R$, (\ref{GGRT}) does not reproduce universal $O(\ell_s^4/R^5)$ term (which are associated with the one loop NG contributions). Hence, the GGRT spectrum can be thought of as a particular resummation of tree level terms in the  $\ell_s/R$ expansion.
The relative success of this resummation suggests that there should exist an alternative perturbative scheme which naturally leads to (\ref{GGRT}) as the leading order approximation.

\subsection{Flux tube spectrum from the world-sheet scattering}
\label{subsec:TBA}

The desired alternative perturbative approach for calculating the finite volume spectrum is based on the observation that the finite volume energies in general depend on several distinct scales. One of them is a compactification scale $R$ and then there also is a scale set by phonon momenta $p$\footnote{Depending on a state, phonon momenta may correspond to several distinct scales. }. The straightforward perturbative expansion mixes all these scales together by replacing the phonon momenta with their free theory values, $p=2\pi N/R$ and then using $\ell_s/R$ as a single expansion parameter. 

An improved perturbative scheme is based on dividing the calculation into two steps. First, one calculates phonon scattering amplitudes in the worldsheet effective theory, using $p\ell_s$ as small expansion parameter.
At the second step one relates the resulting scattering amplitudes to the finite volume spectrum. This second step does not need to be perturbative. For instance, the tree level scattering of the phonons is integrable ({\it i.e.}, no particle production is present) and is characterized by the following flavor independent $2 \to 2$ phase shift\cite{Dubovsky:2012wk},
\begin{equation}
    e^{2i \delta(s)} = e^{is \ell_s^2 /4} \,,
    \label{integrable_phase}
\end{equation}
where $s$ is the standard Mandelstam invariant (the square of the center of mass energy).  Hence, at this leading order the transition between the scattering amplitudes and the finite volume spectrum can be performed using the Thermodynamic Bethe Ansatz (TBA) \cite{zamolodchikov1990thermodynamic,Dorey:1996re}, which allows to exactly calculate the finite volume spectrum of an integrable (1+1)-dimensional theory.

This formalism can also be used to incorporate higher order corrections, at least up to the order where the non-vanishing inelastic amplitudes, such as $2\to 4$, show up. For the worldsheet scattering particle production is absent also in the presence of the next-to-leading order ${\cal O}(\ell_s^4)$ corrections and the corresponding scattering phase shifts take the following form at $D=4$,
\begin{equation}
\begin{split}
    2 \delta_{sym} &= \frac{\ell_s^2 s}{4} -\frac{11 \ell_s^4 s^2}{192 \pi}  +O(s^3) \\
    2 \delta_{anti} &= \frac{\ell_s^2 s}{4} + \frac{11 \ell_s^4 s^2}{192 \pi}  + O(s^3) \\
    2 \delta_{sing} &= \frac{\ell_s^2 s}{4} + \frac{11 \ell_s^4 s^2}{192 \pi} + O(s^3) \,.
    \label{phase_shift}
\end{split}
\end{equation}
where $2\delta_{sym}$ corresponds to the spin 2 (symmetric) channel w.r.t. the $O(2)$ group of transverse rotations, $2\delta_{anti}$ to the antisymmetric (pseudoscalar) channel and $2\delta_{sing}$ to the singlet (scalar) channel.
The leading contribution in all channels coincides with the phase shift of the integrable theory~\eqref{integrable_phase}. The subleading term $\pm\frac{11 \ell_s^4 s^2}{192 \pi}$ (also called the Polchinski-Strominger (PS) amplitude) describes the one-loop scattering. The scattering phase shifts are universal up to this order.

Note that the one-loop phase shifts (\ref{phase_shift}) are the same in the scalar and the pseudoscalar channel. As a consequence, the corresponding scattering is reflectionless in the helicity basis
\begin{equation}
    a_{l(r) \pm}^{\dagger}=a_{l(r) 2}^{\dagger} \pm i a_{l(r) 3}^{\dagger} \,,
\end{equation}
where $a_{l(r)}^\dagger$ are the creation operators for left- (l) and right- (r) moving phonons. Here, we consider a string stretched along the $X^1$ direction, so that $ a_{l(r) 2}^{\dagger}$ and $ a_{l(r) 2}^{\dagger}$ create string oscillations in the $X^2$ and $X^3$ directions.
The TBA for a system of massless excitations (which is our main focus in this paper) with reflectionless scattering can be formulated as a generalized quantization condition around the circle  for each particle  on a string,
\begin{equation}
\begin{aligned}
    p_{l i} R+\sum_j 2 \delta_{a_i a_j}\left(p_{l i}, p_{r j}\right)-i \sum_b \int_0^{\infty} \frac{d q}{2 \pi} \frac{d 2 \delta_{a_i b}\left(i p_{l i}, q\right)}{d q} \ln \left(1-e^{-R \epsilon_r^b(q)}\right)=2 \pi N_{li} \,, \\
    p_{r i} R+\sum_j 2 \delta_{a_j a_i}\left(p_{r i}, p_{l j}\right)+i \sum_b \int_0^{\infty} \frac{d q}{2 \pi} \frac{d 2 \delta_{b a_i}\left(-i p_{r i}, q\right)}{d q} \ln \left(1-e^{-R \epsilon_l^b(q)}\right)=2 \pi N_{ri} \,,
\end{aligned}
    \label{momentum_quantization}
\end{equation}
where $a_i$, $a_j$ label left and right moving excitations of a given string state and $b$ labels all particle species in the (1+1)-dimensional system. In the case of confining strings in $D$ dimensions, these are the left and right moving phonons, with $D-2$ flavors. Here the sum of
phase shifts incorporates local interactions between real excitations of the string. The integral contributions is responsible for non-local winding corrections, associated virtual particles traveling ``around the world". Neglecting the winding corrections reduces TBA to the Asymptotic Bethe Ansatz (ABA). In the language of the straightforward derivative expansion, ABA amounts to a resummation of classical non-linearities, which is main contributor into the improved convergence properties of the TBA.

The winding corrections in the TBA quantizations conditions (\ref{momentum_quantization}) depend on the pseudo-energies $\epsilon_{l(r)}^a$ which satisfy the following integral equations
\begin{equation}
\begin{aligned} 
\epsilon_l^a(q) & =q+\frac{i}{R} \sum_i 2 \delta_{a b_i}\left(q,-i p_{r i}\right)+\frac{1}{2 \pi R} \sum_b \int_0^{\infty} d q^{\prime} \frac{d 2 \delta_{a b}\left(q, q^{\prime}\right)}{d q^{\prime}} \ln \left(1-e^{-R \epsilon_r^b\left(q^{\prime}\right)}\right) \,, \\ \epsilon_r^a(q) & =q-\frac{i}{R} \sum_i 2 \delta_{b_i a}\left(q, i p_{l i}\right)+\frac{1}{2 \pi R} \sum_b \int_0^{\infty} d q^{\prime} \frac{d 2 \delta_{b a}\left(q, q^{\prime}\right)}{d q^{\prime}} \ln \left(1-e^{-R \epsilon_l^b\left(q^{\prime}\right)}\right) \,.
\end{aligned}
\label{pseudoenergy}
\end{equation}

The energy of a state is written as
\begin{equation}
    \Delta E=\sum_i p_{l i}+\sum_i p_{r i}+\frac{1}{2 \pi} \sum_a \int_0^{\infty} d q \ln \left(1-e^{-R \epsilon_l^a(q)}\right)+\frac{1}{2 \pi} \sum_a \int_0^{\infty} d q \ln \left(1-e^{-R \epsilon_r^a(q)}\right) \,.
    \label{energy_TBA}
\end{equation}
Here we denote the energy as $\Delta E$ as an indication that this energy does not include the world-sheet cosmological constant term, which gives rise to the linear piece $R/\ell_s^2$ in the total energy. 
Keeping only the leading order phase shift in~\eqref{phase_shift}, the TBA system reduces to a set of algebraic equations which is straightforward to solve.
As a result one arrives at  the GGRT formula~\eqref{GGRT}. Hence, in the TBA approach the GGRT spectrum indeed comes about as the leading order
approximation, which explains why it provides such a good fit for the majority of the string states.
The PS phase shift gives rise to universal correction to the leading order GGRT spectrum, which lift some of the degeneracies present at the leading order.

Incorporating higher order corrections into the full TBA become more and more challenging both to technical complications  and conceptually.
On a technical side, beyond the leading order TBA equations do not reduce to an algebraic system of equations and one needs to solve a set of integral equations, which in general can only been done numerically. However, for most purposes these technical difficulties can be avoided by neglecting higher order corrections to the phase shift in the winding contributions. This is motivated by the observation that winding corrections are expected to have little UV sensitivity due to exponential suppression of the integral terms in (\ref{momentum_quantization}) and (\ref{pseudoenergy}) at high momenta. In this approximation it follows from (\ref{pseudoenergy}) that pseudoenergies are linear in the particle's momenta. Taking 
 the  two phonon state as an example we obtain
\begin{equation}
    \epsilon_{l(r)}^1(q) = \epsilon_{l(r)}^2(q) = c_{l(r)} q \,,
\end{equation}
and the TBA equations~(\ref{momentum_quantization},\ref{pseudoenergy}) again turn into a system of algebraic equations,
\begin{equation}
\begin{aligned}
    c_l &= 1 + \frac{p_r \ell_s^2}{R} - \frac{\pi \ell_s^2}{6 c_r R^2}  \,, \\
    c_r &= 1 + \frac{p_l \ell_s^2}{R} - \frac{\pi \ell_s^2}{6 c_l R^2}  \,, \\
    p_l R &+ 2 \delta(p_l, p_r) - \frac{\pi \ell_s^2 p_l}{6c_r R} = 2\pi N_L \,, \\
    p_r R &+ 2 \delta(p_r, p_l) - \frac{\pi \ell_s^2 p_r}{6c_l R} = 2\pi N_R \,,
    \label{TBA_equations}
\end{aligned}
\end{equation}
which can be solved with ease, and the energy~\eqref{energy_TBA} becomes
\begin{equation}
    \Delta E = p_l + p_r - \frac{\pi}{6 R c_l} - \frac{\pi}{6 R c_r} \,.
    \label{energy_ABA}
\end{equation}
It was checked in \cite{Dubovsky:2014fma} that for the next-to-leading phase shifts (\ref{phase_shift}) this approximation is practically indistinguishable from 
the actual (numerical) solution of the full TBA system, confirming the expected UV insensitivity of the winding corrections.
Note that these equations can also be used in the opposite direction---to extract the scattering phase shifts as a function of $p_l,p_r$ from the spectrum of 2-phonon excitation states of a confining flux tube.

The conceptual challenge of incorporating higher order corrections is more severe. Indeed, starting at the ${\cal O}(\ell_s^6)$ order the worldsheet S-matrix
exhibit particle production, which leads to the mixing between two- and four-particle states and states with larger number of particles at higher orders.
At the moment there exists no systematic generalization of the TBA (or ABA) approach for relating scattering amplitudes and finite volume spectrum in the presence of these inelasticities. A partial remedy to this problem, called $T\bar{T}$ dressing, was proposed in \cite{Chen:2018keo} and is based on the recently discovered properties of so called $T\bar{T}$ deformed theories \cite{Dubovsky:2013ira,Smirnov:2016lqw,Cavaglia:2016oda}. Namely, the $T\bar{T}$ dressing relates S-matrices of the original and undressed theories in the following way,
\begin{equation}
S_o=e^{\frac{i\ell_s^2}{4}\sum_{i<j}p_i*p_j}S_u\;,
\end{equation}
where $S_o$ is the original S-matrix, $S_u$ is the undressed one, and
$p_i$ are on-shell momenta of the scattered particles cyclically ordered by their rapidities. Then the finite volume spectra of the original and undressed theory are related in the following way 
\begin{equation}
    E(R, \ell_s)=\frac{1}{\mathcal{R}_0} (R+ \frac{\ell_s^2}{2} E(R, \ell_s)) E\left(\mathcal{R}_0, 0\right) +  \frac{\ell_s^2}{2 \mathcal{R}_0} P(R) P\left(\mathcal{R}_0\right) \,,
    \label{TTbar_dressing}
\end{equation}
where $E(R, \ell_s)$ and $E\left(\mathcal{R}_0, 0\right)$ are dressed and undressed energies correspondingly,
$P(R)$ is the total momentum of the state, which is quantized as $P(R) = 2\pi k/R, \,k \in \mathbb{Z}$ and
\begin{equation}
    \mathcal{R}_0 = \sqrt{ (R + \frac{\ell_s^2}{2} E(R, \ell_s))^2 - \frac{ \ell_s^4}{4}P(R)^2 } \,.
\end{equation}
It is important to emphasize that the relation (\ref{TTbar_dressing}) is exact. Consequently, instead of performing a direct 
perturbative calculation of the finite volume spectrum of the dressed theory, one may first calculate the finite volume spectrum of the undressed and then to make use of (\ref{TTbar_dressing}). The choice of the dressing parameter $\ell_s$ is analogous to the choice of a sliding cutoff scale $\mu$ in conventional perturbative calculations. 

The relevance of this procedure for calculating the spectrum of torelons is based on the observation that the leading order (tree level) phase shift  (\ref{integrable_phase})
for the worldsheet scattering exactly coincides with the dressing phase shift for massless particles.
Hence, it is natural to organize a perturbative calculation using the inverse string tension $\ell_s^2$ as an undressing parameter.
Namely, one first calculates the finite volume spectrum of the undressed theory using the straightforward derivative expansion (or ABA, if one works at the order when particle production is still absent), and then improves the result by using the dressing formula. Applying this procedure to the universal 
${\cal O}(\ell_s^4)$ worldsheet phase shifts (\ref{phase_shift}) without the leading order term and restricting to the ABA approximation is equivalent to the earlier prescription summarized by (\ref{TBA_equations}), (\ref{energy_ABA}). Later in Section~\ref{sec:computation} we shall refer to this phase shift by undressed phase shift. The advantage of the $T\bar{T}$ dressing is that it provides a more systematic description of this prescription, which allows to use it as an improvement of the straightforward derivative expansion also at higher orders, in the presence of additional massive particles and for multi-particle states
when the ABA approximation is not applicable in general. In what follows we find this technique to be quite efficient.

\subsection{The Axionic String Ansatz (ASA)}
\label{subsec:ASA}

Universal TBA predictions provide a good description for a large set of confining string excitations, which were measured with high precision using Monte Carlo simulations\cite{Athenodorou:2010cs,Athenodorou:2011rx,Athenodorou:2018sab,Athenodorou:2022pmz} both in $D=4$ and $D=3$.  As expected, for most of the states deviations from the universal predictions increase at shorter radius $R$, indicating the presence of non-universal corrections. In particular, the TBA analysis provides a determination of the first non-trivial Wilson coefficient for $D=3$ confining strings \cite{Dubovsky:2014fma,Chen:2018keo}. However, among 
$D=4$ data one also finds several states which strongly deviate from the universal TBA predictions even at relatively large compactification radius $R$. 
These states indicate the presence of an additional massive excitation on the string worldsheet which was dubbed the string axion \cite{Dubovsky:2013gi,Dubovsky:2014fma}.
The leading order interactions between the worldsheet axion and the translational Goldstones are given by the following action
\begin{equation}
    S_\phi= \int d^2 \sigma \sqrt{-h} \left( -\frac{1}{2} (\partial \phi)^2 - \frac{1}{2} m^2 \phi^2 + \frac{Q_{\phi}}{4} h^{\alpha \beta} \epsilon_{\mu \nu \lambda \rho} \partial_\alpha t^{\mu \nu} \partial_\beta t^{\lambda \rho} \phi \right) \,,
    \label{axion_interaction}
\end{equation}
where $\phi$ is  the pseudoscalar world-sheet axion and
\begin{equation}
    t^{\mu \nu}=\frac{\epsilon^{\alpha \beta}}{\sqrt{-h}} \partial_\alpha X^\mu \partial_\beta X^\nu \,.
\end{equation}
{ We call it ``axion'' because it is a pseudoscalar field which at leading order couples to a worldsheet $\theta$-term constructed from Goldstones, the integration of which is known as the self-intersection number, analogous to how the QCD axion is coupled to a Chern class.} The coupling constant $Q_\phi$ and the axion mass $m$ can be extracted from the Monte-Carlo data of 4d $SU(3)$ Yang-Mills confining flux tube spectrum using the TBA analysis,
\begin{equation}
    Q_{\phi} \approx 0.38 \pm 0.04, \quad m \approx 1.85_{-0.03}^{+0.02} \ell_s^{-1} \,.
\end{equation}
In Section~\ref{sec:subsec_spin0_computation} we will present an determination of these parameters from more precise data provided in Section~\ref{sec:data}. Interestingly, this coupling coincides (within error bars) with the value \cite{Dubovsky:2015zey}
\begin{equation}
\label{Qint}
    Q_{\text{integrable}} = \sqrt{7\over 16 \pi} \approx 0.373 \,,
\end{equation}
for the corresponding coupling in an integrable theory of $D=4$ Goldstones and a massless axion.

This numerical coincidence and several other considerations lead to the proposal of the Axionic String Ansatz for the worldsheet dynamics of the confining strings at large $N$.
It stipulates that both at $D=4$ and $D=3$ the violation of integrability is a transient phenomenon present at intermediate energies of order $\ell_s^{-1}$.
At $D=3$ the only worldsheet degree of freedom is the massless Goldstone mode itself, and at $D=4$ these are two massless Goldstone and a massive axion. One of the goals of the present work is to test that all excitations of the $D=4$ confining string can indeed be described using these worldsheet fields only.

\section{Lattice methodology}
\label{sec:lattice}

In Section~\ref{sec:subsection_lattice_setup_lat} we outline the lattice framework
of our calculations. In Section~\ref{sec:subsection_Q} we discuss the ergodicity of our simulations. In  Section~\ref{sec:subsection_quanta} we describe quantum numbers characterizing string and 
present universal effective string theory predictions for the quantum numbers of the low lying states. 
In Section~\ref{sec:subsection_lattice_operators} we discuss basics of the Generalized Eigenvalue problem method~\cite{Luscher:1984is,Luscher:1990ck},
which is used to determine energies of the excited states and also describe the basis of lattice operators used in our simulation.
 In Section~\ref{sec:subsection_energies_lat} we discuss a number of statistical and systematic uncertainties affecting the extracted energies.
 The main takeaway from this subsection is 
that the uncertainties in the energy determination become larger as  the energy of the state increases\footnote{The latter may increase both when we include higher excited states in our analysis and also when we consider larger values of the compactification radius $R$.}.

\subsection{Lattice Setup}
\label{sec:subsection_lattice_setup_lat}

 We calculate a discrete Euclidean partition function of the $D=4$  gluodynamics
\begin{equation}
Z=\int {\cal{D}}U \exp\{- \beta S[U]\},
\label{eqn_Z}
\end{equation}
where  $U$'s are $SU(N)$ valued gauge fields which live on the links of the lattice and  ${\cal{D}}U$ is the Haar measure.
We work on a hypercubic periodic lattice of size $L_x L_{\perp}^2 L_t$ with lattice spacing $a$. We study flux tubes wound around the $x$ direction, with a physical size $R = a L_x$.

We use the conventional Wilson plaquette action,
\begin{equation}
\beta S = \beta \sum_p \left\{1-\frac{1}{N} {\text{ReTr}} U_p\right\} \,,
\label{eqn_S}
\end{equation}
where $U_p$ is the path-ordered product of link matrices around the plaquette $p$.  
In the continuum limit one gets
\begin{equation}
\int \prod_{l} dU_l e^{-\beta S} 
\stackrel{a\to 0}{\propto}
\int \prod_{x,\mu} dA_\mu(x) 
e^{-\frac{4}{g^2} \int d^4 x Tr F_{\mu\nu} F_{\mu\nu} }\,, 
\end{equation}
where
\[
g^2={2N\over\beta}
\]
is the Yang--Mills coupling constant at the lattice scale.
The 't Hooft  large $N$ limit is approached by sending $N\to\infty $ while  keeping $g^2N$ fixed, 
which implies that as we vary $N$ we need to also vary  $\beta \propto N^2$ in order to keep the lattice scale 
$a$ approximately constant. 
The actual values of $\beta$ which we used are summarized in Table \ref{tab:table_physics_parameters}.

\begin{table}[htb]
\begin{center}
{\color{black}
\begin{tabular}{c|c|c|c|c|c}\hline \hline
$N$ & $\beta$ &   $\frac{1}{N} {\rm Re} {\rm Tr} \langle U_p \rangle $  &  $a\surd\sigma$ & $R_c/a$ & $a m_G$ \\ \hline \hline
3  & 6.338  & 0.6255952(24)   &  0.12902(15)  & 11.99(9) & 0.4276(37) \\
   & 6.0625 & 0.6000332(32)   &  0.19489(16) &  8.00(2) & 0.6365(43)  \\ \hline
5  & 18.375  & 0.6036543(12)  &  0.13047(25) &  12.55(10) & 0.4078(37)  \\
   & 17.630  & 0.5769714(43)  &  0.19707(30) &  8.32(5) & 0.5961(78)  \\\hline
6  & 25.550  & 0.5715594(35)  &  0.20142(27) &  8.30(4) & 0.6112(41) \\  \hline
\end{tabular}}
\caption{ Parameters of our calculations with some corresponding 
properties of the gauge theories: the average plaquette, the string tension, $\sigma$, the
deconfining length, $R_c$, and the mass gap, $m_G$.}
\label{tab:table_physics_parameters}
\end{center}
\end{table}

Physical observables on the lattice are extracted by calculating the expectation value of some functional $\Phi[U]$.
We first generate a set of $n_g$ gauge field configurations $\{U^I\}; I=1,...,n_g$ distributed 
with the Boltzmann-like action factor included in the probability 
measure, i.e. $ dP \propto \prod_l dU_l \exp\{-\beta S[U]\}$.
To achieve this we use a Cabbibo-Marinari algorithm applied to the $N(N-1)/2$
SU(2) subgroups of the SU($N$) matrix, with a mix of standard heat-bath 
and over-relaxation steps. We can now calculate the expectation value of 
an arbitrary functional $\Phi[U]$ of the gauge fields as follows,
\begin{equation}
\langle \Phi \rangle
=
\frac{1}{Z}
\int \prod_{l} dU_l  \Phi[U] e^{-\beta S}
=
\frac{1}{n_g} \sum^{n_g}_{I=1} \Phi[U^I] + 
O\left( \frac{1}{\surd n_g}\right)
\label{eqn_avPhi}
\end{equation}
where the last term is an estimate of  the statistical error.

\subsection{Topological freezing and Ergodicity}
\label{sec:subsection_Q}

In a Monte Carlo sequence of $SU(N)$ lattice gauge fields, the changes in the topological charge $Q$ become
rapidly less frequent as $a(\beta)\to 0$ at fixed $N$ and as $N\to\infty$ at fixed $a(\beta)$.
This precocious critical slowing down raises the question whether our fields provide an adequate
sampling of different topological sectors, although any resulting bias is expected to decrease
as an inverse power of $N$ as $N$ increases. (For a more detailed discussion see, for example,
Section 2.4 of \cite{Athenodorou:2021qvs}.) Here we briefly assess the extent of this `topological freezing'
in our calculations.

In the calculations of this paper we did not calculate the values of $Q$ but we can use the
calculations performed in  \cite{Athenodorou:2021qvs} to estimate how well we sample the different
topological sectors. In that paper we calculated the average number of Monte Carlo sweeps
between changes in $Q$ of 1 unit. Scaling for the lattice volume we choose to estimate the number of
changes in $Q$ on the smallest lattice volumes in each of the 5 relevant calculations, i.e
$16^4$ at the coarser lattice spacing in $SU(3),SU(5)$ and $SU(6)$ and $24^4$ in the
finer lattice calculation in $SU(3)$ and $SU(5)$. Using the results in Tables 42,43 of
\cite{Athenodorou:2021qvs} we estimate the following number of changes in $Q$ in the 5 sequences
used in this paper:
in $SU(3)$ on a $16^4$ lattice at $\beta=6.0625$ roughly 490,000 changes within the sequence
of $6.25\times 10^6$ fields;
in $SU(3)$ on a $24^4$ lattice at $\beta=6.338$ roughly 29,000 changes within the sequence
of $6.25\times 10^6$ fields;
in $SU(5)$ on a $16^4$ lattice at $\beta=17.63$ roughly 1,400 changes within the sequence
of $3.12\times 10^6$ fields; 
in $SU(5)$ on a $24^4$ lattice at $\beta=18.375$ roughly 53 changes within the sequence
of $12.5\times 10^6$ fields; 
in $SU(6)$ on a $16^4$ lattice at $\beta=25.55$ roughly 335 changes within the sequence
of $6\times 10^6$ fields.

In all these cases $\langle Q^2 \rangle \sim 2-3$ so the true distribution
of $Q$ will contain $Q=0,\pm 1,\pm 2$, with some $\pm 3$ and then little else.
So almost all our runs above should be adequate to explore such a distribution
of $Q$, with the case of $SU(5)$ at $\beta=18.375$ being marginal.

\subsection{Quantum Numbers of Confining String States}
\label{sec:subsection_quanta}

 Before proceeding to describe how we measure the spectrum of excited string states, let us explain what are the quantum numbers of these states, both on a lattice and in the continuum limit.
 Let us consider a closed confining flux tube in the fundamental representation of $SU(N)$, which means that in the continuum limit we consider a state created by an operator of the form
 \[
 {\cal O}=\Tr P e^{i\oint_{\cal C} A_{\mu}dx^\mu}\;,
 \]
 where trace is taken in the fundamental representation of the gauge group, and $P$ stands for the path ordering along the closed curve ${\cal C}$.
 For torelons the curve ${\cal C}$ winds around one of the compact directions which in our case is chose to be the $x$-circle
 (the corresponding operators are called the Polyakov loops). On a lattice the path ordered exponent of the gauge field is replaced by a product of the link fields $U$ along the curve ${\cal C}$ on a lattice.
 Torelon states are characterized by the following quantum numbers.
\begin{itemize}
    \item[ ] {\it One-form $\mathbb{Z}_N$ charge $k$}: A flux tube can wind several times around the compactified $x$-direction, and one can also take the trace in different $SU(n)$  representations, producing states of different  ``$N$-ality", {\it i.e.} states carrying different charge $k=0,1,\dots, N-1$
     w.r.t. the $\mathbb{Z}_N$ 1-form center symmetry. In this work we restrict to $k=1$ flux tubes, created by  Polyakov loops in the fundamental representation with a single winding.
    \item[ ] {\it Transverse Momentum $p_{\perp}$}: A flux tube can carry a non-vanishing momentum along the transverse directions  {$p_{\perp} = (p_y, p_z) $} with $p_y = 2 \pi q_y / L_y$ and $p_z = 2 \pi q_z / L_z$, where $q_y$ and $q_z$ are integers (we use the $a=1$ units here). 
    Adding a transverse momentum to a string state amounts to moving this state ``as a whole" in the transverse direction.
    The resulting states have energies related to the energies of the strings at rest by a lattice version of the momentum dispersion relation. 
   There is not much new to learn from these  states and in what follows we restrict to the $p_{\perp}= 0$ sector.
    \item[] {\it Longitudinal Momentum $p$}: A flux tube can carry a non-vanishing momentum along the longitudinal axis defined as the $x$-direction.
    This longitudinal momentum can take values  $p=2\pi q/L_x$, where $q$ is an integer. States with opposite values of $q$ have the same spectrum, so in what follows we restrict to  $q\ge 0$.
    \item[] {\it Angular Momentum  (``Spin") $J$}: A straight confining flux tube in three spatial dimensions is invariant under transverse rotations.
    This implies that its excitations can be characterized by transverse angular momentum  $J$. It can take  values $J=0, \pm 1, ...$. 
    The excitation spectra at opposite values of $J$ are the same\footnote{Strictly speaking this assumes that neither transverse rotations not parity are spontaneously broken by the flux tube ground state. We do not see evidence for such breaking in our data.}.
   
    \item[] {\it Transverse Parity $P_{\perp}$}: Flux tube states  also carry  two different parity quantum numbers. 
    The first parity transformation, called transverse parity $P_{\perp}$, corresponds to the spatial parity in the transverse plane.
    From the viewpoint of the underlying gauge theory, 
  $P_{\perp}$ originates from the spatial parity $P$.
  Without loss of generality it can be defined as $P_{\perp}: (y,z) \to (y,-z)$. Transverse parity does not commute with $SO(2)$ rotations in the transverse plane, so that the full transverse symmetry takes a form of a semi-direct product,
    \[
    O(2)=SO(2) \rtimes P_{\perp}\;.
    \]
  As a result,  $J>0$ states come in pairs of opposite $P_{\perp}$ which combine into $ O(2)$ doublets.

    \item[] {\it Longitudinal Parity $P_{\parallel}$}:   In addition to the transverse parity string excitations are also characterized by their transformation properties w.r.t. the longitudinal parity which act as a reflection along the flux tube, $P_{\parallel}: x\to -x$.  Given that this transformation reverses the color flux direction, this symmetry originates from the $CP$ parity of the underlying gauge theory. Note also that $P_{\parallel}$ flips the longitudinal momentum $q$, so that $q>0$ states are not  eigenstates of  $P_{\parallel}$.

\end{itemize}
In what follows we will use the notation $|J|^{P_{\perp} P_{\parallel}}$ to represent quantum numbers of  $q=0$ states. Similarly,
we will label  $q>0$ states as $(q, |J|^{P_{\perp}})$.

Let us now summarize universal quantum numbers of the states predicted by the effective string theory. All these states are guaranteed to be present at sufficiently large values of $R$. In general, one may expect to find also additional states, associated with massive fields on the string worldsheet, such as the worldsheet axion. Instead, the GGRT spectrum contains only these universal string states.

Effective string theory in the static gauge implies the following construction of the  Hilbert space of torelon states.
One starts with a translationally invariant vacuum state $|0\rangle$, which has zero spin and positive parities, in other words it is a $0^{++}$ state. All excited
states a constructed by acting with phonon creation operators  $a^{\pm}_k$, where $k$ is an integer which determines the longitudinal momentum, $q=2\pi k/L_x$ and $\pm$ determines whether the created phonon carries positive or negative unit  of the angular momentum.
The transformation rules of these operators under the transverse and longitudinal parities are
\begin{eqnarray}
P_\perp:    a^+_k \longleftrightarrow a^-_k\,,\\
P_{\parallel}:  a^\pm_k \longleftrightarrow a^\pm_{-k} \,.
    \end{eqnarray}
For future use we list  in Tables~\ref{tab:table_NGstates_q0}-\ref{tab:table_NGstates_q2} the explicit phonon content of the universal lowest lying states for longitudinal momenta of $q=0,1,2$\footnote{When making the tables of excitations on a string state here and later in Section~\ref{sec:subsec_spin0_computation} we distinguish string states in terms of representations of symmetry in the continuum gauge theory, because physically that is what we are interested. Soon we will mention that symmetries on a lattice and in the continuum spacetime are distinct, and thus give birth to subtly different classifications of representations.  }. In these Tables and in what follows we are making use of the standard notation for the ``levels" $N_{L(R)}$, which count the total number of units of left (right) moving momentum. States at the same levels are degenerate at the leading order in the effective string theory expansion (and also in the GGRT quantization), as a consequence of the tree level integrability of the  Nambu--Goto action.

\begin{table}[htp] 
\begin{center}
\centering{\scalebox{0.85}{
\begin{tabular}{c|c|c|c|r} \hline \hline
$N_L,N_R$ & $J$ & $P_{\perp}$ & $P_{\parallel}$  & \ \ String State \ \  \\ \hline \hline
 $N_L=N_R=0$ & $0$ &  $+$ & $+$ & $| 0 \rangle$ \\ \midrule 
\multirow{4}*{$N_L=N_R=1$} & $0$ &  $+$ & $+$ & $\left( a^{+}_{1} a^{-}_{-1}+a^{-}_{1} a^{+}_{-1} \right) | 0 \rangle$ \\
& $0$ &  $-$ & $-$ & $\left( a^{+}_{1} a^{-}_{-1}-a^{-}_{1} a^{+}_{-1} \right) | 0 \rangle$ \\
& $2$ &  $\pm$ & $+$ & $\left( a^{+}_{1} a^{+}_{-1} \pm a^{-}_{1} a^{-}_{-1} \right) | 0 \rangle$ \\
\midrule 
\multirow{17}*{$N_L=2,N_R=2$} & $0$ &  $+$ & $+$ & $\left( a^{+}_{2} a^{-}_{-2}+a^{-}_{2} a^{+}_{-2} \right) | 0 \rangle$ \\
& $0$ &  $-$ & $-$ & $\left( a^{+}_{2} a^{-}_{-2}-a^{-}_{2} a^{+}_{-2} \right) | 0 \rangle$ \\
& $0$ &  $+$ & $+$ & $\left( a^{+}_{1}a^{+}_{1} a^{-}_{-1} a^{-}_{-1} +a^{-}_{1}  a^{-}_{1} a^{+}_{-1} a^{+}_{-1} \right) | 0 \rangle$ \\
& $0$ &  $-$ & $-$ & $\left( a^{+}_{1}a^{+}_{1} a^{-}_{-1} a^{-}_{-1} -a^{-}_{1}  a^{-}_{1} a^{+}_{-1} a^{+}_{-1} \right) | 0 \rangle$ \\
& $0$ &  $+$ & $+$ & $ a^{+}_{1}a^{-}_{1} a^{+}_{-1} a^{-}_{-1} | 0 \rangle$ \\
& $1$ &  $\pm$ & $+$ & $\big[ ( a^{+}_{1}a^{+}_{1} a^{-}_{-2}+ a^{-}_{2}a^{+}_{-1} a^{+}_{-1}) \pm  ( a^{-}_{1}a^{-}_{1} a^{+}_{-2}+ a^{-}_{2}a^{+}_{-1} a^{+}_{-1}) \big] | 0 \rangle$ \\
& $1$ &  $\pm$ & $-$ & $\big[ ( a^{+}_{1}a^{+}_{1} a^{-}_{-2}- a^{-}_{2}a^{+}_{-1} a^{+}_{-1}) \pm  ( a^{-}_{1}a^{-}_{1} a^{+}_{-2}- a^{-}_{2}a^{+}_{-1} a^{+}_{-1}) \big] | 0 \rangle$ \\
& $1$ &  $\pm$ & $+$ & $\big[ ( a^{+}_{1}a^{-}_{1} a^{+}_{-2}+ a^{+}_{2}a^{-}_{-1} a^{+}_{-1}) \pm  ( a^{-}_{1}a^{+}_{1} a^{-}_{-2}+ a^{-}_{2}a^{+}_{-1} a^{-}_{-1}) \big] | 0 \rangle$ \\ 
& $1$ &  $\pm$ & $-$ & $\big[ ( a^{+}_{1}a^{-}_{1} a^{+}_{-2}- a^{+}_{2}a^{-}_{-1} a^{+}_{-1}) \pm  ( a^{-}_{1}a^{+}_{1} a^{-}_{-2}- a^{-}_{2}a^{+}_{-1} a^{-}_{-1}) \big] | 0 \rangle$ \\ 
& $2$ &  $\pm$ & $+$ & $\left( a^{+}_{2} a^{+}_{-2} \pm a^{-}_{2} a^{-}_{-2} \right) | 0 \rangle$ \\
& $2$ &  $ \pm $ & $+$ &  $\left[   \left( a^{+}_{1}a^{+}_{1} a^{+}_{-1} a^{-}_{-1} \pm a^{-}_{1}a^{-}_{1} a^{-}_{-1} a^{+}_{-1} \right)+\left( a^{+}_{1}a^{-}_{1} a^{-}_{-1} a^{-}_{-1} \pm a^{-}_{1}a^{+}_{1} a^{+}_{-1} a^{+}_{-1} \right) \right] | 0 \rangle$ \\
& $2$ &  $ \pm $ & $-$ &  $\left[   \left( a^{+}_{1}a^{+}_{1} a^{+}_{-1} a^{-}_{-1} \pm a^{-}_{1}a^{-}_{1} a^{-}_{-1} a^{+}_{-1} \right)-\left( a^{+}_{1}a^{-}_{1} a^{-}_{-1} a^{-}_{-1} \pm a^{-}_{1}a^{+}_{1} a^{+}_{-1} a^{+}_{-1} \right) \right] | 0 \rangle$ \\
& $3$ &  $\pm$ & $+$ & $\big[ ( a^{+}_{1}a^{+}_{1} a^{+}_{-2}+ a^{+}_{2}a^{+}_{-1} a^{+}_{-1}) \pm  ( a^{-}_{1}a^{-}_{1} a^{-}_{-2}+ a^{-}_{2}a^{-}_{-1} a^{-}_{-1}) \big] | 0 \rangle$ \\
& $3$ &  $\pm$ & $-$ & $\big[ ( a^{+}_{1}a^{+}_{1} a^{+}_{-2}- a^{+}_{2}a^{+}_{-1} a^{+}_{-1}) \pm  ( a^{-}_{1}a^{-}_{1} a^{-}_{-2}- a^{-}_{2}a^{-}_{-1} a^{-}_{-1}) \big] | 0 \rangle$ \\
& $4$ &  $\pm$ & $+$ & $\left( a^{+}_{1}a^{+}_{1} a^{+}_{-1} a^{+}_{-1} \pm a^{-}_{1}  a^{-}_{1} a^{-}_{-1} a^{-}_{-1} \right) | 0 \rangle$ \\
\midrule 
\end{tabular} }}
\end{center}
\caption{\label{tab:table_NGstates_q0}
Universal low lying string excitations with $q=0$.}
\end{table}

\begin{table}[htp] 
\begin{center}
\centering{\scalebox{0.85}{
\begin{tabular}{c|c|c|c|r} \hline \hline
$N_L,N_R$ & $J$ & $P_{\perp}$ & $P_{\parallel}$  & \ \ \hspace{9cm} String State \ \  \\ \hline \hline
 $N_L=1,N_R=0$ & $1$ &  $\pm$ & $\ $ & $\left( a^{+}_{1} \pm a^{-}_{1} \right) | 0 \rangle$ \\ \midrule 
\multirow{7}*{$N_L=2,N_R=1$} & $0$ &  $+$ & $\ $ & $\left( a^{+}_{2} a^{-}_{-1}+a^{-}_{2} a^{+}_{-1} \right) | 0 \rangle$ \\
& $0$ &  $-$ & $\ $ & $\left( a^{+}_{2} a^{-}_{-1}-a^{-}_{2} a^{+}_{-1} \right) | 0 \rangle$ \\
& $1$ &  $\pm$ & $\ $ & $\left( a^{+}_{1}a^{+}_{1} a^{-}_{-1} \pm  a^{-}_{1}a^{-}_{1}a^{+}_{-1} \right) | 0 \rangle$ \\
& $1$ &  $\pm$ & $\ $ & $\left( a^{+}_{1}a^{-}_{1} a^{-}_{-1} \pm a^{-}_{1}a^{+}_{1} a^{+}_{-1}  \right) | 0 \rangle$ \\
& $2$ &  $\pm$ & $\ $ & $\left( a^{+}_{2} a^{+}_{-1} \pm a^{-}_{2} a^{-}_{-1} \right) | 0 \rangle$ \\
& $3$ &  $\pm$ & $\ $ & $\left( a^{+}_{1}a^{+}_{1} a^{+}_{-1} \pm a^{-}_{1}a^{-}_{1} a^{-}_{-1}  \right) | 0 \rangle$ \\\midrule 
\end{tabular} }}
\end{center}
\caption{\label{tab:table_NGstates_q1}
Universal low lying string excitations with $q=1$.}
\end{table}

\begin{table}[htp] 
\begin{center}
\centering{\scalebox{0.85}{
\begin{tabular}{c|c|c|c|r} \hline \hline
$N_L,N_R$ & $J$ &$P_{\perp}$ & $P_{\parallel}$  & \ \ \hspace{9cm} String State \ \  \\ \hline \hline
\multirow{4}*{$N_L=2,N_R=0$} & $0$ &  $+$ & $\ $ & $ a^{+}_{1} a^{-}_{1} | 0 \rangle$ \\
& $1$ &  $\pm$ & $\ $ & $\left( a^{+}_{2} \pm  a^{-}_{2}  \right) | 0 \rangle$ \\
& $2$ &  $\pm $ & $\ $ & $\left( a^{+}_{1} {a}^{+}_{1} \pm a^{-}_{1}{a}^{-}_{1} \right) | 0 \rangle$ \\
\midrule 
\multirow{16}*{$N_L=3,N_R=1$} & $0$ &  $+$ & $\ $ & $\left( a^{+}_{3} a^{-}_{-1}+a^{-}_{3} a^{+}_{-1} \right) | 0 \rangle$ \\
& $0$ &  $-$ & $\ $ & $\left( a^{+}_{3} a^{-}_{-1}-a^{-}_{3} a^{+}_{-1} \right) | 0 \rangle$ \\
& $0$ &  $+$ & $\ $ & $\left( a^{+}_{1}a^{+}_{1} a^{-}_{1} a^{-}_{-1} +a^{-}_{1}  a^{-}_{1} a^{+}_{1} a^{+}_{-1} \right) | 0 \rangle$ \\
& $0$ &  $-$ & $\ $ & $\left( a^{+}_{1}a^{+}_{1} a^{-}_{1} a^{-}_{-1} -a^{-}_{1}  a^{-}_{1} a^{+}_{1} a^{+}_{-1} \right) | 0 \rangle$ \\
& $1$ &  $\pm$ & $\ $ & $\left( a^{+}_{2}a^{+}_{1} a^{-}_{-1} \pm  a^{-}_{2}a^{-}_{1}a^{+}_{-1} \right) | 0 \rangle$ \\
& $1$ &  $\pm$ & $\ $ & $\left( a^{+}_{2}a^{-}_{1} a^{-}_{-1} \pm a^{-}_{2}a^{+}_{1} a^{+}_{-1}  \right) | 0 \rangle$ \\
& $1$ &  $\pm$ & $\ $ & $\left( a^{-}_{2}a^{+}_{1} a^{-}_{-1} \pm a^{+}_{2}a^{-}_{1} a^{+}_{-1}  \right) | 0 \rangle$ \\
& $2$ &  $\pm $ & $\ $ & $\left( a^{+}_{3} a^{+}_{-1} \pm a^{-}_{3} a^{-}_{-1} \right) | 0 \rangle$ \\
& $2$ &  $\pm $ & $\ $ & $\left( a^{+}_{1}a^{+}_{1} a^{+}_{1} a^{-}_{-1} \pm a^{-}_{1}  a^{-}_{1} a^{-}_{1} a^{+}_{-1} \right) | 0 \rangle$ \\
& $2$ &  $\pm $ & $\ $ & $\left( a^{+}_{1}a^{+}_{1} a^{-}_{1} a^{+}_{-1} \pm a^{-}_{1}  a^{-}_{1} a^{+}_{1} a^{-}_{-1} \right) | 0 \rangle$ \\
& $3$ &  $\pm$ & $\ $ & $\left( a^{+}_{2}a^{+}_{1} a^{+}_{-1} \pm a^{-}_{2}a^{-}_{1} a^{-}_{-1}  \right) | 0 \rangle$ \\
& $4$ &  $\pm $ & $\ $ & $\left( a^{+}_{1}a^{+}_{1} a^{+}_{1} a^{+}_{-1} \pm a^{-}_{1}  a^{-}_{1} a^{-}_{1} a^{-}_{-1} \right) | 0 \rangle$ \\
\midrule 
\end{tabular} }}
\end{center}
\caption{\label{tab:table_NGstates_q2}
Universal low lying string excitations with $q=2$.}
\end{table}

\begin{figure}[htb]
\vspace{1cm}
\centerline{\scalebox{1.00}{\includegraphics[scale=0.20]{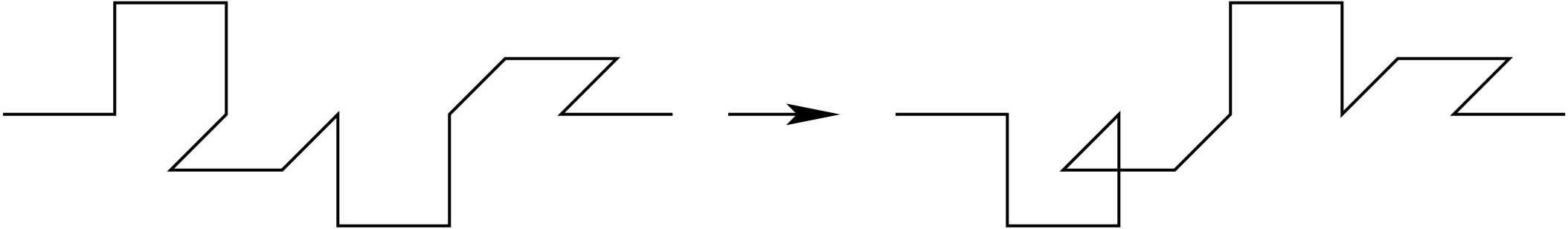} \put(-108,30){${\textcolor{cyan}{P_{\perp}}}$}}
\hspace{1cm}
\scalebox{1.00}{\includegraphics[scale=0.20]{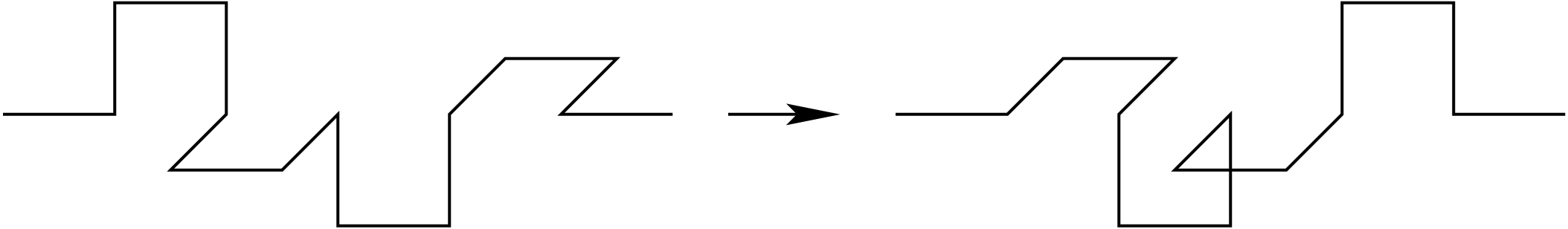} \put(-108,30){$\textcolor{red}{P_{\parallel}}$}}}
\caption{Action of the two parities $P_{\perp}$ parity (left) and $P_{\parallel}$ (right) on a lattice Polyakov loop.}
\label{fig:examplesofparities}
\end{figure}

Our discussion so far applies in the continuum theory. On a lattice states with different quantum numbers are constructed by modifying the transverse shape of the curve ${\cal C}$, which is used to build a Polyakov loop. To illustrate this, in Fig.~\ref{fig:examplesofparities}) we presented how transverse and longitudinal parities act on a sample Polyakov loop.

An obvious important difference between lattice and continuum gauge theory is that on a lattice the continuum $O(2)$  transverse rotation symmetry is broken down to its discrete dihedral (square) symmetry subgroup $D_4$ (see, e.g., \cite{Meyer:2004gx} for a brief review). In particular, only $\pi/4$ rotations survive out of the continuum $SO(2)$ rotations. 
As a result, on a square lattice all states can be decomposed into five distinct irreducible representations (irreps) of the lattice group. The correspondence between these irreps and the continuum $O(2)$ representations can be summarized as follows,
 \begin{itemize}
    \item[$A_1$]: $\, J_{{\rm mod}4}=0,  P_{\perp}=+$
    \item[$A_2$]: $\, J_{{\rm mod}4}=0,  P_{\perp}=-$
    \item[$E$]: $\, |J_{{\rm mod}4}|=1, P_{\perp}=\pm$
    \item[$B_1$]: $\, J_{{\rm mod}4}=2,  P_{\perp}=+$
    \item[$B_2$]: $\, J_{{\rm mod}4}=2,  P_{\perp}=-$
\end{itemize}
In particular, we see that all $D_4$  irreps are one-dimensional except for $E$ which 
is two-dimensional.  This implies that the even $J\neq 0$ parity partners belong to different $D_4$  irreps and in general exhibit $O(a^2)$ splittings.
 As we will see later in Section~\ref{sec:subsec_spin0_computation}, this effect introduces the main source of a systematic error in the determination of the axion coupling.

\subsection{Generalized Eigenvalue Problem and Operator Basis}
\label{sec:subsection_lattice_operators}
{\color{black}
The conventional method to extract excited states is the so called Generalized Eigenvalue problem (GEVP) or Variational Calculation which requires to construct  a large basis of operators for each choice of quantum numbers~\cite{Luscher:1984is,Luscher:1990ck}. GEVP proceeds through the following steps,
\begin{itemize}
    \item One first creates a correlation matrix built out of a basis of operators $\phi_{i}$ with $i=1, \dots, N_{\rm ops}$. The operators need to be created in such a way so that they project onto a given set of quantum numbers; in our case to a specific combination of $\left\{ J, P_{\perp}, P_{\parallel}, q  \right\}$. The correlation matrix has the following index structure,
    \begin{eqnarray}
    C_{i,j}(t)=\langle \phi^{\dagger}_{i}(t) \phi_{j}(0) \rangle \,.
    \end{eqnarray}
    \item One diagonalizes the following matrix 
    \begin{eqnarray}
    C^{-1}(t=0) C(t=a)\,,
    \end{eqnarray}
    and extract the associated eigenvectors $\psi_i$ and corresponding eigenvalues $\lambda_i$.  
    \item The eigenvector with the highest eigenvalue $\lambda_1$ corresponds to the wave-functional of the lowest energy level (ground state). The next highest eigenvalue $\lambda_2$ corresponds to the first excitation level, the next eigenvalue $\lambda_3$ to the second excited state and so forth. In this manner, one constructs  wave functionals of all excitation energy levels. In practice, for reasons that we explain later, one can extract only several low lying energy levels for one selected sector.
\end{itemize}

The simplest operator one can use to project onto the ground state of a torelon is the straight Polyakov loop, which is the path ordered product of link matrices  winding once around the $x$-circle,
\begin{equation}
l_p(n_t) =  \sum_{n_y,n_z} \mathrm{Tr} 
\left\{\prod^{L_x}_{n_x=1} U_x(n_x,n_y,n_z,n_t)\right\} 
\label{eq:eqn_poly}
\end{equation}
 The sum over translations in the transverse  $y$ and $z$ directions projects onto states with zero transverse momentum, $(p_y,p_z) = (0,0)$. This operator is invariant under translations in the longitudinal $x$-direction, so that it also projects on the states with  zero longitudinal momentum $q =0$.
 This operator transforms trivially under all parity transformations and rotations.  This indicates that a simple straight Polyakov loop projects only onto $A_1^+$ states. 
 
 It is important to emphasize that the Polyakov loop and all its possible deformations have zero overlap onto glueball states.
These states are created by path ordered products along the closed loops without any windings.
The orthogonality between torelon excitations and glueballs is guaranteed by the super-selection rules of the $\mathbb{Z}_N$ one-form ``center" symmetry of the $SU(N)$ Yang-Mills theory.

 It is worth pointing out that as the length $R$ of the $x$-circle decreases there is a critical length $R_c=1/T_c$, where $T_c$  is the deconfining temperature, below which the 
Polyakov loop acquires a non-vanishing vacuum expectation value (vev), so that the
one-form center symmetry is spontaneously broken. At shorter radii the theory is in the deconfined phase, where states with and without windings are mixed with each other and stable torelon states no longer exist.
The critical sizes $R_c$  for the Yang-Mills theories considered in our simulations  are summarized in Table~\ref{tab:table_physics_parameters}. In this paper we always restrict to the confined region $R>R_c$.

In order to calculate excited states efficiently we need a large basis of operators which are constructed using lattice paths which are less symmetric than the straight Polyakov loop. The operators we build  consist of combinations of shapes which encode certain values of $J,P_{\perp},P_{\parallel},$ and $q$. We achieve this by constructing linear combination of generalized Polyakov loops build by  path ordered products of links following paths consisting of various transverse deformations along the propagation of the flux. 

To build an operator transforming irreducibly under the parity operations $P_{\perp},P_{\parallel}$, we first create the path-ordered product of links along a Polyakov loop with a particular transverse deformation. Then we create its reflection under the  $P_{\perp}$ parity. By adding (+) or subtracting (-) the traces of the two operators we can construct an operator with $P_{\perp}=+, \ -$ respectively. If the reflection leads to the same shape as that of the original operator, it means that this operator does not incorporate enough asymmetry to probe the $P_{\perp}=-$ sector. Hence, it is characterised trivially by $P_{\perp}=+$. 
We do the same for $P_{\parallel}$ subsequently. 
The above procedure is illustrated in Fig.~\ref{fig:examplesofparities}. Hence, for a generic shape we end up creating four different operators
corresponding to four different combinations of $P_{\perp} = \pm$ and $P_{\parallel}=\pm$. 

To construct operators corresponding to specific $D_4$ irreps one needs to construct linear combinations of operators with  specific values of angular momentum $J$. The procedure reads as follows. One starts  with an arbitrary operator $\phi_{0}$; in our case the operator is a path ordered product along a curve ${\cal C}$ localized on a fixed $t$-slice,  which wraps around the $x$-direction.
Subsequently, one generates $\pi n/2$ rotations of this operator around the $x$-axis, $\phi_{n}$. Finally, one constructs the $\phi_J$ operator defined as 
\begin{equation}
\phi_J= \sum_{n=1}^4 e^{iJ n \frac{\pi}{2}} 
\phi_{n}\;.
\label{eqn_opJ}
\end{equation}
Note also, that as discussed before, we always work with operators which do not carry any momentum in the transverse $(y,z)$ plane. This is achieved by summing over all possible $(y,z)$ translations,
\[
\phi_J( \vec{p}_{\perp}=0) \equiv \sum_{n_y,n_z}\phi_J(n_y, n_z)\;,
\]
where $\phi_J(n_y, n_z)$ is an operator constructed using the curve ${\cal C}_{n_y,n_z}$ obtained by translating the original one by a $(n_y,n_z)$ vector.

Clearly,  the ``spin" $J$, as defined by~\eqref{eqn_opJ} is defined only modulo 4, 
\[
J\sim J+4 n
\]
 for any integer $n$. In particular, for even $J$, states related by reflection, which takes $J\to -J$, are equivalent, because the corresponding spin difference is zero modulo 4, 
\[
\Delta J=2J\equiv 0\mod 4\;.
\]
 On the other hand, for odd $J$ the reflected state is not equivalent to the original ones, which explains why odd spin $D_4$ irreps are two-dimensional rather than one dimensional. For odd $J$ the corresponding states are exactly degenerate also on a lattice, so we calculate energies for only one of them.  The knowledge of  the lattice ``spin" $J$ in general does not allows us to reconstruct the actual continuum spin.

To illustrate this procedure
 in~(\ref{eqn_A12}-\ref{eqn_A14}) we summarized the resulting expressions corresponding to all possible $D_4$ irreps and longitudinal parity assignments for a specific choice of the $\phi_{0}$ operator. Namely, by considering linear combinations of the form 
\begin{eqnarray}
\hspace{-0.25cm} \phi_{A}= {\rm Tr}\left[ \parbox{12.5cm}{\rotatebox{0}{\includegraphics[width=12.5cm]{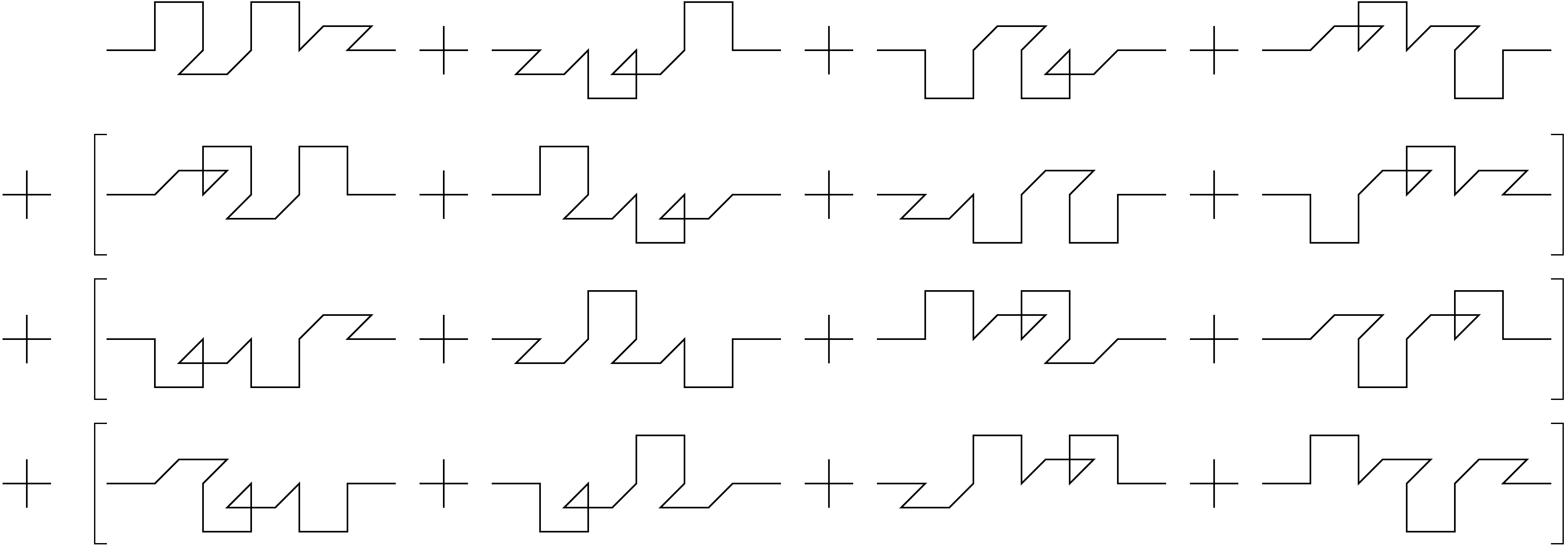}\put(-347,76){ $I$} \put(-347,43){ $L$} \put(-348,10){ $K$}}} \ \right] 
\label{eqn_A12}
\end{eqnarray}
one obtains different $A$ representations for the following choices of the sign factors, $I,J,K$,
\begin{itemize}
    \item For $I=L=K=+1$ the operator projects onto $\{ A_1, P_{\parallel}=+ \}$,
    \item For $I=+1,L=K=-1$ the operator projects onto $\{ A_2, P_{\parallel}=+ \}$,
    \item For $I=-1,L=+1,K=-1$ the operator projects onto $\{ A_1, P_{\parallel}=- \}$,
    \item For $I=L=-1,K=+1$, the operator projects onto $\{ A_2, P_{\parallel}=- \}$.
\end{itemize}
Similarly, linear combinations of the form 
\begin{eqnarray}
\hspace{-0.25cm} \phi_{B}= {\rm Tr}\left[ \parbox{12.5cm}{\rotatebox{0}{\includegraphics[width=12.5cm]{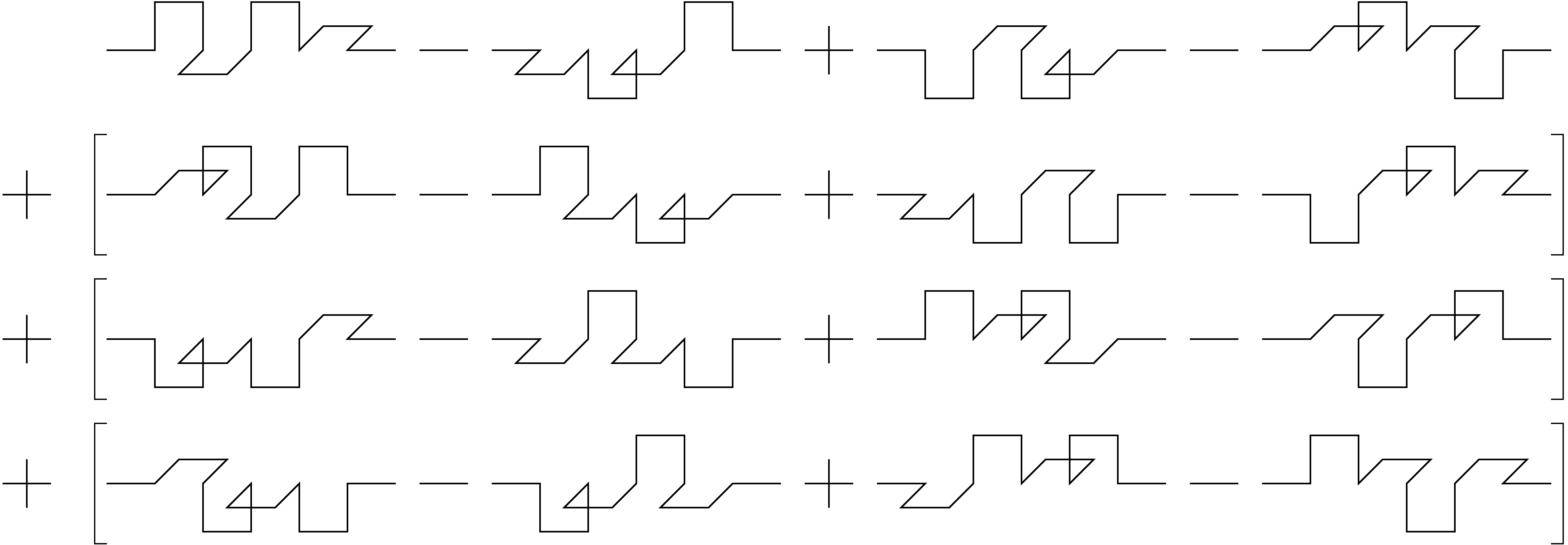}\put(-345,76){$I$} \put(-345,43){$L$} \put(-346,10){$K$}}} \ \right] 
\label{eqn_A13}
\end{eqnarray}
correspond to the $B$ irreps with the following sign choices
\begin{itemize}
    \item For $I=L=K=+1$ the operator projects onto $\{ B_1, P_{\parallel}=+ \}$,
    \item For $I=+1,L=K=-1$ the operator projects onto $\{ B_2, P_{\parallel}=+ \}$,
    \item For $I=-1,L=+1,K=-1$ the operator projects onto $\{ B_1, P_{\parallel}=- \}$,
    \item For $I=L=-1,K=+1$, the operator projects onto $\{ B_2, P_{\parallel}=- \}$.
\end{itemize}
Finally, linear combinations of the form
\begin{eqnarray}
\hspace{-0.25cm} \phi_{E}= {\rm Tr}\left[ \parbox{12.5cm}{\rotatebox{0}{\includegraphics[width=12.5cm]{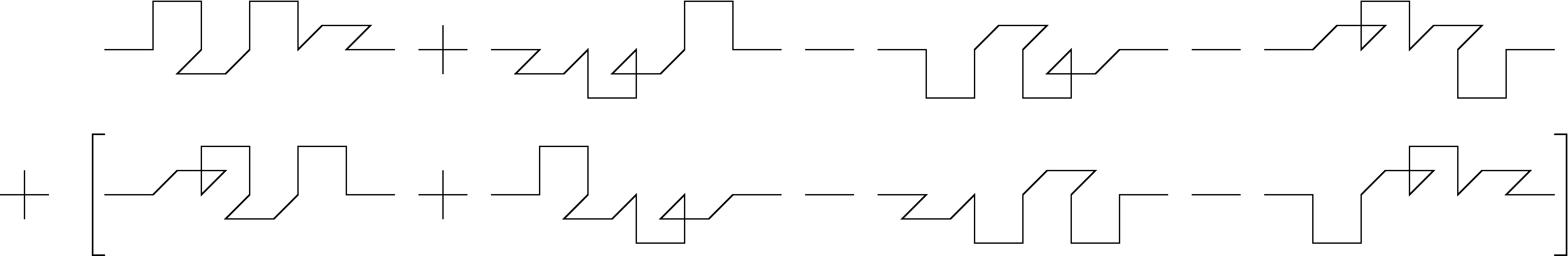} \put(-346,10){$K$}\put(-249,10){\large $i$}\put(-249,43){\large $i$} \put(-74,10){\large $i$}\put(-74,43){\large $i$}}} \ \right] 
\label{eqn_A14}
\end{eqnarray}
project on the $E$ irreps with the following choices,
\begin{itemize}
    \item For $K=+1$ the operator projects onto $\{ E, P_{\parallel}=+ \}$,
    \item For $K=-1$ the operator projects onto $\{ E, P_{\parallel}=- \}$.
\end{itemize}
\begin{figure}[!h]
\begin{center}
    \rotatebox{0}{\includegraphics[width=17cm]{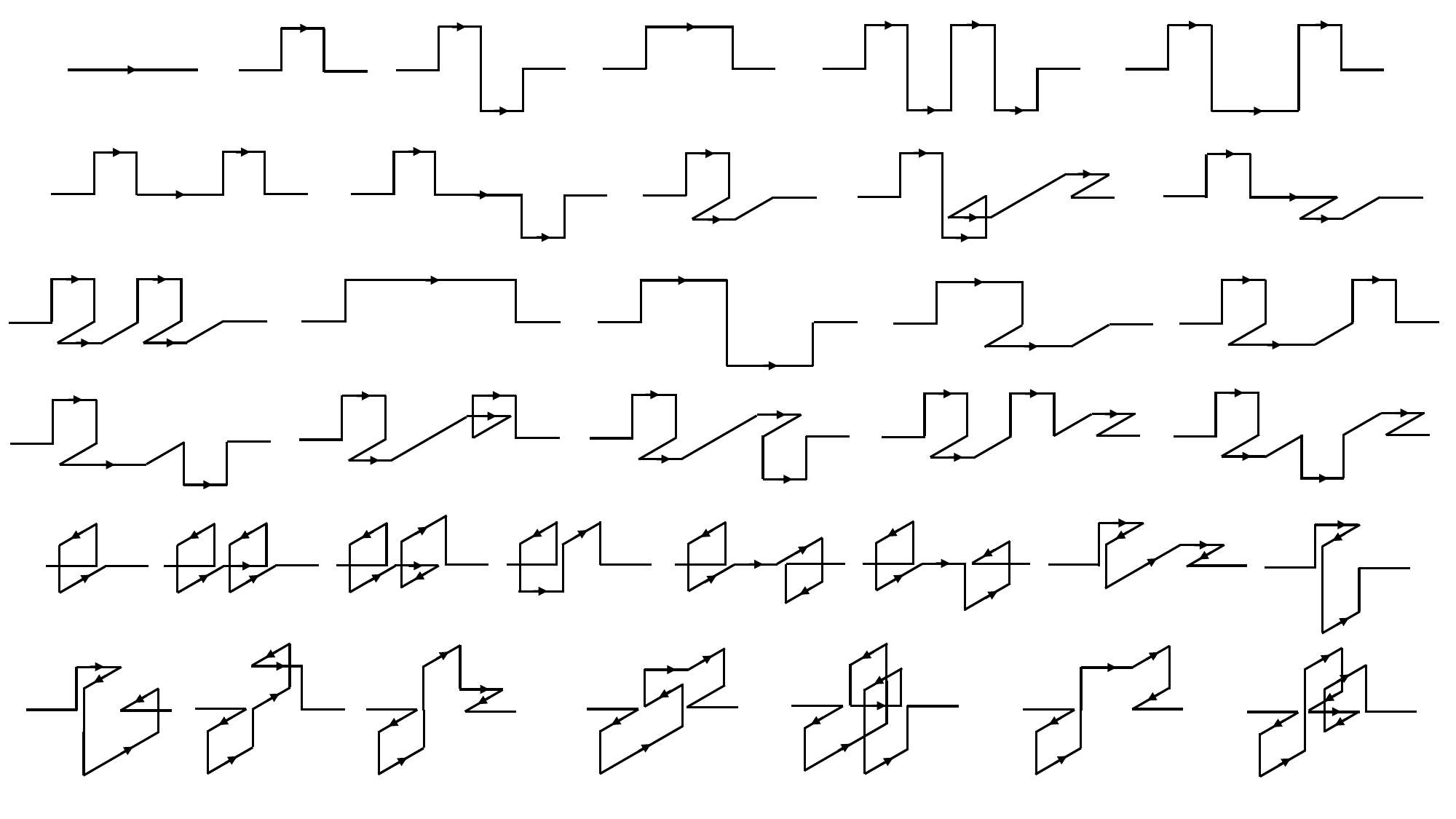}}
\end{center}
\caption{\label{fig:transverse_deformations_use_for_construction} All the transverse deformations used for the construction of the operators.}
\end{figure}

In addition, in order to project on states of a definite momentum $p_{\shortparallel} = 2\pi q/l$ one averages over the longitudinal translation,
\begin{equation}
\phi(p_{\shortparallel}) = \sum_{n_{\shortparallel}}
e^{ip_{\shortparallel} n_{\shortparallel}} \phi_{n_{\shortparallel}}\;,
\label{eqn_opPJ}
\end{equation}
where $\phi_{n_{\shortparallel}}$ stands for an operator $\phi$ translated by a vector $n_{\shortparallel}$ along the $x$ axis.
As discussed before, the longitudinal parity $P_\parallel$ flips $p_{\shortparallel} \to -p_{\shortparallel}$, so it is not defined for states of definite non-vanishing $p_{\shortparallel}$.

To enhance the projection of our operators onto the physical length scales of the states, we apply smearing and blocking techniques~\cite{Lucini:2004my,Teper:1987wt}. Each operator, thus, is built by "super-links" at various smearing and blocking levels. All the paths used for the construction of the operators are presented in Fig.~\ref{fig:transverse_deformations_use_for_construction}; this leads to a basis of around 1200 operators.

In contrast to the basis of operators used in our previous work, the current basis provides enough operators so that we successfully project onto states for each different irreducible representation, with more symmetric representations probed by $\sim 200$ operators and less symmetric by $>50$ operators\footnote{As discussed before, only paths asymmetric enough will contribute to sectors with nontrivial charges. }.

Specifying to the  $q=0$ case as an example, one can check that the operators built for the purposes of this work, 
indeed project irreducibly on $(D_4,P_{\parallel})$  representations. 
This can be done by constructing the full correlation matrix 
consisting of a basis of operators which includes all linear combinations created. 
This should have a block diagonal form with all off-diagonal entries corresponding 
to different irreducible representations being equal to zero within statistics. 
This is illustrated in~\eqref{eq:full_corr_matrix} where each diagonal entry $C_{R_i, R_i}$ represents a block diagonal matrix with a size equal to the number of operators with a given choice of  $(D_4,P_{\parallel})$ quantum numbers. We have checked numerically that the correlation matrix indeed takes this form. 

{\scriptsize
\begin{eqnarray}
C_{R^i, R^j}=
\begin{pmatrix}
C_{A^+_1, A^+_1} & 0 & 0 & 0  & 0 & 0 & 0 & 0  & 0 & 0     \\
0 & C_{A^-_1, A^-_1} & 0 & 0  & 0 & 0 & 0 & 0  & 0 & 0    \\
0 & 0 &  C_{A^+_2, A^+_2} & 0  & 0 & 0 & 0 & 0  & 0 & 0   \\
0 & 0 &   0    & C_{A^-_2, A^-_2} & 0 & 0 & 0 & 0  & 0 & 0 \\
0 & 0 &   0    & 0 & C_{E^+, E^+} & 0  & 0 & 0  & 0 & 0\\
0 & 0 &   0    & 0 & 0 &  C_{E^-, E^-} & 0 & 0  & 0 & 0 \\
0 & 0 &   0    & 0 & 0 &  0 & C_{B^+_1, B^+_1} & 0  & 0 & 0 \\
0 & 0 &   0    & 0 & 0 &  0 & 0 & C_{B^-_1, B^-_1}  & 0 & 0 \\
0 & 0 &   0    & 0 & 0 &  0 & 0 & 0  & C_{B^+_2, B^+_2} & 0 \\
0 & 0 &   0    & 0 & 0 &  0 & 0 & 0  & 0 & C_{B^-_2, B^-_2}
\end{pmatrix} 
\label{eq:full_corr_matrix}
\end{eqnarray}}\\

}

\subsection{Extraction of Torelon Energies and Related Uncertainties}
\label{sec:subsection_energies_lat}
In this subsection we will discuss several sources of statistical and systematic uncertainties affecting the extraction of torelon energies from the lattice data.

Energies of colour singlet states in lattice gauge theories are calculated using the spectral decomposition of a Euclidean two point correlator of some operator $\phi(t)$ with sufficiently high  overlap onto physical states,
\begin{eqnarray}
\langle \phi^\dagger(t=an_t)\phi(0) \rangle
& = &
\langle \phi^\dagger e^{-Han_t} \phi \rangle
=
\sum_i |c_i|^2 e^{-aE_in_t}\,,
\label{eqn_cortomass}
\end{eqnarray} 
where the energies are ordered, $E_{i+1}\geq E_i$, so that at large $t$ the correlator is dominated by the state with the lowest energy $E_0$. Only states
with non-zero overlaps
\[
c_j = \langle vac | \phi^\dagger | j \rangle \neq 0
\]
contribute to the sum (\ref{eqn_cortomass}).
 Therefore, we need to match quantum numbers of the operator $\phi$ to those of the state we are interested in. In this work we study torelons---excitations of a  flux tube wound around the $x$ direction.

\subsubsection{Effective Energies}
\label{sec:Effectibe energies}
As explained above, in order to extract the energies one first constructs linear combinations $\psi_i$ corresponding to approximate energy eigenstates by diagonalizing the correlation matrix at $t=a$. Then one measures the two point correlation function  $\langle \psi_i^\dagger(t)\psi_i(0) \rangle$ at later times $t$ and extracts the corresponding energy eigenvalue from its exponential falloff. 

Let us first discuss the simplest case when the goal is to measure the energy of the ground state in a given symmetry channel, which is approximated by $\psi_0$. The statistical error of the measurement of the $\langle \psi_0^\dagger(t)\psi_0(0) \rangle$ correlator is roughly independent of the time $t$, so that  the signal-to-noise ratio decreases exponentially with $t$. 
This implies that the energy extraction needs to be performed at the shortest possible times $t$. The complication is that in practice  $\psi_0$ is not an exact energy eigenstate, so that the correlator is really a sum of several exponents, 
\begin{equation}
\langle \psi_0^\dagger(t)\psi_0(0) \rangle
=
\sum_k |c_{0k}|^2 e^{-aE_k n_t} \;,
\label{eq:eqn_cortomass2}
\end{equation}
where $c_{0k}$ is an overlap between the approximate ground state $\psi_0$ and the $k$-th exact energy eigenstate in its symmetry sector. 
Hence, in order to be able to perform an accurate energy determination one needs the overlap $c_{00}$ between the approximate and exact ground state to be large enough so that subleading exponents in the sum (\ref{eq:eqn_cortomass2}) have a chance to become negligible at sufficiently short times, when the correlator can still be distinguished from the statistical noise.

To implement this procedure it is convenient to work with an effective energy $E_{i,eff}(n_t)$ defined as\footnote{To account for finite volume corrections, in practice we use a cosh function instead of an exponential function for fitting. See discussions in Section~\ref{sec:finite_volume_effects}. } 
\begin{equation}
e^{-aE_{i,eff}(n_t)}
=
\frac{\langle \psi_i^\dagger(n_t)\psi_i(0) 
\rangle}{\langle \psi_i^\dagger(n_t-1)\psi_i(0) \rangle}
\label{eqn_Eeff}
\end{equation}
If $|i\rangle$ is the lightest state within a given symmetry channel  then as $n_t$ grows the effective energy $a E_{i,eff}(n_t)$ decreases and approaches $a E_i$.
In practice, we measure $a E_i$ when the values of  $a E_{i,eff}(n_t)$ form a plateau as a function of time $n_t$. Given that the plateau is identified at finite $n_t$, there will always be some leftover effects from higher excited states which lead to systematic over-estimation of the energy. On the other hand, when the plateau is too short, the statistical error can also be underestimated. Because of the exponential falloff of the signal-to-noise ratio, for heavier states we are forced to identify the plateaux at smaller $n_t$ to avoid being completely dominated by noise, so this effect is more pronounced in the spectrum of heavier states, as we shall often see later in this paper. 

All these considerations apply also in the case when  $\psi_i$ is not a ground state.
An additional subtlety present in this case is that $\psi_i$ now also has a (small) overlap with lower energy state so that  
at larger $n_t$ the effective energy will decrease to the corresponding lower plateau. This provides yet another reason to measure the plateau as early as it becomes pronounced.

This discussion indicates that we need to maximize overlaps between the approximate eigenstates obtained by solving the GEVP and the true energy eigenstates.
The quality of these overlaps depend on the basis of operators which one uses.
In terms of this choice of operators, the present work should be regarded
as exploratory. Not surprisingly we find that the overlaps are
not as good as in our earlier $D=2+1$ study. However for a number
of states we do have overlaps $|c|^2 \geq 0.9$, at which point we can
regard the difficult-to-estimate systematic errors as being under control.
Of course, for heavier states our identification of an `asymptotic'
exponential behaviour becomes less reliable, and therefore so does our
estimate of the overlap. On the other hand, in $D=3+1$ one has more 
channels with different quantum numbers than in $D=2+1$, and each of
these channels has its own ground state(s) which corresponds to some
excitation of the absolute ground state. Since these ground states are
afflicted by fewer systematic errors, this allows us to obtain
relatively reliable results for a number of string excitation even 
with our poorer overlaps.

\begin{figure}[htb]
  \begin{center} 
 \rotatebox{0}{\includegraphics[width=17cm]{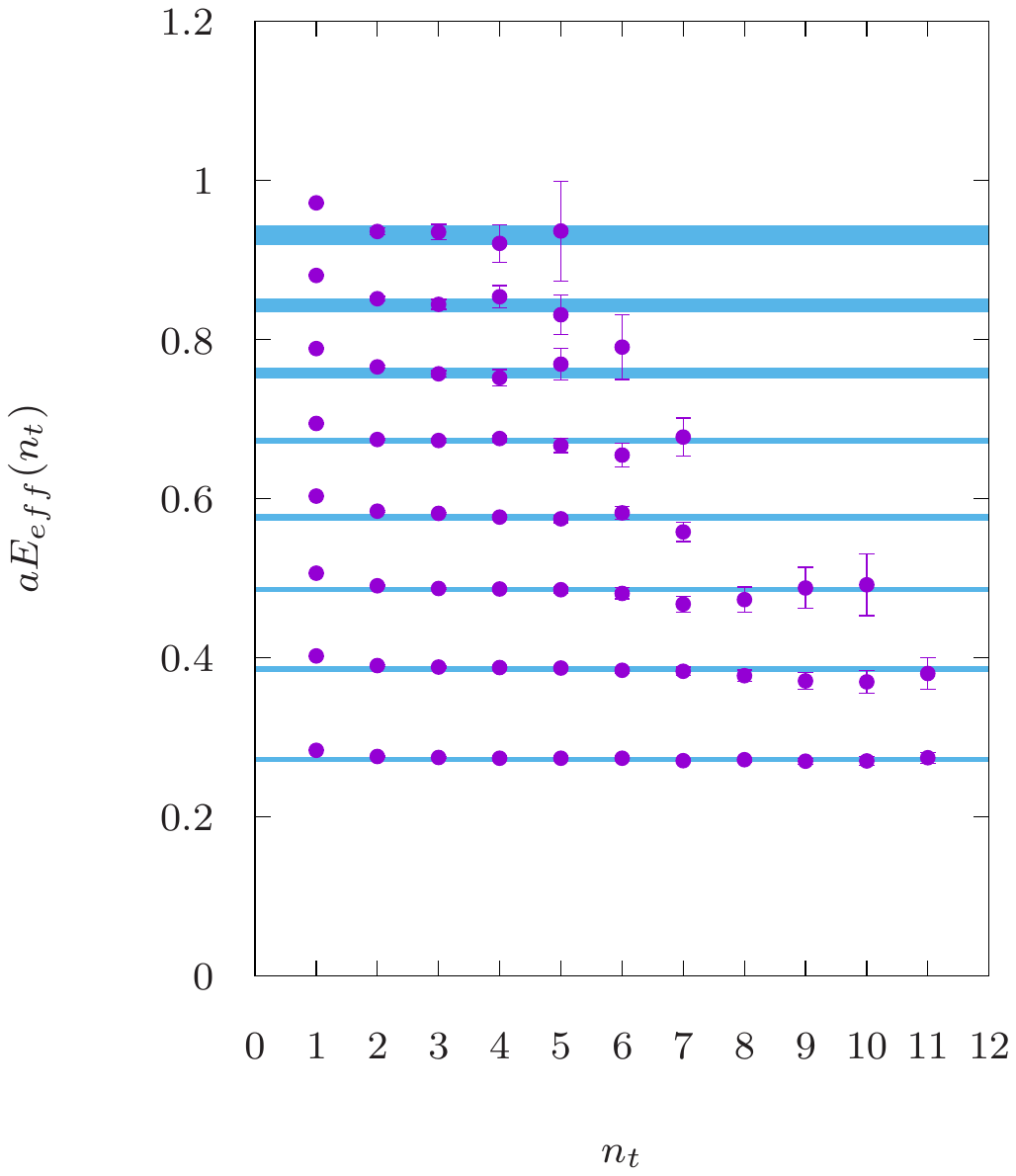}}
 \vspace{-4cm}
\caption{\label{fig:ground_states} Effective energies extracted from the correlator $C(t=an_t)$ 
using eqn(\ref{eqn_Eeff}). The absolute ground state
of a flux tube of length $R/a=10,12,14,16,18,20,22,24$ denoted as $\circ$ in ascending order for $SU(6)$ at $\beta=25.55$.}
  \end{center}
\end{figure}

Let us now illustrate these general considerations with several concrete examples from our simulations.
In Fig.~\ref{fig:ground_states} we present the values of $aE_{eff}(n_t)$ for ground states from our $SU(6)$ calculation at $\beta=25.55$. The effective masses of the absolute ground state ($q= 0$, $0^{++}$ ground state) are denoted with solid circles for flux tube lengths $R/a= 10,12,14,16,18,20,22,24$. For all but the largest values of the energy $E$, the statistical errors are invisible on this plot except at large values of $n_t$. The horizontal red bands correspond to the extracted energies with the associated statistical uncertainty. We note how once the errors become large, at larger $n_t$, the points have a tendency to drift away from the plateau value. Nonetheless, even for $R= 24a$ where the plateau is shorter, it is clear that the calculation of $a E_0(R)$ is unambiguous and is under good control. This is aided by the fact that the plateau begins at very small $n_t$. As a result the overlaps are close to 100\%.

\begin{figure}[htb]
  \begin{center} 
 \rotatebox{0}{\includegraphics[width=19cm]{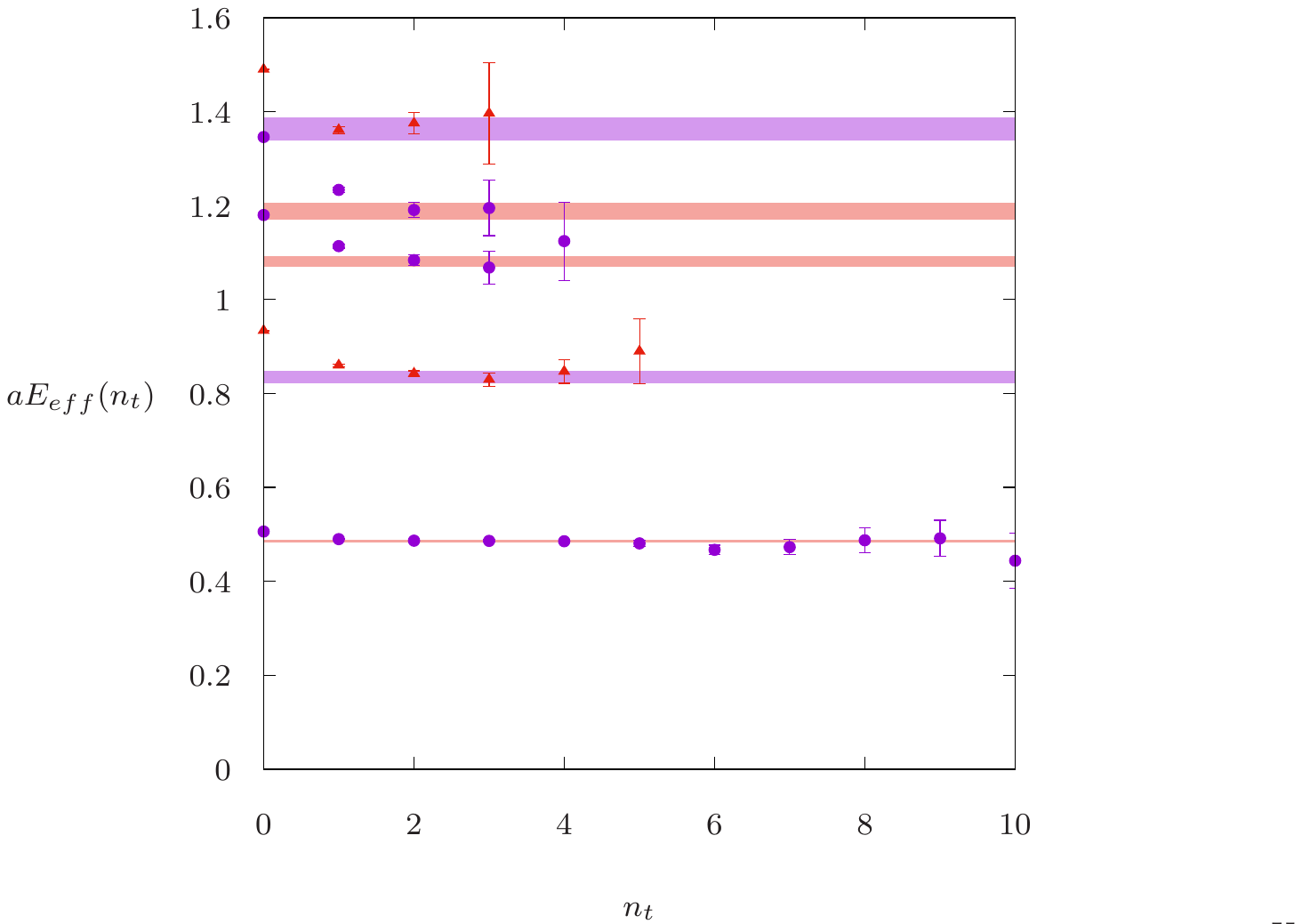}}
\caption{\label{fig:excited_states_} Effective energies extracted from the correlator $C(t=an_t)$ 
using eqn(\ref{eqn_Eeff}) for ground as well as excited states. In $\circ$ and in ascending order we present results for the ground, first excited state as well as second excited states with for torelons with quantum numbers $0^{++}$ while in $\vartriangle$ we present results for the ground as well as first excited state for torelons with quantum numbers $0^{--}$. The results have been extracted for length $R/a=14$ and $SU(6)$ at $\beta=25.55$.}
  \end{center}
\end{figure}

Moving now to Fig.~\ref{fig:excited_states_}, with solid circles we represent the ground as well as the $1^{\rm st}$ and $2^{\rm nd}$ excited states of a flux tube with $q= 0$, $0^{++}$ and with a length $R=14a$. The horizontal red bands correspond to the extracted energies with the associated error. The ground state is well determined, but the two excited states demonstrate the joint problem of a less good overlap and larger energy making the identification of a plateau less clear-cut. To get an idea how good the projection onto these states is, one can look at the normalized overlaps. In fact, the normalized overlap of the first excited state is $|c_{1}|^2 \sim 0.88$ while for the second excited state $|c_{1}|^2 \sim 0.82$; both overlaps are strikingly lower compared to that for the ground state with $|c_{1}|^2 \sim 0.97$. At the same figure, in solid triangles we represent the ground as well as the first excited states of a flux tube with quantum numbers $q= 0$, $0^{--}$. The horizontal purple lines correspond to the best estimates for the energy of each state bound by the associated statistical uncertainty. Once again we see that the ground state within a sector of given quantum numbers can be estimated with better accuracy compared to the excitations. The first excited state demonstrates the difficulty of providing a good energy estimate due the problem of an overlap being not good  enough at the larger energy. Note that the normalized overlaps are  $|c_{i}|^2 \sim 0.90$ for both states in this case.

As can be seen from the cases above, if the overlap is smaller, there is a greater contribution from higher excited states at smaller values of $n_t$, so that the effective energy at those time separations appears larger. With very short plateaux, the statistical errors are sometimes underestimated, due to lack of enough data. If the overlap is very small then the ``signal" is overwhelmed by the statistical noise long before we reach large enough times to see an energy plateau, and we are then left with what appears to be an ill-defined but highly excited state. As a general remark we mention that our experience showed that it is very hard to identify states with an overlap of less than $|c_{i}|^2 \sim 0.5$ while the energy extraction typically becomes extremely difficult to pursue for $|c_{i}|^2 \leq 0.75$\footnote{We note in passing that this is directly related to the appearance of states contained within multi-trace operators, such as scattering states between torelons as well as between torelons and glueballs. Any states that involve multi-trace operators are expected to have a much smaller overlap onto our single trace operators than this, and to be completely invisible within the  generalized eigenvalue problem calculation; such effects are expected to be further suppressed as moving towards to the large-$N$ limit. }.

What the above paragraphs tell us by demonstrating data for our largest value of $N$, is that our results are mostly under good control. This control begins to slip for the states with highest energies, tainted by systematic error that is biased upward and not systematically estimated, especially when such a state is not the ground state of a given set of quantum numbers.  

\subsubsection{Finite volume Effects}
\label{sec:finite_volume_effects}

Another potential source of errors in energy determination is associated to finite volume effects.
Namely, when calculating the spectrum of flux tubes in a hyper-cubic box with dimensions of $L_x \times L_y \times L_z \times L_{t}$, it is essential to ensure that corrections arising from both the finite transverse spatial size, denoted as $L_{\perp}$ (where $L_{\perp} = L_y = L_z$), and from the finite temporal extent $L_t$, are negligible.

Thus, the finite volume corrections are of two kinds: those that arise from
the finite spatial extent and those that arise from the finite temporal extent. Finite spatial extent effects can be caused by the emission of a virtual glueball by a flux tube which propagates around one of the orthogonal spatial directions with compactification length $L_{\perp}$ and then it is being reabsorbed. This will cause a shift in the flux tube energy by an amount of $\Delta E \propto {\tilde g}^2 e^{-a m_G L_{\perp}}$ where $m_G$ is the bulk mass gap, namely the mass of the lightest glueball and ${\tilde g}$ is the closed string coupling, and hence adding a handle to the string world-sheet results in a factor of $\tilde{g}^2$. If the string length $L_{x}$ is too small then the lightest state of emission will be a combination of a flux loop and a conjugate loop and in that case the leading large-$L_{\perp}$ contribution will be received from its propagation around an orthogonal compact direction, giving $\Delta E \sim {\tilde g}^2 e^{-2a E_0(R) L_{\perp}}$. So we expect that as we reduce $L_x$ we will have to have a larger spatial volume to minimise these corrections. However, as we move towards the large-$N$ limit such effects should affect less and less our calculation\footnote{The finite volume effects that we listed above include those that play a role in our computations, and they are all suppressed in the large $N$ limit. However we do not know whether all kinds of finite volume corrections should vanish when $N \to \infty$. }.

We now move on to discuss the systematic effects induced by finite temporal extent $L_t$. There are two sources of this kind. The first one is that on a circle, the two-point correlator is not given by a exponential function of the shortest distance, but it should also include propagations from the other direction and with windings. As a first order correction, which is good enough for our lattice systems, we should replace the expression for effective energy~\eqref{eqn_Eeff} by a cosh function. 

The second source comes from the contribution of thermal particles. Imposing a finite temporal extent $L_t$ means that the partition function $Z$ in  the denominator of~\eqref{eqn_avPhi} receives contributions not just  from the vacuum but from other states propagating along the temporal torus. 
The same state will be contributing to the path integral which is present in the numerator of~\eqref{eqn_avPhi}. In the limit of $N \to \infty$, states which are colour singlets do not interact, and, thus, this additional contribution will cancel between the numerator and the denominator. Of course, at finite value of $N$ a state propagating on the back of the time torus will interact with the torelon which propagates between the Wilson line operators in the numerator implying an insufficient cancellation, so there will be a shift in the energy of the torelon. Typically these contributions are of order $\Delta E \sim O(e^{-a m L_t})$ where 
$m$ is the lightest energy. For small string lengths $R$, the lightest state is the winding 
flux tube, so that $m = E_0(R)$, and we must make $l_T$ larger as $R$ decreases, so as to maintain  $E_0(R)l_t \gg 1$.

\begin{table}[htb]
\begin{center}
\begin{tabular}{c|c|c|c|c}\hline \hline
\multicolumn{5}{c}{SU(3) ; $\beta=6.0625$} \\ \hline \hline
$R/a$ & $L_{\perp} \times L_{\perp} \times L_T$ & $aE^{\rm gs}_{0^{++}}(R)$ &  $aE^{\rm f.ex}_{0^{++}}(R)$ & $aE^{\rm gs}_{0^{--}}(R)$ \\ \hline
12 & $16\times 16\times 24$ &  0.3479(18)  &  0.9694(68)   &  0.7386(64) \\ 
   & $20\times 20\times 24$ &   0.3549(21) &  1.0076(71)   &  0.7425(56) \\ 
   & $24\times 24\times 24$ &   0.3544(17) &  1.0078(72)   &  0.7316(80) \\ \hline
14 & $16\times 16\times 24$ &   0.4467(18) &  1.0333(88)   &  0.8143(77) \\ 
   & $20\times 20\times 24$ &   0.4489(18) &  1.0548(96)   &  0.8221(91) \\ \hline
16 & $16\times 16\times 16$ &   0.5387(17) &  1.0701(117)  &  0.8847(155) \\
   & $20\times 20\times 24$ &   0.5388(18) &  1.0900(111)  &  0.8943(150) \\ \hline
\end{tabular}
\caption{\label{tab:finite_volume} Energies of the ground and first excited states for quantum numbers $0^{++}$ as well as the ground state for $0^{--}$ 
of a closed flux tube of length $R$, on various spatial volumes. We have considered $SU(3)$ gauge theory at $\beta=6.0625$ and for each measurement we performed 6.25 million Monte-Carlo sweeps.}
\end{center}
\end{table}

As a standard procedure we monitor the finite volume corrections by performing calculations for various transverse and temporal extents for each different value of the flux tube length $R$. A detailed study for $SU(6)$ can be found in \cite{Meyer:2004hv}, however this was done for a basis of operators which consists of the simple line operator. As a matter of fact in many of our previous papers we have described tests of such corrections in some detail, and the volumes used in this paper have been chosen accordingly. However most of  those  tests  were  done  with  a  small  basis  of  operators,  which  allowed  us  to calculate  the absolute  ground  state  but  did  not  allow  an  accurate  determination  of  excited states. Since(some) excited states will have a larger total ‘width’ than the ground state, and hence might be more sensitive to the transverse boundaries (the temporal extent is not a problem here), more attention should be invested also on the low-lying excitation spectrum.

One, of course, could choose large enough transverse and temporal lattice extents so that such systematics are vanishing; unfortunately increasing lattice size increases computational cost and, thus, one should opt wisely the selection of the lattice volumes. Nevertheless, since it is not too expensive to make just one lattice torus very large, we choose to do so for the time torus and then we just vary the spatial extents to check the associated corresponding finite spatial volume corrections. At $N \to \infty$, corrections due to finite transverse lattice extent are vanishing because of the $\tilde{g}^2$ factor, thus one should search for those at smallest possible values of $N$. Thus, we choose to investigate $SU(3)$ at $\beta=6.0625$ since $\beta = 6.338$ corresponds to smaller lattice spacing and would require the usage of much larger lattice volumes. Similar study has been carried out in our previous investigation for which we used just the simple line operator. A good idea would be to start our investigation from the largest lattice volumes used in the above study. The values of $R$ and lattice sizes for the $SU(3)$ calculation are listed in Table~\ref{tab:finite_volume}. It would only make sense to investigate such effects on the energy levels on which we have good control. Thus, we focus on the three lightest states of torelons with $p=0$. Namely, we calculate the ground and first excited state energies for torelons with quantum numbers $0^{++}$ as well as the ground state of a torelon with $0^{--}$. 

In Table~\ref{tab:finite_volume} we provide a comparison of energy levels for different lattice volumes of flux tubes with lengths $R/a = 12, 14, 16$. We can see that the finite volume corrections on the absolute ground state are negligible. In point of fact only the smallest volume for the shortest flux tube experiences such an effect with deviations of no more than $\sim 2 \% $. However, because we are interested in identifying leading and  subleading corrections to the linear behaviour of $E_0(R)$, these  corrections would play an important role; we, thus, choose to work with the largest lattice extent for flux tubes with length of $R/a =12$. For our shortest flux tube with length $R/a = 10$ we choose to simply increase the transverse lattice extents of our previous calculation in \cite{Athenodorou:2010cs} from $l_{\perp}/a= 20$ to  $l_{\perp}/a= 24$ and observed that the ground state remains the same. The leading lattice large volume correction decreases exponentially with the flux tube size. For $R/a=16$ we decided to use transverse size equal to the size of the tube. Hence, we expect that the energy values extracted for flux tubes with larger lengths than those presented in Table~\ref{tab:finite_volume} will experience no finite volume corrections if we set the transverse lattice size equal to the longitudinal.

Concerning the first excited state with quantum numbers $0^{++}$ as well as the ground state with $0^{--}$, it appears that no important finite volume effects, for the chosen volumes, are present for the given statistical accuracy. As a matter of fact the only statistically significant discrepancy appears for the shortest flux tube with length $R/a=12$ and the smallest transverse lattice extent with deviations of no more than $\sim 2 \% $ from the results extracted for the two larger lattice transverse extents. The adoption of the largest lattice volume for $R/a = 12$ suggests that there is no reason to worry for finite volume effects in the low-lying excitation spectrum. We report that for the $R/a = 12, 14$ we performed calculations for the largest transverse lattice extents reported in Table~\ref{tab:finite_volume} while for $R/a=16$ we keep all lattice extents the same.

\section{Summary of lattice data}
\label{sec:data}

{\color{black} We  performed calculations of the excitation spectrum of a closed flux tube carrying a fundamental color flux, which winds once around a compactified spatial direction of length $R$, which corresponds to the $\mathbb{Z}_N$ charge $k=1$. We consider values of $N$ and inverse bare couplings, $\beta=2N/g^2$ which are listed in Table~\ref{tab:table_physics_parameters}. For each choice of $N$ and $\beta$ we provide the average plaquette value as well as the string tension of  $a \sqrt{\sigma}$ extracted from our measurement of the absolute ground state string energy $E_0(R)$. In addition we list some basic physical properties of each lattice gauge theory such as the critical compactification radius below which the theory deconfines and the flux tube dissolves (for $N > 2$), as well as the mass of the lightest scalar glueball. To extract the value of $a^2\sigma$ we fit the ground state data using as a fitting function the GGRT expression with an additional term proportional to $O(1/R^7)$. We observe that the string tension is not sensitive to  the choice of the correction term.
 In principle, one can extract the string tension just by fitting the ground state with the universal Nambu-Goto prediction for $R\sqrt{\sigma} >  2.5$ with an acceptable $\chi^2$. 
 
 The dimensionless quantity $a \sqrt{\sigma}$  enables us to express the lattice spacing $a$ in physical units by setting the string tension $\sigma$ to its experimentally determined value  $\surd\sigma \sim 440 {\mathrm{MeV}}$. Of course this involves some level of ambiguity because the $SU(N)$ gauge theory and the real world QCD are two different theories and there can be no single rescaling that will match their physics. }

{\color{black} The values we provide for $R_c$ have been obtained by extrapolating and interpolating in $\beta$ results reported in the works~\cite{Lucini:2002ku,Lucini:2003zr,Lucini:2005vg}. Our investigation will go down to flux tube lengths $R$ which are very close to the critical length $R_c$. At these lengths the flux tube should be about as wide as it is long. In practice at such lengths, the flux tube looks nothing like a thin string but rather as a blob.}

We have performed high statistics calculations with a large number of operators for all choices of gauge groups; lattice sizes and the number of MC sweeps used in this work are reported in Table~\ref{table_param}. We start our investigation by simulating $SU(3)$ at $\beta=6.0625$ and extracting the energy spectrum for 12 different flux tube lengths from $R \sqrt{\sigma} \sim 1.9$ to $\sim 7$. We then move to larger values $N=5,6$ while keeping  $a \sqrt{\sigma}$ approximately fixed, $a \sqrt{\sigma}\sim 0.2$, which corresponds to  inverse couplings of $\beta = 17.630$ and $\beta = 25.55$ respectively. For both gauge groups we considered 8 values of the length $R$ starting from $R \sqrt{\sigma} \sim 1.9$ to $\sim 4.7$. Next, we move closer to the continuum limit by investigating $SU(3)$ at $\beta = 6.338$ and $SU(5)$ at $\beta = 18.375$. For both gauge groups these values correspond to lattice spacing $a \sqrt{\sigma} \sim 0.13$. By expressing $a$ in a more intuitive `fermi' unit the above values become $a\sim  0.09 {\mathrm{fm}}$  and $a\sim  0.06 {\mathrm{fm}}$ respectively. This illustrates our strategy of investigating possible finite lattice spacing artifacts. Namely, we kept $N=3,5$ fixed and reduced the lattice spacing by a factor of $\sim 2/3$. 
For $SU(3)$, $\beta=6.338$ we extract the spectrum for 14 values of the flux tube length ($R \sqrt{\sigma} \sim ~ 1.8 - 4.7$) while for $SU(5)$, $\beta = 17.63$ for 5 values of $R$ $(R \sqrt{\sigma} \sim 2.6 - 3.6)$.

\begin{figure}[htb]
  \begin{center} 
 \rotatebox{0}{\hspace{-2.0cm}\includegraphics[width=14cm]{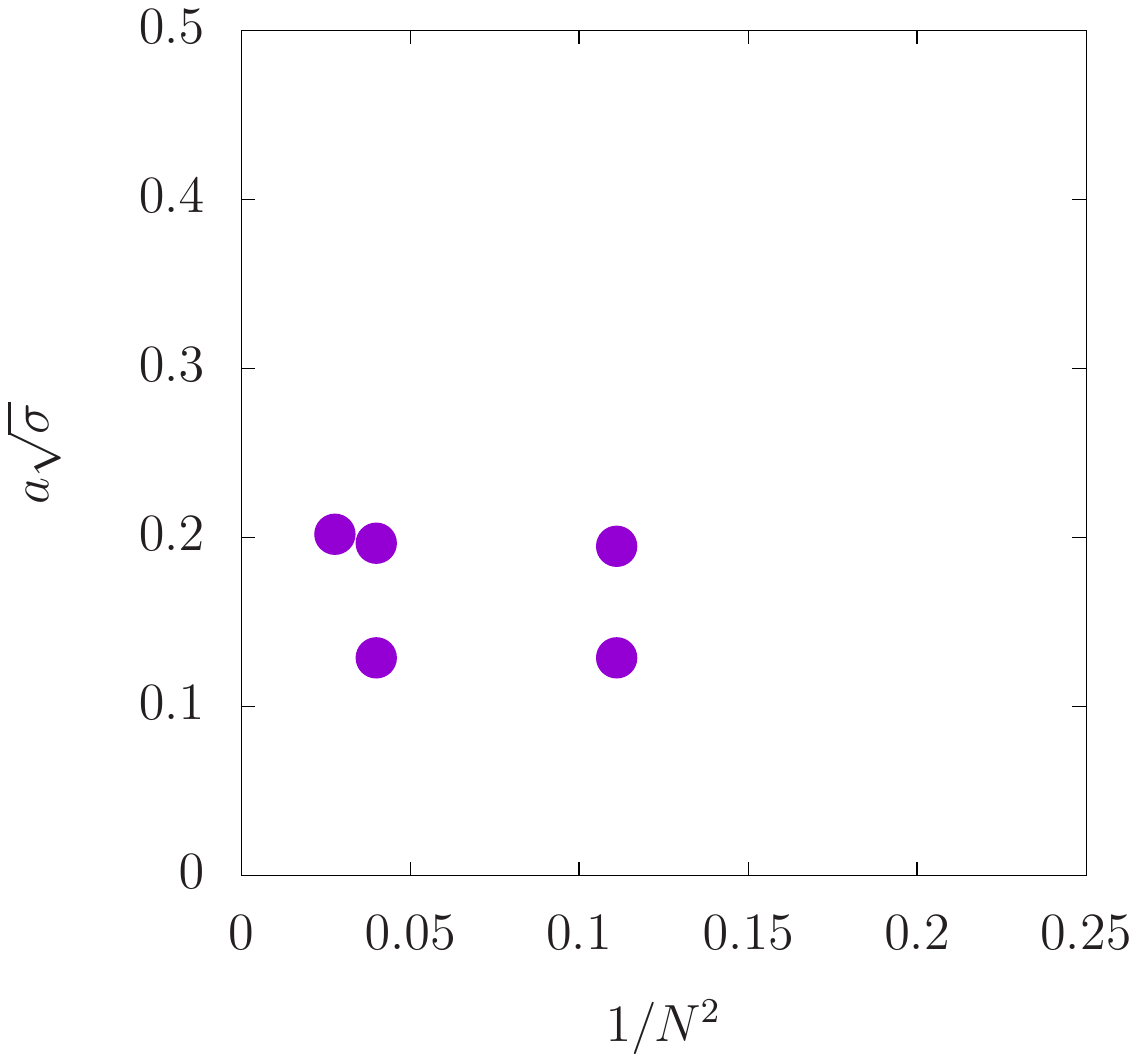}}
\caption{\label{fig:strategy} A graphic representation of our choice of the parameters $N$ and $\beta$ on the plane $a \sqrt{\sigma}$ vs. $1/N^2$. As can be seen in the plot, we investigate possible large-$N$ corrections by fixing the value of $a \sqrt{\sigma}$ and altering $N$, and finite lattice spacing effects by keeping $N$ constant and reducing $a \sqrt{\sigma}$.}
  \end{center}
\end{figure}

 Note that as a consequence of changing the lattice spacing the masses of states in lattice units also change which  affects  possible systematic over-estimations of these masses, as discussed earlier, especially for heavy states. Within our computations, it is hard to tell between this effect and the genuine lattice dependence originating from the discretization effects. A graphic illustration of our choices of $N$ and $\beta$ values is provided in Fig.~\ref{fig:strategy}.

In what follows we start presenting our results with the absolute ground state $E_0(R)$; this is the ground state with quantum numbers $0^{++}$. We then move to the low-lying spectrum of states  within each different irreducible representation $R^{P_{\parallel}}$ and zero longitudinal momentum $p_{\shortparallel}=2\pi q/R=0$; our large basis of operators enables us to extract states within each single irreducible representation $R^{P_{\parallel}}$. 
A special attention is given to the ground state with quantum numbers $0^{--}$ which corresponds to a massive mode, namely the world-sheet axion. 

Subsequently, we move to the low-lying spectrum of flux tubes that carry a non-vanishing longitudinal momentum $p_{\shortparallel}=2\pi q/R$, for $q=1,2$. Once again, the large basis of operators provides access to excitation levels within all the irreducible representations $R$. 
In the absence of massive excitations, the low-lying states are expected to be predicted by the GGRT formula~\eqref{GGRT}, at least in the long string limit. 
However we shall see that  many states deviate drastically from the GGRT predictions. However, all these exotic states  can be incorporated in the simple framework with a single massive worldsheet axion in the effective string action, as we shall explain later in Section~\ref{sec:computation}. In this section, we will summarize the data and see  many examples of the breakdown of the GGRT formula.

\subsection{The absolute ground state}
\label{subsection_groundq0}

In this section, we examine how the absolute ground state energy depends on the parameter $R$. This energy level corresponds to the ground state of a closed flux tube that wraps around the spatial circle once and possesses trivial quantum numbers of $0^{++}$. As a result, we anticipate encountering this state as the ground state within the $A_1^{+}$ irreducible representation. It is worth noting that our primary focus in this study is the investigation of the excitation spectrum, rather than an in-depth analysis of the ground state.  Nevertheless, our analysis can still yield some insights into the universal corrections to the GGRT spectrum.

In Table~\ref{tab:table_absolute_ground_state}, we present our measurements of the ground state energy $E_0$ in the $0^{++}$ channel for  $SU(3)$, $SU(5)$, and $SU(6)$ gauge groups. To start with  let us  see  what this data tells us about the value of the leading universal 
 $1/R$ correction, often referred to as the L\"uscher term. Specifically, we need to test whether $E_0(R)$
 has an asymptotic form~\cite{Luscher:2004ib}
\begin{equation}
E_0(R) \stackrel{R\to\infty}{=} \sigma R - \frac{\pi}{3R}+{\cal O}(1/R^3)\;.
\label{eq:Luscher_term}
\end{equation}
We tackle this  question  by extracting the effective parameter denoted as "$c_{eff}$," as it is a key component in this expression,
\begin{equation}
E_0(R) = \sigma_{eff} R - c_{eff}\frac{1}{R}\,,
\label{eq:eqn_MLceff}
\end{equation}
where, $\sigma_{eff}$ represents the effective string tension. To investigate this, we compute the energies $E_0(R)$ for various lengths, starting from $R = R_{1}, R_2, \ldots$, where each subsequent length $R_{i+1}$ is greater than or equal to the previous one ($R_1 = R_{\text{min}}$). Through a fitting process using (\ref{eq:eqn_MLceff}) as an ansatz for $E_0(R)$ 
we determine the parameters $c_{eff}$ and $\sigma_{eff}$ for each set of energy values, beginning with $E_0(R_{\text{min}})$.

It is worth noting that we anticipate that for smaller values of $R_{\text{min}}$, the value of $c_{eff}$ may exhibit significant deviations from its limit at large $R_{\text{min}}$, owing to additional corrections introduced by higher-order terms in $1/R$ and also due to a possible breakdown of $1/R$ expansion caused by massive modes. Consequently, our objective is to address whether $c_{eff}$ tends toward $\pi/3$ as $R_{\text{min}}$ approaches infinity.

\begin{figure}[htb]
  \begin{center} 
 \rotatebox{0}{\put(350,300){$\pi/3$}\hspace{-3.5cm}\includegraphics[width=24cm]{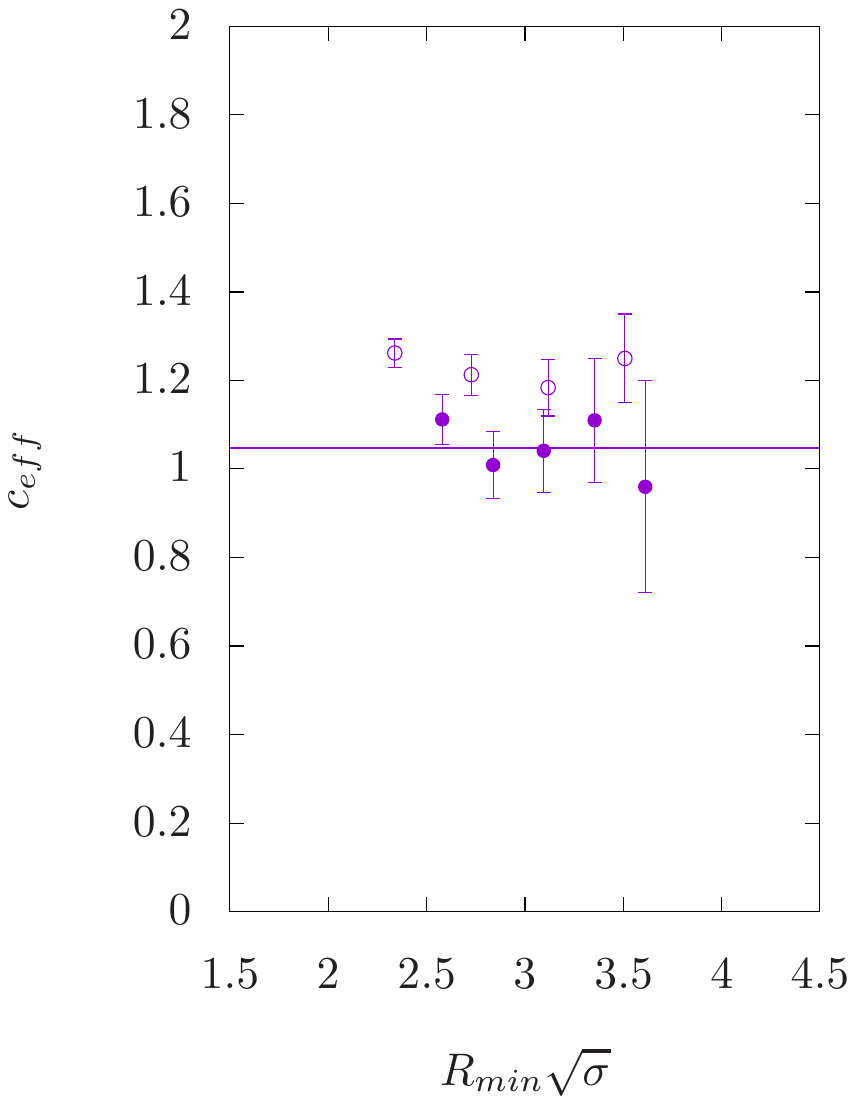}}
\caption{\label{fig:ceff_Luscher} Fits to flux tube energies for $R\ge R_{min}$ with $E(R) = \sigma R - c/R$
  in $SU(3)$ at $\beta=6.338$ ($\bullet$) and $\beta=6.0625$ ($\circ$). Solid line
  is the expected universal (L\"uscher) value of $\pi/3$. }
  \end{center}
\end{figure}

In Fig.~\ref{fig:ceff_Luscher} we present our results for $c_{eff}$ as a function of $R_{min} \sqrt{\sigma}$ for $SU(3)$ and $\beta=6.338$ as well as $\beta=6.0625$. The results for $\beta=6.338$ appear to exhibit perfect agreement with the expectation of the L\"uscher correction for $R \sqrt{\sigma} \geq 2.6$ while the data for $\beta=6.0625$ exhibits a minor discrepancy of 2-3 sigmas which could be attributed to discretization effects.

\begin{figure}[htb]
  \begin{center} 
 \rotatebox{0}{\hspace{-2cm}\includegraphics[width=20cm]{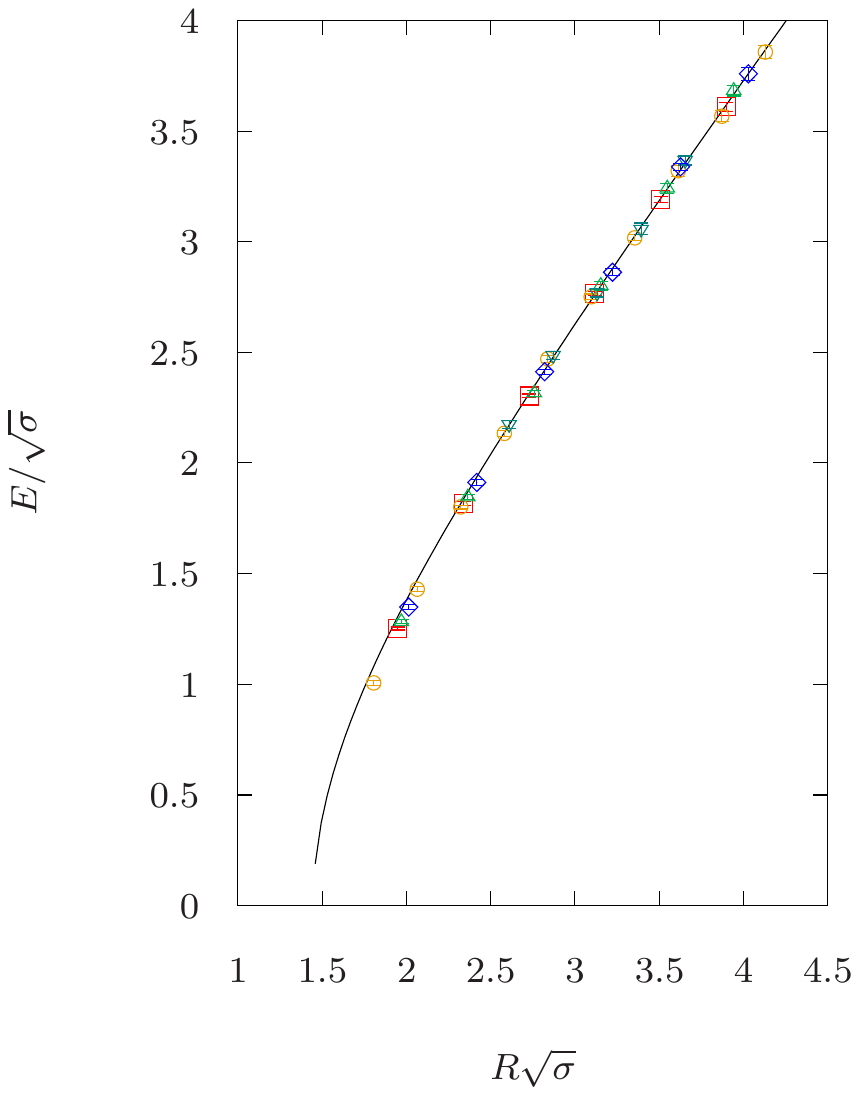}\put(-320,120){\small $N_L=N_R=0$}}
\caption{\label{fig:absolute_ground} The absolute ground state of the confining string with quantum numbers $0^{++}$ and $q=0$. The line presents the GGRT level with $N_L=N_R=0$. The representation of the different gauge groups goes as follows: $SU(3)$, $\beta=6.0625$ is represented by $\square$, $SU(3)$, $\beta=6.338$ by $\circ$, $SU(5)$, $\beta=17.630$ by $\triangle$, $SU(5)$, $\beta=18.375$ by $\triangledown$ and $SU(6)$, $\beta=25.55$ by  $\diamond$.} 
  \end{center}
\end{figure}

As was observed in earlier study~\cite{Athenodorou:2011rx} the ground state data can be described remarkably well using the GGRT expression for the energy. This agreement is visually evident in Fig.~\ref{fig:absolute_ground}. This plot also indicates that for the shortest flux tube lengths, there are subtle deviations from the GGRT prediction. As reviewed in Section~\ref{sec:theory} the leading deviation from the GGRT formula scales as $1/R^7$ for the flux tube  ground state\footnote{Recall that the universal $1/R^5$ Polchinski--Strominger correction vanishes for the ground state~\cite{Aharony:2009gg}.}.

\begin{figure}[htb]
  \begin{center} 
 \rotatebox{0}{\hspace{0.0cm}\includegraphics[width=20cm]{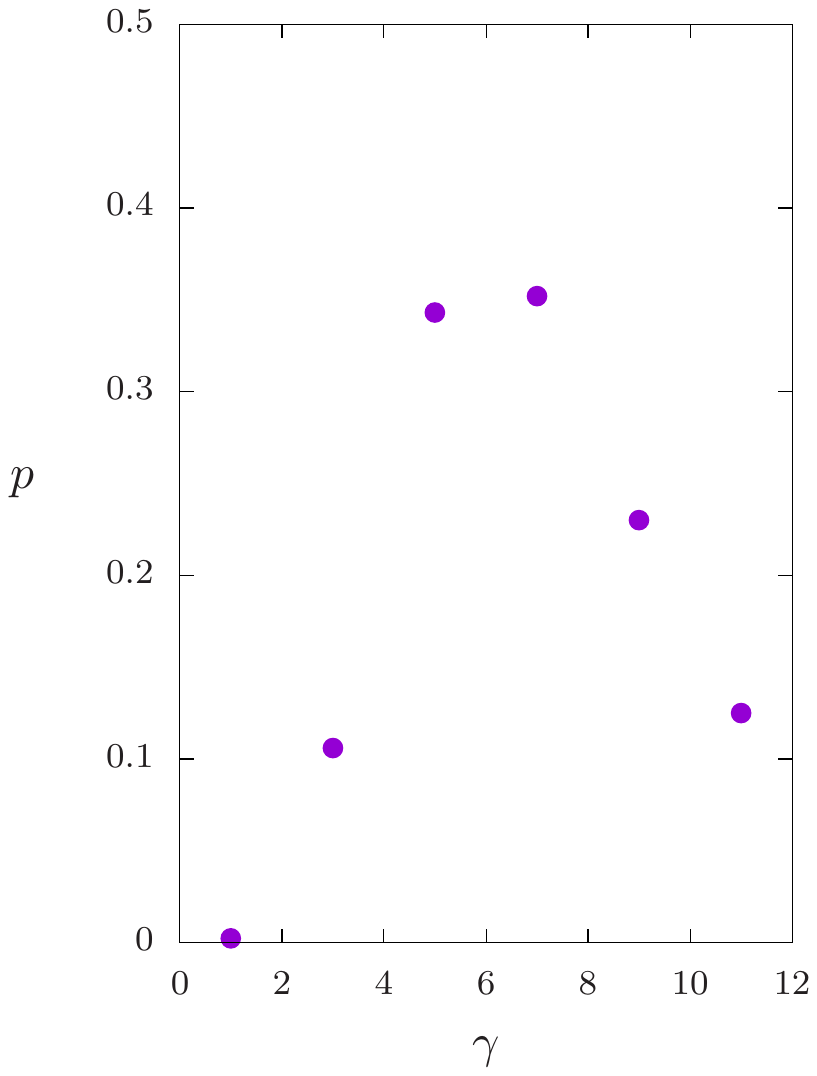}}
\caption{\label{fig:gamma} Standard $p$-value for the power of the correction to GGRT:
  $E(R) = E_{\rm GGRT}(R) + c\surd\sigma/\{R\surd\sigma\}^{\gamma}$. For
  the flux loop in $SU(3)$ at $\beta=6.338$.}
  \end{center}
\end{figure}

Therefore, a valuable exercise involves examining the correction to the ground state spectrum of the confining flux tube and assessing its consistency with theoretical predictions. To accomplish this, we perform data fitting employing the following ansatz
\begin{eqnarray}
    E(R)=E_{\rm GGRT}(R) - \frac{\tilde{c}\sqrt{\sigma}}{(R \sqrt{\sigma})^\gamma}\,.
\end{eqnarray}
At the first stage we would like to confirm that our data agrees with the theoretical prediction $\gamma=7$.
Here we focus on the $SU(3)$, $\beta = 6.338$ data set as the most precise one and extending to the shortest values of $R$.
In Fig.~\ref{fig:gamma} we present the standard $p$-value for the power of the correction to the GGRT string. It is evident that the dominant powers in the fit are $\gamma=5$ and $\gamma=7,$ with a slight preference for the latter. This aligns well with the theoretical preference for $\gamma=7$. We note that if we replace $E_{GGRT}(R)$  in the fits by the sum of the universal terms for the ground state energy, then the conclusion remains the same except that gamma=3 becomes much less likely.

At the second stage we set the power to $\gamma=7$ and direct our attention toward the precise determination of the correction prefactor $\tilde{c}$. The results of the fits can be found in Table~\ref{tab:table_coefficient_7}. It's evident that we can obtain values for $\tilde{c}$ for all the ensembles except for $SU(5)$ at $\beta=18.374$, where the absence of data for the shortest flux tubes hinders the extraction of results. However, one should be conservative when taking this fitted non-universal coefficient at face value, because it is based on the assumption that the deviation only has one term of power $\gamma=7$, but we know there should an infinite series of higher order corrections. Given the precision and the corresponding string lengths, it is hard to eliminate, for instance, the effect of the $O(\sigma^4/R^9)$ term, so there should be a large systematic error that we fail to estimate for $\tilde{c}$. Also the $1/R$ expansion may break down for short strings, so that a reliable extraction of $\tilde{c}$ would require scanning over the short $R$ cutoff similarly to how we did for the L\"uscher correction. 

\begin{table}[htb]
\centering
\begin{tabular}{cc|cc} \hline \hline
  $SU(N)$  & $\beta$  &  $\tilde{c}$   &  $a^2\sigma$   \\ \hline \hline
 $SU(3)$   & 6.3380 & 5.16(77) & 0.016659(28)  \\
 $SU(3)$   & 6.0625 & 5.88(85) & 0.037995(54)  \\ \hline
 $SU(5)$   & 18.375 &  --      & 0.017040(50)  \\
 $SU(5)$   & 17.630 & 5.77(1.34) & 0.03872(10)  \\ \hline
 $SU(6)$   & 25.550 & 6.94(1.54) & 0.04055(10)  \\ \hline
\end{tabular}
\caption{Fits to flux tube energies of the form
  $E(R)=E_{\rm GGRT}(R) - \tilde{c}\surd\sigma/\{R\surd\sigma\}^7$
  for the lattice fields shown.}
\label{tab:table_coefficient_7}
\end{table}

\subsection{Excited states of the confining flux tubes with longitudinal momentum $q=0$}
\label{sec:results_for_q=0}

\subsubsection{The $N_L=N_R=1$ states and the axion}
\label{subsec:Nl=Nr=1}

We proceed with presenting our findings regarding the $q=0$ longitudinal momentum sector, specifically focusing on data related to the GGRT level with $N_L=N_R=1$. We compare the measured energy spectrum  with the predictions of the GGRT string model. 
The phonon composition of such states is detailed in the second column of Table~\ref{tab:table_NGstates_q0}.

According to this Table~\ref{tab:table_NGstates_q0} the $N_L=N_R=1$ should be four-fold degenerate consisting of a  $0^{++}$ state with phonon content $\left( a^{+}_{1} a^{-}_{-1}-a^{-}_{1} a^{+}_{-1} \right) | 0 \rangle$, a $0^{--}$ state  $\left( a^{+}_{1} a^{-}_{-1}-a^{-}_{1} a^{+}_{-1} \right) | 0 \rangle$,  and a spin 2 doublet 
consisting of a $2^{++}$ state  $\left( a^{+}_{1} a^{+}_{-1}+a^{-}_{1} a^{-}_{-1} \right) | 0 \rangle$ as well as a $2^{-+}$ state  $\left( a^{+}_{1} a^{+}_{-1}-a^{-}_{1} a^{-}_{-1} \right) | 0 \rangle$.

\begin{figure}[htb]
  \begin{center} 
  \vspace{-2.0cm}
 \rotatebox{0}{\hspace{-2.0cm}\includegraphics[width=15cm]{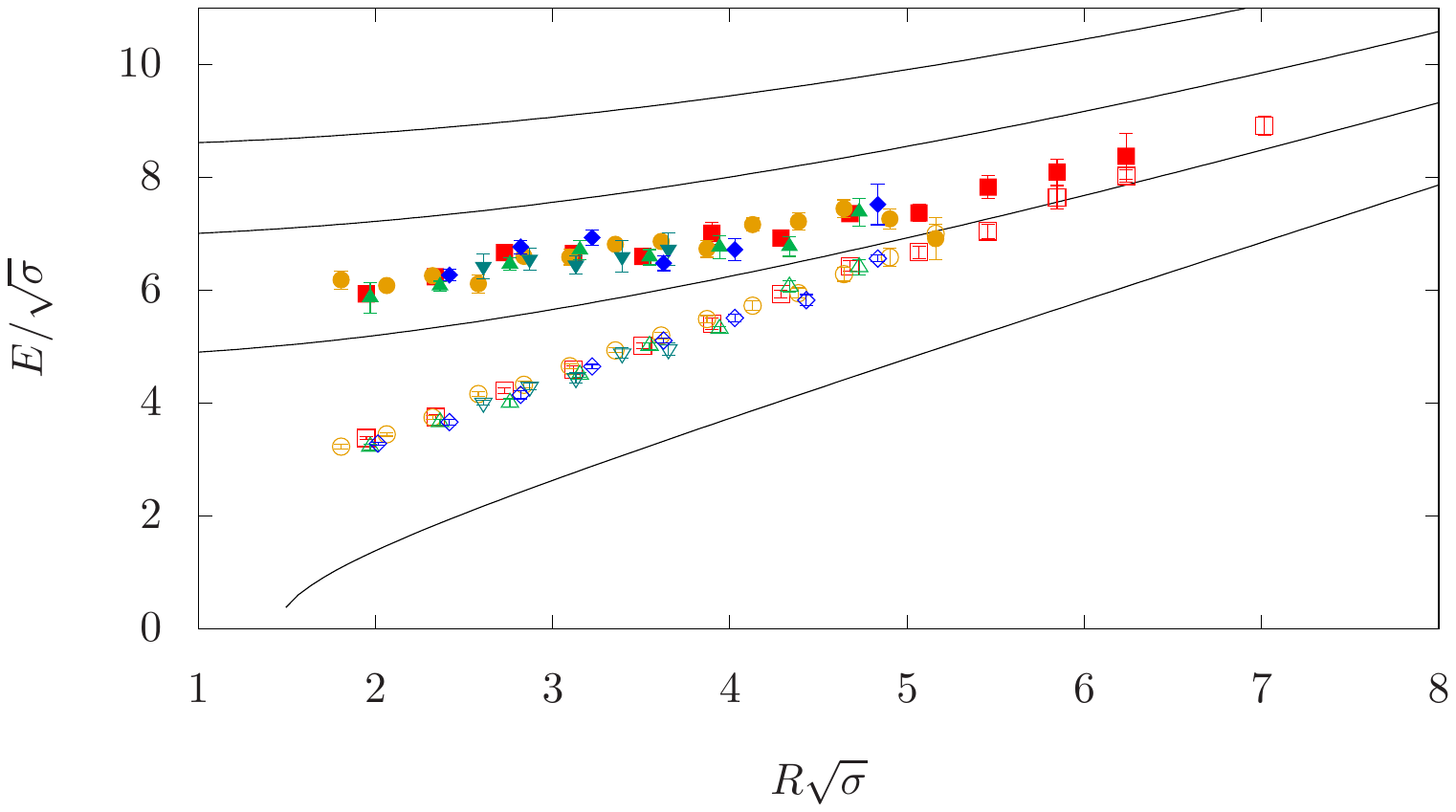}\put(-20,224){\scriptsize $N_L=N_R=0$}\put(-20,247){\scriptsize $N_L=N_R=1$}\put(-20,270){\scriptsize $N_L=N_R=2$}}
 
 \vspace{-3.0cm}
 \rotatebox{0}{\hspace{-2.0cm}\includegraphics[width=15cm]{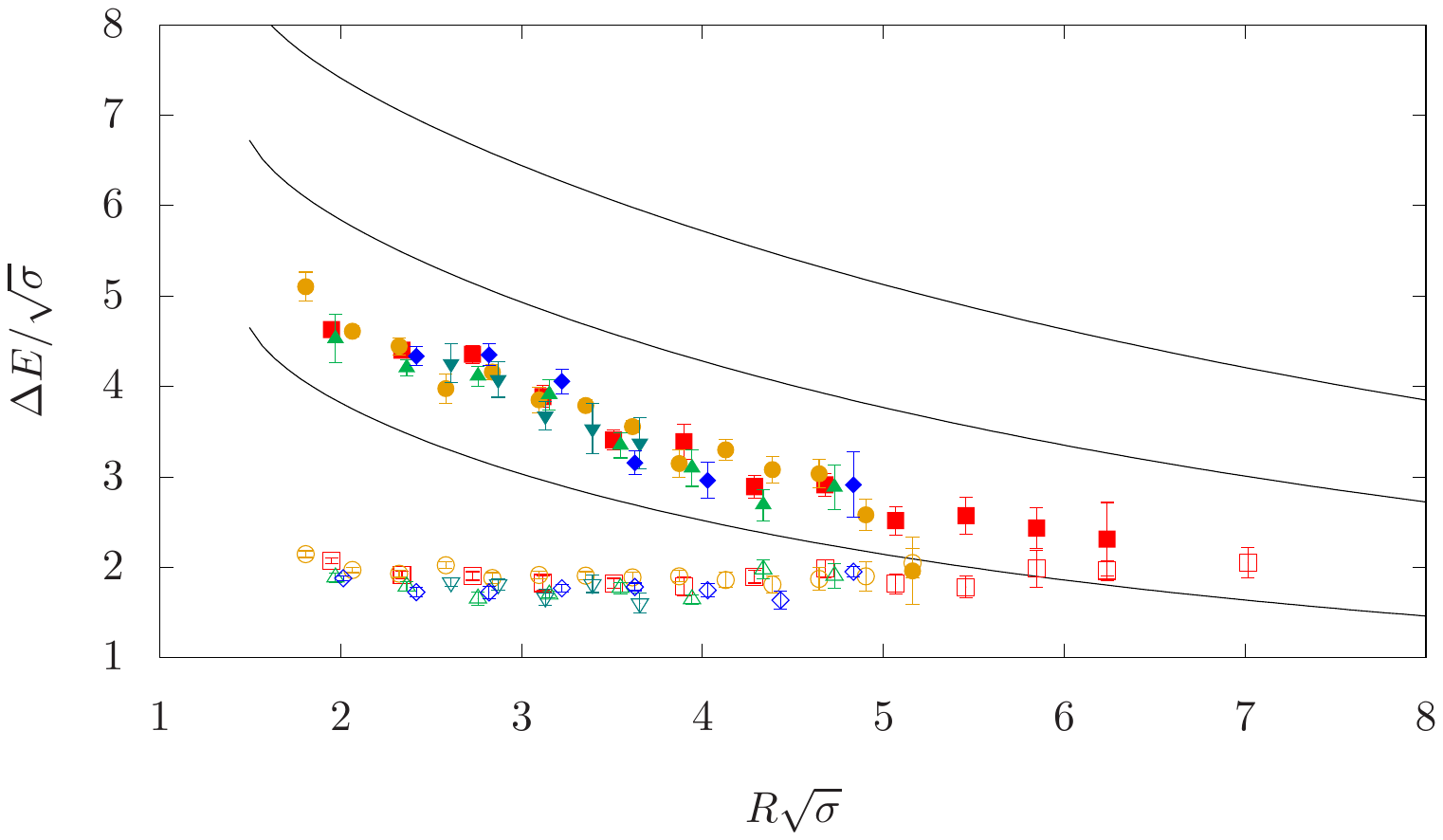}\put(-20,114){\scriptsize $N_L=N_R=1$}\put(-20,143){\scriptsize $N_L=N_R=2$}\put(-20,170){\scriptsize $N_L=N_R=3$}}

 \vspace{-1.0cm}
\caption{\label{fig:plot_J0_Pp-_Pr-_Q0} Results for the confining string with quantum numbers $0^{--}$. On the upper plot we visualize the energy $E/\sqrt{\sigma}$ while on the lower plot the energy minus the absolute ground GGRT level $\Delta E/\sqrt{\sigma}$. The representation of the different gauge groups goes as follows: $SU(3)$, $\beta=6.0625$ is represented by $\square$ ($\blacksquare$), $SU(3)$, $\beta=6.338$ by $\circ$ ($\bullet$), $SU(5)$, $\beta=17.630$ by $\triangle$ ($\blacktriangle$), $SU(5)$, $\beta=18.375$ by $\triangledown$ ($\blacktriangledown$) and $SU(6)$, $\beta=25.55$ by  $\diamond$ ($\blacklozenge$) for ground (first excited) state.}
  \end{center}
\end{figure}

In Fig.~\ref{fig:plot_J0_Pp+_Pr+_Q0} we provide results on the $0^{++}$ first excited state, while in Figs.~\ref{fig:plot_J0_Pp-_Pr-_Q0},~\ref{fig:plot_J2_Pp+_Pr+_Q0},~\ref{fig:plot_J2_Pp-_Pr+_Q0} on the $2^{++}$, $2^{-+}$ and $0^{--}$ ground states for all five $SU(N)$ ensembles. We compare the above data with the GGRT prediction for $N_R=N_L=1$, which dictates them to be four-fold degenerate. We observe that $0^{++}$, $2^{++}$ and $2^{-+}$ flux tube excitations exhibit mild deviations for short values of $R\sqrt{\sigma}$ and for longer strings become consistent with GGRT. On the other hand the $0^{--}$ ground state exhibits much larger deviations from the GGRT string
In Fig.~\ref{fig:plot_J0_Pp-_Pr-_Q0} we plot the ground state and also the first excited state with quantum numbers $0^{--}$ for all gauge groups considered in this work. It appears that both states are only mildly affected by lattice effects. In addition, our data indicates that this state does not exhibit 
strong   $N$ dependence.

The anomalous $0^{--}$ ground state have  clear characteristics of a massive resonance. Namely, the corresponding mass gap is nearly independent of the compactification 
size $R$ for a broad range of $R$.
This can be made more manifest by  subtracting the absolute flux tube ground state energy $E_0(R)$ or, almost equivalently,  the GGRT ground state energy $E_{\rm GGRT}(0,0)$
\begin{eqnarray}
    \Delta E = E^{0^{--}}_{\rm gs} - E_{\rm GGRT}(0,0) \,,
\end{eqnarray}
which makes it easy to see the  plateau. We present $\Delta E$ in the bottom panel of Fig.~\ref{fig:plot_J0_Pp-_Pr-_Q0} for all data sets.
We see a clear plateau which indicates the presence of a massive  axion excitation on the flux tube worldsheet \cite{Dubovsky:2013gi,Dubovsky:2014fma}.
We will further discuss the properties of this state in~\ref{sec:subsec_spin0_computation} and will present the extraction of  the axion mass and coupling
with highest precision thus far using the data presented here, focusing on the case of $SU(3)$ at $\beta=6.338$.

\subsubsection{The second excited $0^{++}$ state}
\label{sec:second_exscited_state_0++}

 In the preceding two sections, we observed that the ground and first excited  $0^{++}$
can be interpreted as the $N_L=N_R=0$ and $N_L=N_R=1$ GGRT states, respectively. Therefore, we now turn to the second excited $0^{++}$ state and investigate whether it aligns with the next GGRT excitation level with $N_L=N_R=2$.

\begin{figure}[htb]
  \begin{center} 
  \vspace{-2.0cm}
 \rotatebox{0}{\hspace{-2.0cm}\includegraphics[width=15cm]{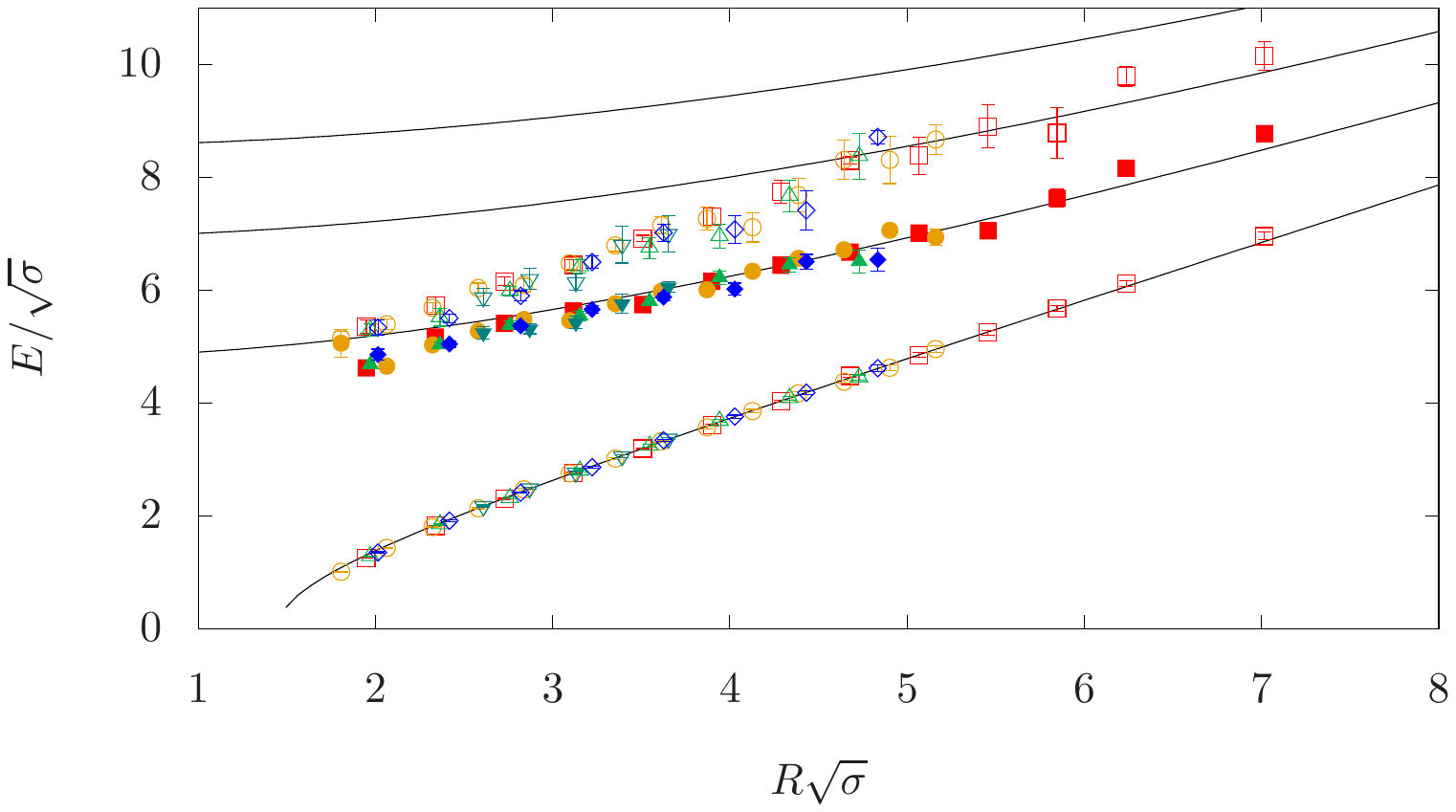} \put(-20,224){\scriptsize $N_L=N_R=0$}\put(-20,247){\scriptsize $N_L=N_R=1$}\put(-20,270){\scriptsize $N_L=N_R=2$}}
 
 \vspace{-3.0cm}
 \rotatebox{0}{\hspace{-2.0cm}\includegraphics[width=15cm]{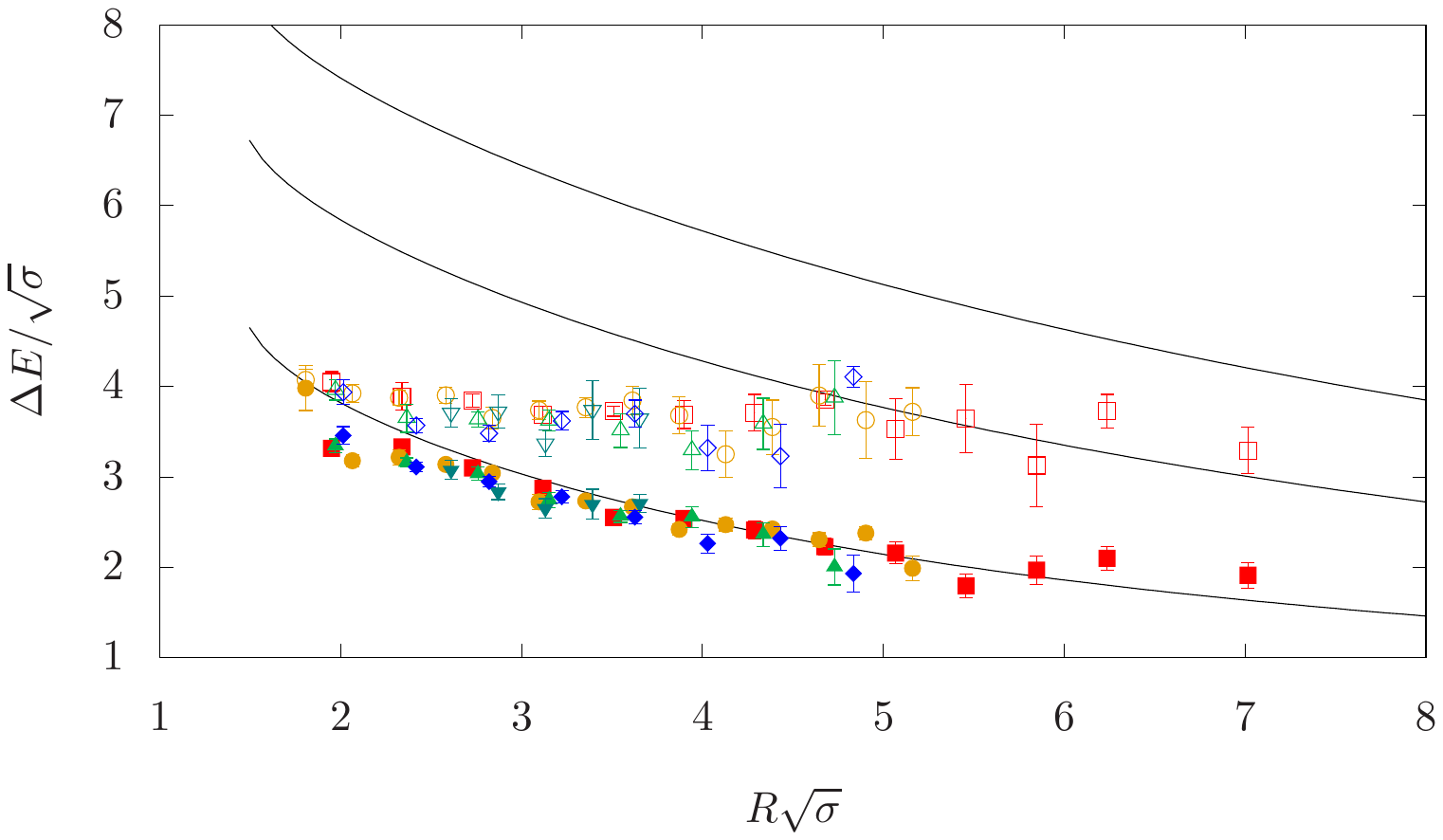}\put(-20,114){\scriptsize $N_L=N_R=1$}\put(-20,143){\scriptsize $N_L=N_R=2$}\put(-20,170){\scriptsize $N_L=N_R=3$}}

 \vspace{-1.0cm}
\caption{\label{fig:plot_J0_Pp+_Pr+_Q0} Results for the confining string with quantum numbers $0^{++}$. On the upper plot we visualize the energy $E/\sqrt{\sigma}$ while on the lower plot the energy with the ground GGRT level being subtracted i.e. $\Delta E/\sqrt{\sigma}$. The representation of the different gauge groups goes as follows: $SU(3)$, $\beta=6.0625$ is represented by $\square$ ($\blacksquare$), $SU(3)$, $\beta=6.338$ by $\circ$ ($\bullet$), $SU(5)$, $\beta=17.630$ by $\triangle$ ($\blacktriangle$), $SU(5)$, $\beta=18.375$ by $\triangledown$ ($\blacktriangledown$) and $SU(6)$, $\beta=25.55$ by  $\diamond$ ($\blacklozenge$) for ground (first excited) state.}
  \end{center}
\end{figure}

In Table~\ref{tab:0++_second_excited} and Fig.~\ref{fig:plot_J0_Pp+_Pr+_Q0}, we present the second excited $0^{++}$ state. It is important to note that beyond this energy level, we encounter a multitude of $0^{++}$ states that reflect the considerable degeneracy of the GGRT string when $N_L = N_R = 2$. Interestingly, this state exhibits the same resonance behavior as the $0^{--}$ ground state, with the corresponding energy gap being largely independent of $R$ in a broad range of $R$.
This is illustrated   in the lower part of Fig.~\ref{fig:plot_J0_Pp+_Pr+_Q0}. 
Furthermore the magnitude of this gap is close to twice the value of the axion mass. This makes it natural to interpret this state as two axions at rest, so its existence provides a nice self-consistency check for the axion explanation of the anomalous $0^{--}$ state.
A more comprehensive TBA analysis of this state  will be presented in Section~\ref{sec:subsec_spin0_computation}, with the focus on the case of $SU(3)$ at $\beta=6.338$.

\subsubsection{ $0^{+-}$ ground state}
\label{sec:results_0+-_channel}
\begin{figure}[htb]
  \begin{center} 
  \vspace{-2.0cm}
 \rotatebox{0}{\hspace{-2.0cm}\includegraphics[width=15cm]{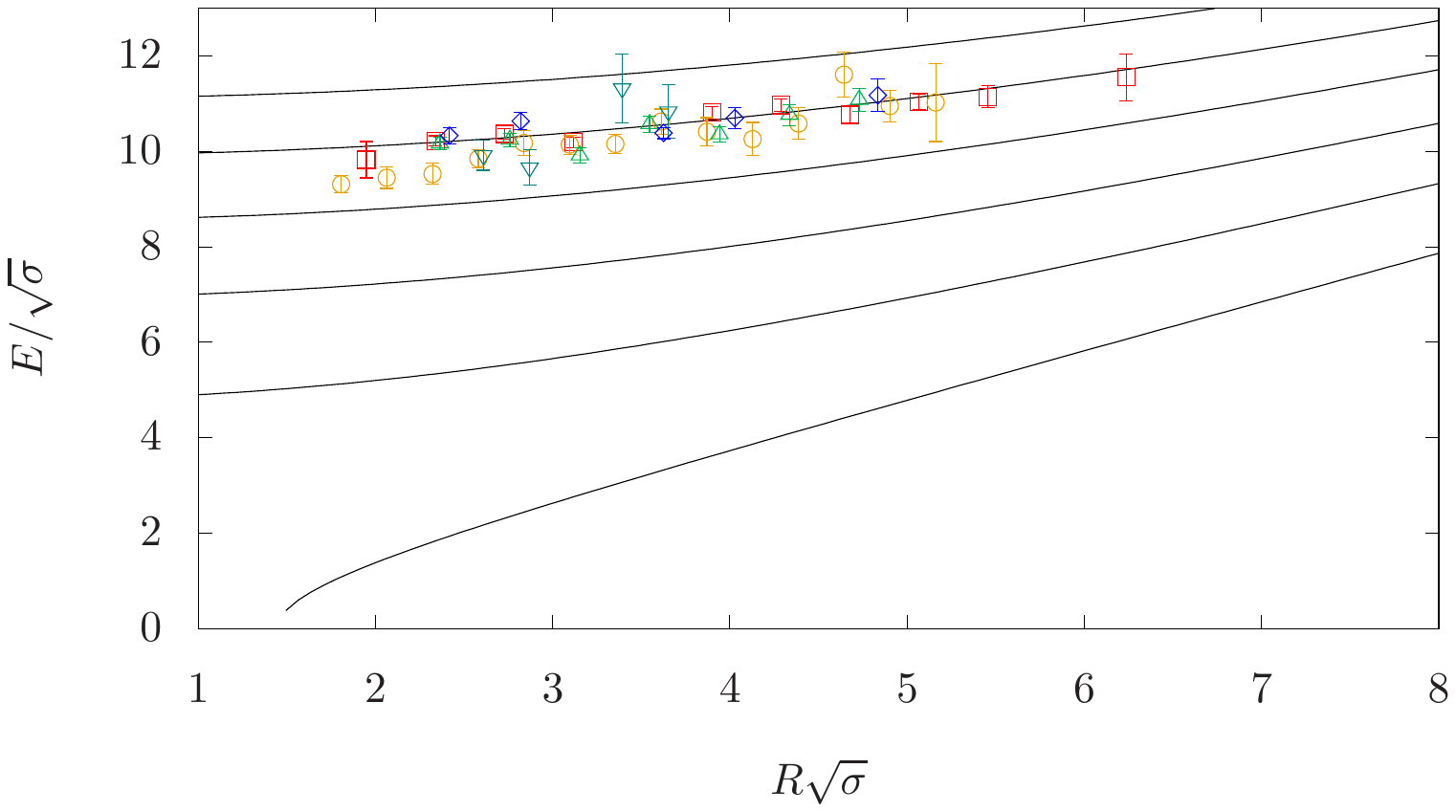}\put(-20,207){\scriptsize $N_L=N_R=0$}\put(-20,225){\scriptsize $N_L=N_R=1$}\put(-20,240){\scriptsize $N_L=N_R=2$}\put(-20,257){\scriptsize $N_L=N_R=3$}\put(-20,270){\scriptsize $N_L=N_R=4$}}
 
 \vspace{-3.0cm}
 \rotatebox{0}{\hspace{-2.0cm}\includegraphics[width=15cm]{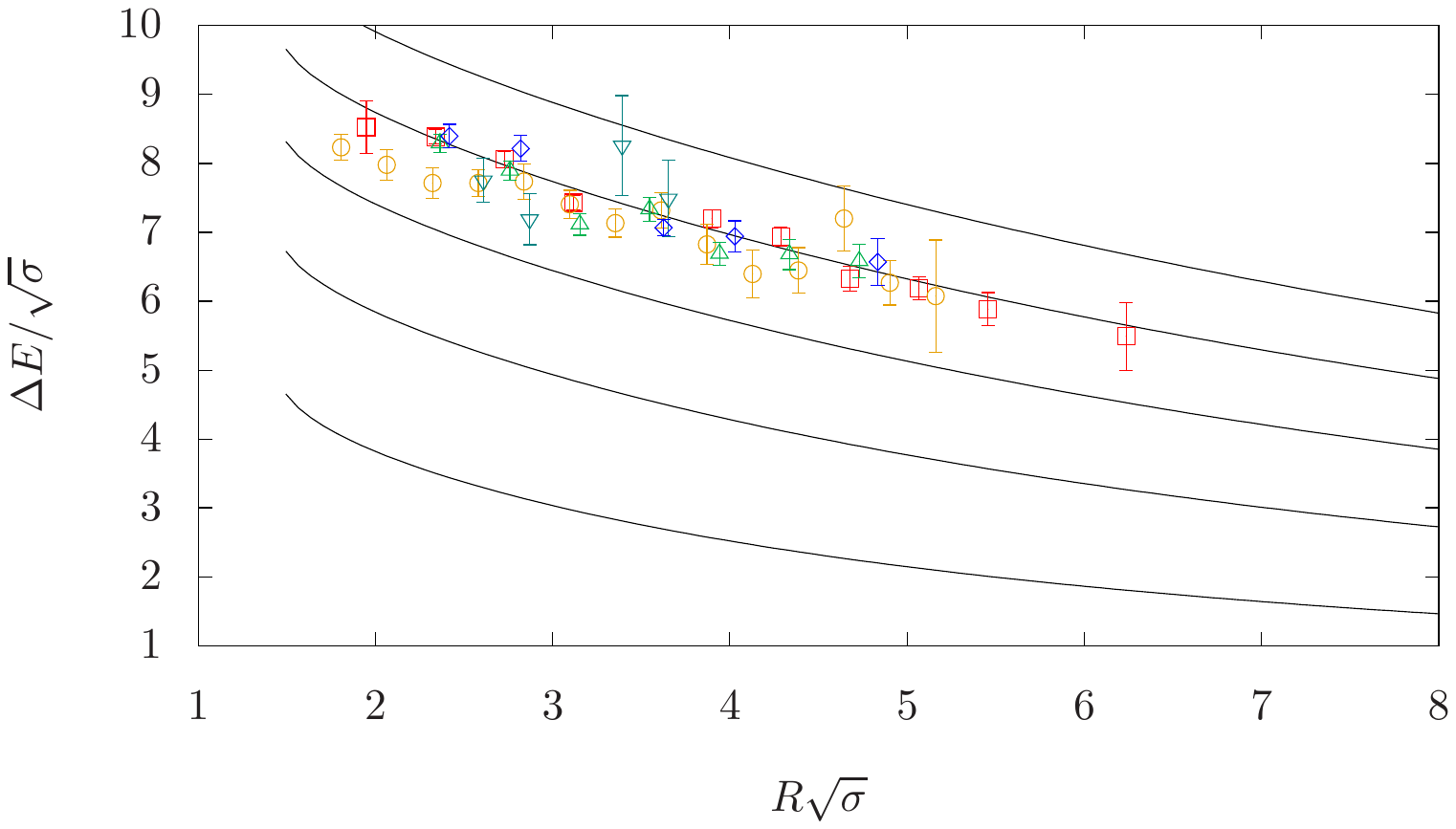}\put(-20,114){\scriptsize $N_L=N_R=1$}\put(-20,136){\scriptsize $N_L=N_R=2$}\put(-20,157){\scriptsize $N_L=N_R=3$}\put(-20,176){\scriptsize $N_L=N_R=4$}\put(-20,195){\scriptsize $N_L=N_R=5$}}

 \vspace{-1.0cm}
\caption{\label{fig:plot_J0_Pp+_Pr-_Q0} Results for the confining string with quantum numbers $0^{+-}$. On the upper plot we visualize the energy $E/\sqrt{\sigma}$ while on the lower plot the energy minus the absolute ground GGRT level $\Delta E/\sqrt{\sigma}$. The representation of the different gauge groups goes as follows: $SU(3)$, $\beta=6.0625$ is represented by $\square$, $SU(3)$, $\beta=6.338$ by $\circ$, $SU(5)$, $\beta=17.630$ by $\triangle$, $SU(5)$, $\beta=18.375$ by $\triangledown$ and $SU(6)$, $\beta=25.55$ by  $\diamond$ for ground state.}
  \end{center}
\end{figure}

We now turn our attention to the  ground  $0^{+-}$ state of the confining flux tube. One does not find such a state in Table~\ref{tab:table_NGstates_q0} because it first appears 
at the $N_L=N_R=3$ level in the GGRT spectrum and can be written as 
\[
[(a_1^+ a_1^+ a^-_1 a_{-3}^- -a_{-1}^+ a_{-1}^+ a^-_{-1}a_{3}^-) + (a_1^- a_1^- a^+_1 a_{-3}^+ -a_{-1}^- a_{-1}^- a^+_{-1}a_{3}^+)]|0\rangle
\]
in terms of the oscillators. 
  This state is therefore expected to be quite heavy and, as a result,  to be subject to large systematic uncertainties.
Our most reliable results emerge from the $SU(3)$ data at $\beta=6.338$, which exhibit better plateaus. Nonetheless, it is noteworthy that we have acquired results from all five different ensembles. In Table~\ref{tab:0+-_ground_state} and Fig.~\ref{fig:plot_J0_Pp+_Pr-_Q0}, we present our findings concerning the $0^{+-}$ ground state. The results generally overshoot the $N_L=N_R=3$ GGRT level, which signals a large systematic over-estimation of such heavy states.

\subsubsection{$0^{-+}$ ground state}
\label{sec:results_0-+_channel}
Let us discuss now the $0^{-+}$ ground state.
As follows from Table~\ref{tab:table_NGstates_q0}, the lowest GGRT state in this sector appears at the $N_L=N_R=2$ level,
\[
\left( a^{+}_{1}a^{+}_{1} a^{+}_{-1} a^{+}_{-1} -a^{-}_{1} a^{-}_{1} a^{-}_{-1} a^{-}_{-1} \right) | 0 \rangle\;.
\]
Note that in the continuum theory  this state has   $4^{-+}$ quantum numbers. However, as discussed before, on a lattice  $4^{-+}$ and $0^{-+}$ states fall in the same representation, so they should both appear in the same sector.

\begin{figure}[htb]
  \begin{center} 
  \vspace{-2.0cm}
 \rotatebox{0}{\hspace{-2.0cm}\includegraphics[width=15cm]{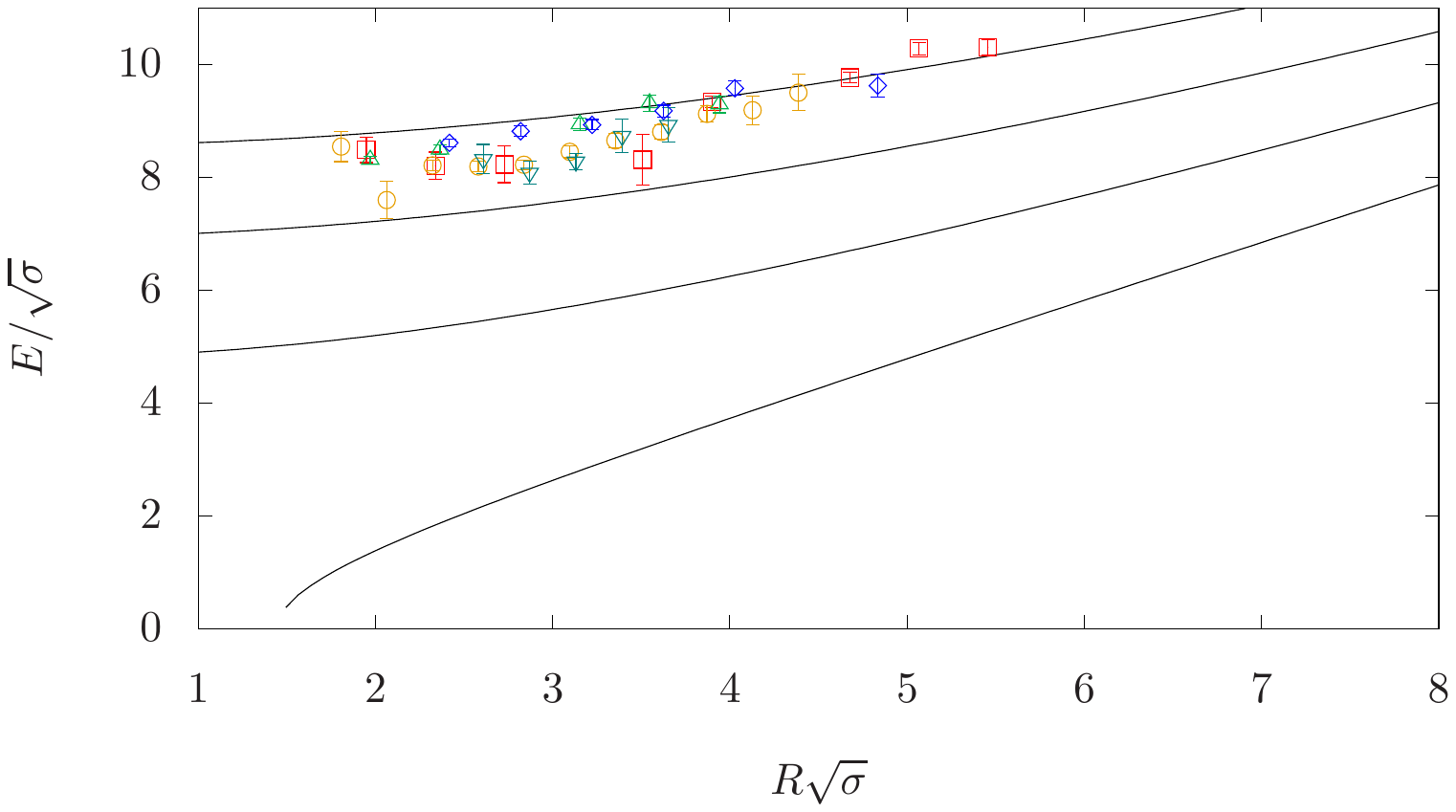}\put(-20,224){\scriptsize $N_L=N_R=0$}\put(-20,247){\scriptsize $N_L=N_R=1$}\put(-20,270){\scriptsize $N_L=N_R=2$}}
 
 \vspace{-3.0cm}
 \rotatebox{0}{\hspace{-2.0cm}\includegraphics[width=15cm]{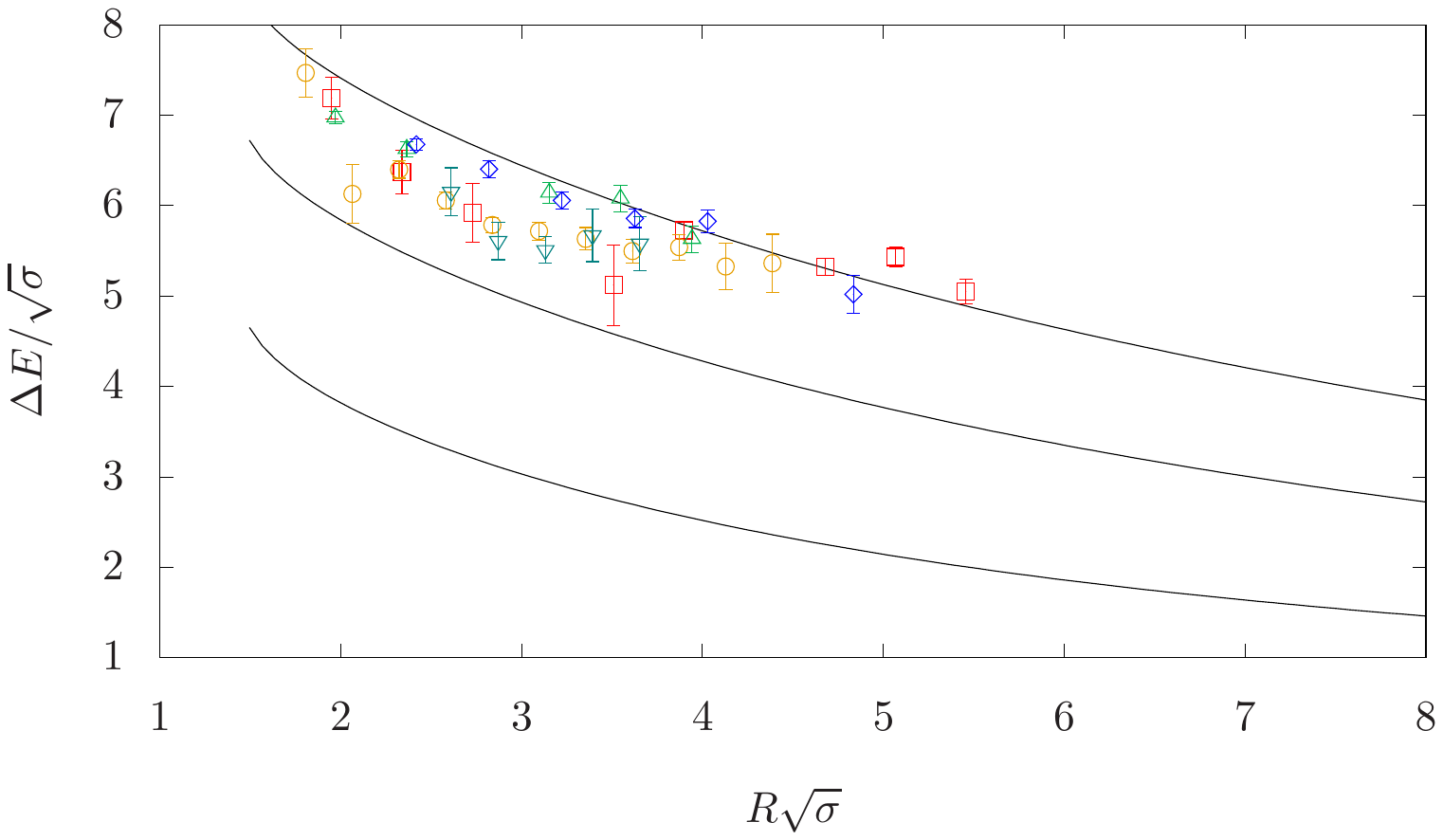}\put(-20,114){\scriptsize $N_L=N_R=1$}\put(-20,143){\scriptsize $N_L=N_R=2$}\put(-20,170){\scriptsize $N_L=N_R=3$}}

 \vspace{-1.0cm}
\caption{\label{fig:plot_J0_Pp-_Pr+_Q0} Results for the ground state of the confining string with quantum numbers $0^{-+}$. On the upper plot we visualize the energy $E/\sqrt{\sigma}$ while on the lower plot the energy minus the absolute ground GGRT level $\Delta E/\sqrt{\sigma}$. The representation of the different gauge groups goes as follows: $SU(3)$, $\beta=6.0625$ is represented by $\square$, $SU(3)$, $\beta=6.338$ by $\circ$, $SU(5)$, $\beta=17.630$ by $\triangle$, $SU(5)$, $\beta=18.375$ by $\triangledown$ and $SU(6)$, $\beta=25.55$ by  $\diamond$.} 
  \end{center}
\end{figure}

In Table~\ref{tab:0-+_ground_state} and Fig.~\ref{fig:plot_J0_Pp-_Pr+_Q0}, we present the $0^{-+}$ ground state for all five data sets. The relevant energy levels appear to approach the $N_L=N_R=3$ GGRT level for long strings rather than the expected $N_L=N_R=2$. The  $SU(5)$ data 
exhibit noticeable discretization effects, which can also be attributed to the systematic over-estimation of the spectrum at $\beta=17.630$, where the ground state is extremely heavy in the lattice unit. The same can also affect the  $SU(6)$ data.
Nevertheless, no  $N_L=N_R=2$ state shows up in the spectrum.  Most likely this indicates that our basis has a poor overlap with the $4^{-+}$ state.
We will further investigate the observed $0^{-+}$ state in Section~\ref{sec:subsec_spin0_computation} and will argue that it is more likely to be
described as a  massive axionic excitation with two additional phonons rather than a GGRT state.

\subsubsection{$1^{+}$ ground state}
\label{sec:results_1+_channel}
Let us now present results for the ground characterized with odd spin $|J_{\rm mod4}|=1$ and positive longitudinal parity  $P_{||}=+$.
As discussed in  Section~\ref{sec:subsection_quanta} such states appear on a lattice as doublet representations of the lattice group, which combine two states of different transverse parities $P_{\perp}={\pm}$, so in what follows we often omit the superscript of $P_{\perp}$ for odd-spin sectors. 
The lightest GGRT states with these quantum numbers arise at  $N_L=N_R=2$. Namely, in Table~\ref{tab:table_NGstates_q0} one finds  two $J=1$ doublets
\[ \big[ ( a^{+}_{1}a^{+}_{1} a^{-}_{-2}+ a^{-}_{2}a^{+}_{-1} a^{+}_{-1}) \pm ( a^{-}_{1}a^{-}_{1} a^{+}_{-2}+ a^{-}_{2}a^{+}_{-1} a^{+}_{-1}) \big] | 0 \rangle\]
 and \[\big[ ( a^{+}_{1}a^{-}_{1} a^{+}_{-2}+ a^{+}_{2}a^{-}_{-1} a^{+}_{-1}) \pm ( a^{-}_{1}a^{+}_{1} a^{-}_{-2}+ a^{-}_{2}a^{+}_{-1} a^{-}_{-1}) \big] | 0 \rangle\] with $J=1$, as well as a $J=3$ doublet \[\big[ ( a^{+}_{1}a^{+}_{1} a^{+}_{-2}+ a^{+}_{2}a^{+}_{-1} a^{+}_{-1}) \pm ( a^{-}_{1}a^{-}_{1} a^{-}_{-2}+ a^{-}_{2}a^{-}_{-1} a^{-}_{-1}) \big] | 0 \rangle\;.\]

\begin{figure}[htb]
  \begin{center} 
  \vspace{-2.0cm}
 \rotatebox{0}{\hspace{-2.0cm}\includegraphics[width=15cm]{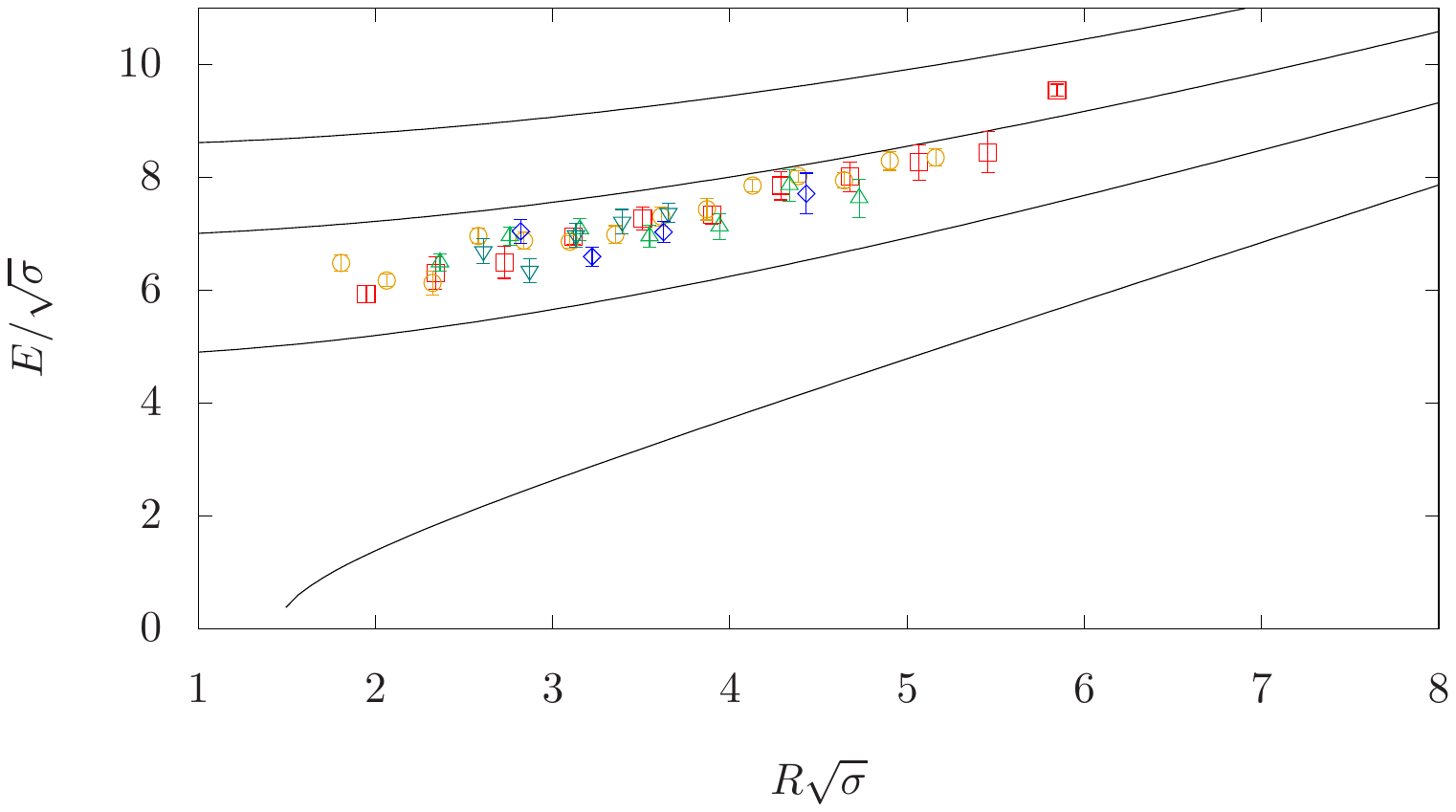}\put(-20,224){\scriptsize $N_L=N_R=0$}\put(-20,247){\scriptsize $N_L=N_R=1$}\put(-20,270){\scriptsize $N_L=N_R=2$}}
 
 \vspace{-3.0cm}
 \rotatebox{0}{\hspace{-2.0cm}\includegraphics[width=15cm]{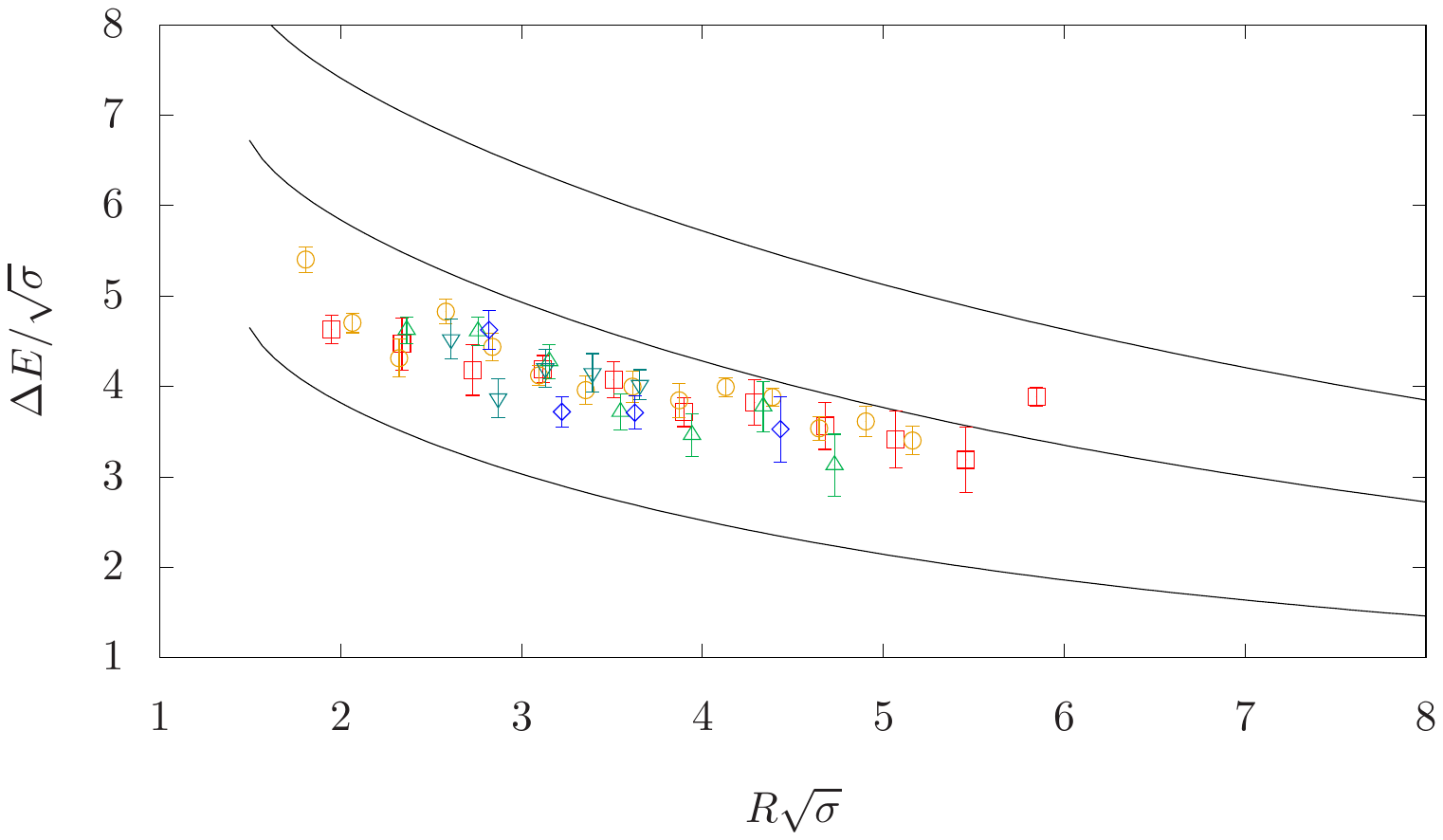}\put(-20,114){\scriptsize $N_L=N_R=1$}\put(-20,143){\scriptsize $N_L=N_R=2$}\put(-20,170){\scriptsize $N_L=N_R=3$}}

 \vspace{-1.0cm}
\caption{\label{fig:plot_J1_Pr+_Q0} Results for the ground state of the confining string with quantum numbers $1^+$, $q=0$. On the upper plot we visualize the energy $E/\sqrt{\sigma}$ while on the lower plot the energy minus the absolute ground GGRT level $\Delta E/\sqrt{\sigma}$. The representation of the different gauge groups goes as follows: $SU(3)$, $\beta=6.0625$ is represented by $\square$, $SU(3)$, $\beta=6.338$ by $\circ$, $SU(5)$, $\beta=17.630$ by $\triangle$, $SU(5)$, $\beta=18.375$ by $\triangledown$ and $SU(6)$, $\beta=25.55$ by  $\diamond$.} 
  \end{center}
\end{figure}

In our simulation we managed to extract the corresponding ground state and observe also a couple of low lying excitations albeit with not so convincing plateaus.
Our results for the ground state are presented in Fig.~\ref{fig:plot_J1_Pr+_Q0} as well as in Table~\ref{tab:1+_ground_state}. The ground state appears to approach the $N_L=N_R=2$ GGRT energy level for the longest strings with $R\sigma^{1/2}\gtrsim 5 $, and exhibits significant deviations for shorter flux tubes. Using the TBA analysis,
we will argue  in Section~\ref{subsec:q0jnonzero_computation} that this state corresponds to a string state with a massive axion excitation and an additional  phonon excitation. 

\subsubsection{$1^{-}$ ground state}
\label{sec:results_1-_channel}
\begin{figure}[htb]
  \begin{center} 
  \vspace{-2.0cm}
 \rotatebox{0}{\hspace{-2.0cm}\includegraphics[width=15cm]{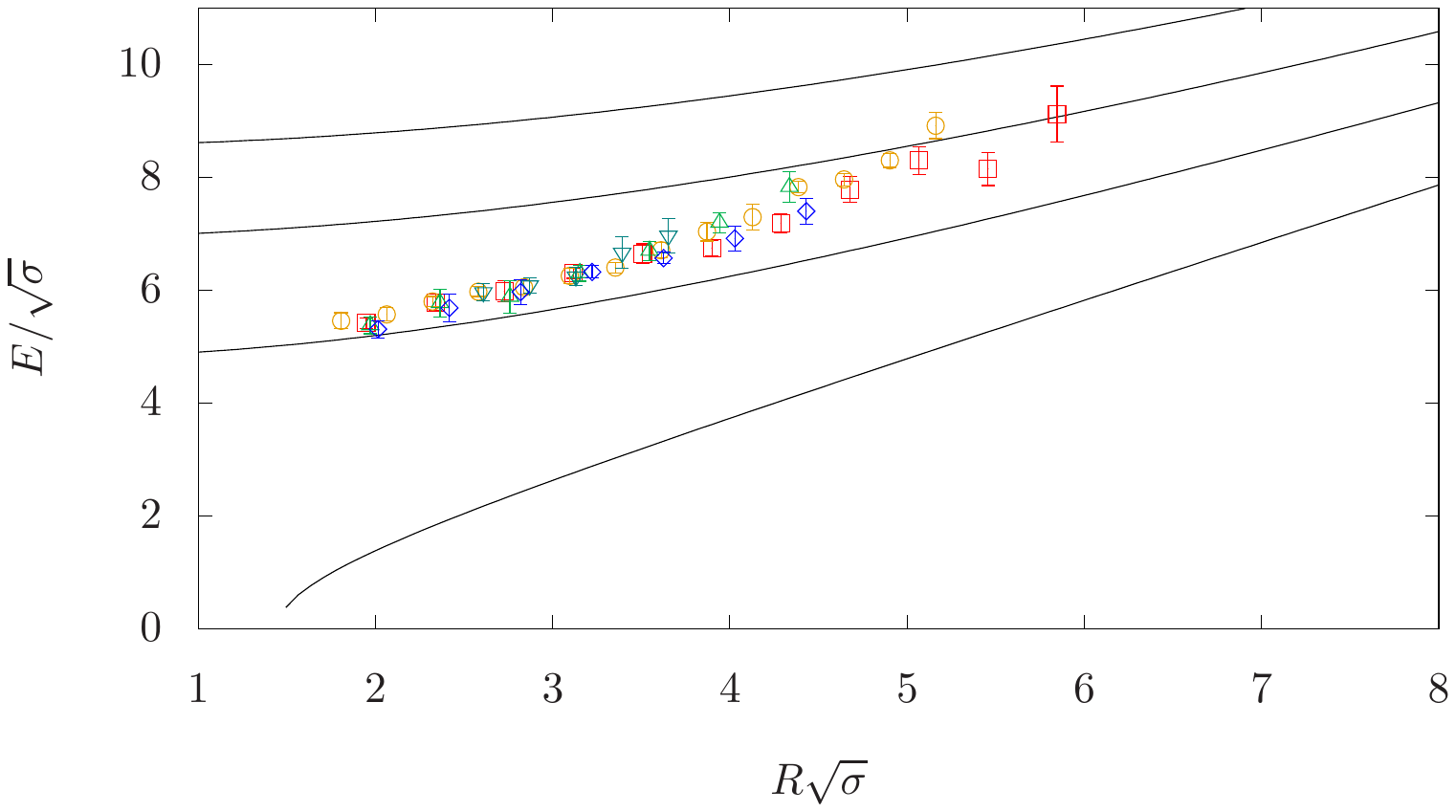}\put(-20,224){\scriptsize $N_L=N_R=0$}\put(-20,247){\scriptsize $N_L=N_R=1$}\put(-20,270){\scriptsize $N_L=N_R=2$}}
 
 \vspace{-3.0cm}
 \rotatebox{0}{\hspace{-2.0cm}\includegraphics[width=15cm]{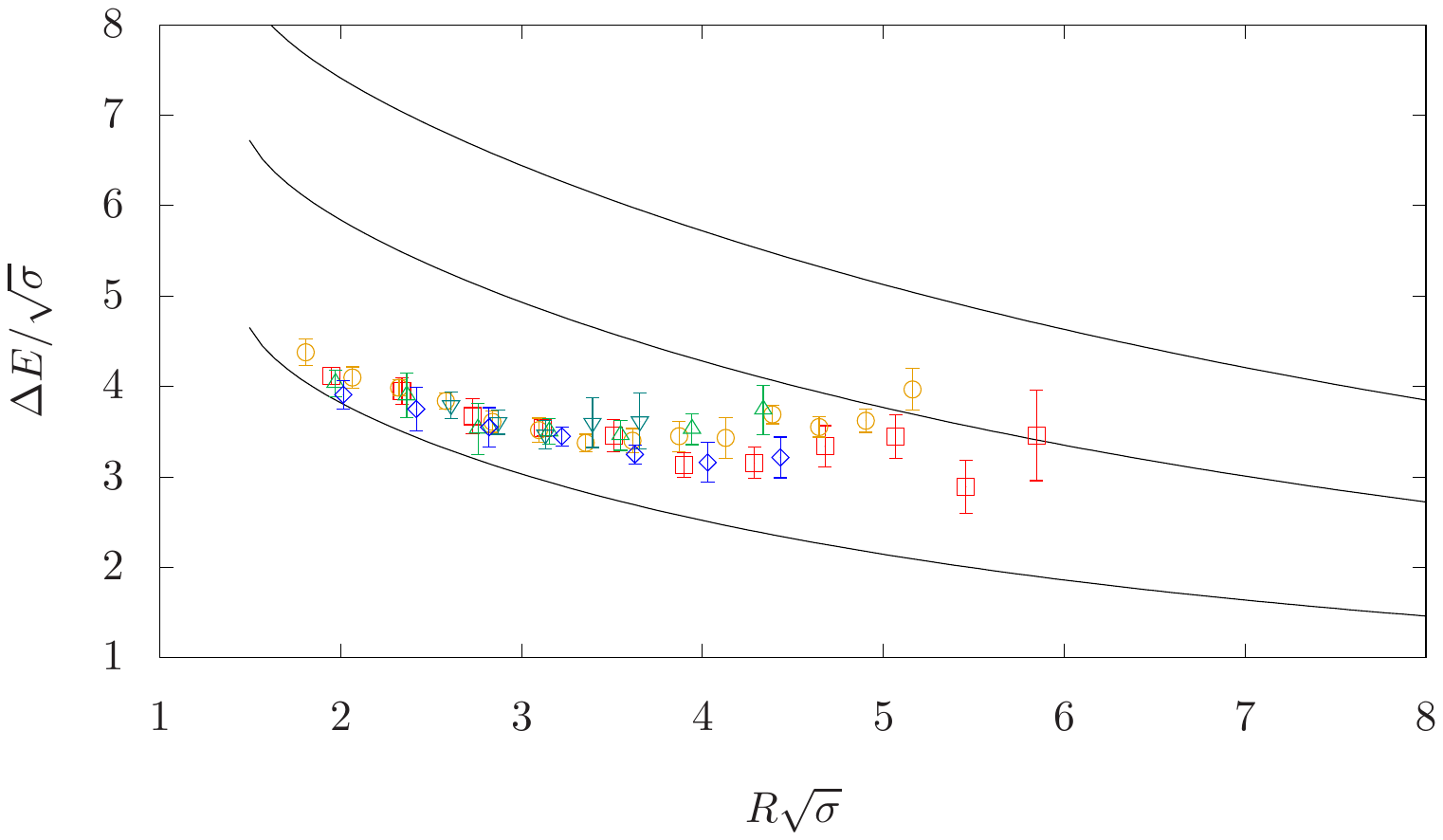}\put(-20,114){\scriptsize $N_L=N_R=1$}\put(-20,143){\scriptsize $N_L=N_R=2$}\put(-20,170){\scriptsize $N_L=N_R=3$}}

 \vspace{-1.0cm}
\caption{\label{fig:plot_J1_Pr-_Q0} Results for the ground state of the confining string with quantum numbers $1^-$, $q=0$. On the upper plot we visualize the energy $E/\sqrt{\sigma}$ while on the lower plot the energy minus the absolute ground GGRT level $\Delta E/\sqrt{\sigma}$. The representation of the different gauge groups goes as follows: $SU(3)$, $\beta=6.0625$ is represented by $\square$, $SU(3)$, $\beta=6.338$ by $\circ$, $SU(5)$, $\beta=17.630$ by $\triangle$, $SU(5)$, $\beta=18.375$ by $\triangledown$ and $SU(6)$, $\beta=25.55$ by  $\diamond$.}
  \end{center}
\end{figure}

The situation for odd spin $|J_{\rm mod4}|=1$ states with negative  longitudinal parity $P_{||}=-$ is very similar. In Table~\ref{tab:table_NGstates_q0} one again finds two $J=1$ doublets and one $J=3$ at the $N_L=N_R=2$ level. Our results 
 for the ground state in this sector are listed in Table~\ref{tab:1-_ground_state} and plotted in Fig.~\ref{fig:plot_J1_Pr-_Q0}. Again the ground state gradually approaches the  $N_L=N_R=2$ GGRT energy, while displaying significant deviations for short flux tubes. The TBA analysis in Section~\ref{subsec:q0jnonzero_computation} again strongly suggests that this ground state
 describes a two-particle state containing a massive axion and a phonon.
 
 \subsubsection{$2^{++}$ and  $2^{-+}$ ground and first excited states}
\label{sec:results_2pm+}
Let us discuss now first excitations in the $2^{++}$ and  $2^{-+}$ sectors. It is natural to discuss these states together because they should be exactly degenerate in the continuum theory. They correspond to the different representation of the lattice group though. There is no obvious low lying candidate axionic state like this, so one expects these states to be approximated by the GGRT spectrum reasonably well. By inspecting Table~\ref{tab:table_NGstates_q0} one finds two such states at the $N_L=N_R=2$
level. The corresponding oscillator content is given by \[\left( a^{+}_{2} a^{+}_{-2}\pm a^{-}_{2} a^{-}_{-2} \right) | 0 \rangle\] and \[\left[   \left( a^{+}_{1}a^{+}_{1} a^{+}_{-1} a^{-}_{-1} \pm a^{-}_{1}a^{-}_{1} a^{-}_{-1} a^{+}_{-1} \right)+\left( a^{+}_{1}a^{-}_{1} a^{-}_{-1} a^{-}_{-1} \pm a^{-}_{1}a^{+}_{1} a^{+}_{-1} a^{+}_{-1} \right) \right] | 0 \rangle\;.\]
 
\begin{figure}[htb]
  \begin{center} 
  \vspace{-2.0cm}
 \rotatebox{0}{\hspace{-2.0cm}\includegraphics[width=15cm]{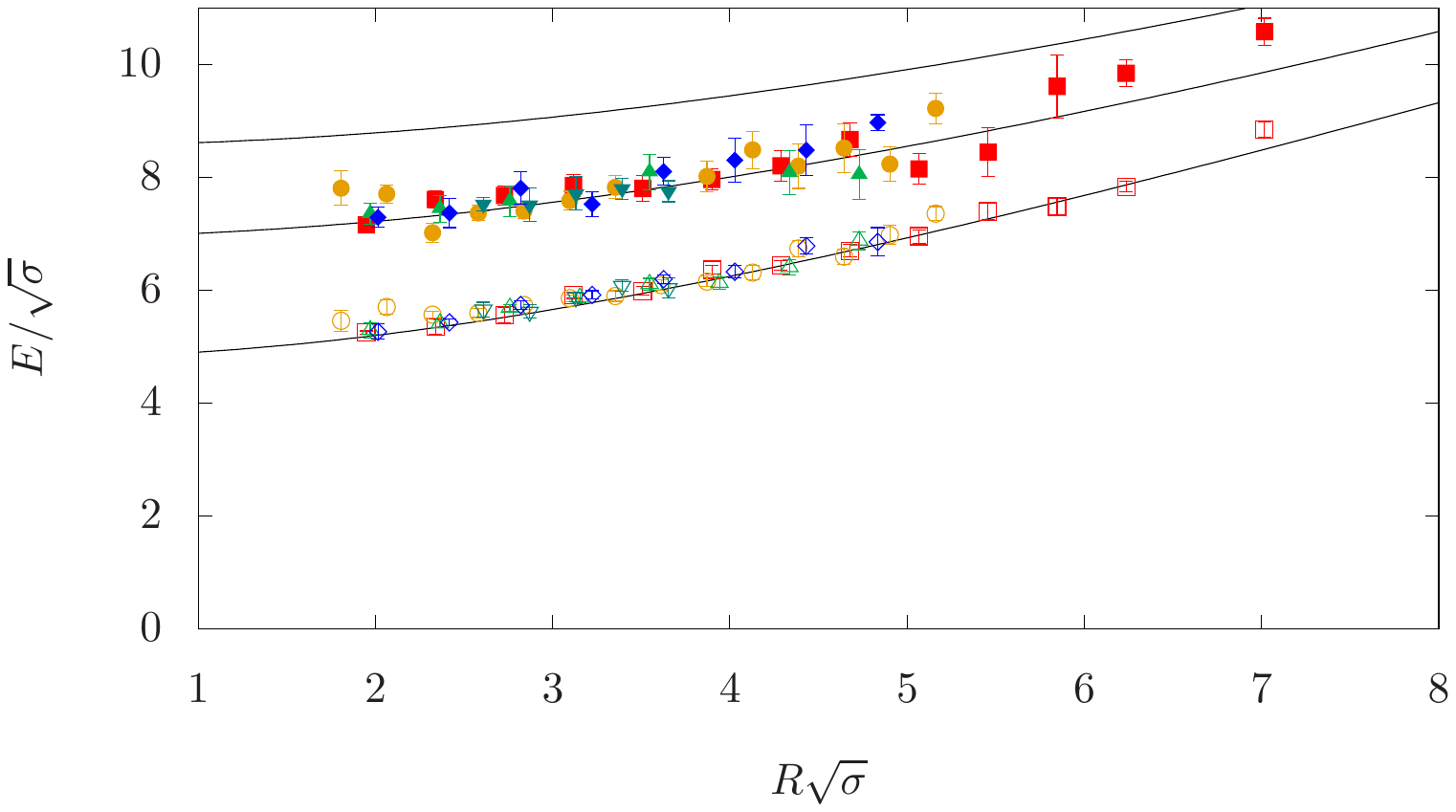}\put(-20,247){\scriptsize $N_L=N_R=1$}\put(-20,270){\scriptsize $N_L=N_R=2$}}
 
 \vspace{-3.0cm}
 \rotatebox{0}{\hspace{-2.0cm}\includegraphics[width=15cm]{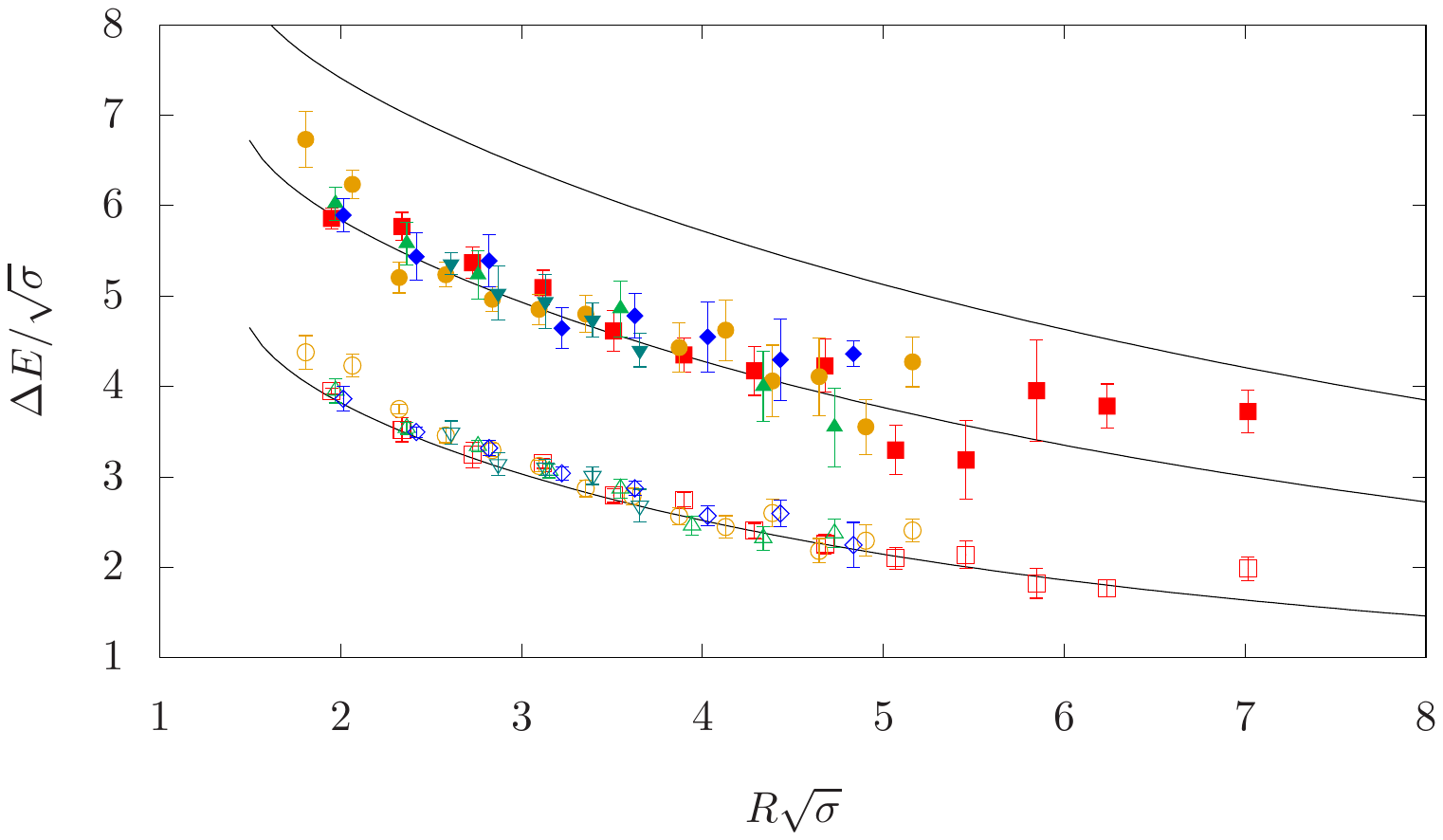}\put(-20,114){\scriptsize $N_L=N_R=1$}\put(-20,143){\scriptsize $N_L=N_R=2$}\put(-20,170){\scriptsize $N_L=N_R=3$}}

 \vspace{-1.0cm}
\caption{\label{fig:plot_J2_Pp+_Pr+_Q0} Results for the confining string with quantum numbers $2^{++}$. On the upper plot we visualize the energy $E/\sqrt{\sigma}$ while on the lower plot the energy with the ground GGRT level being subtracted $\Delta E/\sqrt{\sigma}$. The representation of the different gauge groups goes as follows: $SU(3)$, $\beta=6.0625$ is represented by $\square$ ($\blacksquare$), $SU(3)$, $\beta=6.338$ by $\circ$ ($\bullet$), $SU(5)$, $\beta=17.630$ by $\triangle$ ($\blacktriangle$), $SU(5)$, $\beta=18.375$ by $\triangledown$ ($\blacktriangledown$) and $SU(6)$, $\beta=25.55$ by  $\diamond$ ($\blacklozenge$) for ground (first excited) state.}
  \end{center}
\end{figure}

\begin{figure}[htb]
  \begin{center} 
  \vspace{-2.0cm}
 \rotatebox{0}{\hspace{-2.0cm}\includegraphics[width=15cm]{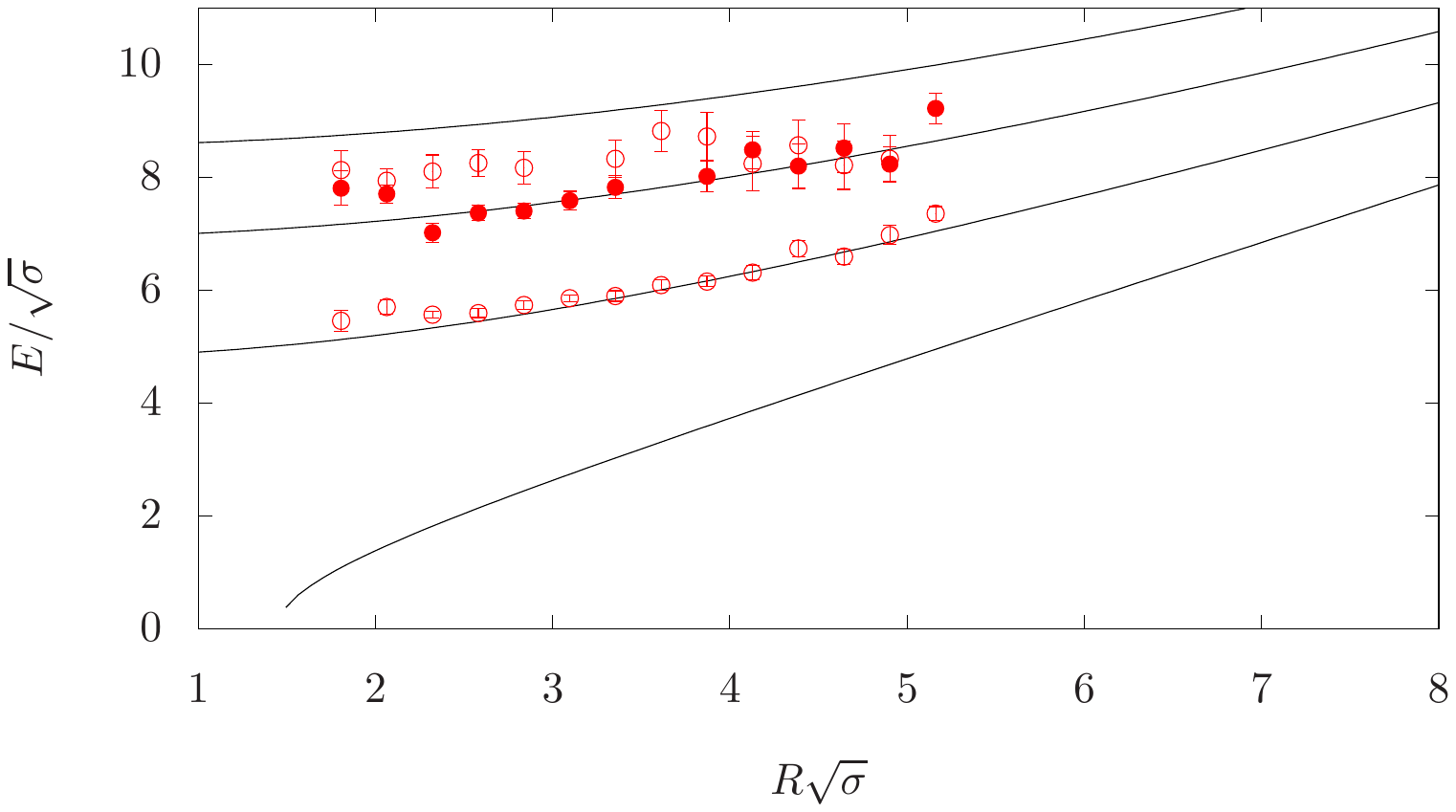}\put(-20,224){\scriptsize $N_L=N_R=0$}\put(-20,247){\scriptsize $N_L=N_R=1$}\put(-20,270){\scriptsize $N_L=N_R=2$}}
 
 \vspace{-3.0cm}
 \rotatebox{0}{\hspace{-2.0cm}\includegraphics[width=15cm]{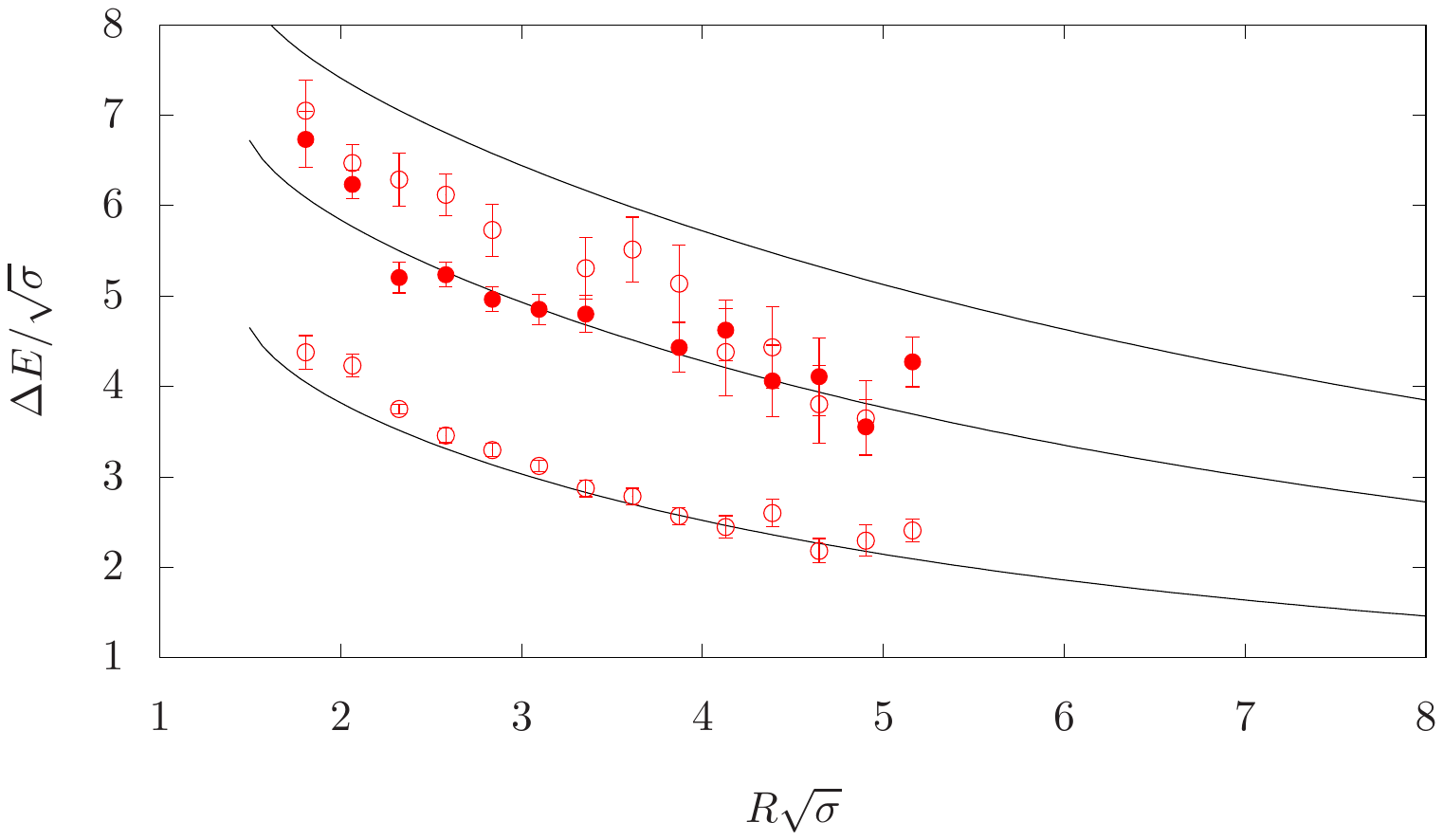}\put(-20,114){\scriptsize $N_L=N_R=1$}\put(-20,143){\scriptsize $N_L=N_R=2$}\put(-20,170){\scriptsize $N_L=N_R=3$}}

 \vspace{-1.0cm}
\caption{\label{fig:plot_J2_Pp+_Pr+_Q0_2nd} Results for the confining string with quantum numbers $2^{++}$ for $SU(3)$, $\beta=6.338$. On the upper plot we visualize the energy $E/\sqrt{\sigma}$ while on the lower plot the Energy with the ground GGRT level being subtracted $\Delta E/\sqrt{\sigma}$. The representation of the data points goes as follows. The ground state is represented by $\circ$, the first excited by $\bullet$ and the second excited state by $\circ$.}
  \end{center}
\end{figure}

 In  Fig.~\ref{fig:plot_J2_Pp+_Pr+_Q0} we present the ground and first excited $2^{++}$ states for all five data sets. The data are also listed in Tables~\ref{tab:2++_ground_state} and \ref{tab:2++_first_excited_state}. Clearly, it is evident that the first excited state is entirely in line with the $N_L=N_R=2$ GGRT level, considering the provided statistical accuracy, while lattice  and large-$N$ effects appear to be negligible.
 Unfortunately, due to statistical limitations we could only extract reliable results for the second excited state just for the $SU(3)$ data set with $\beta=6.338$. 
 The results are presented in Fig.~\ref{fig:plot_J2_Pp+_Pr+_Q0_2nd} demonstrating that although the second excited state appears to exhibit large deviations from the $N_L=N_R=2$ GGRT level for short strings, it finally converges for long strings and $R \sqrt{\sigma} > 4$ suggesting that both first and second excited states are string-like and in accordance with the degeneracy pattern prescribed in Table~\ref{tab:table_NGstates_q0}.
 
\begin{figure}[htb]
  \begin{center} 
  \vspace{-2.0cm}
 \rotatebox{0}{\hspace{-2.0cm}\includegraphics[width=15cm]{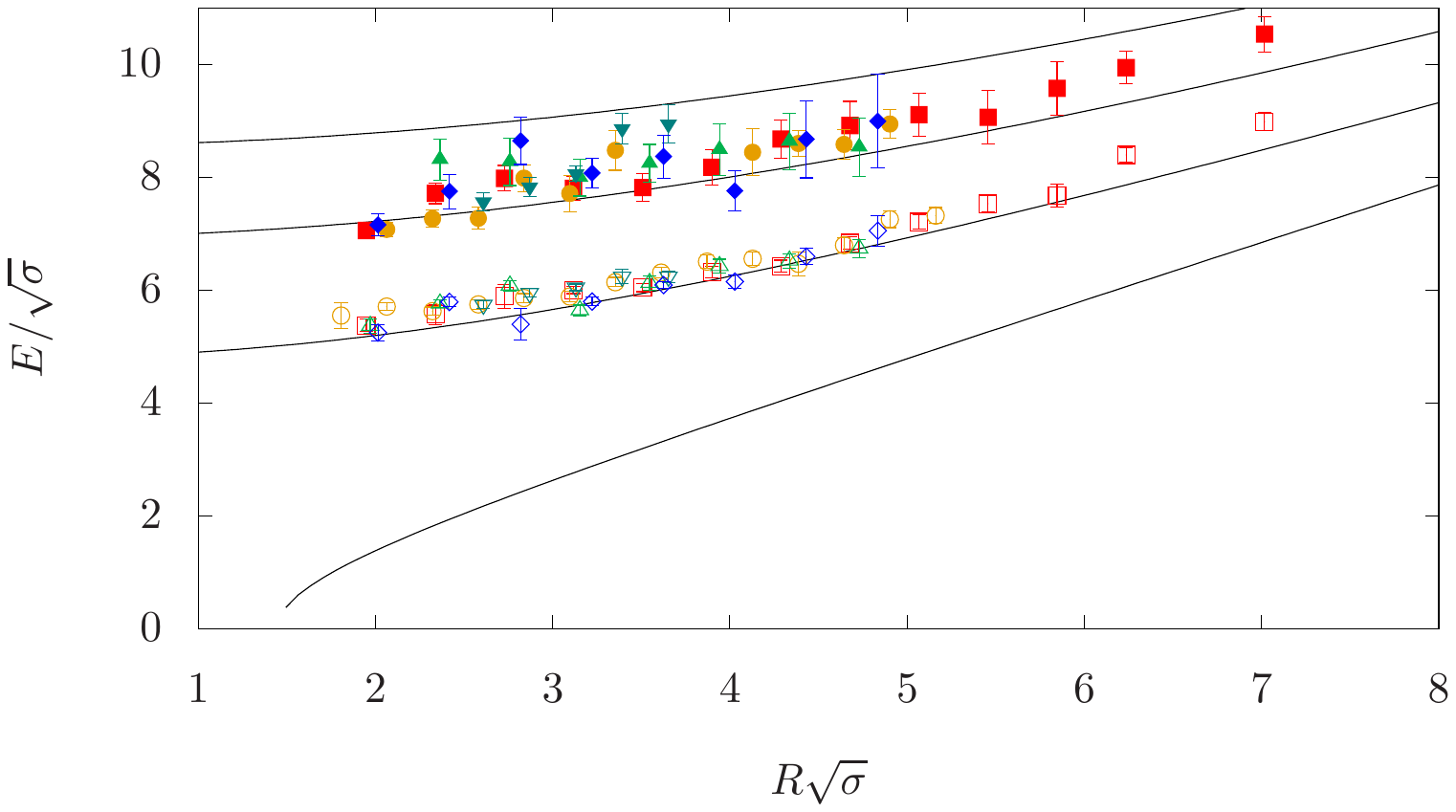}\put(-20,224){\scriptsize $N_L=N_R=0$}\put(-20,247){\scriptsize $N_L=N_R=1$}\put(-20,270){\scriptsize $N_L=N_R=2$}}
 
 \vspace{-3.0cm}
 \rotatebox{0}{\hspace{-2.0cm}\includegraphics[width=15cm]{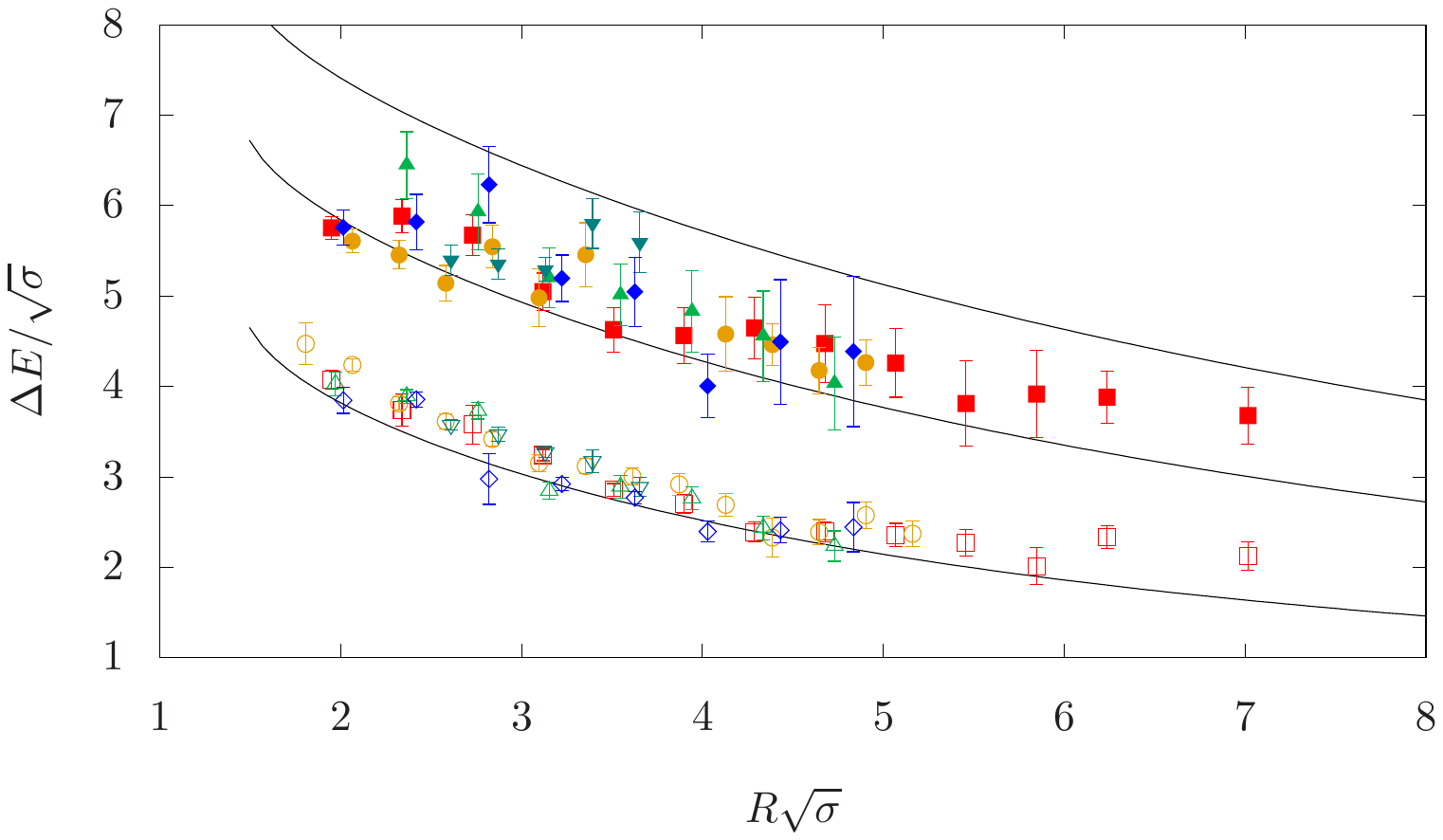}\put(-20,114){\scriptsize $N_L=N_R=1$}\put(-20,143){\scriptsize $N_L=N_R=2$}\put(-20,170){\scriptsize $N_L=N_R=3$}}

 \vspace{-1.0cm}
\caption{\label{fig:plot_J2_Pp-_Pr+_Q0} Results for the confining string with quantum numbers $2^{-+}$. On the upper plot we visualize the energy $E/\sqrt{\sigma}$ while on the lower plot the energy with the ground GGRT level being subtracted $\Delta E/\sqrt{\sigma}$. The representation of the different gauge groups goes as follows: $SU(3)$, $\beta=6.0625$ is represented by $\square$ ($\blacksquare$), $SU(3)$, $\beta=6.338$ by $\circ$ ($\bullet$), $SU(5)$, $\beta=17.630$ by $\triangle$ ($\blacktriangle$), $SU(5)$, $\beta=18.375$ by $\triangledown$ ($\blacktriangledown$) and $SU(6)$, $\beta=25.55$ by  $\diamond$ ($\blacklozenge$) for ground (first excited) state.}
  \end{center}
\end{figure}

  In Fig.~\ref{fig:plot_J2_Pp-_Pr+_Q0}, we display the ground and first excited $2^{-+}$ states for all five data sets. The data are also listed in Tables~\ref{tab:2-+_ground_state} and \ref{tab:2-+_first_excited_state}. It's apparent that the first excited state aligns entirely with the $N_L=N_R=2$ GGRT level, given the provided statistical precision. Additionally, lattice effects and significant  finite $N$ corrections seem to be negligible. Due to limitations in statistical data, it was not possible to extract estimations for higher excitations. 

\begin{figure}[htb]
  \begin{center} 
  \vspace{-2.0cm}
 \rotatebox{0}{\hspace{-2.0cm}\includegraphics[width=15cm]{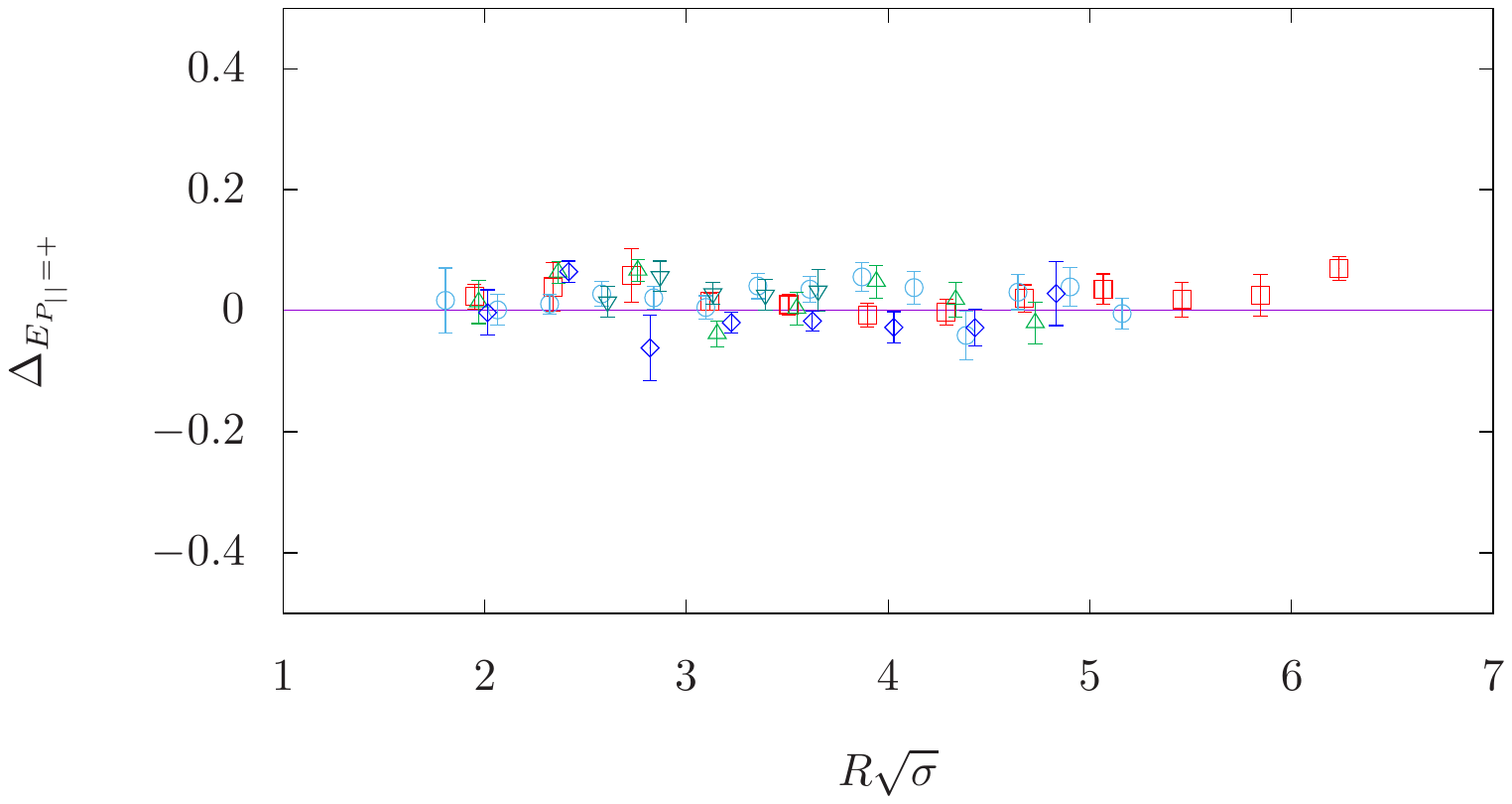}}
 
 \vspace{-3.0cm}
 \rotatebox{0}{\hspace{-2.0cm}\includegraphics[width=15cm]{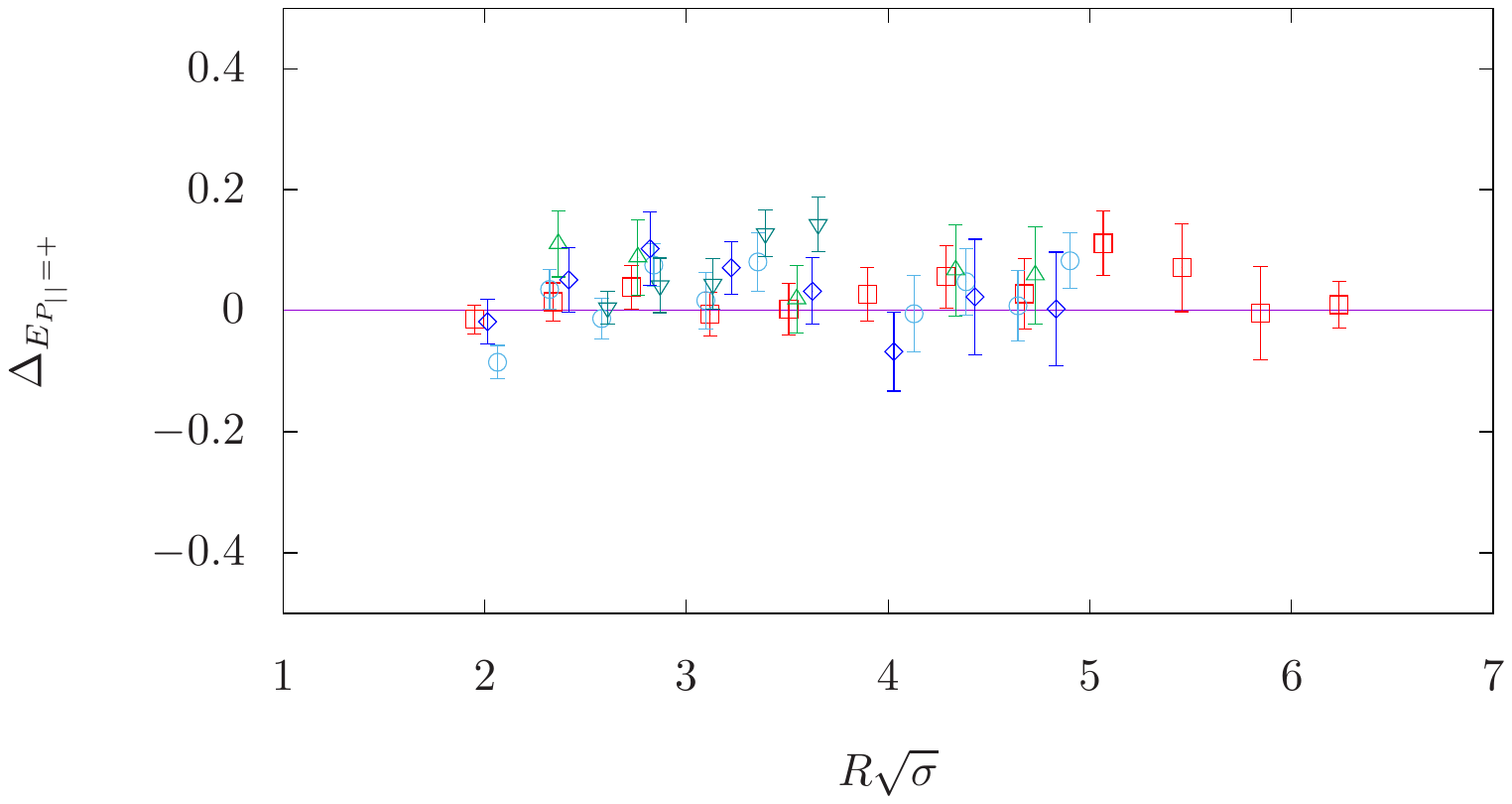}}

 \vspace{-1.0cm}
\caption{\label{fig:splitting_J2Pp} The relative splitting between $2^{-+}$ and $2^{++}$ ground (upper plot) and first excited states (lower plot) at $q=0$. The representation of the different gauge groups goes as follows: $SU(3)$, $\beta=6.0625$ is represented by $\square$ ($\blacksquare$), $SU(3)$, $\beta=6.338$ by $\circ$ ($\bullet$), $SU(5)$, $\beta=17.630$ by $\triangle$ ($\blacktriangle$), $SU(5)$, $\beta=18.375$ by $\triangledown$ ($\blacktriangledown$) and $SU(6)$, $\beta=25.55$ by  $\diamond$ ($\blacklozenge$) for ground (first excited) state. }
  \end{center}
\end{figure}

Due to parity doubling, excitations between the $2^{++}$ and $2^{-+}$ states are expected to be exactly degenerate in the continuum theory at all levels. However, at finite lattice spacing, there may be a splitting between states in these two representations due to lattice artifacts. To investigate the extent of this deviation and the existence for such degeneracy, we present the relative splittings for the ground and first excited states in Fig.~\ref{fig:splitting_J2Pp}. Both plots show that the states exhibit agreement within a couple of standard deviations at most, indicating that a clear determination of these splittings would require a significant increase in statistical data. This question could be addressed in future investigations.

\subsubsection{$2^{+-}$ and $2^{--}$  ground states}
\label{sec:results_2pm-}
Finally, let us discuss the $2^{+-}$ and $2^{--}$ symmetry channel. Again, these two states are degenerate in the continuum theory, so it is natural to discuss them together.
As follows from Table~\ref{tab:table_NGstates_q0}, the corresponding GGRT states first show up at the $N_L=N_R=2$ and the corresponding phonon content is
\[\left[   \left( a^{+}_{1}a^{+}_{1} a^{+}_{-1} a^{-}_{-1} \pm a^{-}_{1}a^{-}_{1} a^{-}_{-1} a^{+}_{-1} \right)-\left( a^{+}_{1}a^{-}_{1} a^{-}_{-1} a^{-}_{-1} \pm a^{-}_{1}a^{+}_{1} a^{+}_{-1} a^{+}_{-1} \right) \right] | 0 \rangle\;.\] However, unlike for the first excited  $2^{\pm+}$ state, in this case there is also a natural candidate axionic state, obtained by adding an axion excitation to the ground $2^{\pm+}$ state.

\begin{figure}[htb]
  \begin{center} 
  \vspace{-2.0cm}
 \rotatebox{0}{\hspace{-2.0cm}\includegraphics[width=15cm]{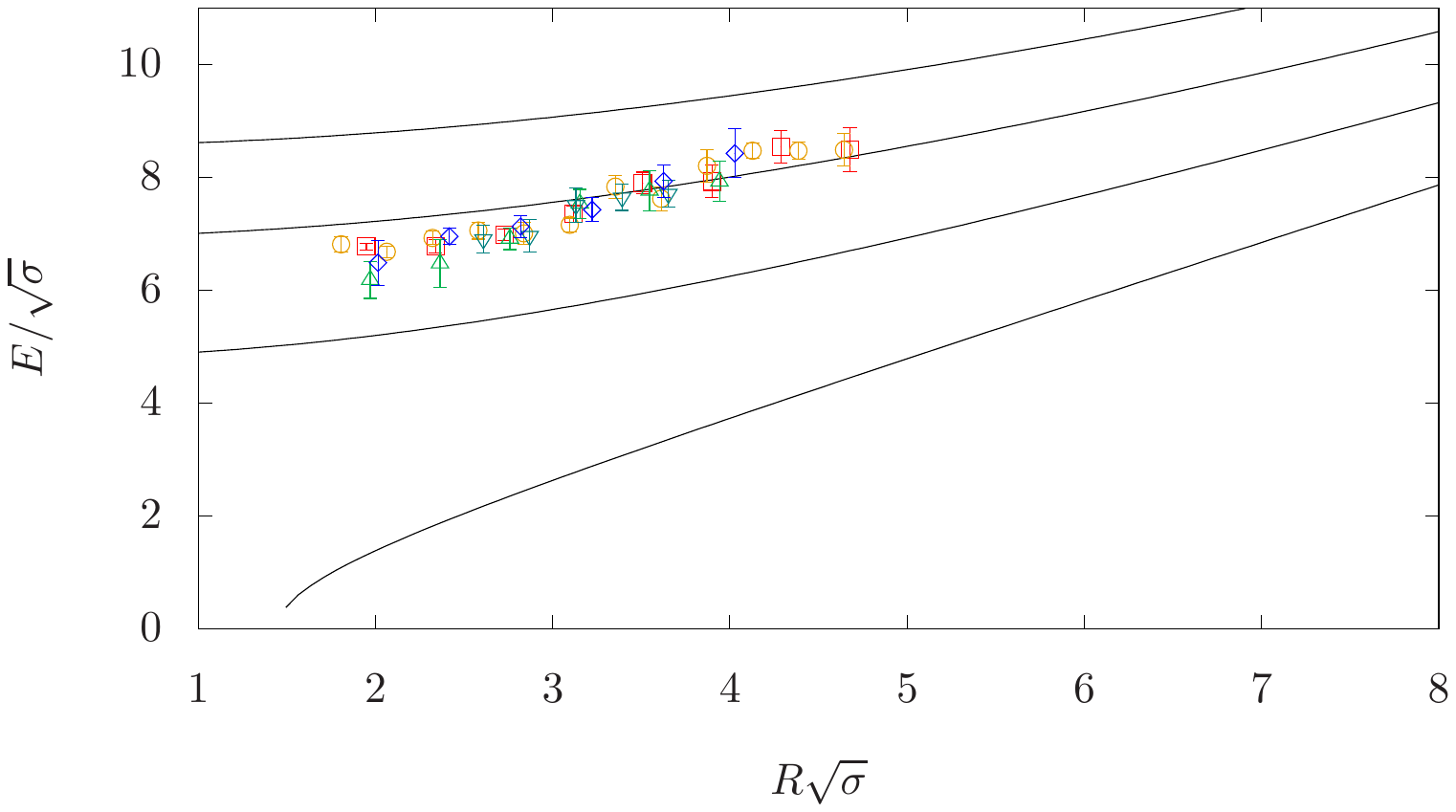}\put(-20,224){\scriptsize $N_L=N_R=0$}\put(-20,247){\scriptsize $N_L=N_R=1$}\put(-20,270){\scriptsize $N_L=N_R=2$}}
 
 \vspace{-3.0cm}
 \rotatebox{0}{\hspace{-2.0cm}\includegraphics[width=15cm]{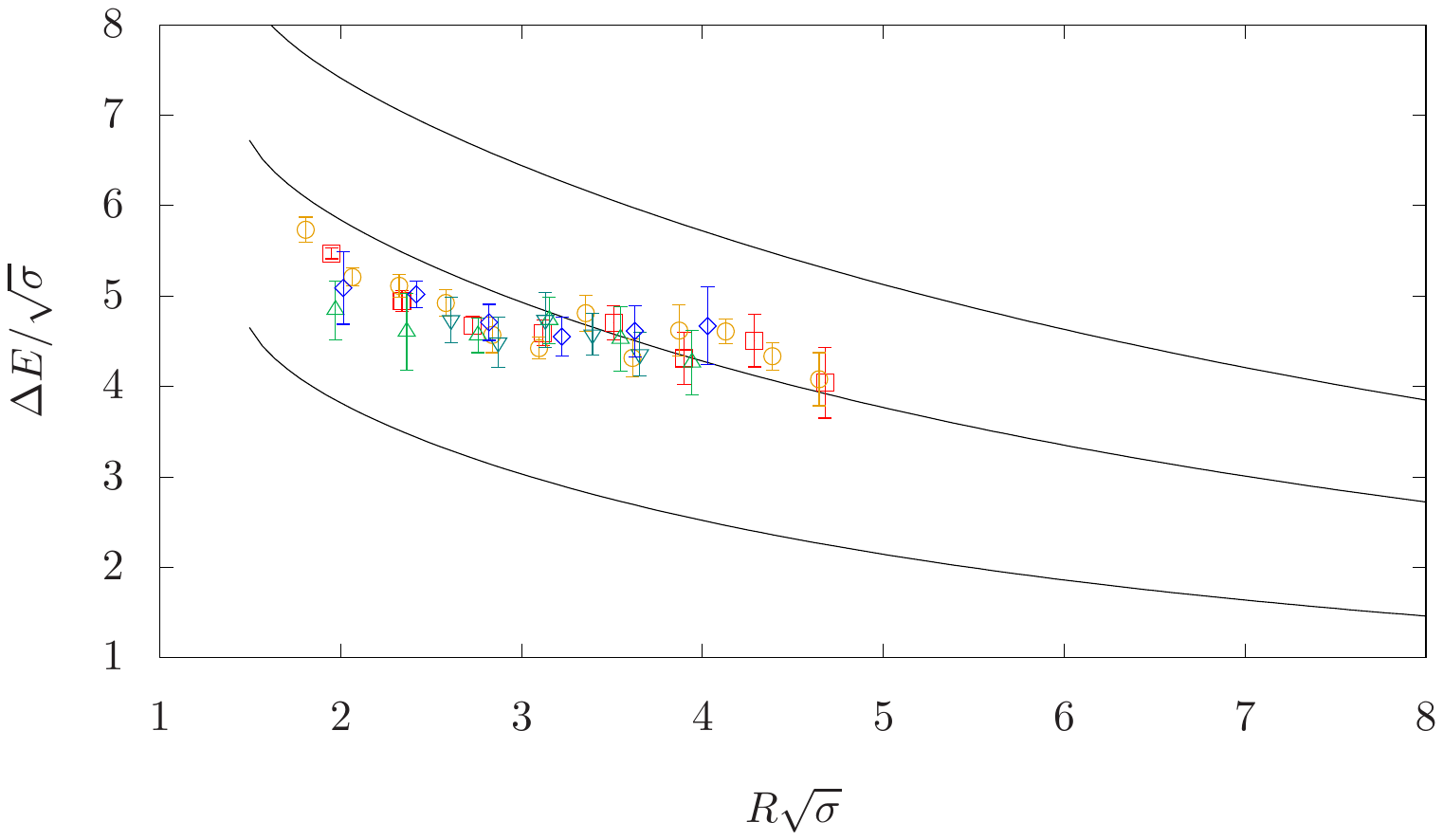}\put(-20,114){\scriptsize $N_L=N_R=1$}\put(-20,143){\scriptsize $N_L=N_R=2$}\put(-20,170){\scriptsize $N_L=N_R=3$}}

 \vspace{-1.0cm}
\caption{\label{fig:plot_J2_Pp+_Pr-_Q0} Results for the ground state of the confining string with quantum numbers $2^{+-}$. On the upper plot we visualize the energy $E/\sqrt{\sigma}$ while on the lower plot the energy with the ground GGRT level being subtracted $\Delta E/\sqrt{\sigma}$. The representation of the different gauge groups goes as follows: $SU(3)$, $\beta=6.0625$ is represented by $\square$, $SU(3)$, $\beta=6.338$ by $\circ$, $SU(5)$, $\beta=17.630$ by $\triangle$, $SU(5)$, $\beta=18.375$ by $\triangledown$ and $SU(6)$, $\beta=25.55$ by $\diamond$.}
  \end{center}
\end{figure}

\begin{figure}[htb]
  \begin{center} 
  \vspace{-2.0cm}
 \rotatebox{0}{\hspace{-2.0cm}\includegraphics[width=15cm]{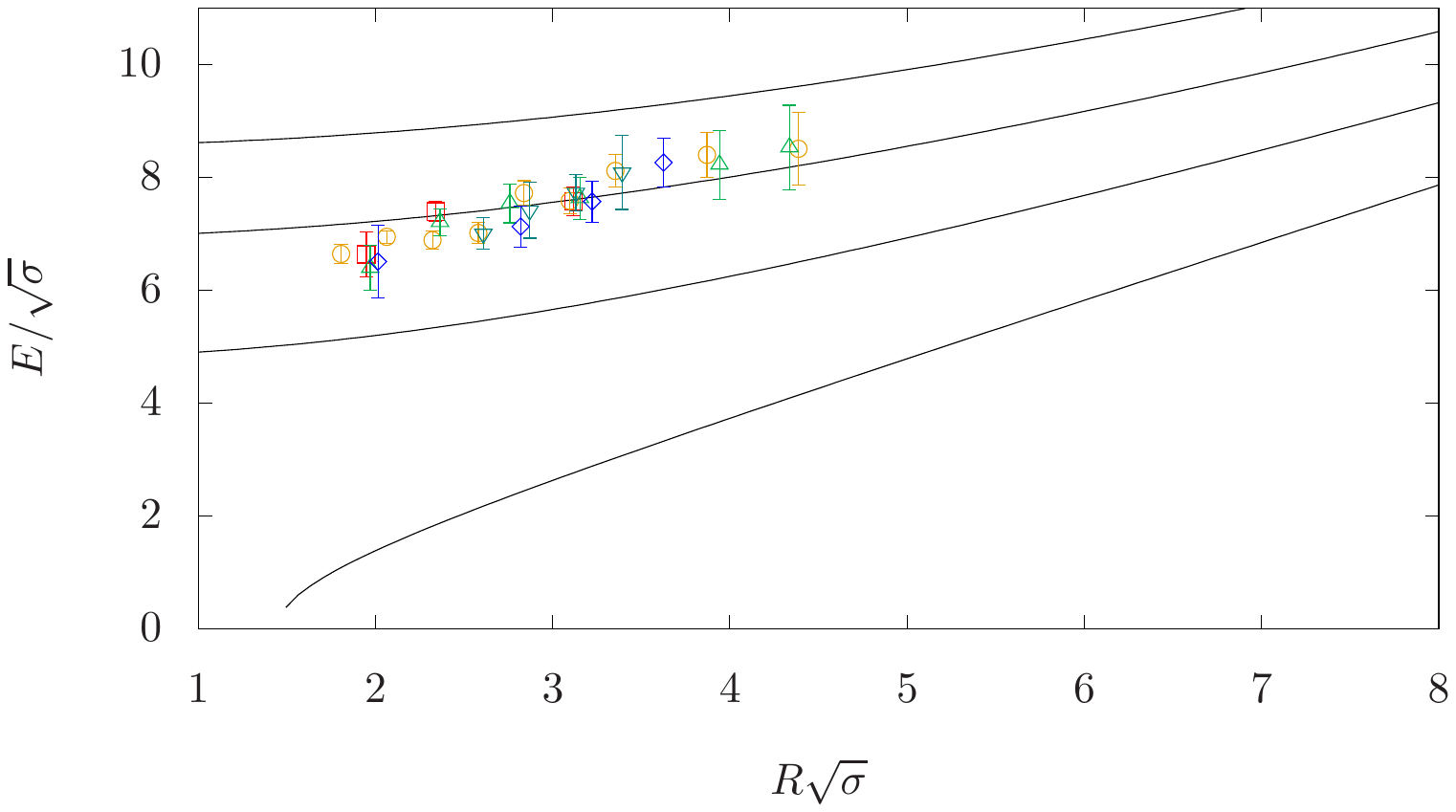}\put(-20,224){\scriptsize $N_L=N_R=0$}\put(-20,247){\scriptsize $N_L=N_R=1$}\put(-20,270){\scriptsize $N_L=N_R=2$}}
 
 \vspace{-3.0cm}
 \rotatebox{0}{\hspace{-2.0cm}\includegraphics[width=15cm]{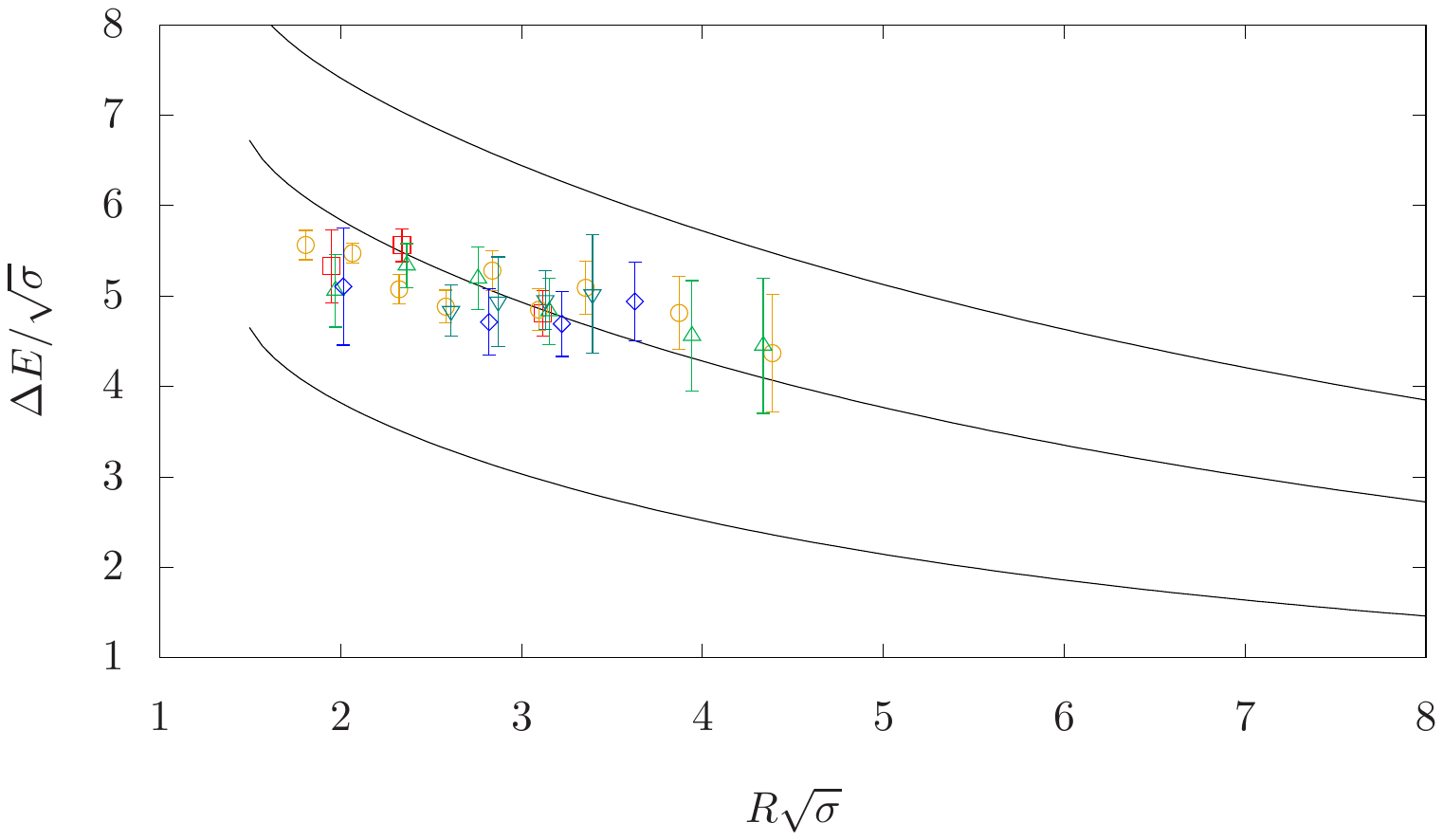}\put(-20,114){\scriptsize $N_L=N_R=1$}\put(-20,143){\scriptsize $N_L=N_R=2$}\put(-20,170){\scriptsize $N_L=N_R=3$}}

 \vspace{-1.0cm}
\caption{\label{fig:plot_J2_Pp-_Pr-_Q0} Results for the ground state of the confining string with quantum numbers $2^{--}$. On the upper plot we visualize the energy $E/\sqrt{\sigma}$ while on the lower plot the energy with the ground GGRT level being subtracted $\Delta E/\sqrt{\sigma}$. The representation of the different gauge groups goes as follows: $SU(3)$, $\beta=6.0625$ is represented by $\square$, $SU(3)$, $\beta=6.338$ by $\circ$, $SU(5)$, $\beta=17.630$ by $\triangle$, $SU(5)$, $\beta=18.375$ by $\triangledown$ and $SU(6)$, $\beta=25.55$ by  $\diamond$.}
  \end{center}
\end{figure}

In Fig.~\ref{fig:plot_J2_Pp+_Pr-_Q0} and Table~\ref{tab:2+-_ground_state} we present the ground state of the $2^{+-}$ sector across all five data sets, while we do the same for the $2^{--}$ sector in Fig.~\ref{fig:plot_J2_Pp-_Pr-_Q0} and Table~\ref{tab:2--_ground_state}. Notably, these states exhibit substantial mass, rendering the estimation of their effective masses a challenging endeavor. As outlined in Table~\ref{tab:2--_ground_state}, we were able to obtain reliable mass values for only a few data sets for $2^{--}$\footnote{As discussed in Section~\ref{sec:subsection_lattice_operators}, the basis of operators in the $2^{--}$ sector is generically smaller than those in $2^{+-}$, essentially because only less symmetric lattice paths can contribute to $2^{--}$. As a result, the spectrum extraction is usually worse in the $2^{--}$ sector. }. Note that for both sectors, the next excitation levels emerge with significantly higher masses, suggesting that the ground states of the flux tube are non-degenerate. 

We should see in Section~\ref{subsec:q0jnonzero_computation} that both states should be interpreted as a string state with a massive excitation and two phonon excitations, so we defer the comparison to theory till then. There are no detectable large-$N$ effects or lattice spacing dependence in the ground state given the large errors here.

\begin{figure}[htb]
  \begin{center} 
  \vspace{-2.0cm}
 \rotatebox{0}{\hspace{-2.0cm}\includegraphics[width=15cm]{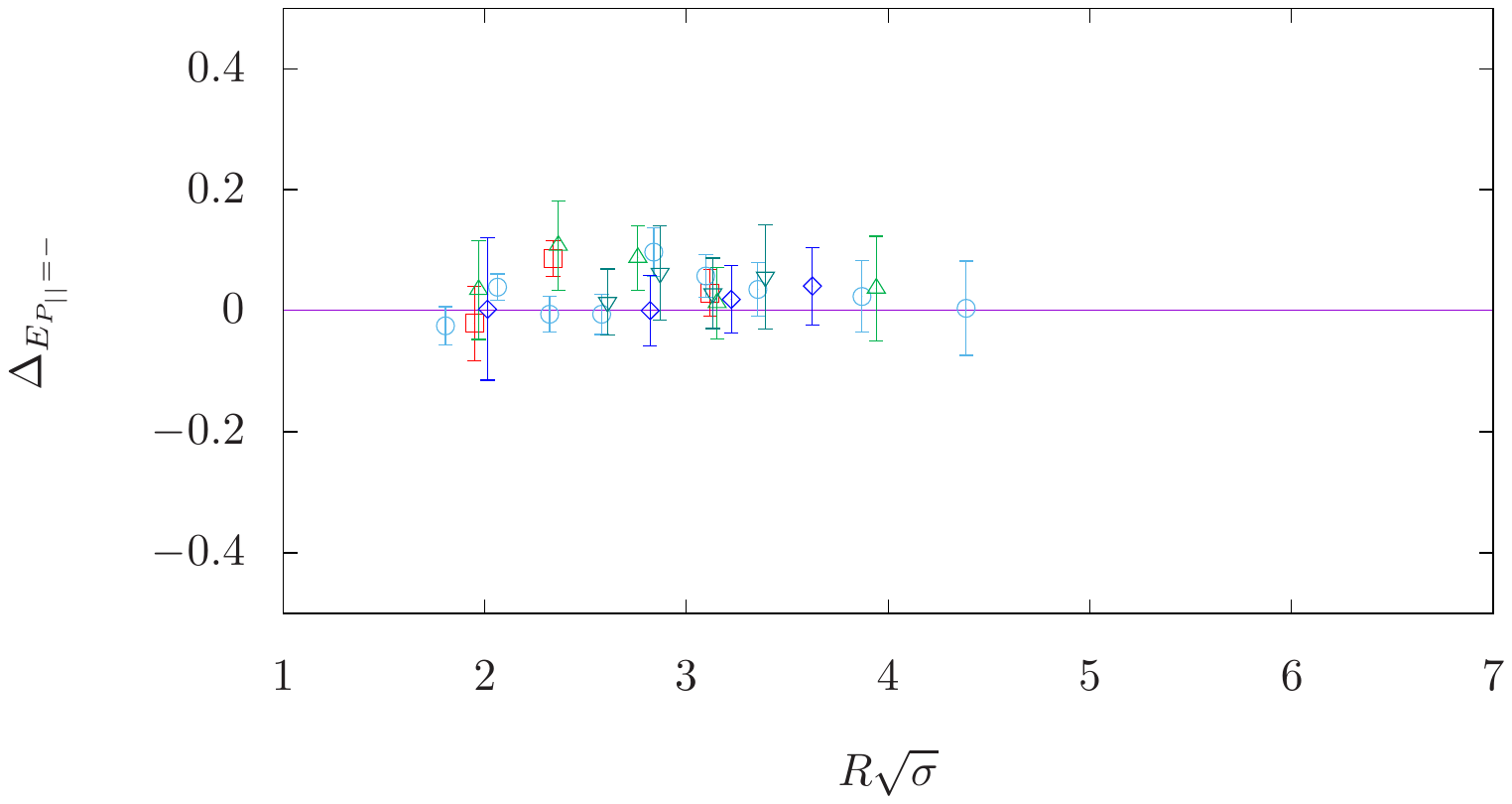}}

 \vspace{-1.0cm}
\caption{\label{fig:splitting_J2Pm} The relative splitting between $2^{--}$ and $2^{+-}$ ground state at $q=0$. The representation of the different gauge groups goes as follows: $SU(3)$, $\beta=6.0625$ is represented by $\square$ ($\blacksquare$), $SU(3)$, $\beta=6.338$ by $\circ$ ($\bullet$), $SU(5)$, $\beta=17.630$ by $\triangle$ ($\blacktriangle$), $SU(5)$, $\beta=18.375$ by $\triangledown$ ($\blacktriangledown$) and $SU(6)$, $\beta=25.55$ by  $\diamond$ ($\blacklozenge$) for ground (first excited) state. }
  \end{center}
\end{figure}

{
Similarly as in Section~\ref{sec:results_2pm+}, the $2^{+-}$ and $2^{--}$ states are expected to be perfectly degenerate in the continuum theory but at nonzero lattice spacing, lattice artifacts may introduce a splitting. To explore this potential discrepancy, we present the relative splittings for the ground state in Fig.~\ref{fig:splitting_J2Pm}. The plots reveal that the states are consistent within only a few standard deviations, implying again that a more precise determination of these splittings would necessitate a substantial increase in the statistical dataset.}

\subsection{Confining flux tube with $q=1$ longitudinal momentum}
\label{sec:results_for_q=1}
In this section, we present our results on the spectrum of flux tube excitations carrying one unit of longitudinal momentum $q=1$, where $q$ is related to the longitudinal momentum as $p=\frac{2 \pi q}{ R}$. Let us recall that string excitations with nonzero longitudinal momentum are characterized by their spin $|J|$ and the transverse parity $P_{\perp}$, while the longitudinal parity ceases to be a good quantum number because it does not commute with $p$ at $p\neq 0$.

\subsubsection{$q=1$ ground state}
\label{sec:J1_q1_ground_state}
We begin the presentation of the $q \neq 0$ results with the lightest state. This is the ground state in the $J=1$, $q=1$ sector. We present this state for the five $SU(N)$ data sets in Fig.~\ref{fig:plot_J1_Q1} together with  the GGRT prediction for $N_L=1, N_R=0$. The data are also listed in Table~\ref{tab:Q1_J1_ground_state}. According to Table~\ref{tab:table_NGstates_q1}, the GGRT lightest $q=1$ state consists of a single phonon with momentum $2 \pi / R$. Our data are in a good agreement  with this GGRT level.  

\subsubsection{$0^+$, $q=1$ ground state}
\label{sec:J0_Pp+_q1_ground_state}
\begin{figure}[htb]
  \begin{center} 
  \vspace{-2.0cm}
 \rotatebox{0}{\hspace{-2.0cm}\includegraphics[width=15cm]{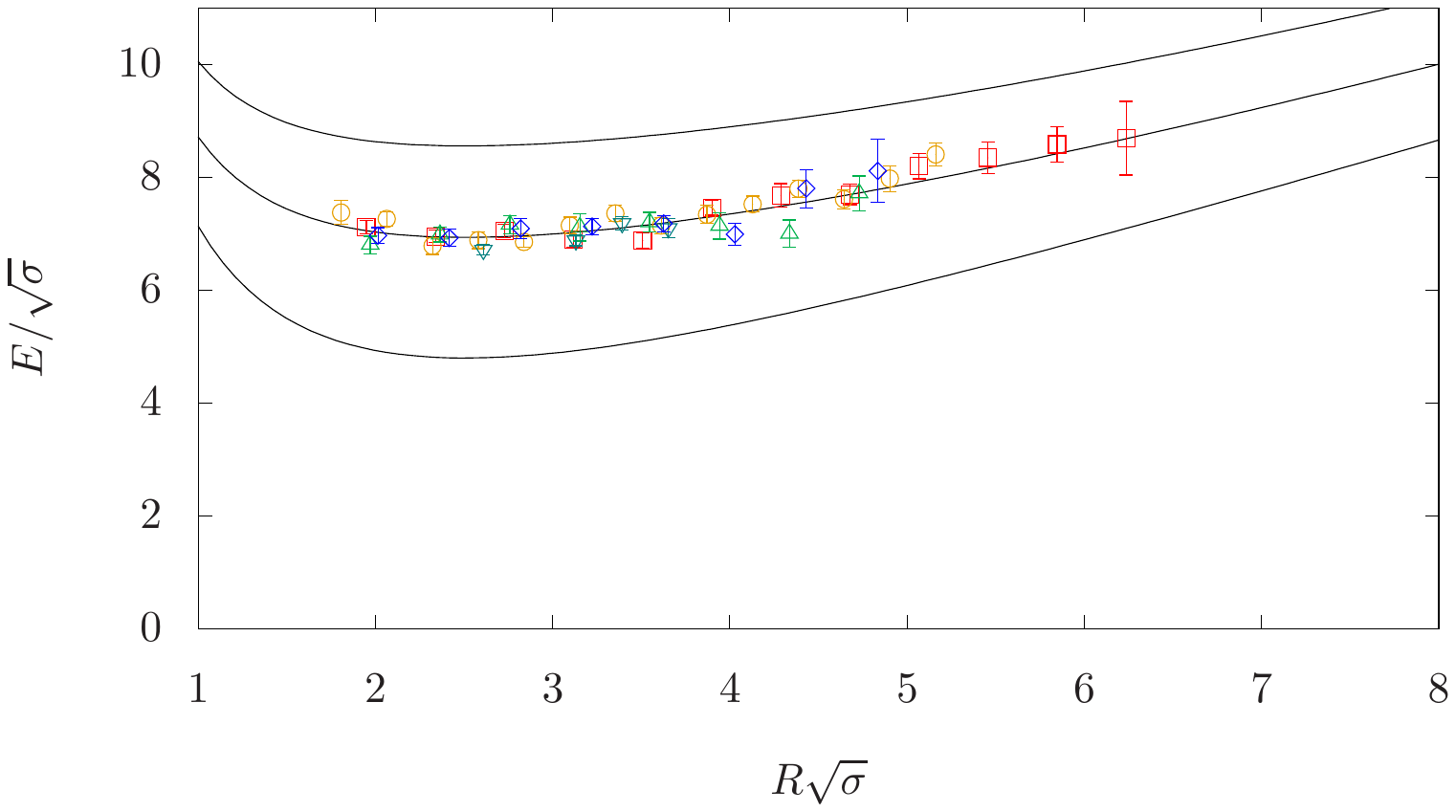}\put(-20,237){\scriptsize $N_L=1, N_R=0$}\put(-20,258){\scriptsize $N_L=2,N_R=1$}\put(-20,275){\scriptsize $N_L=3, N_R=2$}}
 
 \vspace{-3.0cm}
 \rotatebox{0}{\hspace{-2.0cm}\includegraphics[width=15cm]{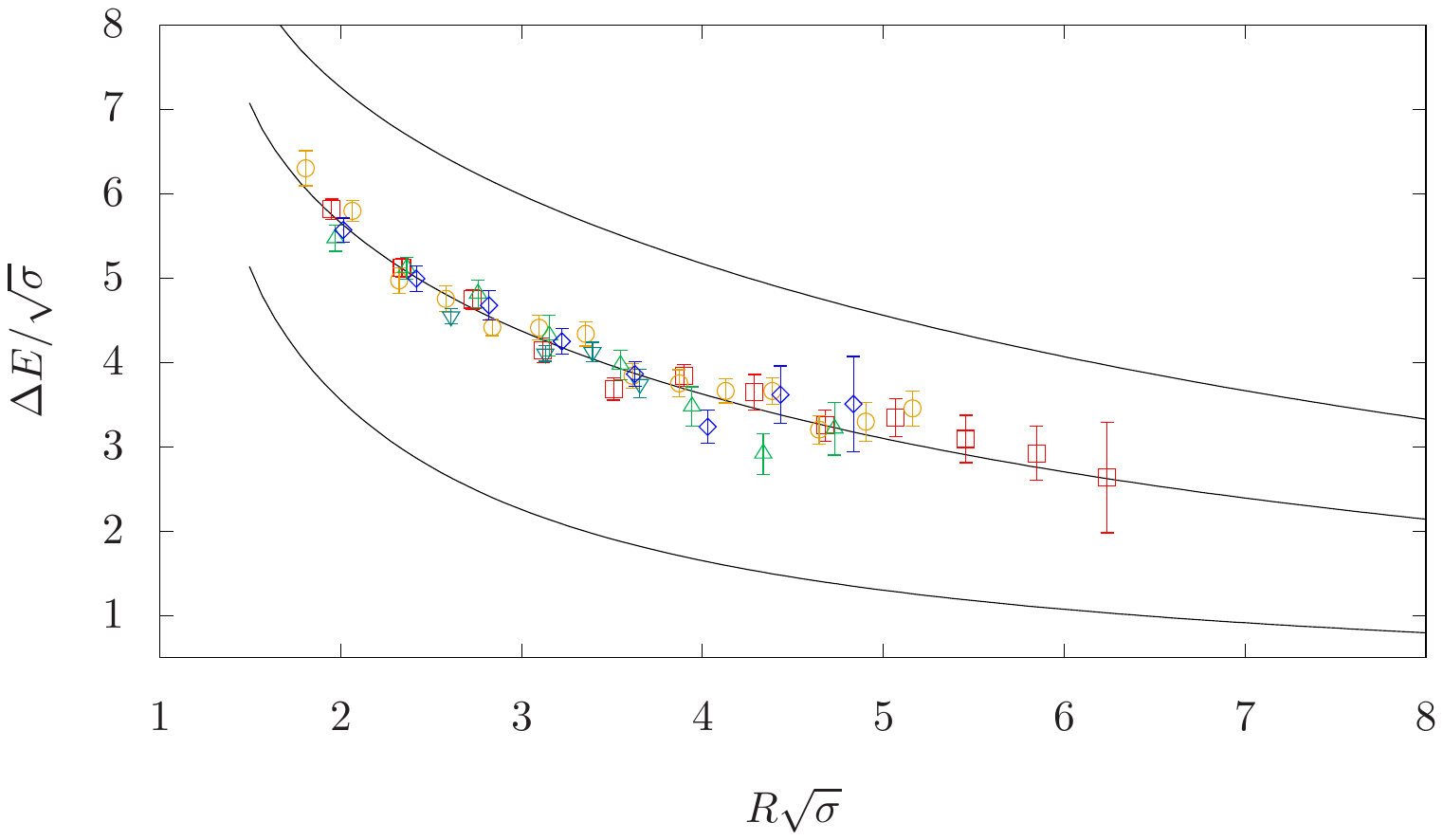}\put(-20,108){\scriptsize $N_L=1, N_R=0$}\put(-20,140){\scriptsize $N_L=2,N_R=1$}\put(-20,165){\scriptsize $N_L=3, N_R=2$}}

 \vspace{-1.0cm}
\caption{\label{fig:plot_J0_Pp+_Q1} Results for the confining string with quantum numbers $0^+$, $q=1$. On the upper plot we visualize the energy $E/\sqrt{\sigma}$ while on the lower plot the energy minus the ground GGRT level $\Delta E/\sqrt{\sigma}$. The representation for the different gauge groups goes as follows: $SU(3)$, $\beta=6.0625$ is represented by $\square$, $SU(3)$, $\beta=6.338$ by $\circ$, $SU(5)$, $\beta=17.630$ by $\triangle$, $SU(5)$, $\beta=18.375$ by $\triangledown$ and $SU(6)$, $\beta=25.55$ by  $\diamond$ for the ground state.}
  \end{center}
\end{figure}

According to Table~\ref{tab:table_NGstates_q1}, the lowest GGRT $0^+$, $q=1$ arises at $N_L=2$ and $N_R=1$ 
 with the oscillator content given by 
\[\left( a^{+}_{2} a^{-}_{-1}+a^{-}_{2} a^{+}_{-1} \right) | 0 \rangle\;.
\] 
In Fig.~\ref{fig:plot_J0_Pp+_Q1} and Table~\ref{tab:J0_Pp+_Q1_ground_state} we provide  results for the $0^+$, $q=1$ ground state across all five different data sets. 
The data  indeed demonstrate a good agreement with the $N_L=2$ and $N_R=1$ GGRT prediction. 
The first excited  $0^+$, $q=1$ level is noticeably heavier than the ground state (although the corresponding  mass plateaus are not very well pronounced, so that the accurate mass determination is problematic). This again agrees with the GGRT prediction that there is a single $0^+$, $q=1$ state with $N_L=2$ and $N_R=1$.

\subsubsection{$0^-$, $q=1$ states and the axion}
\label{sec:J0_Pp-_q1_state}
\begin{figure}[htb]
  \begin{center} 
  \vspace{-2.0cm}
 \rotatebox{0}{\hspace{-2.0cm}\includegraphics[width=15cm]{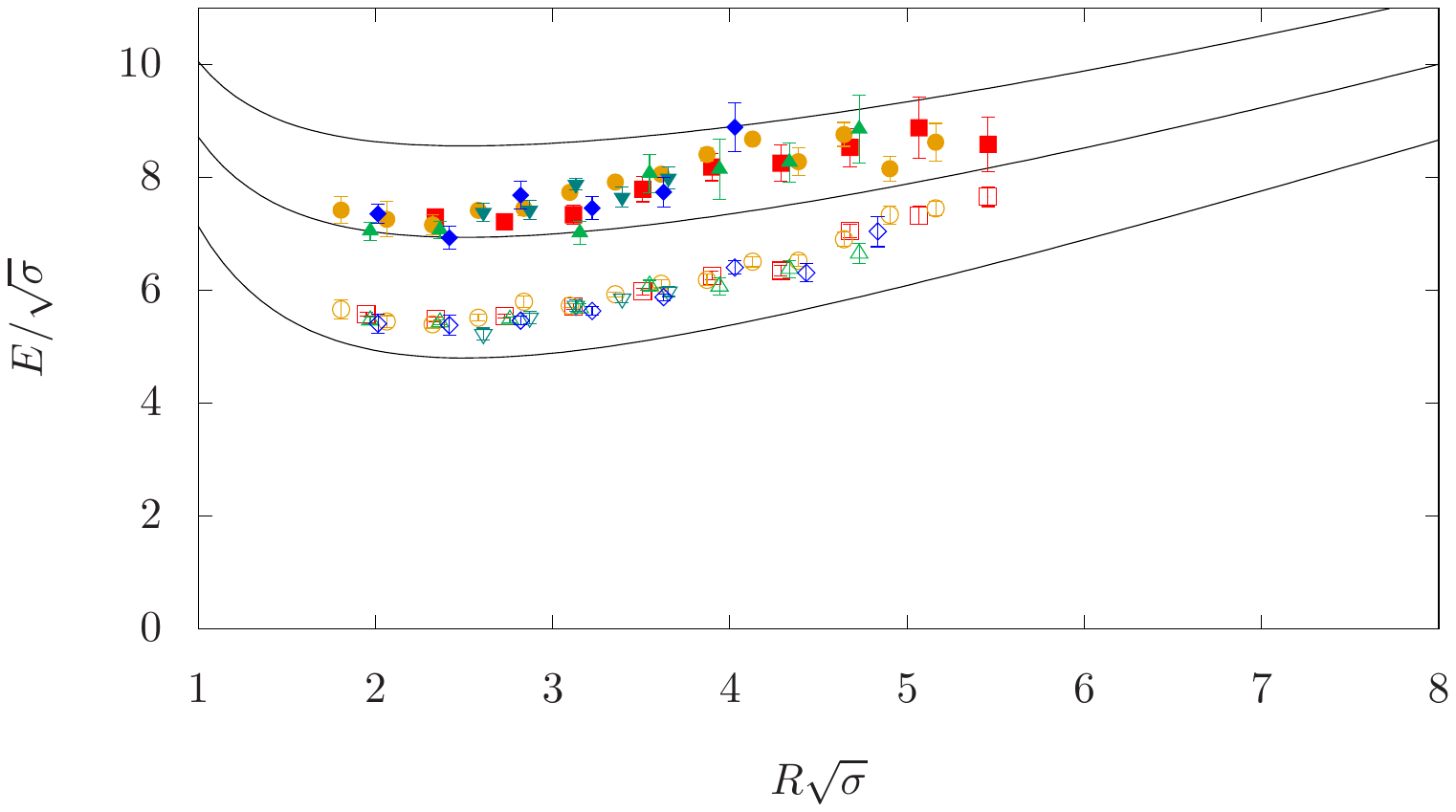}\put(-20,237){\scriptsize $N_L=1, N_R=0$}\put(-20,258){\scriptsize $N_L=2,N_R=1$}\put(-20,275){\scriptsize $N_L=3, N_R=2$}}
 
 \vspace{-3.0cm}
 \rotatebox{0}{\hspace{-2.0cm}\includegraphics[width=15cm]{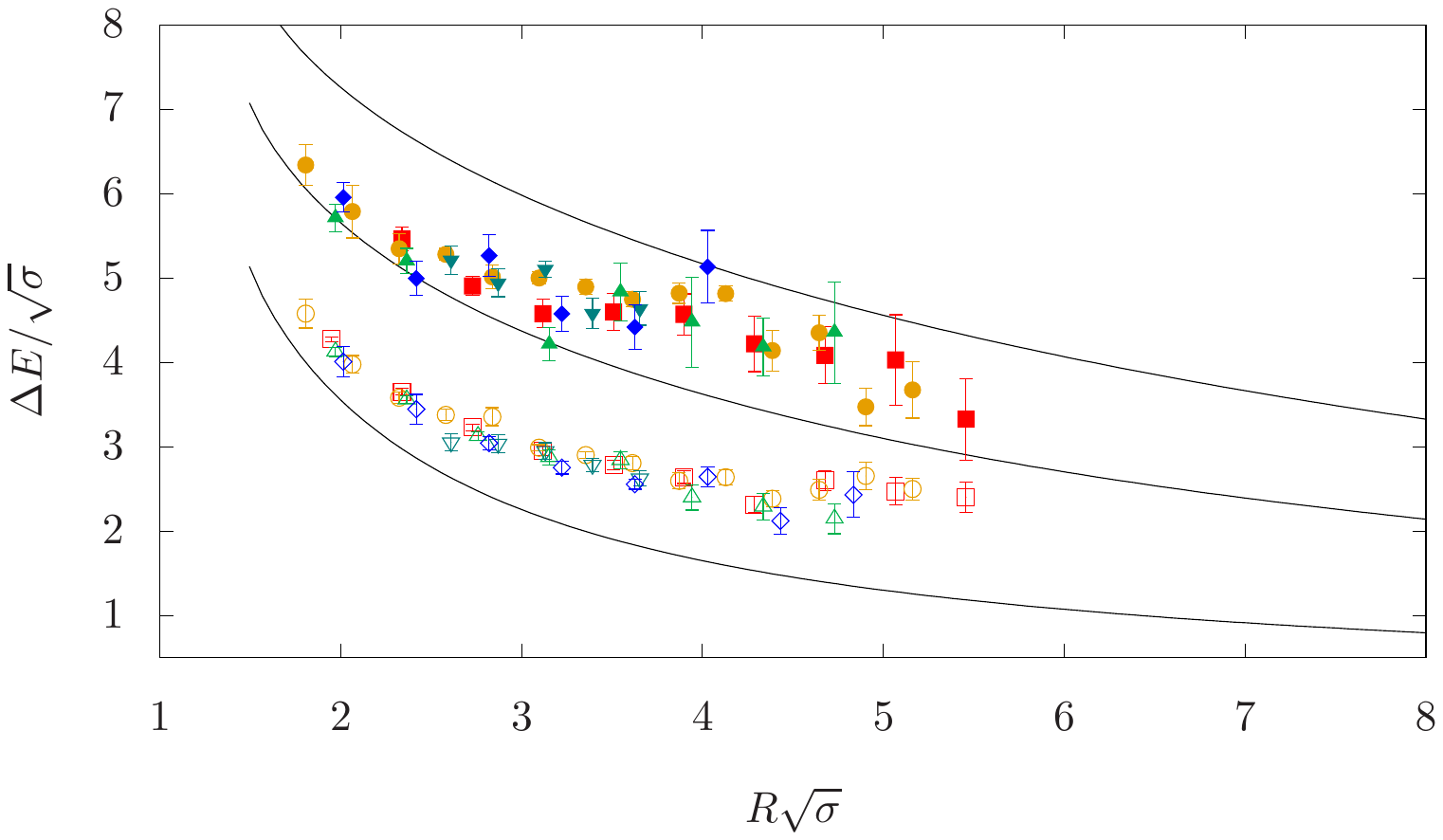}\put(-20,108){\scriptsize $N_L=1, N_R=0$}\put(-20,140){\scriptsize $N_L=2,N_R=1$}\put(-20,165){\scriptsize $N_L=3, N_R=2$}}

 \vspace{-1.0cm}
\caption{\label{fig:plot_J0_Pp-_Q1} Results for the confining string with quantum numbers $0^-$,  $q=1$. On the upper plot we visualize the energy $E/\sqrt{\sigma}$ while on the lower plot the energy minus the ground GGRT level $\Delta E/\sqrt{\sigma}$. The representation of the different gauge groups goes as follows: $SU(3)$, $\beta=6.0625$ is represented by $\square$ ($\blacksquare$), $SU(3)$, $\beta=6.338$ by $\circ$ ($\bullet$), $SU(5)$, $\beta=17.630$ by $\triangle$ ($\blacktriangle$), $SU(5)$, $\beta=18.375$ by $\triangledown$ ($\blacktriangledown$) and $SU(6)$, $\beta=25.55$ by  $\diamond$ ($\blacklozenge$) for ground (first excited) state.}
  \end{center}
\end{figure}

This channel is expected to have very different properties, because in addition to the $N_L=2$, $N_R=1$ GGRT state one can get a $0^-$, $q=1$ state also by boosting an axion.
In Fig.~\ref{fig:plot_J0_Pp-_Q1} we provide the energies for the $0^-$, $q=1$ ground state as well as for the first excited state in the same sector. The data are summarized in Tables~\ref{tab:J0_Pp-_Q1_ground_state} and \ref{tab:J0_Pp-_Q1_first_excited_state}. We observe that the ground state exhibits large deviations from the GGRT predictions in all range of measured $R$.
The TBA analysis in Section~\ref{subsec:q1computation} confirms that  this state indeed corresponds to a boosted massive excitation.

\subsubsection{$1$, $q=1$ excited states}
\label{sec:J1_q1_excited_states}
As we just discussed in Section~\ref{sec:J1_q1_ground_state}, the ground state of the confining flux tube with quantum numbers $J=1$, $q=1$ agrees well with the GGRT  $N_L=1$, $N_R=0$ state, which corresponds to a single phonon excitation of a flux tube. The first GGRT excited states in this channel arise at the $N_L=2$, $N_R=1$ level. Namely, by inspecting Table~\ref{tab:table_NGstates_q1} one finds there two $J=1$ doublets 
\[\left( a^{+}_{1}a^{+}_{1} a^{-}_{-1} \pm  a^{-}_{1}a^{-}_{1}a^{+}_{-1} \right) | 0 \rangle\]
and
\[\left( a^{+}_{1}a^{-}_{1} a^{-}_{-1} \pm a^{-}_{1}a^{+}_{1} a^{+}_{-1}  \right) | 0 \rangle\]
and also an additional $J=3$ doublet (which is indistinguishable from $J=1$ states on a lattice),
\[
\left( a^{+}_{1}a^{+}_{1} a^{+}_{-1} \pm a^{-}_{1}a^{-}_{1} a^{-}_{-1}  \right) | 0 \rangle\;.
\]
In addition, one also expects to find a non-GGRT state, obtained by adding a massive axion to the ground $J=1$, $q=1$ state.

\begin{figure}[htb]
  \begin{center} 
  \vspace{-2.0cm}
 \rotatebox{0}{\hspace{-2.0cm}\includegraphics[width=15cm]{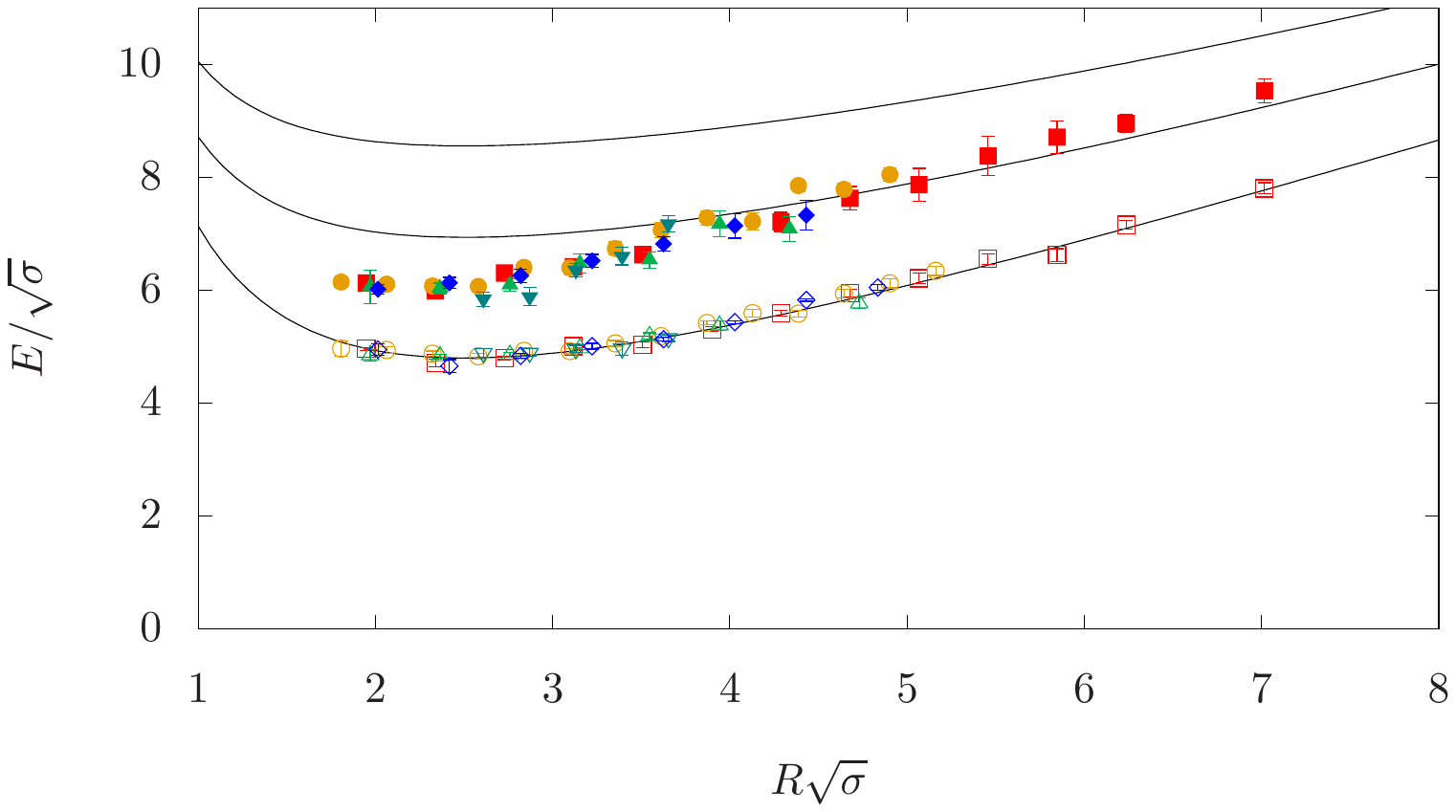}\put(-20,237){\scriptsize $N_L=1, N_R=0$}\put(-20,258){\scriptsize $N_L=2,N_R=1$}\put(-20,275){\scriptsize $N_L=3, N_R=2$}}
 
 \vspace{-3.0cm}
 \rotatebox{0}{\hspace{-2.0cm}\includegraphics[width=15cm]{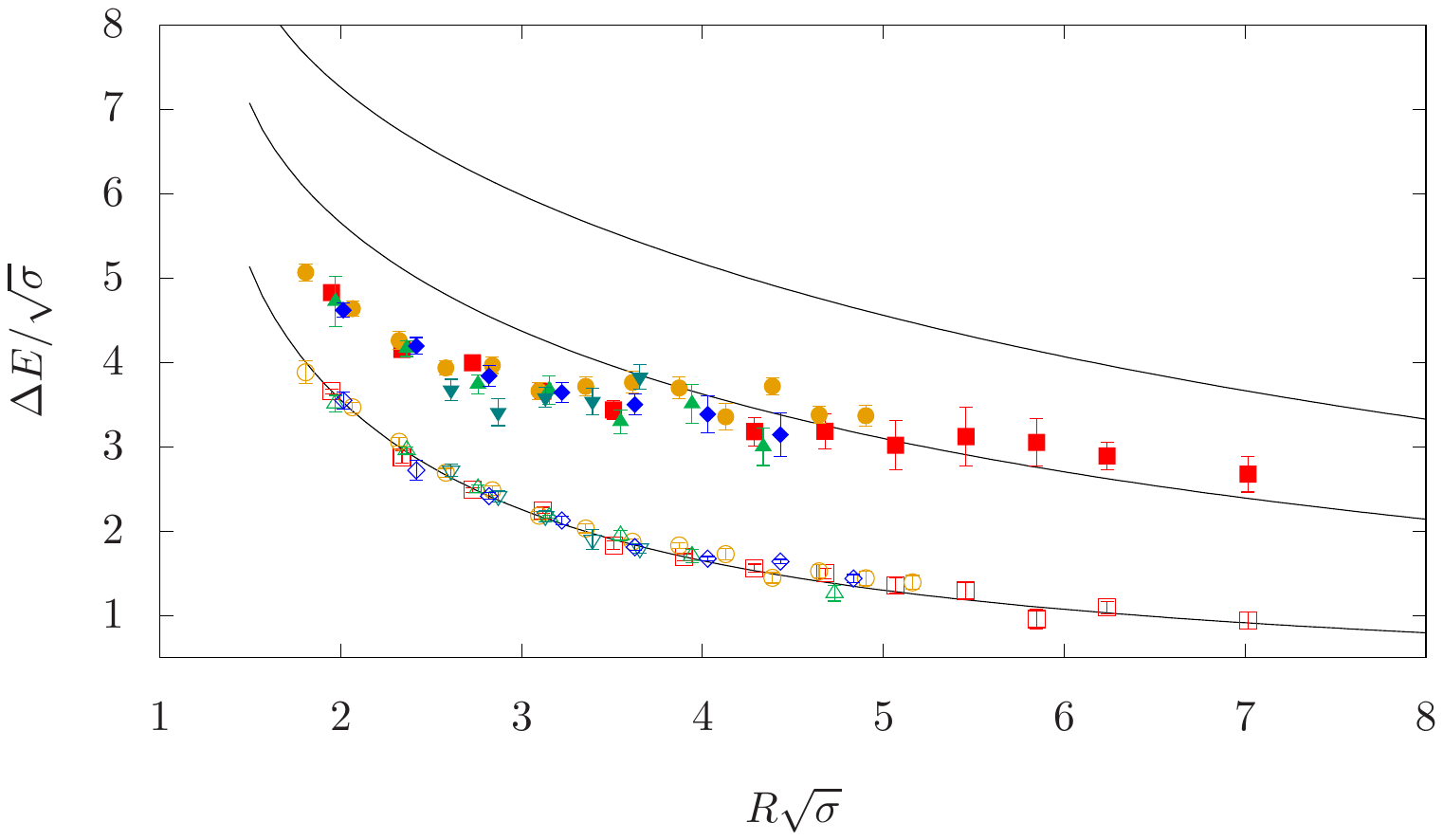}\put(-20,108){\scriptsize $N_L=1, N_R=0$}\put(-20,140){\scriptsize $N_L=2,N_R=1$}\put(-20,165){\scriptsize $N_L=3, N_R=2$}}

 \vspace{-1.0cm}
\caption{\label{fig:plot_J1_Q1} Results for the confining string with quantum numbers $|J|=1$, $q=1$. On the upper plot we visualize the energy $E/\sqrt{\sigma}$ while on the lower plot the energy minus the ground GGRT level $\Delta E/\sqrt{\sigma}$. The representation of the different gauge groups goes as follows: $SU(3)$, $\beta=6.0625$ is represented by $\square$ ($\blacksquare$), $SU(3)$, $\beta=6.338$ by $\circ$ ($\bullet$), $SU(5)$, $\beta=17.630$ by $\triangle$ ($\blacktriangle$), $SU(5)$, $\beta=18.375$ by $\triangledown$ ($\blacktriangledown$) and $SU(6)$, $\beta=25.55$ by  $\diamond$ ($\blacklozenge$) for ground (first excited) state.}
  \end{center}
\end{figure}

  In Fig.~\ref{fig:plot_J1_Q1} we present the ground as well as the first excited $J=1$, $q=1$ state. The data are listed in Tables~\ref{tab:Q1_J1_ground_state} and \ref{tab:Q1_J1_first_excited_state}. 
 We observe that the first excited state we observe exhibits significant deviations from the $N_L=2$, $N_R=1$ GGRT excitation level for $R\sqrt{\sigma} \le 4$ and approaches it for $R\sqrt{\sigma} \geq 4$. In Section~\ref{subsec:q1computation}, we provide TBA analysis that precisely captures such deviation, with the interpretation that  it represents  a string state with a massive excitation and a phonon excitation.

\begin{figure}[htb]
  \begin{center} 
  \vspace{-2.0cm}
 \rotatebox{0}{\hspace{-2.0cm}\includegraphics[width=15cm]{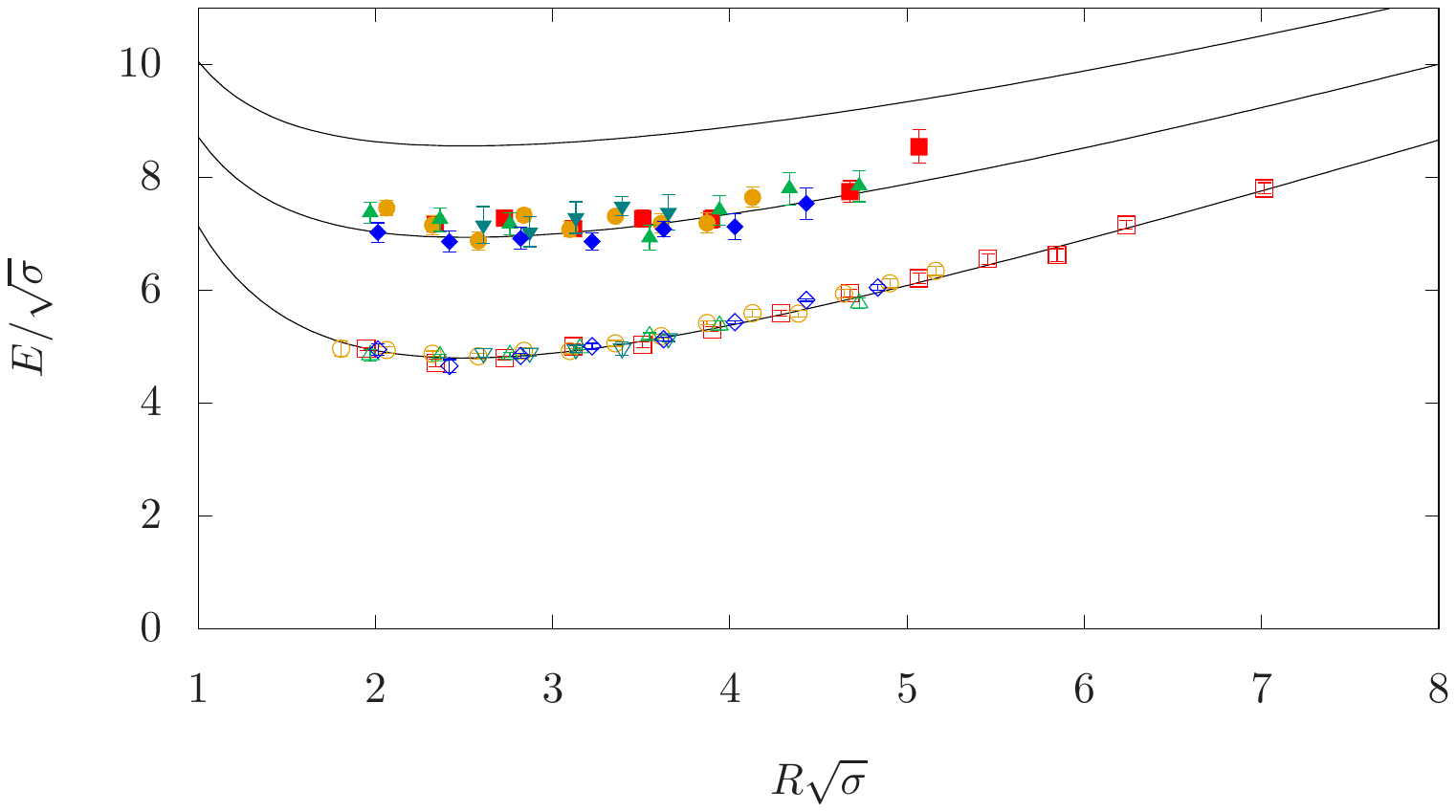}\put(-20,237){\scriptsize $N_L=1, N_R=0$}\put(-20,258){\scriptsize $N_L=2,N_R=1$}\put(-20,275){\scriptsize $N_L=3, N_R=2$}}
 
 \vspace{-3.0cm}
 \rotatebox{0}{\hspace{-2.0cm}\includegraphics[width=15cm]{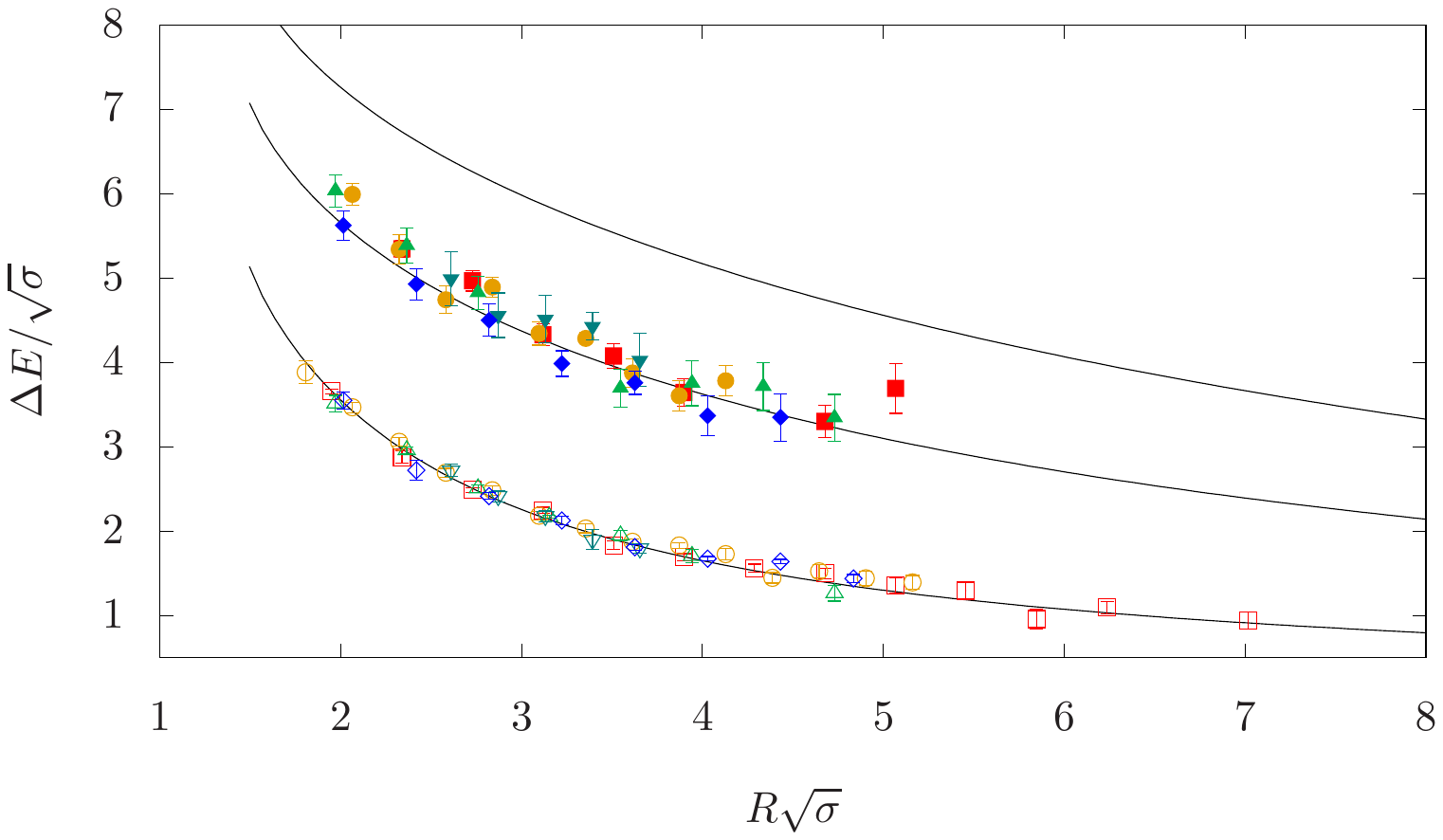}\put(-20,108){\scriptsize $N_L=1, N_R=0$}\put(-20,140){\scriptsize $N_L=2,N_R=1$}\put(-20,165){\scriptsize $N_L=3, N_R=2$}}

 \vspace{-1.0cm}
\caption{\label{fig:plot_J1_Q1_2nd} Results for the confining string with quantum numbers $|J|=1$, $q=1$. On the upper plot we visualize the energy $E/\sqrt{\sigma}$ while on the lower plot the energy minus the ground GGRT level $\Delta E/\sqrt{\sigma}$. The representation of the different gauge groups goes as follows: $SU(3)$, $\beta=6.0625$ is represented by $\square$ ($\blacksquare$), $SU(3)$, $\beta=6.338$ by $\circ$ ($\bullet$), $SU(5)$, $\beta=17.630$ by $\triangle$ ($\blacktriangle$), $SU(5)$, $\beta=18.375$ by $\triangledown$ ($\blacktriangledown$) and $SU(6)$, $\beta=25.55$ by  $\diamond$ ($\blacklozenge$) for ground (second excited) state.}
  \end{center}
\end{figure}

Turning to the next excitation level, in Fig.~\ref{fig:plot_J1_Q1_2nd} we present the second excited  $J=1$, $q=1$ state for the five different data sets. The data are listed in Table~\ref{tab:Q1_J1_second_excited_state}. Our data demonstrate an adequate agreement with the $N_L=2$, $N_R=1$ GGRT level. Unfortunately, with the current statistics we are not able to 
  extract reliably  higher excitations, so that  the expected degeneracy of the $N_L=2$, $N_R=1$  level cannot be confirmed.

\subsubsection{$2^{\pm}$, $q=1$ ground state}
\label{sec:J2_Pp+_q1_state}

\begin{figure}[htb]
  \begin{center} 
  \vspace{-2.0cm}
 \rotatebox{0}{\hspace{-2.0cm}\includegraphics[width=15cm]{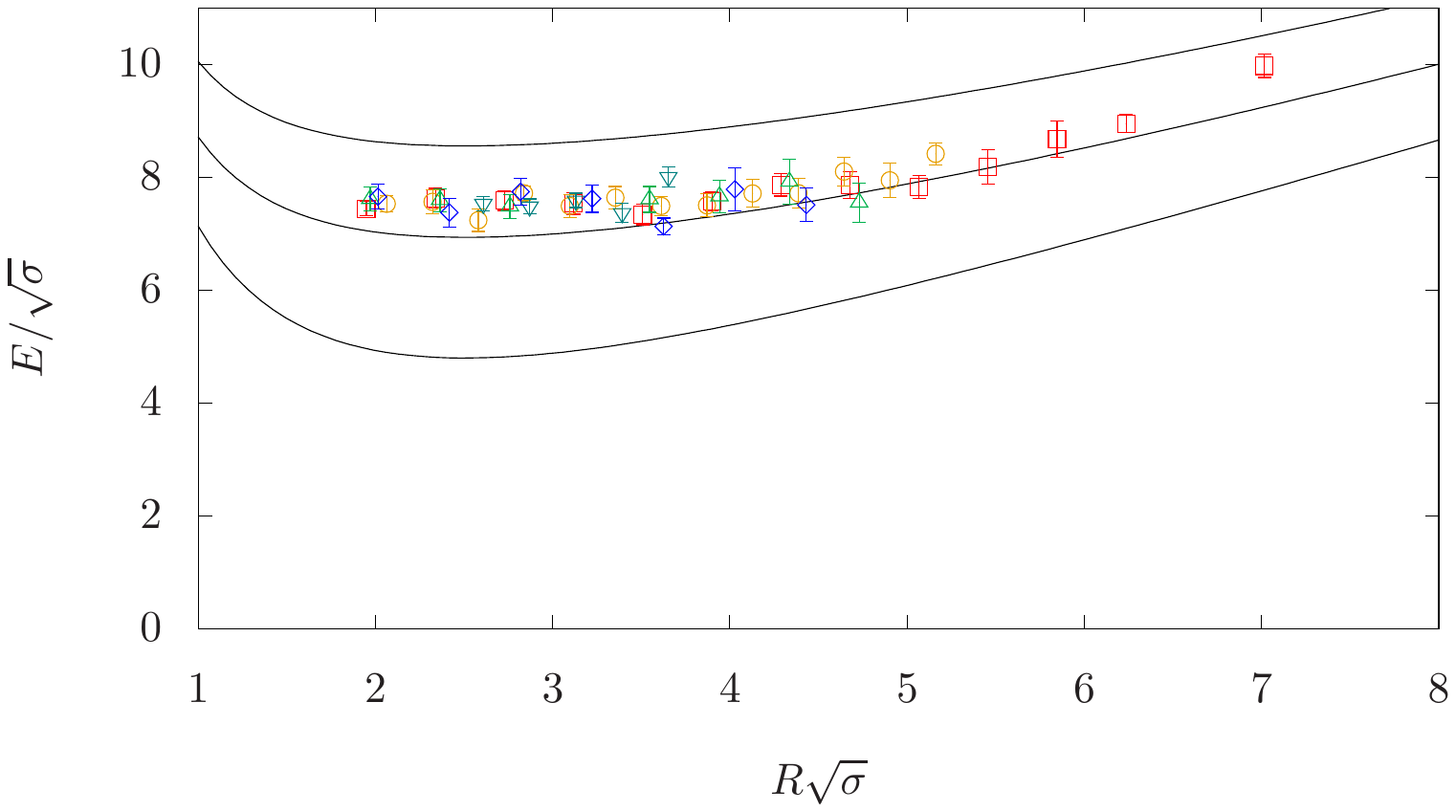}\put(-20,237){\scriptsize $N_L=1, N_R=0$}\put(-20,258){\scriptsize $N_L=2,N_R=1$}\put(-20,275){\scriptsize $N_L=3, N_R=2$}}
 
 \vspace{-3.0cm}
 \rotatebox{0}{\hspace{-2.0cm}\includegraphics[width=15cm]{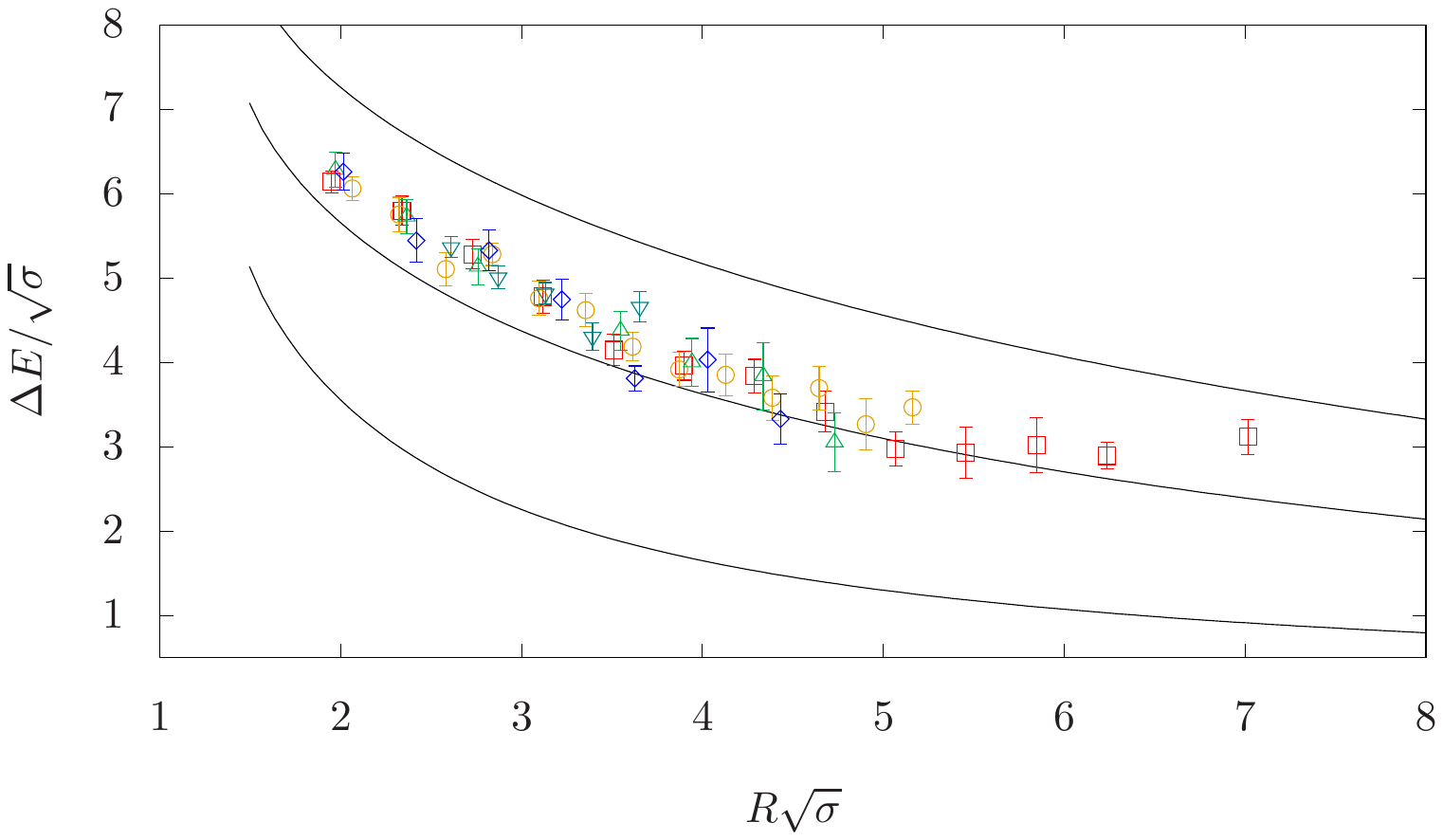}\put(-20,108){\scriptsize $N_L=1, N_R=0$}\put(-20,140){\scriptsize $N_L=2,N_R=1$}\put(-20,165){\scriptsize $N_L=3, N_R=2$}}

 \vspace{-1.0cm}
\caption{\label{fig:plot_J2_Pp+_Q1} Results for the ground state of the confining string with quantum numbers $2^+$, $q=1$. On the upper plot we visualize the energy $E/\sqrt{\sigma}$ while on the lower plot the energy minus the ground GGRT level $\Delta E/\sqrt{\sigma}$. The representation for the different gauge groups goes as follows: $SU(3)$, $\beta=6.0625$ is represented by $\square$, $SU(3)$, $\beta=6.338$ by $\circ$, $SU(5)$, $\beta=17.630$ by $\triangle$, $SU(5)$, $\beta=18.375$ by $\triangledown$ and $SU(6)$, $\beta=25.55$ by  $\diamond$.}
  \end{center}
\end{figure}

\begin{figure}[htb]
  \begin{center} 
  \vspace{-2.0cm}
 \rotatebox{0}{\hspace{-2.0cm}\includegraphics[width=15cm]{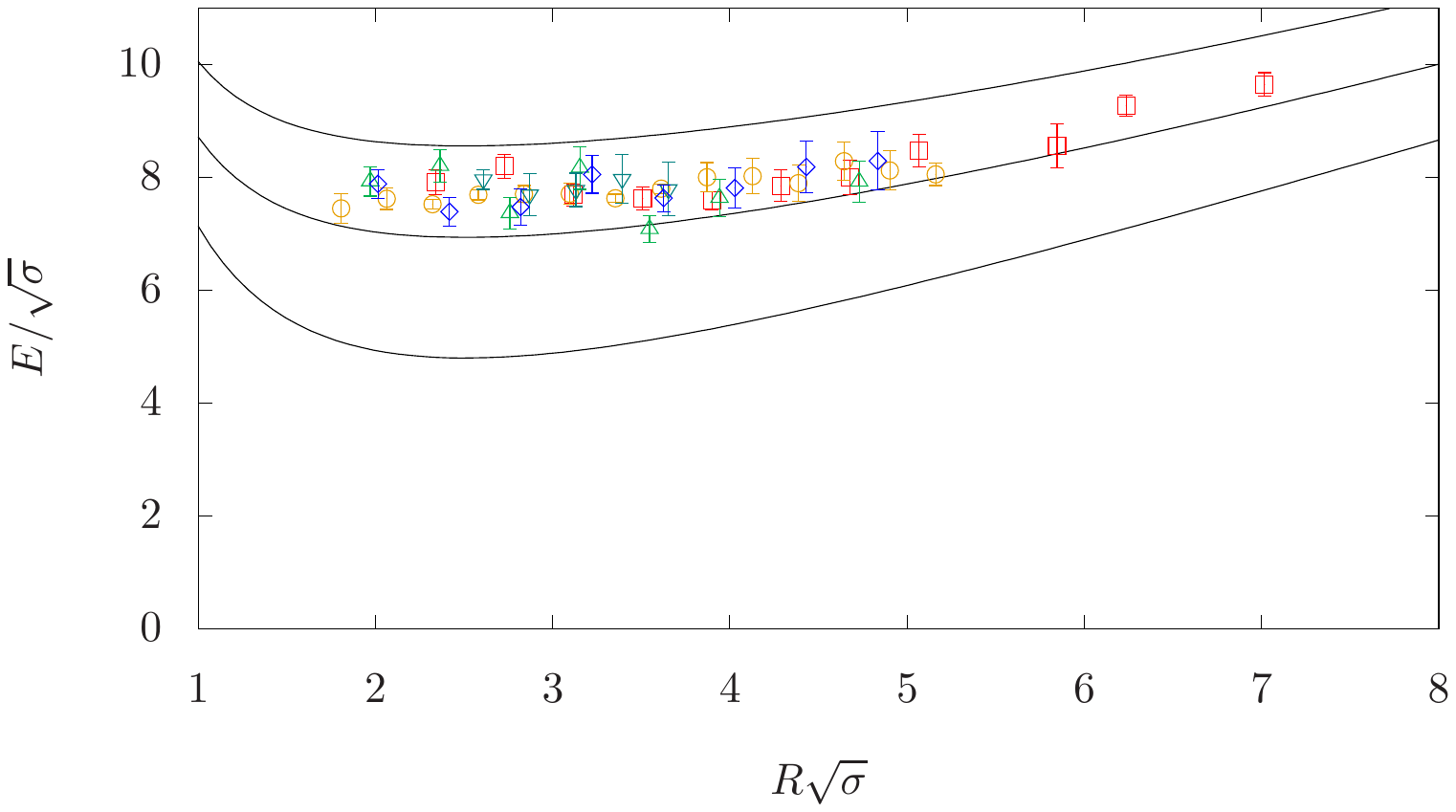}\put(-20,237){\scriptsize $N_L=1, N_R=0$}\put(-20,258){\scriptsize $N_L=2,N_R=1$}\put(-20,275){\scriptsize $N_L=3, N_R=2$}}
 
 \vspace{-3.0cm}
 \rotatebox{0}{\hspace{-2.0cm}\includegraphics[width=15cm]{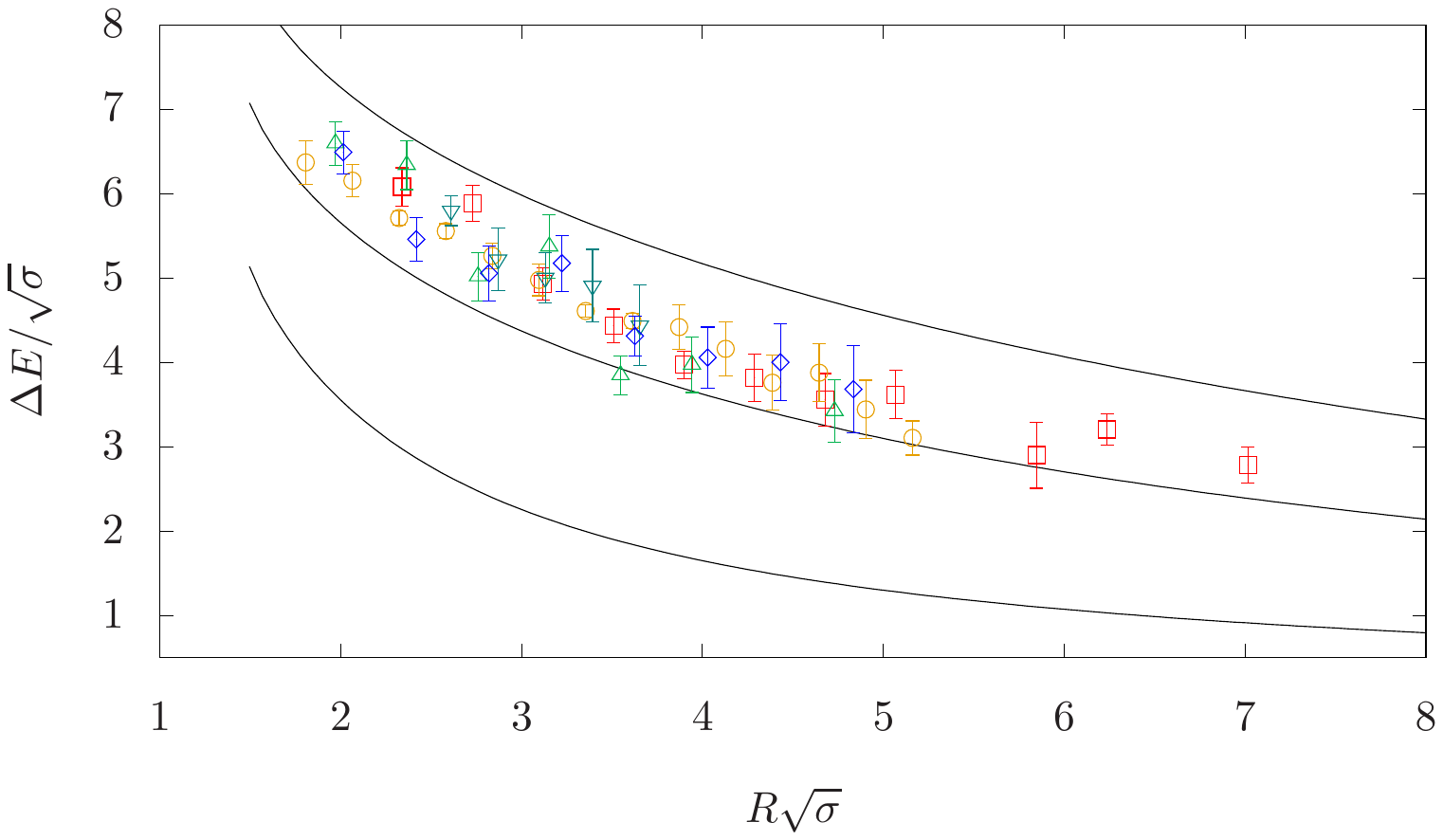}\put(-20,108){\scriptsize $N_L=1, N_R=0$}\put(-20,140){\scriptsize $N_L=2,N_R=1$}\put(-20,165){\scriptsize $N_L=3, N_R=2$}}

 \vspace{-1.0cm}
\caption{\label{fig:plot_J2_Pp-_Q1} Results for the ground state of the confining string with quantum numbers $2^-$, $q=1$. On the upper plot we visualize the energy $E/\sqrt{\sigma}$ while on the lower plot the energy minus the ground GGRT level $\Delta E/\sqrt{\sigma}$. The representation for the different gauge groups goes as follows: $SU(3)$, $\beta=6.0625$ is represented by $\square$, $SU(3)$, $\beta=6.338$ by $\circ$, $SU(5)$, $\beta=17.630$ by $\triangle$, $SU(5)$, $\beta=18.375$ by $\triangledown$ and $SU(6)$, $\beta=25.55$ by  $\diamond$.}
  \end{center}
\end{figure}

Finally, let us discuss the $2^{\pm}$, $q=1$ ground state.  These two states are expected to be approximated by the GGRT states with the oscillator content given by
\[
\left( a^{+}_{2} a^{+}_{-1}\pm a^{-}_{2} a^{-}_{-1} \right) | 0 \rangle\;.
\]
We presented the corresponding data in Figs.~\ref{fig:plot_J2_Pp+_Q1} and~\ref{fig:plot_J2_Pp-_Q1} for all five different data sets. The data are listed in Tables~\ref{tab:J2_Pp+_Q1_ground_state} and \ref{tab:J2_Pp-_Q1_ground_state}. Within the current accuracy, the data points do not appear to exhibit significant $N$ dependence or finite lattice effects. Our results demonstrate that the ground states deviate from the $N_L=2$, $N_R=1$  for $R \sqrt{\sigma} \leq 4$ and gradually approaches the GGRT prediction for longer strings. Also, within error bars, the two states are degenerate. As we will see later, the TBA analysis shows that the deviation from the GGRT spectrum is described, within error bars, by the Polchinski-Strominger interaction and the axion contribution~\eqref{res_phase} to the symmetric channel scattering phase shift, as illustrated in Fig.~\ref{fig:q1j2}.

Unfortunately, the present statistical accuracy does not allow the extraction of heavier states within this sector.

\subsection{Confining flux tube with $q=2$ longitudinal momentum}
\label{sec:results_for_q=2}
As a final set of data, let us  present our findings concerning excitations with two units of longitudinal momentum, $q=2$. Similarly to the preceding section, the quantum numbers that characterize these states  are the spin $J$ and the transverse parity $P_{\perp}$. 
The lowest energy $q=2$ excitations are expected to be the $N_L=2$, $N_R=0$ GGRT states. One finds five such states in Table~\ref{tab:table_NGstates_q2}.
First there is a two phonon $J=0$ and $P_{\perp}=+$ state of the form 
\[
 a^{+}_{1} a^{-}_{1} | 0 \rangle\;.
 \]
Second there is a doublet of single phonon $J^\pm$ states,
\[
\left( a^{+}_{2} \pm a^{-}_{2} \right) | 0 \rangle\;.
\]
Finally, there is a pair of two phonon $2^\pm$ states,
\[
\left( a^{+}_{1} {a}^{+}_{1}\pm a^{-}_{1}{a}^{-}_{1} \right) | 0 \rangle\;.
\]
Let us check whether the lattice data conforms with these predictions and what it say about excited states.

\subsubsection{$0^+$, $q=2$ states}
\begin{figure}[htb]
  \begin{center} 
  \vspace{-2.0cm}
 \rotatebox{0}{\hspace{-2.0cm}\includegraphics[width=15cm]{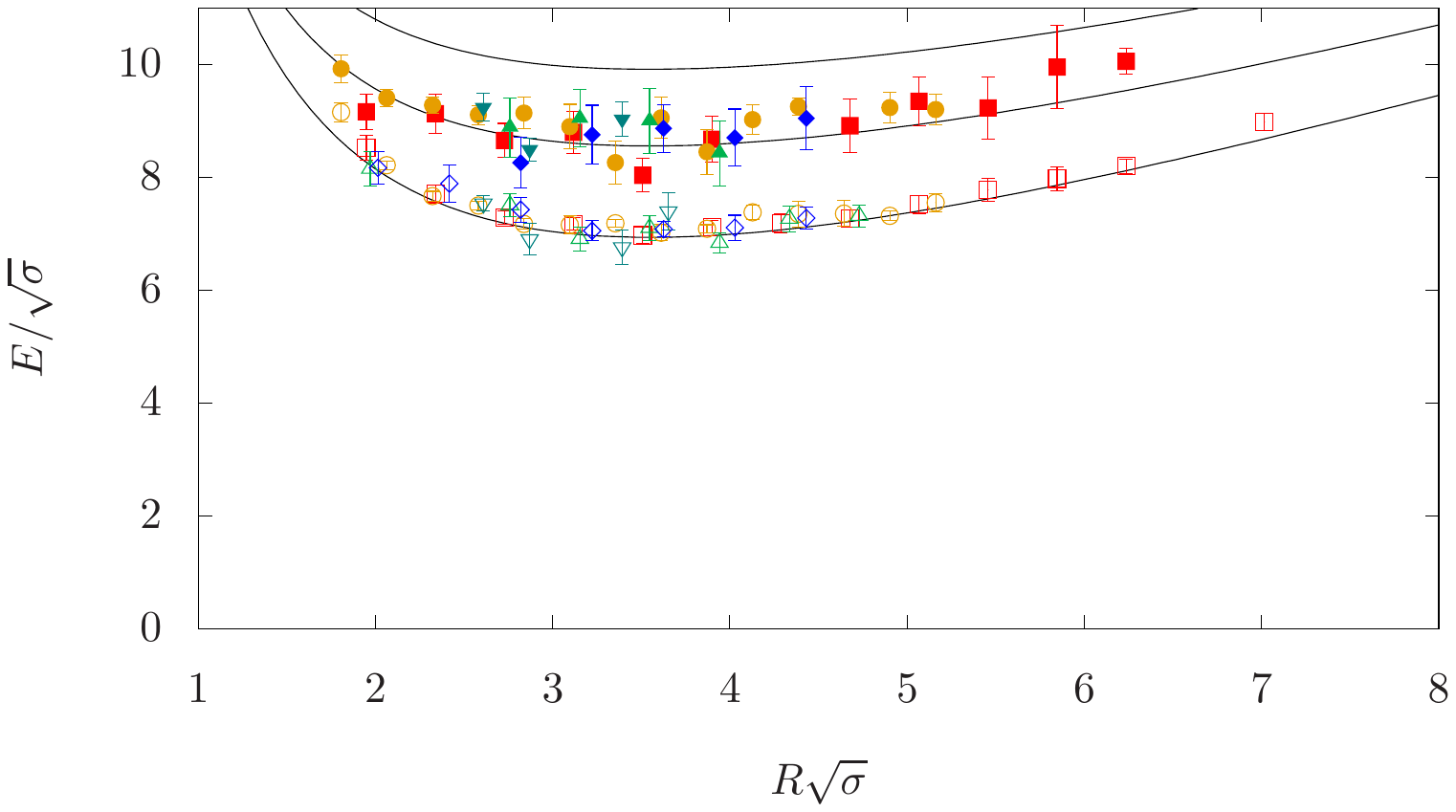}\put(-20,248){\scriptsize $N_L=2, N_R=0$}\put(-20,268){\scriptsize $N_L=3,N_R=1$}}
 
 \vspace{-3.0cm}
 \rotatebox{0}{\hspace{-2.0cm}\includegraphics[width=15cm]{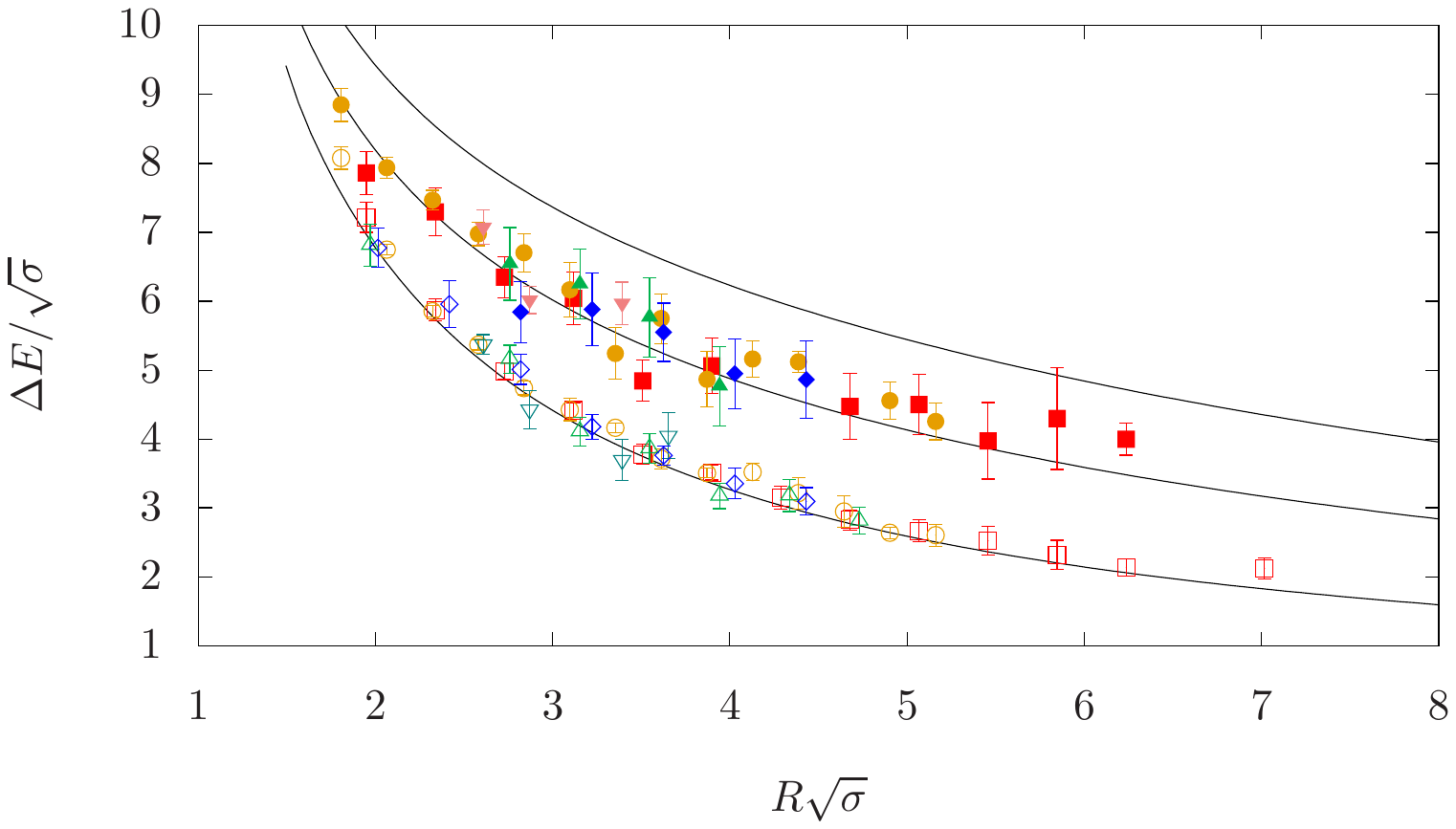}\put(-20,115){\scriptsize $N_L=2, N_R=0$}\put(-20,138){\scriptsize $N_L=3,N_R=1$}\put(-20,160){\scriptsize $N_L=4, N_R=2$}}

 \vspace{-1.0cm}
\caption{\label{fig:plot_J0_Pp+_Q2} Results for the confining string with quantum numbers $0^+$,  $q=2$. On the upper plot we visualize the energy $E/\sqrt{\sigma}$ while on the lower plot the energy minus the ground GGRT level $\Delta E/\sqrt{\sigma}$. The representation of the different gauge groups goes as follows: $SU(3)$, $\beta=6.0625$ is represented by $\square$ ($\blacksquare$), $SU(3)$, $\beta=6.338$ by $\circ$ ($\bullet$), $SU(5)$, $\beta=17.630$ by $\triangle$ ($\blacktriangle$), $SU(5)$, $\beta=18.375$ by $\triangledown$ ($\blacktriangledown$) and $SU(6)$, $\beta=25.55$ by  $\diamond$ ($\blacklozenge$) for ground (first excited) state.}
  \end{center}
\end{figure}

Energies of observed $0^+$, $q=2$ states are presented in Fig.~\ref{fig:plot_J0_Pp+_Q2} for all five different data sets. The data are summarized in Tables~\ref{tab:J0_Pp+_Q2_ground_state},\ref{tab:J0_Pp+_Q2_first_excited_state}. The corresponding ground state 
 exhibits a good agreement with the $N_L=2$, $N_R=0$ GGRT level. This agreement holds for long down to short strings with $R \sqrt{\sigma} \leq 2$. Shifting out attention to the first excited state in this sector , we observe that the extracted data have an approximate agreement with next GGRT level characterized by $N_L=3$ and $N_R=1$. We do not observe any states with massive excitations in this sector. 

\subsubsection{$0^-$, $q=2$ states}
\begin{figure}[htb]
  \begin{center} 
  \vspace{-2.0cm}
 \rotatebox{0}{\hspace{-2.0cm}\includegraphics[width=15cm]{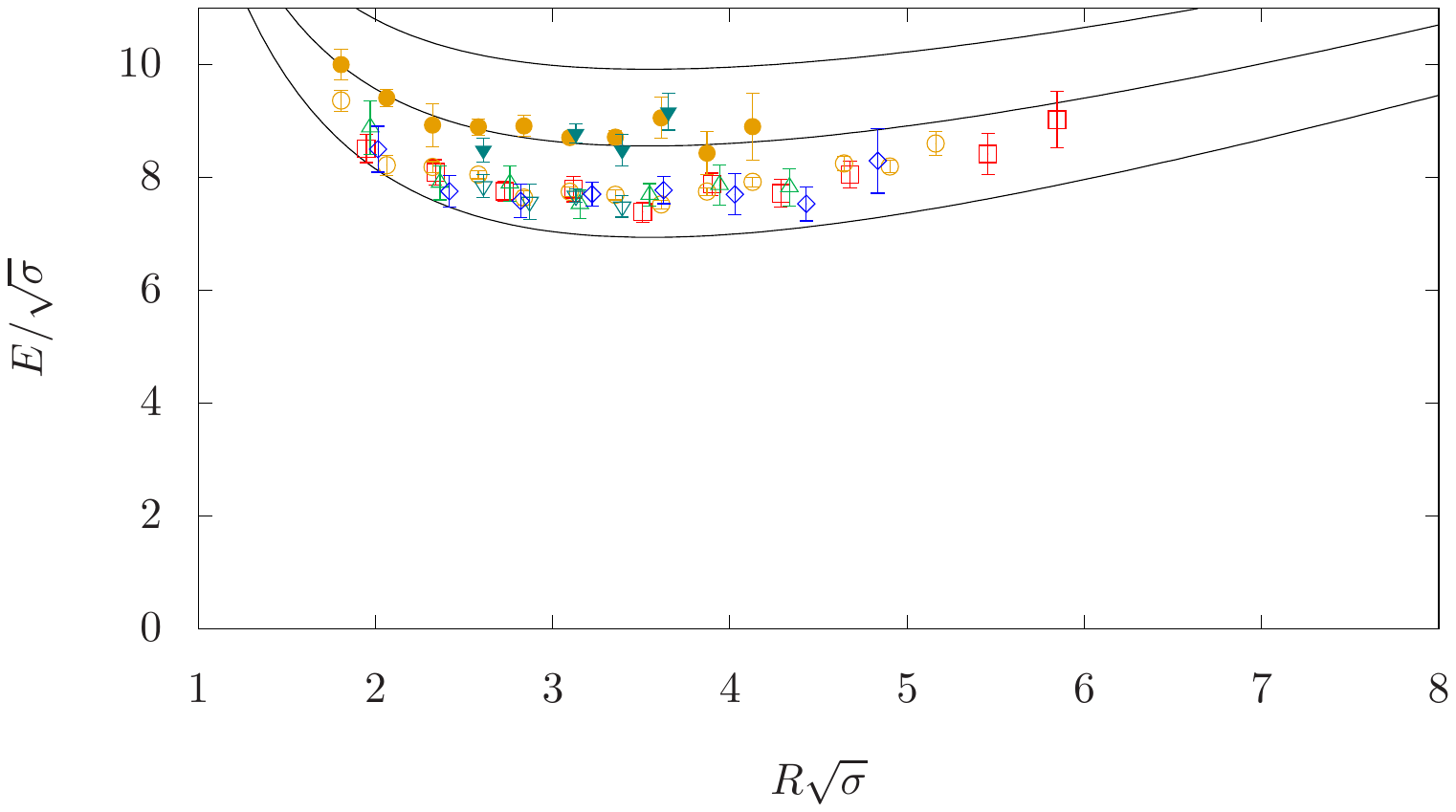}\put(-20,248){\scriptsize $N_L=2, N_R=0$}\put(-20,268){\scriptsize $N_L=3,N_R=1$}}
 
 \vspace{-3.0cm}
 \rotatebox{0}{\hspace{-2.0cm}\includegraphics[width=15cm]{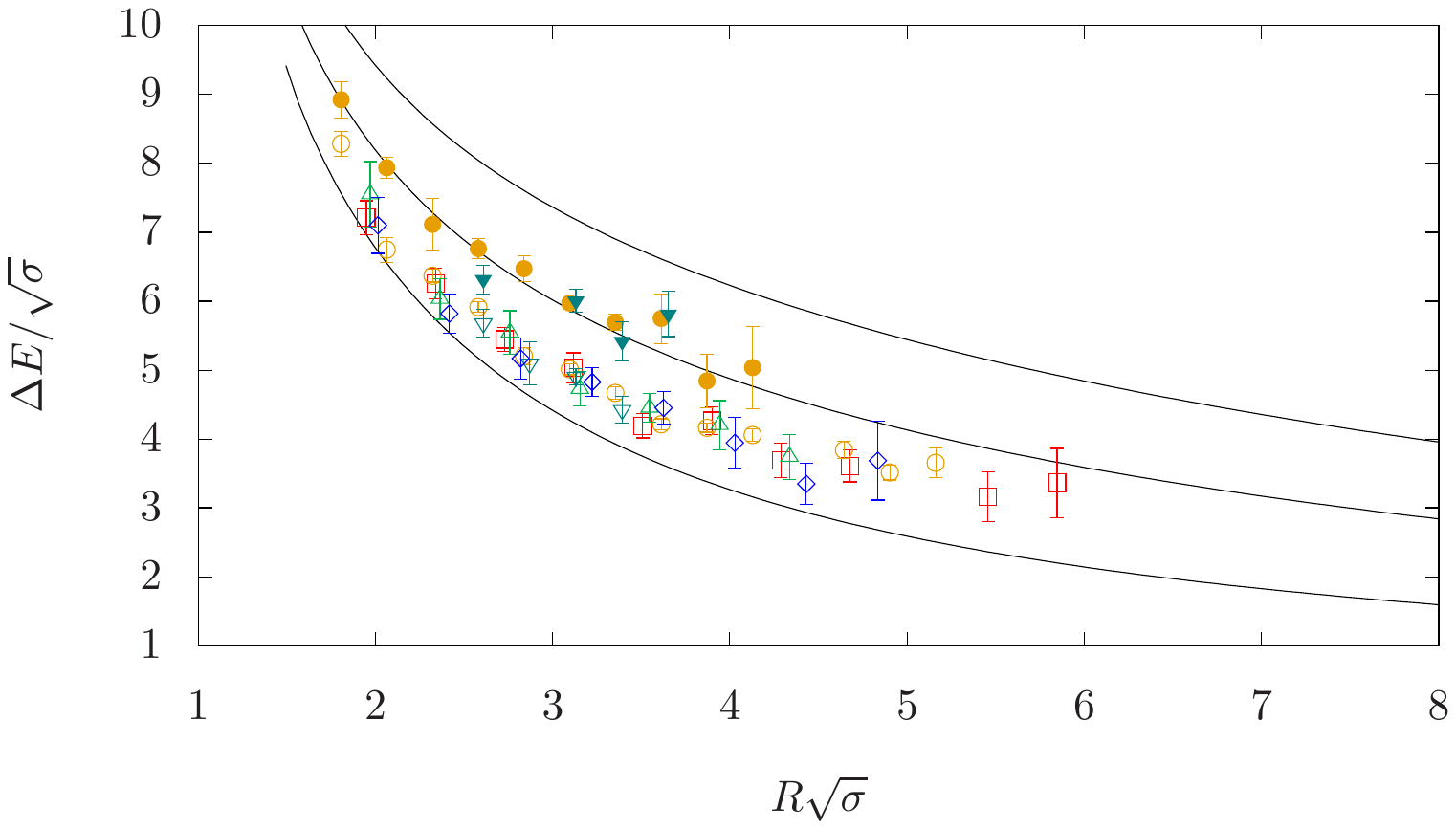}\put(-20,115){\scriptsize $N_L=2, N_R=0$}\put(-20,138){\scriptsize $N_L=3,N_R=1$}\put(-20,160){\scriptsize $N_L=4, N_R=2$}}

 \vspace{-1.0cm}
\caption{\label{fig:plot_J0_Pp-_Q2} Results for the confining string with quantum numbers $0^-$,  $q=2$. On the upper plot we visualize the energy $E/\sqrt{\sigma}$ while on the lower plot the energy minus the ground GGRT level $\Delta E/\sqrt{\sigma}$. The representation of the different gauge groups goes as follows: $SU(3)$, $\beta=6.0625$ is represented by $\square$ ($\blacksquare$), $SU(3)$, $\beta=6.338$ by $\circ$ ($\bullet$), $SU(5)$, $\beta=17.630$ by $\triangle$ ($\blacktriangle$), $SU(5)$, $\beta=18.375$ by $\triangledown$ ($\blacktriangledown$) and $SU(6)$, $\beta=25.55$ by  $\diamond$ ($\blacklozenge$) for ground (first excited) state.}
  \end{center}
\end{figure}

Turning now to the  $0^-$, $q=2$ states, Fig.~\ref{fig:plot_J0_Pp-_Q2} indicates  that there is no flux tube states associated with the GGRT  $N_L=2$, $N_R=0$ level.
This is in agreement with Table~\ref{tab:table_NGstates_q2}, which predicts the first GGRT $0^-$, $q=2$ state to appear at the $N_L=3$, $N_R=1$ level (in fact, there are three such states). The observed  ground state appears to exhibit large deviations from the  $N_R=3$, $N_L=1$ GGRT level. This is also expected given that in addition to the GGRT states there should also be  a boosted axion state similarly to what was observed at $q=1$. On the other hand, the first excited $0^-$, $q=2$ state agrees well with the $N_R=3$, $N_L=1$ GGRT level. Unfortunately, the second excited state cannot be extracted due to statistical limitations. The data are summarized in Tables~\ref{tab:J0_Pp-_Q2_ground_state},\ref{tab:J0_Pp-_Q2_excited_state}. 
We present the TBA analysis of these states in  Section~\ref{subsec:q2computation}. 

\subsubsection{$1$, $q=2$ states}
\begin{figure}[htb]
  \begin{center} 
  \vspace{-2.0cm}
 \rotatebox{0}{\hspace{-2.0cm}\includegraphics[width=15cm]{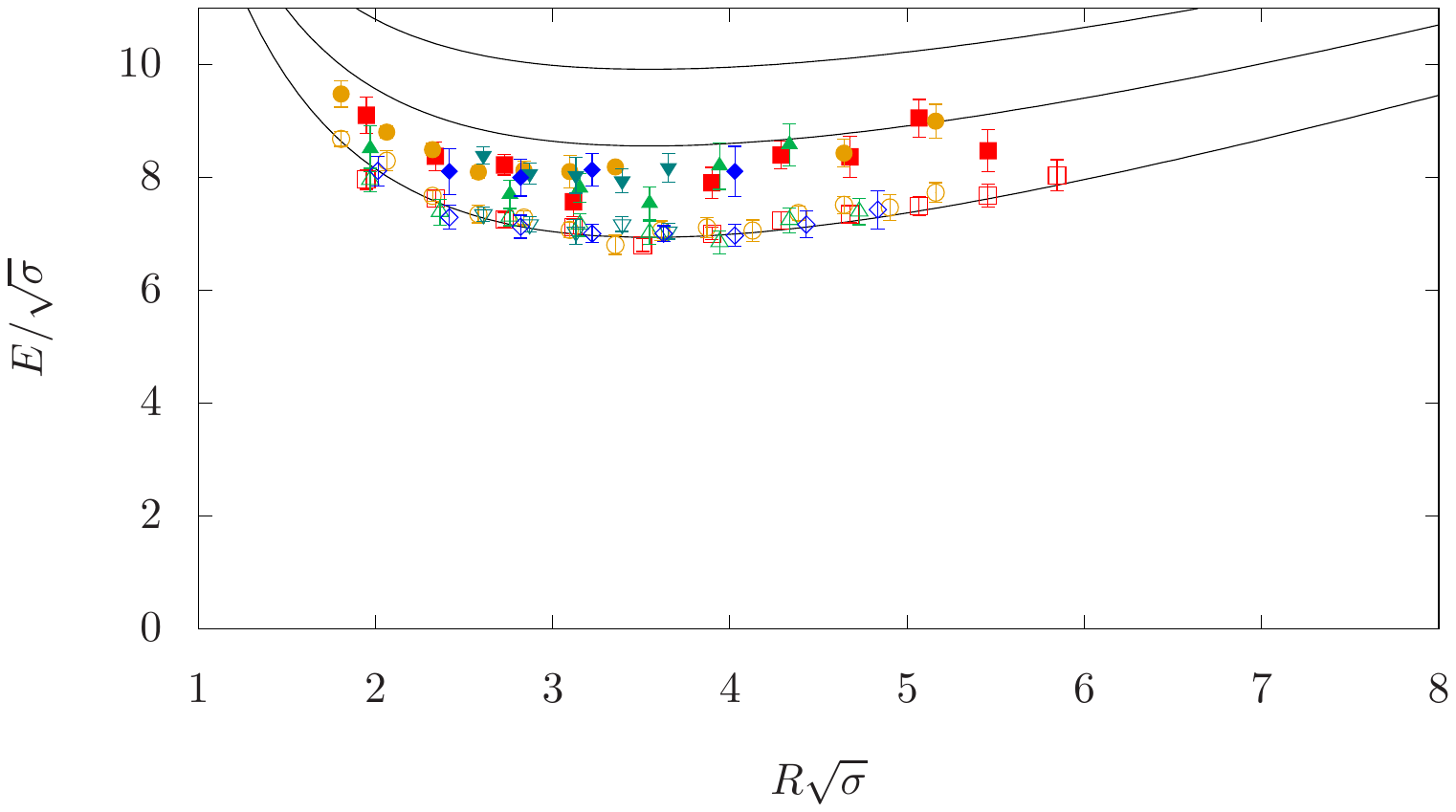}\put(-20,248){\scriptsize $N_L=2, N_R=0$}\put(-20,268){\scriptsize $N_L=3,N_R=1$}}
 
 \vspace{-3.0cm}
 \rotatebox{0}{\hspace{-2.0cm}\includegraphics[width=15cm]{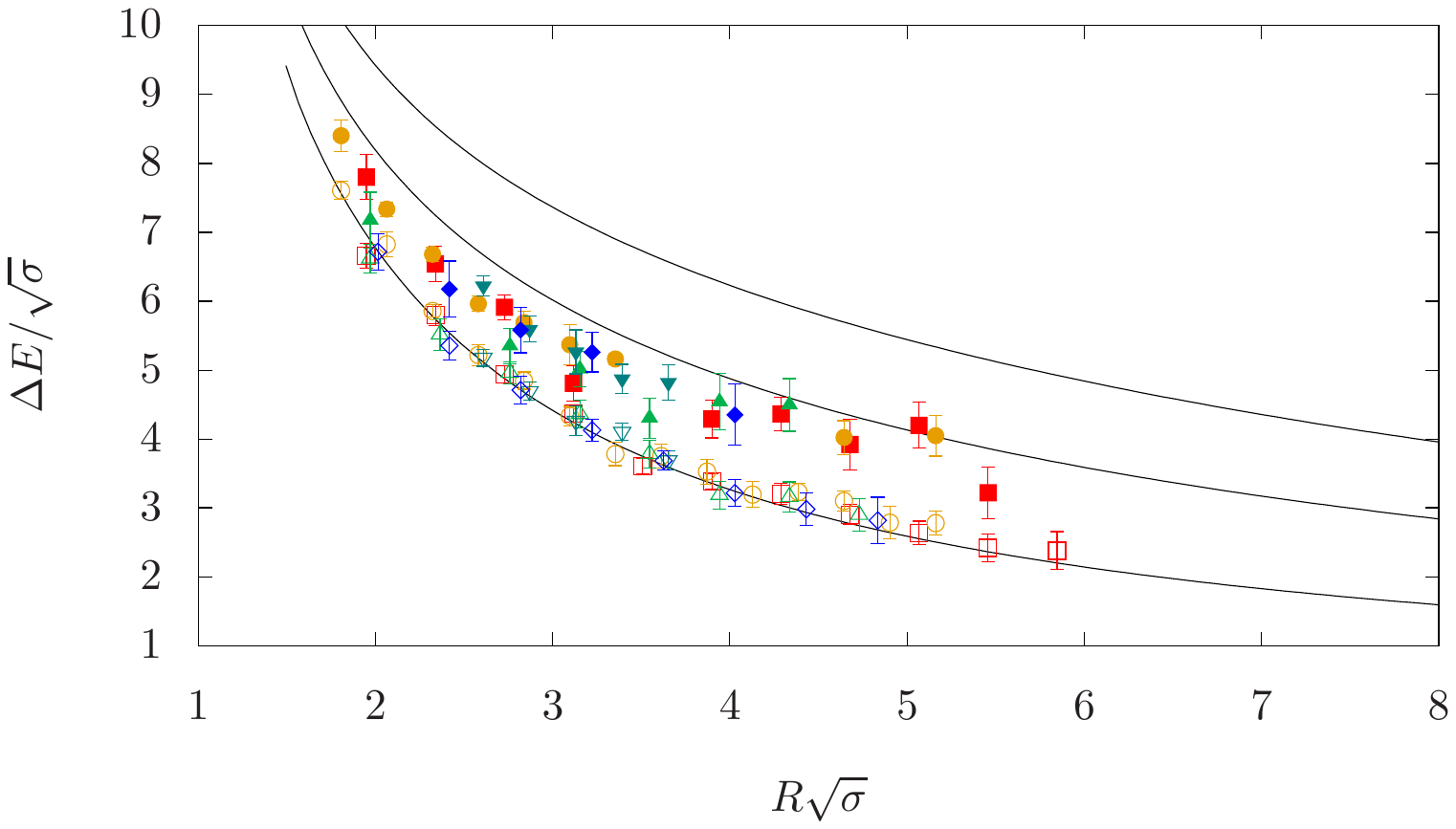}\put(-20,115){\scriptsize $N_L=2, N_R=0$}\put(-20,138){\scriptsize $N_L=3,N_R=1$}\put(-20,160){\scriptsize $N_L=4, N_R=2$}}

 \vspace{-1.0cm}
\caption{\label{fig:plot_J1_Q2} Results for the confining string with quantum numbers $|J|=1$, $q=2$. On the upper plot we visualize the energy $E/\sqrt{\sigma}$ while on the lower plot the energy minus the ground GGRT level $\Delta E/\sqrt{\sigma}$. The representation of the different gauge groups goes as follows: $SU(3)$, $\beta=6.0625$ is represented by $\square$ ($\blacksquare$), $SU(3)$, $\beta=6.338$ by $\circ$ ($\bullet$), $SU(5)$, $\beta=17.630$ by $\triangle$ ($\blacktriangle$), $SU(5)$, $\beta=18.375$ by $\triangledown$ ($\blacktriangledown$) and $SU(6)$, $\beta=25.55$ by  $\diamond$ ($\blacklozenge$) for ground (first excited) state.}
  \end{center}
\end{figure}

$|J_{\rm mod4}|=1$, $q=2$ states are presented in Fig.~\ref{fig:plot_J1_Q2} and Tables~\ref{tab:J1_Q2_ground_state},\ref{tab:J1_Q2_first_excited_state}. Similar to the $J=0$, $P_{\perp}=+$ case, the $J=1$ ground state demonstrates an almost perfect agreement with the GGRT $N_L=2$, $N_R=0$ state for all five data sets. Also the data exhibits a clear gap between this ground state and the first excited one.
For the latter, the GGRT spectrum gives rise to four $N_L=3$, $N_R=1$ states with $|J_{\rm mod4}|=1$. Three of these states, namely, $\left( a^{+}_{2}a^{+}_{1} a^{-}_{-1} \pm  a^{-}_{2}a^{-}_{1}a^{+}_{-1} \right) | 0 \rangle$, $\left( a^{+}_{2}a^{-}_{1} a^{-}_{-1} \pm a^{-}_{2}a^{+}_{1} a^{+}_{-1}  \right) | 0 \rangle$ and $\left( a^{-}_{2}a^{+}_{1} a^{-}_{-1} \pm a^{+}_{2}a^{-}_{1} a^{+}_{-1}  \right) | 0 \rangle$, correspond to the continuum $J=1$ states, while the remaining one, $\left( a^{+}_{2}a^{+}_{1} a^{+}_{-1} \pm a^{-}_{2}a^{-}_{1} a^{-}_{-1}  \right) | 0 \rangle$, corresponds to $J=3$. In addition, one also expects to find a state with a massive axion added to the $1$, $q=2$ ground state.
The observed first excited state exhibits a considerable deviation from the $N_L=3$ and $N_R=1$ GGRT level for $R \sqrt{\sigma} \leq 4.5$. The TBA analysis presented in 
Section~\ref{subsec:q2computation} indicates that this is indeed an axionic state.

\subsubsection{$2^\pm$, $q=2$ states}
\begin{figure}[htb]
  \begin{center} 
  \vspace{-2.0cm}
 \rotatebox{0}{\hspace{-2.0cm}\includegraphics[width=15cm]{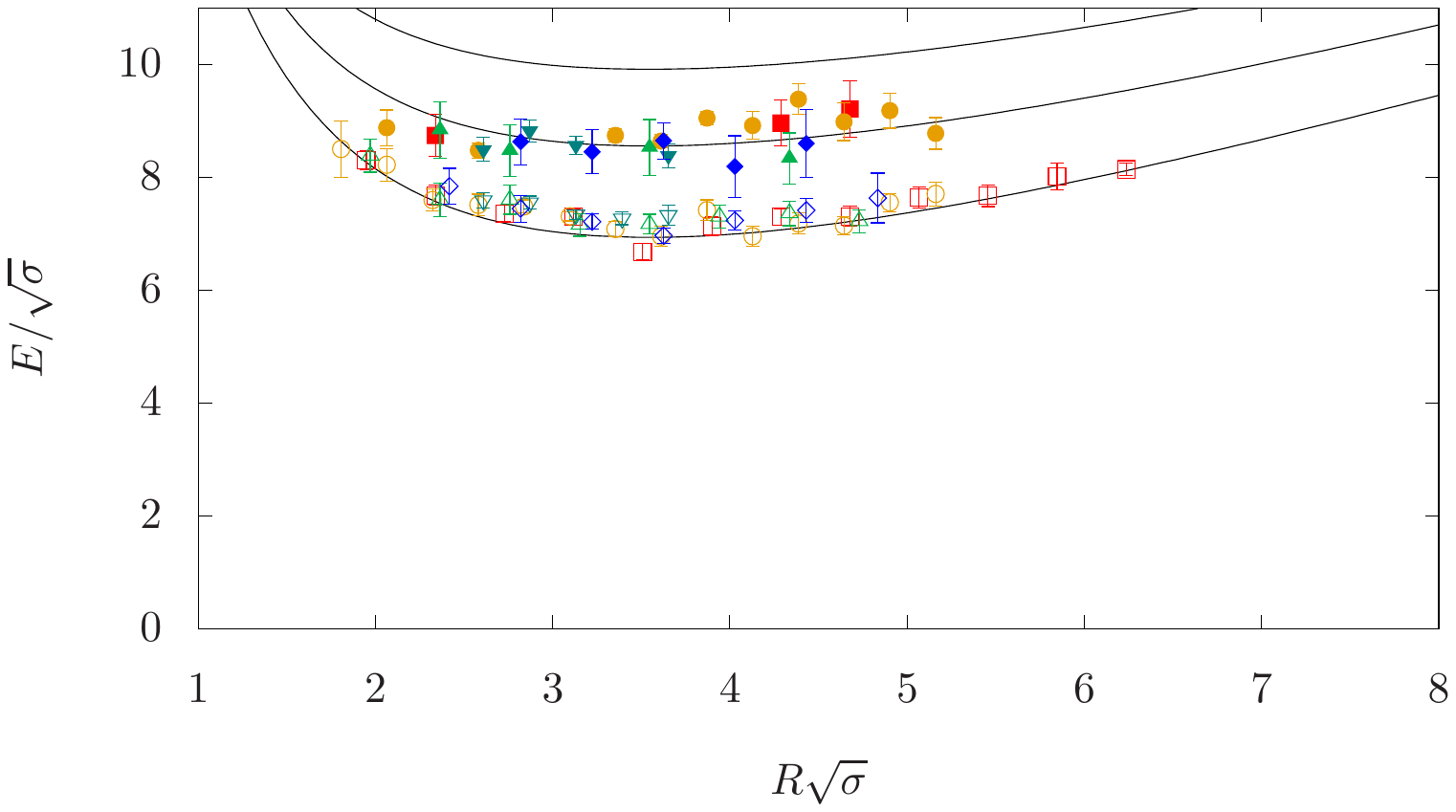}\put(-20,248){\scriptsize $N_L=2, N_R=0$}\put(-20,268){\scriptsize $N_L=3,N_R=1$}}
 
 \vspace{-3.0cm}
 \rotatebox{0}{\hspace{-2.0cm}\includegraphics[width=15cm]{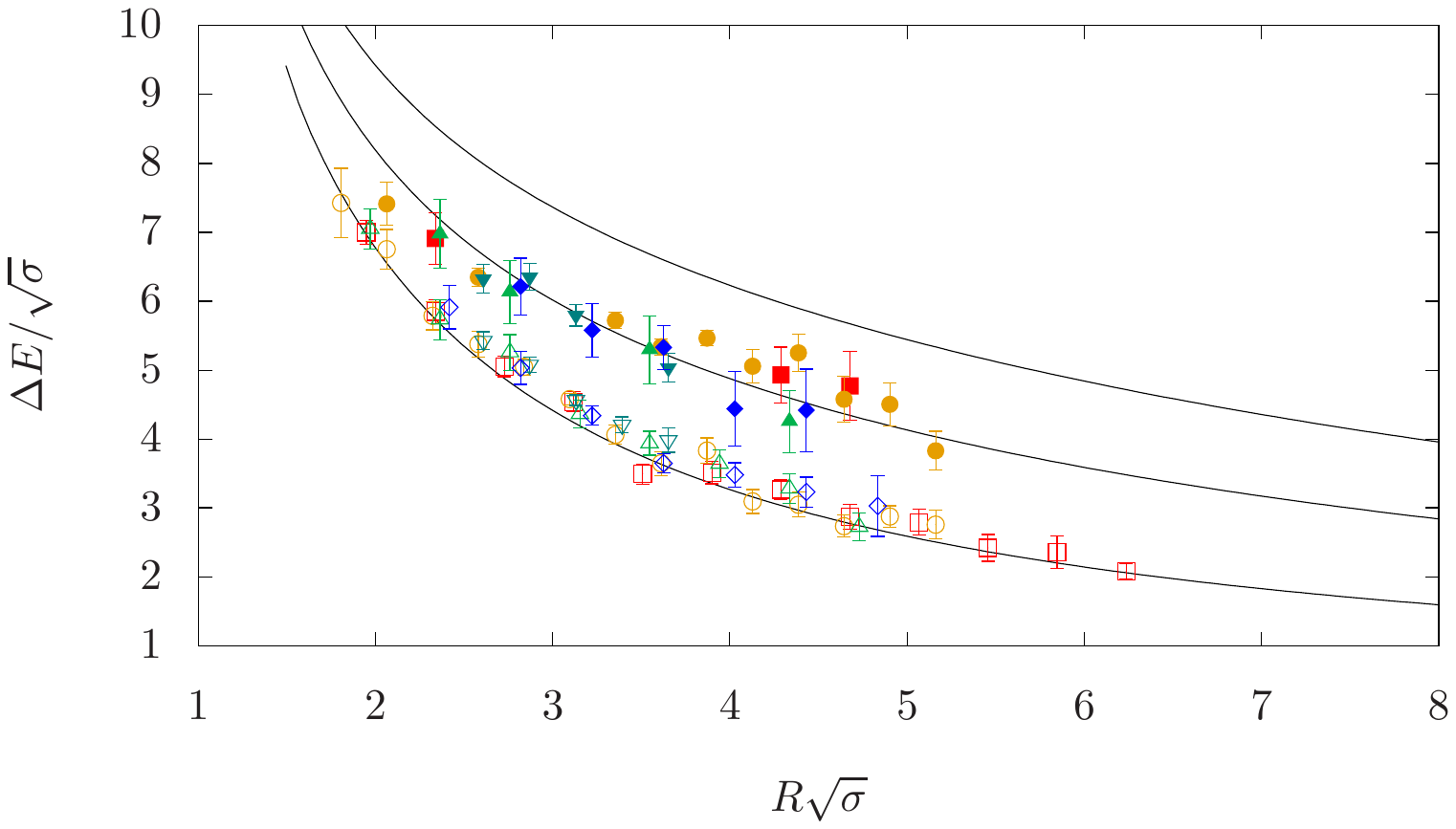}\put(-20,115){\scriptsize $N_L=2, N_R=0$}\put(-20,138){\scriptsize $N_L=3,N_R=1$}\put(-20,160){\scriptsize $N_L=4, N_R=2$}}

 \vspace{-1.0cm}
\caption{\label{fig:plot_J2_Pp+_Q2} Results for the confining string with quantum numbers $2^+$,  $q=2$. On the upper plot we visualize the energy $E/\sqrt{\sigma}$ while on the lower plot the energy minus the ground GGRT level $\Delta E/\sqrt{\sigma}$. The representation of the different gauge groups goes as follows: $SU(3)$, $\beta=6.0625$ is represented by $\square$ ($\blacksquare$), $SU(3)$, $\beta=6.338$ by $\circ$ ($\bullet$), $SU(5)$, $\beta=17.630$ by $\triangle$ ($\blacktriangle$), $SU(5)$, $\beta=18.375$ by $\triangledown$ ($\blacktriangledown$) and $SU(6)$, $\beta=25.55$ by  $\diamond$ ($\blacklozenge$) for ground (first excited) state.}
  \end{center}
\end{figure}

\begin{figure}[htb]
  \begin{center} 
  \vspace{-2.0cm}
 \rotatebox{0}{\hspace{-2.0cm}\includegraphics[width=15cm]{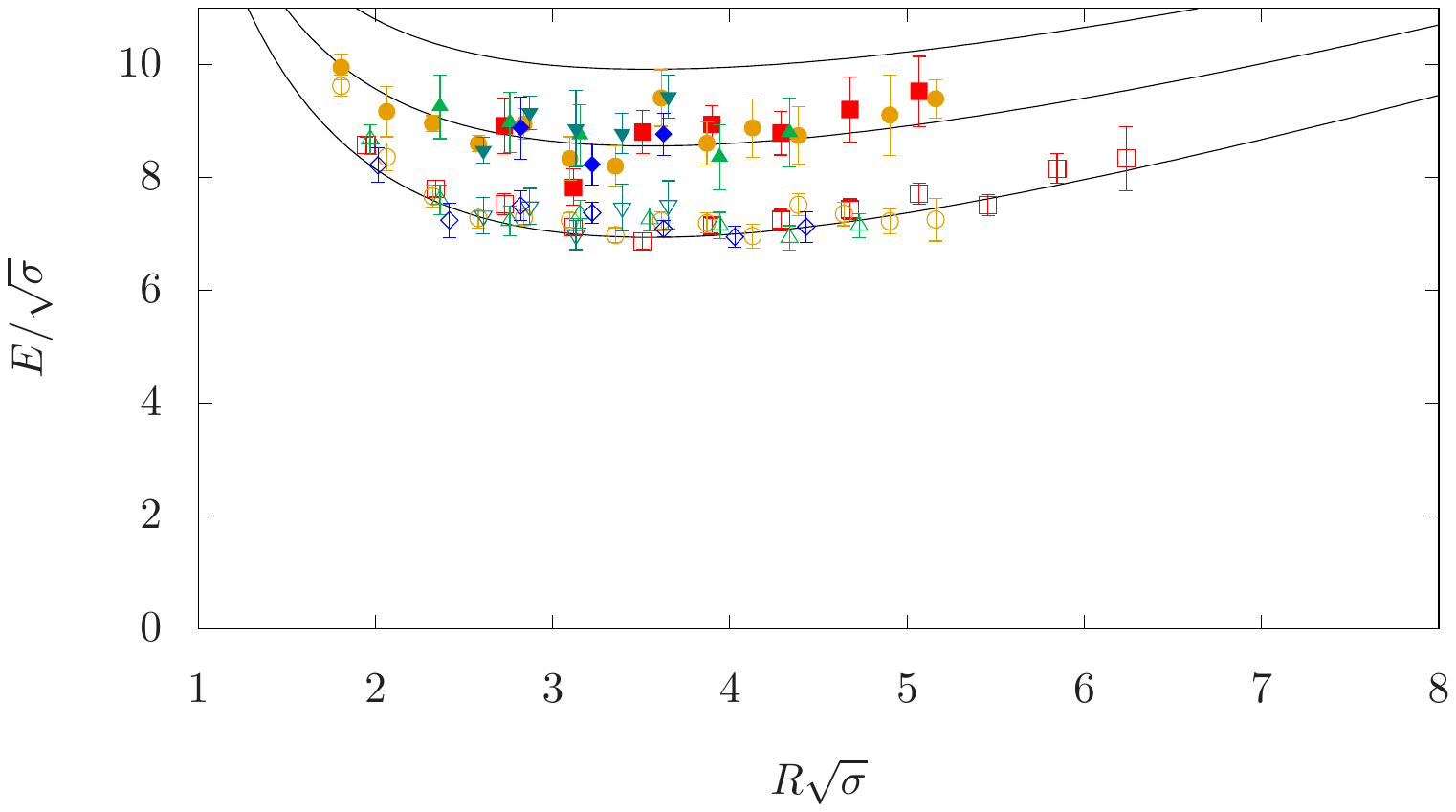}\put(-20,248){\scriptsize $N_L=2, N_R=0$}\put(-20,268){\scriptsize $N_L=3,N_R=1$}}
 
 \vspace{-3.0cm}
 \rotatebox{0}{\hspace{-2.0cm}\includegraphics[width=15cm]{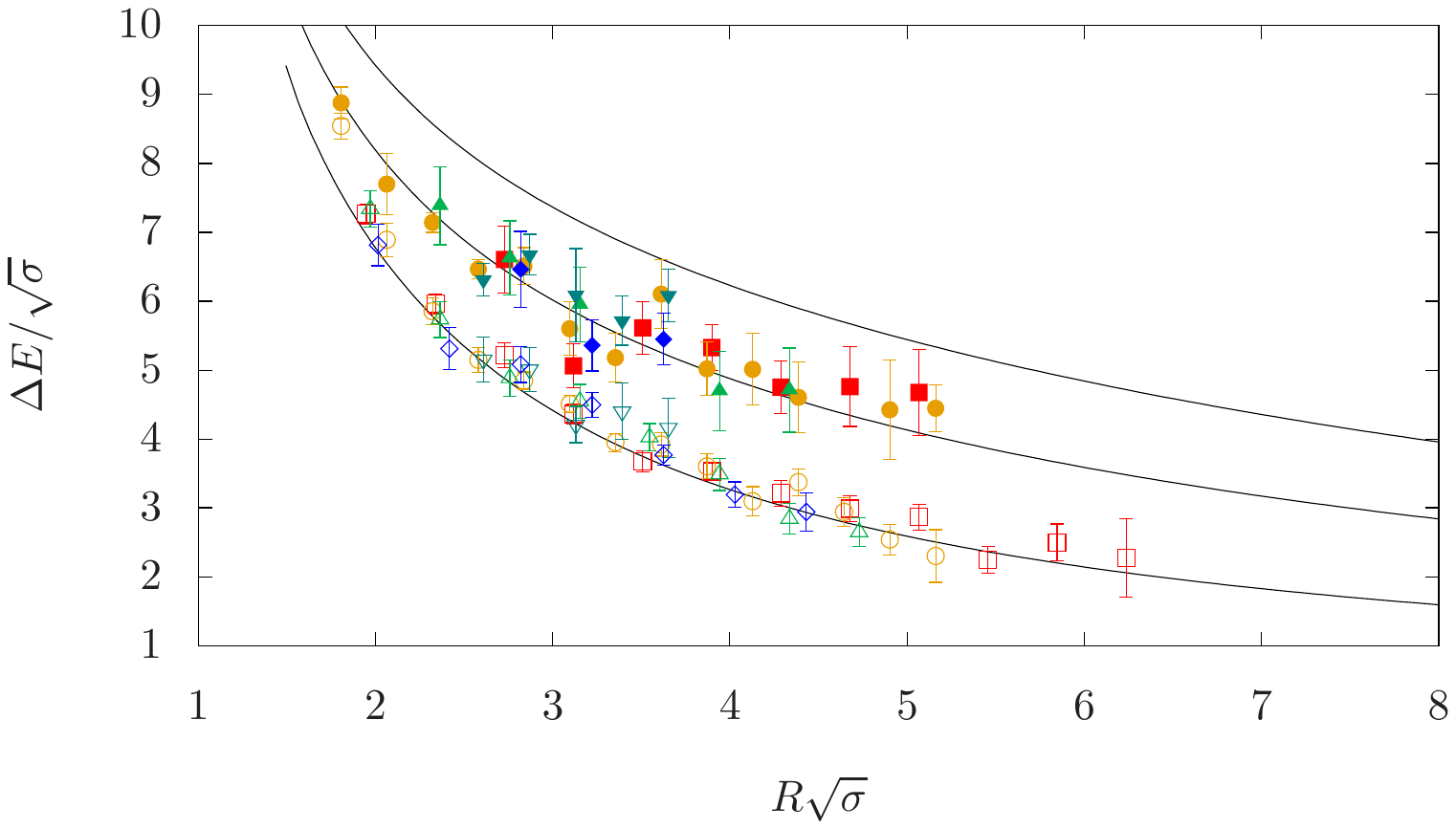}\put(-20,115){\scriptsize $N_L=2, N_R=0$}\put(-20,138){\scriptsize $N_L=3,N_R=1$}\put(-20,160){\scriptsize $N_L=4, N_R=2$}}

 \vspace{-1.0cm}
\caption{\label{fig:plot_J2_Pp-_Q2} Results for the confining string with quantum numbers $2^-$,  $q=2$. On the upper plot we visualize the energy $E/\sqrt{\sigma}$ while on the lower plot the energy minus the ground GGRT level $\Delta E/\sqrt{\sigma}$. The representation of the different gauge groups goes as follows: $SU(3)$, $\beta=6.0625$ is represented by $\square$ ($\blacksquare$), $SU(3)$, $\beta=6.338$ by $\circ$ ($\bullet$), $SU(5)$, $\beta=17.630$ by $\triangle$ ($\blacktriangle$), $SU(5)$, $\beta=18.375$ by $\triangledown$ ($\blacktriangledown$) and $SU(6)$, $\beta=25.55$ by  $\diamond$ ($\blacklozenge$) for ground (first excited) state.}
  \end{center}
\end{figure}

The very final set of $q=2$ states are the $2^\pm$ states, presented in Figs.~\ref{fig:plot_J2_Pp+_Q2} and~\ref{fig:plot_J2_Pp-_Q2}. The data are summarized in Tables~\ref{tab:J2_Pp+_Q2_ground_state}-\ref{tab:J2_Pp-_Q2_first_excited_state}. We observe that the states of opposite parity agree with each other, as expected given that they belong to the same doublet in the continuum theory. In both cases the ground state exhibits a good agreement with 
the $N_L=2$, $N_R=0$ GGRT level. The first excited state agrees well with the $N_L=3$, $N_R=1$ GGRT state, although the TBA analysis presented in Section~\ref{subsec:q2computation} suggests that it contains also an admixture of a state with a massive axion.

\section{$T\bar{T}$ dressing analysis of the spectrum}
\label{sec:computation}

In this section, we provide a theoretical analysis of our Monte--Carlo data on the excitation spectrum of confining strings summarized in Section~\ref{sec:data}. 
The analysis is based on the TBA and $T\bar{T}$ dressing techniques which were briefly reviewed in Section~\ref{subsec:TBA}. We verify 
that all low-lying excited states of confining flux tube accessible in our lattice simulations can be described by an effective long string theory with an additional massive pseudoscalar (the worldsheet axion), in agreement with the ASA.   

More specifically, we performed computations of the excitation spectrum  for states that can be interpreted as string states with some number of  massive excitations and up to two additional phonon excitations. The main focus of this analysis is on testing whether the expected axionic states are indeed present in the data, and also on confirming that all ``anomalous" states ({\it i.e.}, the ones exhibiting pronounced deviations from the GGRT spectrum) can be understood in this way.

The computation uses as an input the $2 \to 2$ phonon scattering phase shifts extracted from two-particle states as discussed in Section~\ref{sec:subsec_spin0_computation}. No new parameters are introduced to describe other states, so the success of this approach is quite non-trivial.

We will focus on the gauge group $SU(3)$ at the coupling $\beta = 6.338$, because this case offers the best statistics for higher excited states. Also, as we saw in Section~\ref{sec:data}  the $N$-dependence and lattice spacing effects are not significant to affect our conclusions. In this section, we assume that the  $SU(3)$, $\beta = 6.338$ is considered unless indicated otherwise.

\subsection{$q=0$ two particle states: extraction of the  worldsheet axion parameters}
\label{sec:subsec_spin0_computation}

The absolute ground state of the confining flux tube is given by the ground state in the $0^{++}$ sector. It has no phonon excitations on the string, Its energy can be approximated using the TBA approach with the tree level NG phase shift, which results in the GGRT expression. Alternatively, it can also be described by the sum of the universal terms in the  $l_s/R$ expansion. With the current data, these two approaches produce practically indistinguishable  results. Then, as explained  in Section~\ref{subsection_groundq0},  by fitting the ground state data we can extract the string tension, 
\begin{equation}
    a^2 \sigma = a^2/\ell_s^2 =  0.01665(4) \,.
\end{equation}
In what follows, we will be expressing all  physical quantities in the physical (string) units. 

The ground states in $0^{--}$, $2^{++}$, $2^{-+}$ sectors and the first excited state in the $0^{++}$ sector are more nontrivial. From Table~\ref{tab:table_NGstates_q0} we  see that they correspond to the GGRT level $N_L=N_R=1$. However, as was first observed in \cite{Athenodorou:2010cs},  the corresponding lattice data exhibits  notable deviations from the GGRT formula~\eqref{GGRT}, especially for the $0^{--}$ state. As discussed in Section~\ref{subsec:Nl=Nr=1} these results are well reproduced with the current more precise data. 

This data can be described by
 the effective string theory~\eqref{effective_string} with an additional massive pseudoscalar (``worldsheet axion'') field\cite{Dubovsky:2013gi,Dubovsky:2014fma}.
 The leading axion-Goldstone interaction is given by~\eqref{axion_interaction}. The most direct evidence for the existence of the axion is provided  by the shape of the $0^{--}$ ground state. The corresponding energy is largely independent of the compactification radius $R$, indicative of a presence of a massive particle at rest on a string, as discussed in Section~\ref{subsec:Nl=Nr=1}.

 More generally, the axion shows up as a relatively sharp resonance in the pseudoscalar channel of the two-phonon scattering. Its dominant decay channel is into two phonons and the corresponding decay width is controlled by the coupling constant $Q_{\phi}$ in \eqref{axion_interaction}. It also contributes to the phase shifts in other channels through $t$-channel diagrams. Overall, the leading axion contribution into two phonon scattering phase shifts is given by the following expression \cite{Dubovsky:2014fma},
\begin{equation}
    2 \delta_{\rm res}(p)=2 \sigma_2 \tan ^{-1}\left(\frac{8 Q_{\phi}^2 \ell_s^4 p^6}{m^2-4 p^2}\right)+\sigma_1 \frac{8 Q_{\phi}^2 \ell_s^4 p^6}{m^2+4 p^2} \,,
    \label{res_phase}
\end{equation}
where $\sigma_1 = (-1,1,1)$, $\sigma_2 = (0,0,1)$ stand for singlet, symmetric and anti-symmetric channels. 
Recall that in terms of two particle $q=0$ phonon states summarized in Table~\ref{tab:table_NGstates_q0}, the singlet channel corresponds to $0^{++}$ states, 
the symmetric channel corresponds to  $2^{\pm+}$ states, and the pseudoscalar (or equivalently, antisymmetric) channel to $0^{--}$ states.
Using these phase shift combined with the universal Nambu-Goto contributions we can apply TBA or the 
$T\bar{T}$ dressing technique (as reviewed in Section~\ref{subsec:TBA}) to extract the finite volume spectrum of the corresponding excited states.

\begin{figure}[htb]
\scalebox{1}{\includegraphics[width=0.77\textwidth]{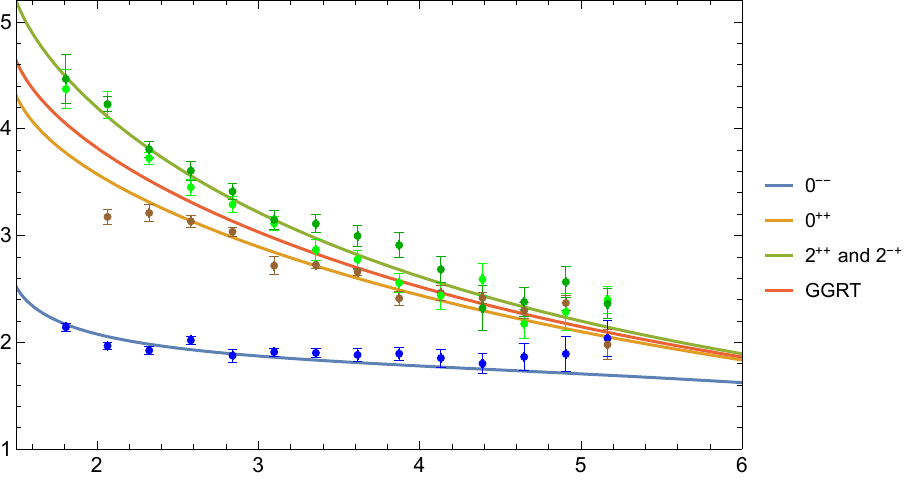}\put(-380,200){ \small $\Delta E\ell_s$}\put(-60,0){\small $R/\ell_s$}}
\centering
\caption{The energy at the level $N_L=N_R=1$ deducted by the GGRT ground state $\Delta E = E - E_{\rm GGRT}(0,0)$ as a function of the string length $R$. Here the solid lines represent the theoretical predictions of the spectrum of the two-phonon excitation states deriving from the $2 \to 2$ phonon anti-symmetric(blue), singlet(brown) and symmetric(green) scattering channels with the inclusion of a massive pseudo-scalar field, using $T\bar{T}$ deformation. The points with corresponding colors are the lattice data. There are two sectors with spin-2, where the light green dots are in $2^{++}$ sector, and the dark green dots are in $2^{-+}$ sector. The solid red line represents the $N_L=N_R=1$ GGRT spectrum. }
\label{fig:p0n1}
\end{figure}

By fitting the ground states in the $0^{--}$, $2^{++}$, $2^{-+}$ sectors and the first excited state in the $0^{++}$ sector one extracts two available free parameters---the axion mass $m$ and the coupling $Q_{\phi}$,
\begin{equation}
    m = 1.812(16) \ell_s^{-1}, \quad Q_{\phi} = 0.365(5) \,,
    \label{fitted_resonance}
\end{equation}  
which agrees well with previous results\cite{Dubovsky:2014fma,Chen:2018keo}\footnote{The convention in \cite{Dubovsky:2014fma} is related to our convention by $Q_{\phi} = \frac{\alpha}{8\pi}$. }. The best fit theoretical predictions and the lattice data are displayed in Fig.~\ref{fig:p0n1}\footnote{Note that everywhere in this section when plotting energies of the states we always subtract GGRT ground state, {\it i.e.} we plot $\Delta E = E - E_{\rm GGRT}(0,0)$. }. Note that the axion mass is mostly determine by the data in the pseudoscalar channel, while two other channels fix the $Q_{\phi} $
value.

We note that the errors presented here are purely statistical. 
 The presented best fit values correspond to $\chi^2/{\rm d.o.f.} = 1.86$. This $\chi^2$-value is somewhat large, which is mostly caused by the systematic splitting between two spin-2 states, which are supposed to be degenerate in the continuum limit.
To quantify the systematic error we performed separate fits using $2^{++}$ or $2^{-+}$ data only. The resulting best fit values are presented Table~\ref{table_fit_resonance}.
We included here also the corresponding results for the $SU(5)$ gauge group at $\beta = 17.63$ and for the $SU(6)$ gauge group. 
The systematic uncertainties for $N_c=5, 6$ are significantly larger because the corresponding lattice spacings are larger. 
We see that these systematic errors mostly affect the determination of the $Q_{\phi}$ coupling. Also we observe that  the resonance mass $m$ exhibits a mild $N_c$ dependence (which has been extrapolated to approach a non-vanishing finite value in the infinite $N_c$  limit  in~\cite{Athenodorou:2017cmw}).

We comment that another way to circumvent the systematic error originating from the lattice splitting, is to extract $Q_{\phi}$ using merely the scalar channel. With the current data, this does not lead to a significantly better result and we leave this approach for the future work. In principle, it should be computationally more manageable to perform higher precision Monte-Carlo computations for this state, than to go to finer lattice spacing.

\begin{table}[htb]
\begin{center}
\begin{tabular}{c|c|c|c}
\hline \hline & $S U(3)$ & $S U(5)$ & $SU(6)$ \\
\hline \hline $2^{++}$ & & & \\
$m \ell_s$ & $1.812(16) $ & $1.647(23) $ & $1.653(17) $ \\
$Q_{\phi}$ & $0.377(7) $ & $0.387(10)$ & $0.385(10)$ \\
\hline $2^{-+}$ & & & \\
$m \ell_s$ & $1.811(16)$ & $1.648(23) $ & $1.656(17)$ \\
$Q_{\phi}$ & $0.354(6)$ & $0.346(7) $ & $0.337(10)$ \\
\hline
\end{tabular}
\end{center}
\caption{The best fit values and statistical errors for the resonance mass $m$ and axion coupling $Q_{\phi}$ for different $N_c$ and different spin-2 channels. }
\label{table_fit_resonance}
\end{table}

Now we get back to the fitting result. Interestingly, within error bars, the extracted values of the axion coupling agree with the one  in the integrable model, given by (\ref{Qint}).
Furthermore, a pseudoscalar resonance with  the same value of mass and coupling is   
also found on one of the boundaries of the allowed region in the S-matrix bootstrap~\cite{EliasMiro:2019kyf}.

 From Fig.~\ref{fig:p0n1} we see also some other systematic errors affecting this theoretical analysis. First, the data are systematically higher than the prediction for very long strings $R \gtrsim 4 l_s$, which could be due to the over-estimation of the energy of very heavy states. 
 Second, the energy of the shortest  strings $R \lesssim 2.3 l_s$ in the $0^{++}$ sector is significantly below the theoretical prediction. One may be tempted to say that this originates from the effect of higher order corrections in the low energy phase shift. However, this interpretation is disfavored by  the data with $P_{\perp}=+$ and nonzero longitudinal momentum, which probes the same range of c.o.m energy
 $s$.  
 Also note that a similar effect has also been observed in 3D Yang-Mills theory~\cite{Dubovsky:2014fma}. It appears more likely to be due to some lattice effect, instead of the high momentum corrections in the effective string theory. Indeed, it is worth noticing that the  plateau for this data point ($R=16a$) is quite poor, which also explains
 why we do not have a data point at $R=14a$ in this sector. The same phenomenon happens also for $SU(5)$ $\beta = 6.0625$ data. It is unclear why this happens only in the $0^{++}$ sector and it will be good to address this puzzle in more detail in the future. 

\begin{figure}
    \centering
    \scalebox{1}{\includegraphics[width=0.65\linewidth]{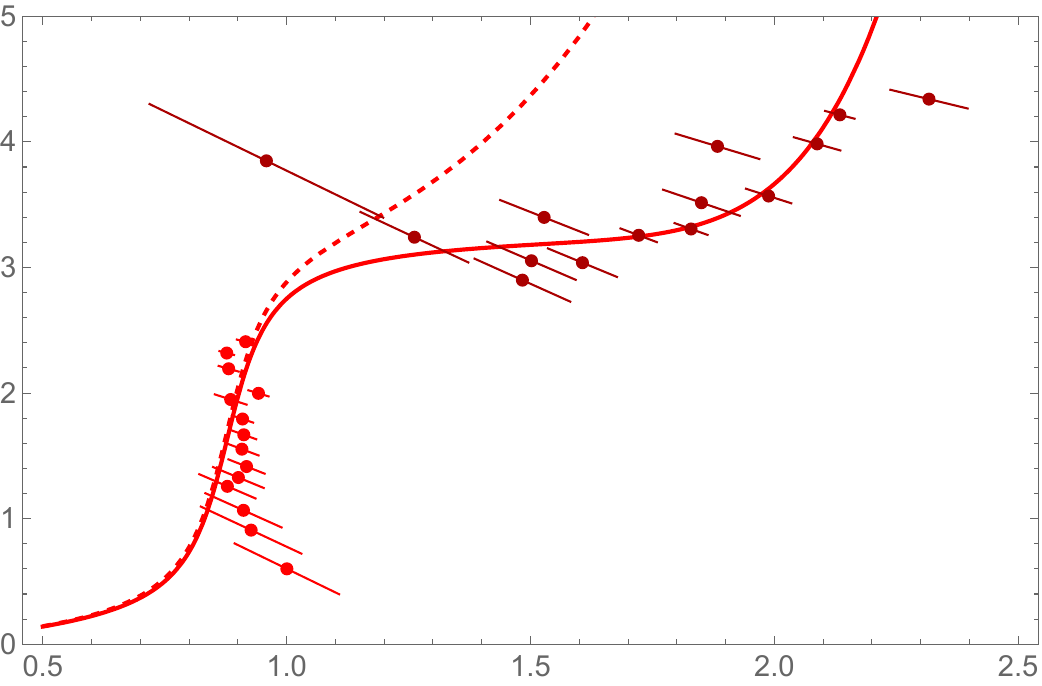}\put(-315,205){ \small $\delta$}\put(4,5){\small $p\ell_s$}}
    \caption{The phase shift of the pseudoscalar channel of the $2 \to 2$ scattering. The solid and dashed red line are the predictions with and without the higher order corrections in the low energy expansion. The data for the ground state (bright red) and the first excited state (dark red) are obtained from TBA. }
    \label{fig:0mmphase}
\end{figure}

It is straightforward to extend this analysis to include also the first excited state in the $0^{--}$ sector. This state probes the pseudoscalar phase shift above the resonance.
The corresponding data are shown in Fig.~\ref{fig:0mmenergy} as dark red dots. This data can still be analyzed with the TBA method using the $2 \to 2$ scattering phase shift, but the momenta of the phonons are higher, so we have to take into account higher order corrections in the low energy expansion. Even more appropriately, one may consider this analysis as a measurement of the scattering phase shift from the lattice data. Namely, we parametrize the phase shift as 
\begin{equation}
    2\delta_{\rm anti}(s) = \frac{s}{4} + \frac{11 s^2}{192 \pi} + 2\delta_{\rm res,anti} + \frac{a_3}{(2\pi)^2} \left( \frac{s}{4} \right)^3 + \frac{a_4}{(2\pi)^3} \left( \frac{s}{4} \right)^4 + \cdots \,,
    \label{phase_expand}
\end{equation} 
and determine the $a_3$, $a_4$ coefficients by fitting the finite volume spectrum using the ABA with dressing.
The best fit values are
\begin{equation}
    a_3 = -12.2(5), \quad a_4 = 11.0(9) \,.
    \label{higher_order}
\end{equation}
and the corresponding finite volume energy  is shown by the dark red solid line in Fig.~\ref{fig:0mmenergy}. Alternatively, one may directly extract the phase shift from the TBA equations, with the result shown in Fig.~\ref{fig:0mmphase}.

Note that most likely it would be inappropriate to identify the fitted coefficients $a_3$, $a_4$ with the non-universal Wilson coefficients in the effective string action~\eqref{effective_string}. Indeed, the fit is performed at high momenta above the resonance mass, where the effective string theory is not expected to be applicable.
Instead this should be treated as a phenomenological fit that determines the phase shift in the whole 
$s$-range where we have the Monte-Carlo data.

Extending the TBA analysis to even further higher excited states in $0^{--}$ and other symmetry channels is rather challenging, because one expects significant mixing between states with different number of phonon excitations due to the lack of integrability. To get around this challenge we adopt the following somewhat ad hoc approach, which can be considered as a variant of the undressing technique. Namely, we will model  the worldsheet theory as a $T\bar{T}$-dressing of  NG phonons interacting through the undressed phase shift and a free massive axion. This corresponds to treating the axion as a very narrow resonance. A priori, one does not expect this approach to fit the data with high precision. However, assuming no additional massive states are present on the worldsheet, one may expect to find at least the qualitative agreement between this model and data. Also, this approach is well motivated by the ASA, according to which the worldsheet scattering in a certain sense approaches the $T\bar{T}$ dressed form in the high energy limit, when the axion mass can be neglected. In fact, we find this approach works surprisingly well at the quantitative level for most of the spectrum we have studied in this work. To illustrate this point, in what follows in addition to the dressed spectrum we also plot the spectrum of GGRT phonons with an additional free axion. We refer to the latter as the ``undressed spectrum'' in the captions.

Within this approach in addition to the GGRT states  listed in Tables~\ref{tab:table_NGstates_q0}-\ref{tab:table_NGstates_q2} one expects to find additional states containing massive axion excitations. Namely, if 
denoted by $A_k$
an axionic creation operator with a longitudinal momentum $\frac{2\pi k}{R}$, one can add to any GGRT state an arbitrary number of  axionic states with all possible momenta. In terms of quantum numbers, adding each additional axion does not affect the spin of the state, and flips both parities\footnote{To be precise, it flips the transverse parity. For the longitudinal one, it gets flipped under the action of $A_k+A_{-k}$ and stays the same if one acts with $A_k-A_{-k}$.}.

Let us check to what extent this expectation is supported by the lattice data discussed in the current work. From a theoretical viewpoint the most natural way to organize 
a discussion of different axionic states may be to fix the ``parent" GGRT state and to discuss possible axionic states which can be obtained from it. However, this is not very convenient for the purpose of this paper, because it would force us to keep changing the quantum numbers of the discussed states. Instead, we organize our discussion of
axionic states similarly to Section~\ref{sec:data}, {\it i.e.} according to the quantum numbers of these states.

\subsection{Axionic $q=0$  states}
\label{subsec:q0jnonzero_computation}
Let us first discuss  axionic $q=0$ states in different symmetry sectors.

\subsubsection{ Second excited $0^{++}$ state: two axions} 
\label{sec:2ax}
One of the most straightforward cross-check for the axionic model is the existence of multi-axion states without any phonon excitations. The simplest such state describes two axions at rest, $A_0 A_0 |0\rangle$, and is expected to show up in the $0^{++}$ sector. Indeed, as illustrated by Fig.~\ref{fig:0pp}, the second excited $0^{++}$ state plays exactly this role. Namely, the corresponding gap has a very mild dependence on the radius, and is approximately equal to twice the axion mass $2m$, as one would expect for a state describing two weakly interacting particles at rest. In addition to this qualitative support of the axionic model, this plot provides a surprisingly impressive quantitative confirmation of the $T\bar{T}$ ansatz for axion self-interactions. Namely, we observe that the data is very well described by the $T\bar{T}$ dressing (the solid blue curve) over the whole range of $R$. 
To illustrate the effect of dressing, we have also shown the corresponding undressed energy (the dashed blue curve). It is clear that dressing provides a significant improvement, which is achieved without introducing any new parameters. 
Clearly, $T\bar{T}$ dressing provides a much better description of the data, indicating that it correctly captures the axion self-interaction, which is repulsive, as follows from the plot.

\subsubsection{$1^{\pm}$ ground states: axion-phonon collision}
\label{sec:q01}
\begin{figure}[htb]
\scalebox{1}{\includegraphics[width=0.65\textwidth]{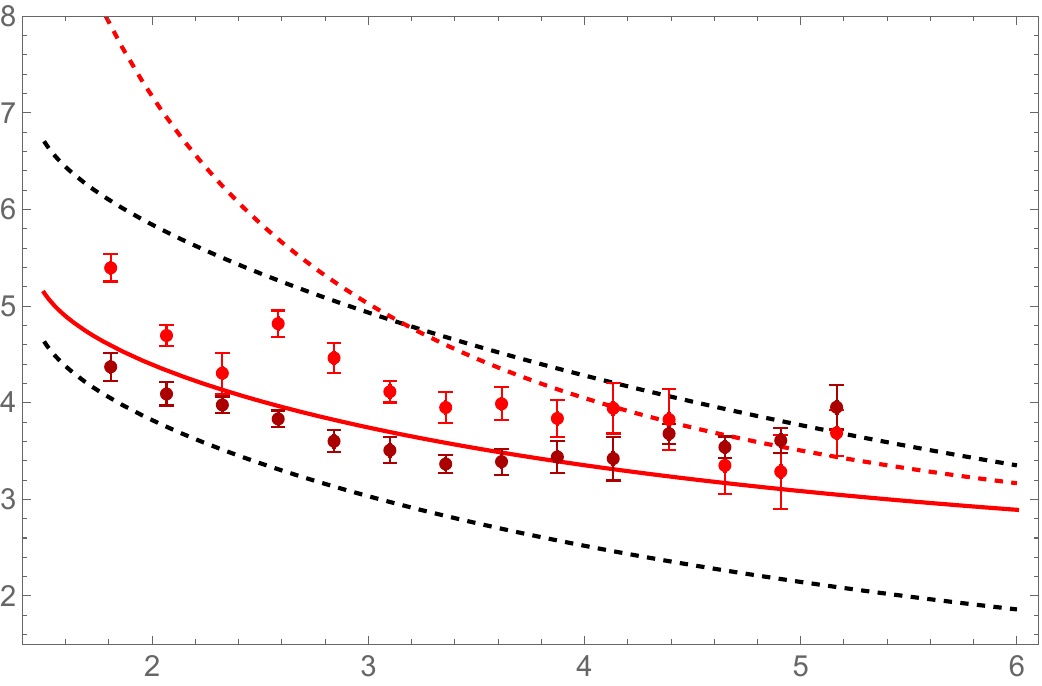}\put(-330,205){ \small $\Delta E\ell_s$}\put(0,0){\small $R/\ell_s$}\put(0,15){\scriptsize $N_L=N_R=1$}\put(0,65){\scriptsize $N_L=N_R=2$}\put(0,50){\scriptsize\color{red} Axion+1 phonon}}
\centering
\caption{The lowest-lying two states in the $1^{+}$ and $1^-$ sectors, with total longitudinal momentum quanta $q=0$. Red and darker red dots represent the ground states in the $1^{+}$ and $1^-$ sectors respectively. The solid and dashed red lines are the theoretical predictions for $ \left[ \left( A_1 a_{-1}^+ \mp A_1 a_{-1}^- \right) 
\mp \left( A_{-1} a_{1}^+ \mp A_{-1} a_{1}^- \right) \right] | 0 \rangle$ with and without doing $T\bar{T}$ dressing respectively. The dashed black lines are GGRT spectra at levels $N_L=N_R=1$ and $N_L=N_R=2$ from bottom to top. }
\label{fig:q0j1}
\end{figure}

As discussed in Sections \ref{sec:results_1+_channel} and \ref{sec:results_1-_channel}, the lowest GGRT states in these sectors correspond to the $N_L=N_R=2$ level.
Instead, one expects to find lighter axionic states, obtained by adding an axion with a single unit of momentum to the lowest single phonon state,
\[ \left[ \left( A_1 a_{-1}^+ \mp A_1 a_{-1}^- \right) 
\mp \left( A_{-1} a_{1}^+ \mp A_{-1} a_{1}^- \right) \right] | 0 \rangle\;.
\] 
This states are particularly interesting because they provide the cleanest probe of the axion-phonon scattering, similarly to the two phonon states which were used to discover the axion.
Within the simple $T\bar{T}$ dressing model these two states are expected to be degenerate, and the corresponding energy is shown in Fig.~\ref{fig:q0j1}. The lattice data indeed reveals states significantly below $N_L=N_R=2$ GGRT level in both cases. 
However, these states exhibit a substantial splitting. As a result the $1^+$ ground state is well described by the $T\bar{T}$ dressing, while the description for the $1^-$ ground state is not so successful. Note however, that this state exhibits  surprisingly large up and down variations in its energy as a function of $R$, which may be indicative of an unaccounted systematics. Overall, we believe that the axionic model works well at the qualitative level in both of these sectors. Also, the leading order axion-phonon interaction can be seen to be described qualitatively well if we compare the dressed and undressed spectrum. Note, that the axion-phonon interaction resulting from the $T\bar{T}$ dressing is attractive (unlike the repulsive interaction between two massive axions at rest discussed earlier). It will be interesting to see whether additional axion-phonon interactions, not accounted by the dressing, are able to describe the splitting between these two states also at the quantitative level (we will discuss this in more detail in section \ref{sec:TTcomments}).

\subsubsection{ Second excited $0^{--}$ state} 
\begin{figure}[htb]
\centering
\scalebox{1}
{\includegraphics[width=0.65\textwidth]{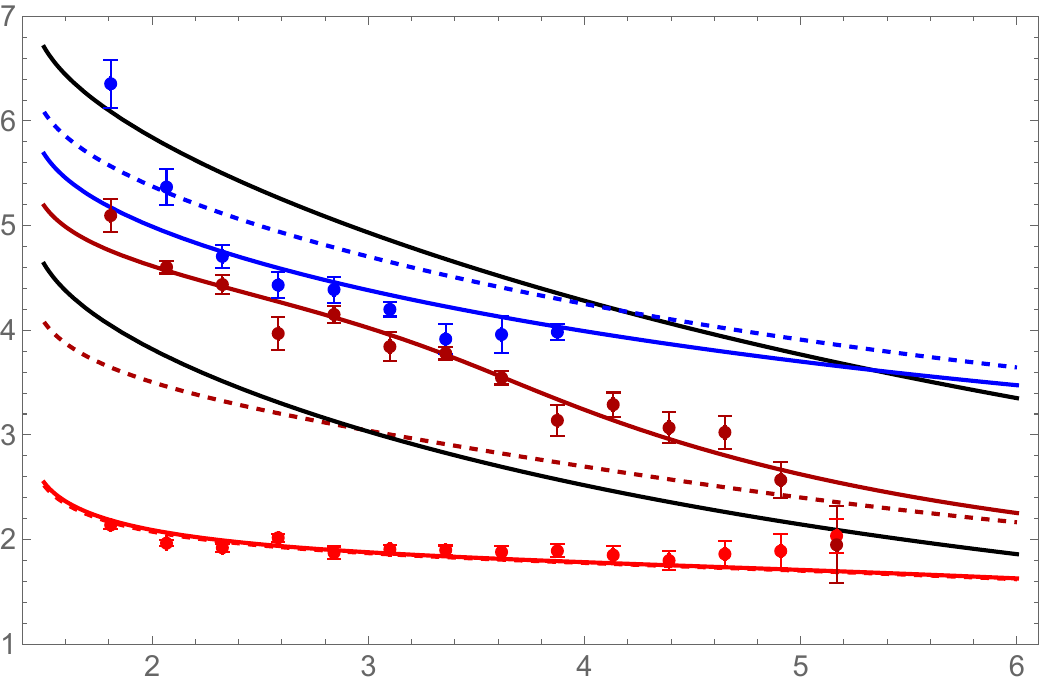}\put(-320,210){\small $\Delta E\ell_s$}\put(0,5){\small $R/\ell_s$}\put(0,38){\scriptsize $N_L = N_R = 1$}\put(0,80){\scriptsize $N_L = N_R = 2$}\put(0,28){\scriptsize\color{red} Axion}\put(0,56){\scriptsize\color{purple} $N_L = N_R = 1$ with}\put(0,48){\scriptsize\color{purple} (higher order) corrections}\put(0,90){\scriptsize\color{blue} Axion+2 phonons}}
\caption{The spectrum of the ground (bright red dots), the first excited states (dark red dots) and the second excited states (blue dots) in the $0^{--}$ sector with total longitudinal momentum quanta $q=0$, plotted as a function of the string length $R$. We also display theoretical predictions from $T\bar{T}$ dressing for these states, respectively interpreted as one axion, two phonons and one axion plus two phonons. In the last case, we also plot the undressed spectrum as a dashed blue curve, for comparison. Note that for two-phonon states, we need to add higher order corrections (distinguished by whether the line is solid or not) to match the data. The solid black lines are the GGRT spectrum at levels $N_L=N_R=1$ and $N_L=N_R=2$. }
\label{fig:0mmenergy}
\end{figure}

The $0^{--}$ sector is the one where the axion was identified in the first place, by observing the single axion state $A_0|0\rangle$.
Another relatively low lying axionic state in this sector can be obtained by adding an axion to the first excited $0^{++}$ state,
\[
A_0 \left( a_1^+ a_{-1}^- + a_1^- a_{-1}^+ \right) | 0 \rangle \;.
\]
 The second excited $0^{--}$ state, shown with blue dots in left Fig.~\ref{fig:0mmenergy}, fits this role very well.
Indeed, to compute the energy of this axionic states, we apply  the $T\bar{T}$ dressing equation to the state of two phonons and an axion at rest. Here the undressed phonon state is described using ABA with the undressed scalar channel phase shift, which is given by a sum of the universal PS contribution and $t$ channel axion exchange.  
The result, shown as a blue solid line in Fig.~\ref{fig:0mmenergy}, provides a good prediction of this state. Comparing to the undressed energy, it indicates that the $T\bar{T}$ dressing provides also a good description of the phonon-axion interactions. As follows from  Fig.~\ref{fig:0mmenergy}, this description ceases to be accurate for the shortest strings, corresponding to the two data points with $R\lesssim 2\ell_s$, which is hardly surprising given that these points correspond to very short strings.

\subsubsection{$2^{\pm -}$ states} 
\begin{figure}[htb]
\scalebox{1}{\includegraphics[width=0.65\textwidth]{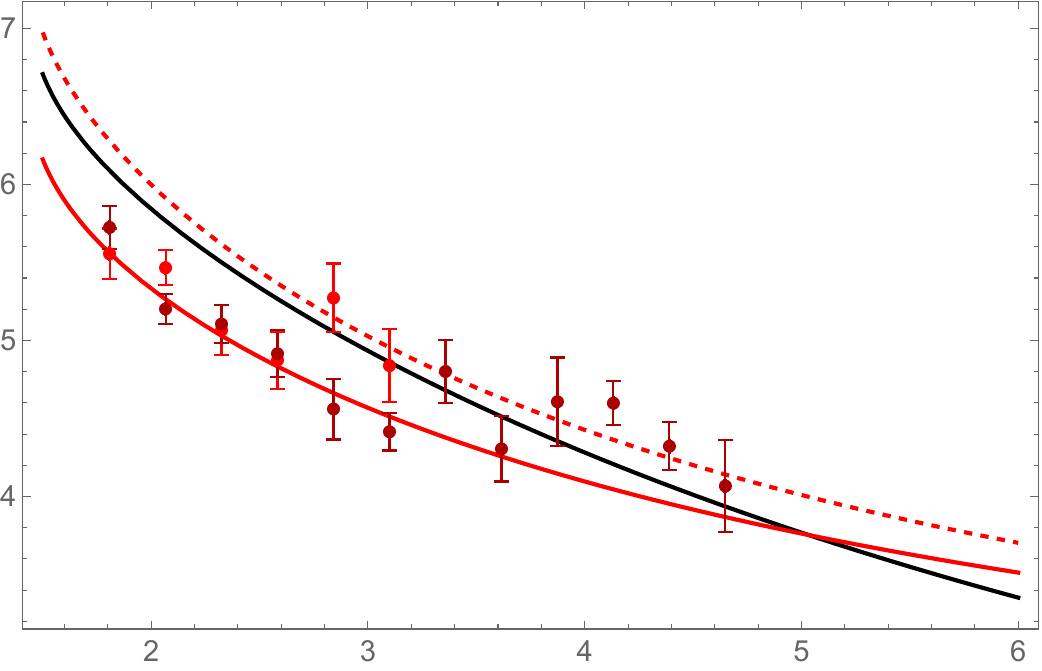}\put(-330,205){ \small $\Delta E\ell_s$}\put(0,0){\small $R/\ell_s$}\put(0,20){\scriptsize $N_L=N_R=2$}\put(0,30){\scriptsize\color{red} Axion+2 phonons}}
\centering
\caption{The ground states in the $2^{+-}$ (darker red dots) and $2^{--}$ (bright red dots) sectors, with total longitudinal momentum quanta $q=0$. The solid and red lines are the theoretical predictions for $A_0 \left( a_1^+ a_{-1}^+ \mp a_1^- a_{-1}^- \right) | 0 \rangle$ with and without $T\bar{T}$ dressing respectively. The solid black lines is GGRT spectrum at level $N_L=N_R=2$. }
\label{fig:q0j2}
\end{figure}

Turning to the spin 2 states, one expects to find a low lying $2^-$ doublet which is obtained by adding an axion at rest to the GGRT ground $2^+$ state, similarly to how the second excited $0^{--}$ state was obtained by adding an axion to the first excited $0^{++}$. The oscillator content of this doublet is 
\[
A_0 \left( a_1^+ a_{-1}^+ \mp a_1^- a_{-1}^- \right) | 0 \rangle\;.
\]
The dressing calculation of the energy of these states proceeds similarly to the second  excited $0^{--}$ state. Namely, one starts with the free massive axion at rest and two phonons interacting through the 
undressed spin 2 phase shift in the ABA approximation and dresses them up. The result is presented in Fig.~\ref{fig:q0j2}. Note that for these states the axionic prediction is quite close to the $N_L=N_R=2$ GGRT level. Nevertheless, for relatively short strings, $R\lesssim 3\ell_s$, where the data quality is good, the axionic prediction is definitely preferred to the GGRT energy, and reproduces the data very well. Moreover, the axion-phonon interaction is also well captured by the dressing procedure at the quantitative level . For longer strings the quality of the data is not good enough to distinguish between the two. Also, for $R\lesssim 3\ell_s$ the energies of two different polarizations ($2^{+-}$ and $2^{--}$) agree with each other within error bars. For longer strings, we only plot the $2^{+-}$
energies, because the energies of $2^{--}$ were not reliably measured there.

\subsubsection{$0^{-+}$ states}
\label{sec:TBA0mp}
\begin{figure}[htb]
\scalebox{1}{\includegraphics[width=0.65\textwidth]{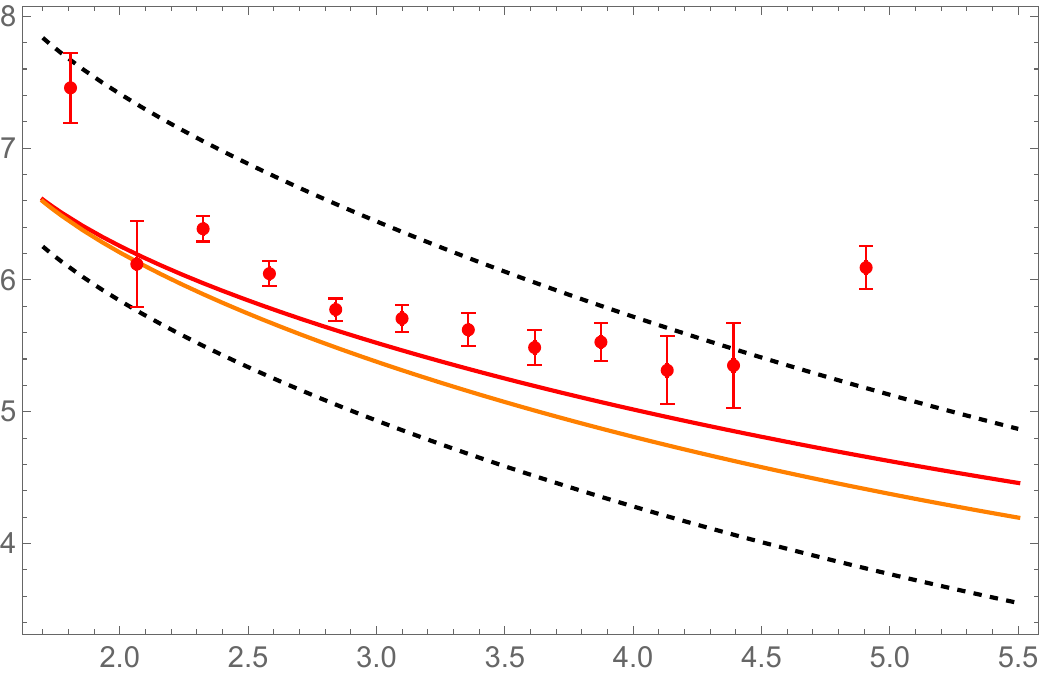}\put(-330,205){ \small $\Delta E\ell_s$}\put(0,0){\small $R/\ell_s$}\put(0,20){\scriptsize $N_L=N_R=2$}\put(0,0){\small $R/\ell_s$}\put(0,70){\scriptsize $N_L=N_R=3$}\put(0,0){\small $R/\ell_s$}\put(0,55){\scriptsize{\color{red} Two }{\color{orange} states }{\color{red} w/}}\put(0,45){\scriptsize{\color{orange} Axion+2 }{\color{red} phonons}}}
\centering
\caption{The ground state in the $0^{-+}$ sector with total longitudinal momentum quanta $q=0$, labeled as red dots. Red and orange lines represent the predictions of $ \left[ A_{-1} \left( a_2^+ a_{-1}^- + a_2^- a_{-1}^+ \right) - A_{1} \left( a_{-2}^+ a_{1}^- + a_{-2}^- a_{1}^+ \right) \right] | 0\rangle$ and $ \left( A_{2}  a_{-1}^+ a_{-1}^- - A_{-2} a_1^+ a_{1}^- \right) | 0\rangle$ respectively from $T\bar{T}$ dressing. The black dashed lines are GGRT spectra at levels $N_L=N_R=2$ and $N_L=N_R=3$ from bottom to top. }
\label{fig:0mp}
\end{figure}

 Finally, let us take a look at the $0^{-+}$ sector. In the continuum theory the lowest states in this sector are expected to be axionic.
In fact, there are two relatively close lightest axionic states with these quantum numbers, which are 
\[ \left[ A_{-1} \left( a_2^+ a_{-1}^- + a_2^- a_{-1}^+ \right) - A_{1} \left( a_{-2}^+ a_{1}^- + a_{-2}^- a_{1}^+ \right) \right] | 0\rangle\]
and 
\[ \left( A_{2}  a_{-1}^+ a_{-1}^- - A_{-2} a_1^+ a_{1}^- \right) | 0\rangle\;.\]
We plotted their energies calculated using the dressing approximation in Fig.~\ref{fig:0mp}. We see that these energies  agree with the measured ground state energy in this sector at the qualitative level. However, the measured energies are all somewhat heavier than the theoretical ones. This can be indicative of a systematic energy overestimation for this state, which is not surprising given that this state is by far the heaviest one that we have analyzed. The bigger issue, however, is that on the lattice one expects to see the $N_L=N_R=2$
$4^{-+}$ GGRT state listed in Table~\ref{tab:table_NGstates_q0} as the ground state in this sector, considering that the lattice symmetry does not distinguish $0^{-+}$ and $4^{-+}$ 
states. However, we did not observe such a state in the simulation. Most likely this indicates that our set of operators does not have a good overlap onto high spin states.

\subsection{Axionic $q=1$ states}
\label{subsec:q1computation}
Let us now discuss axionic $q=1$ states.

\subsubsection{$0^-$, $q=1$  states} 
\begin{figure}[htb]
\centering
\scalebox{1}
{\includegraphics[width=0.65\textwidth]{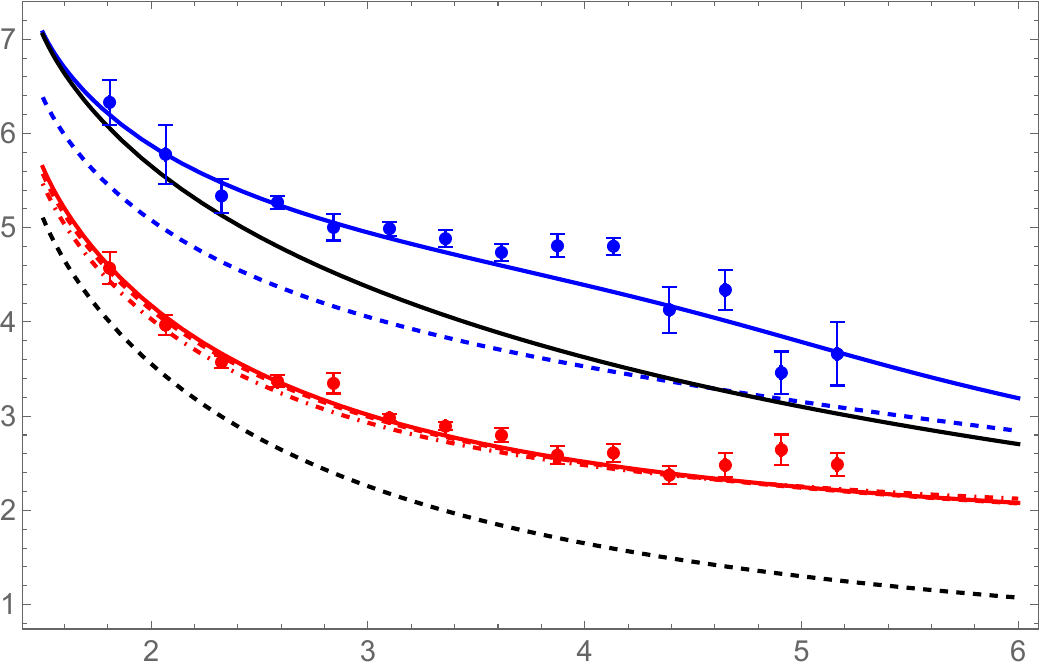}\put(-320,205){ \small $\Delta E\ell_s$}\put(0,5){\small $R/\ell_s$}\put(0,20){\scriptsize $N_L=1,N_R=0$}\put(0,65){\scriptsize $N_L=2,N_R=1$}\put(0,50){\scriptsize\color{red} Axion}\put(0,85){\scriptsize\color{blue} $N_L=2,N_R=1$}\put(0,77){\scriptsize\color{blue} with corrections}}
\caption{The ground (red dots) and first excited states (blue dots) in the $0^-$,  $q=1$ sector. The red and blue dashed lines are the theoretical prediction of the ground state and the first excited state considering them as two phonon excitations, while the dot-dashed red line is the theoretical prediction of the ground state considering it as one massive excitation. The red and blue solid lines are the prediction including higher order corrections in the low energy expansion of the phase shift. The solid and dashed black lines are GGRT spectra at levels $N_L=2,N_R=1$ and $N_L=1,N_R=0$ respectively. }
\label{fig:q1j0}
\end{figure}

Physically, these states can be thought as a boosted version of the $0^{--}$, $q=0$  sector, where the axion has been discovered. In particular, the ground state in this sector is supposed to represent a boosted axion $A_1|0\rangle$. Its energy can be calculated in two ways. First one can solve the TBA equations for the pseudoscalar $q=1$ two phonon state 
\be
\label{q1}
 (a_2^+ a_{-1}^- - a_2^- a_{-1}^+) |0\rangle\;,
\ee
using the phase shift, which was previously extracted from the pseudoscalar $q=0$ states ({\it i.e.}, the phase shift given by the sum of the universal terms, resonance and higher order corrections~\eqref{higher_order}). Alternatively, one may apply the $T\bar{T}$ dressing to a massive one particle state $A_1|0\rangle$. As follows from Fig.~\ref{fig:q1j0} the two approaches are practically indistinguishable for the ground state. 
Both agree fairly well with the data, with the two phonon calculation providing a marginally better agreement.
This agreement between the two theoretical calculations indicates that this state probes the resonant region of the phase shift, which is equivalent to saying that it represents a one axion state.

\begin{figure}
    \centering
    \scalebox{1}{\includegraphics[width=0.65\linewidth]{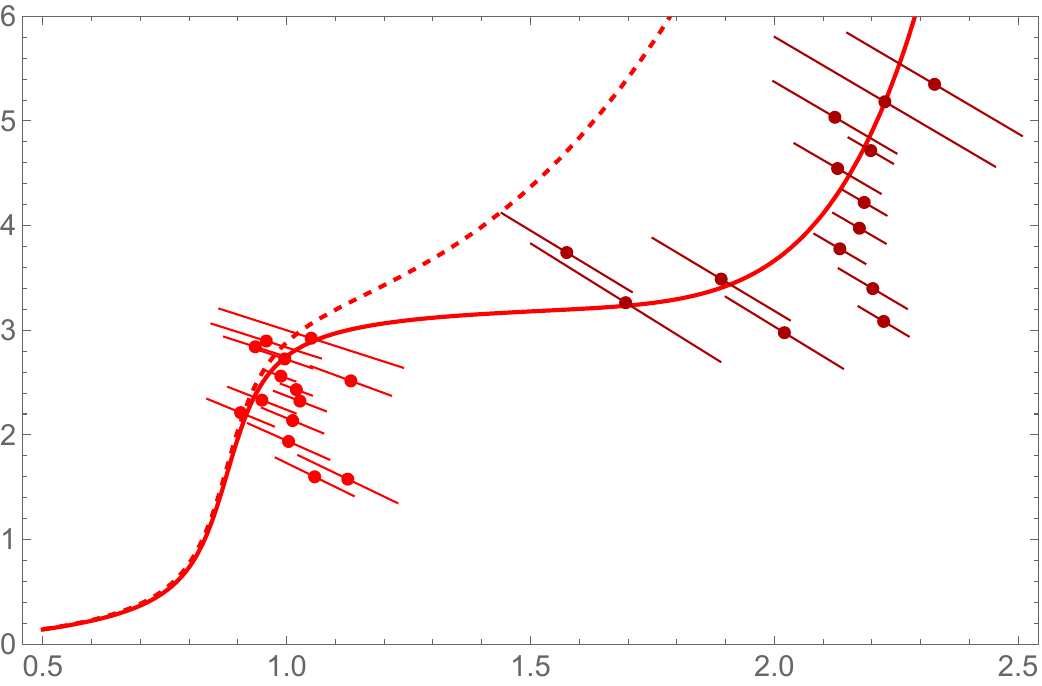}\put(-315,205){ \small $\delta$}\put(4,5){\small $p\ell_s$}}
    \caption{The low energy phase shift with and without $O(s^3)$ and $O(s^4)$ corrections as solid and dashed red lines. The phase shift data extracted from the spectrum of $q=1, \;0^-$ sector for the ground and first excited states are denoted as red and darker red dots. }
    \label{fig:q1j0phase}
\end{figure}

The first excited   $0^-$, $q=1$ state is not axionic, but rather a GGRT-like two phonon state. Its energy can be again calculated using the TBA equation with the same phase shift leading to a good agreement  with the data, as also shown in Fig.~\ref{fig:q1j0}. In the same plot we also presented the result obtained neglecting the higher order corrections to the phase shift. We see that the higher order corrections practically do not affect the ground state, but crucial for the correct prediction of the first excited one.
The same points can be illustrated by extracting the scattering phase shift from the data using the TBA equations~\eqref{TBA_equations} as shown in Fig.~\ref{fig:q1j0phase}.

The first axionic $q=1$  $0^+$ state is expected to be the two axion state $A_1A_0|0\rangle$ which is too heavy to be extracted from the present data.

\subsubsection{$1^{\pm}$, $q=1$ sates: another axion-phonon collision} 
\label{sec:anaxph}
\begin{figure}[htb]
\scalebox{1}{\includegraphics[width=0.65\textwidth]{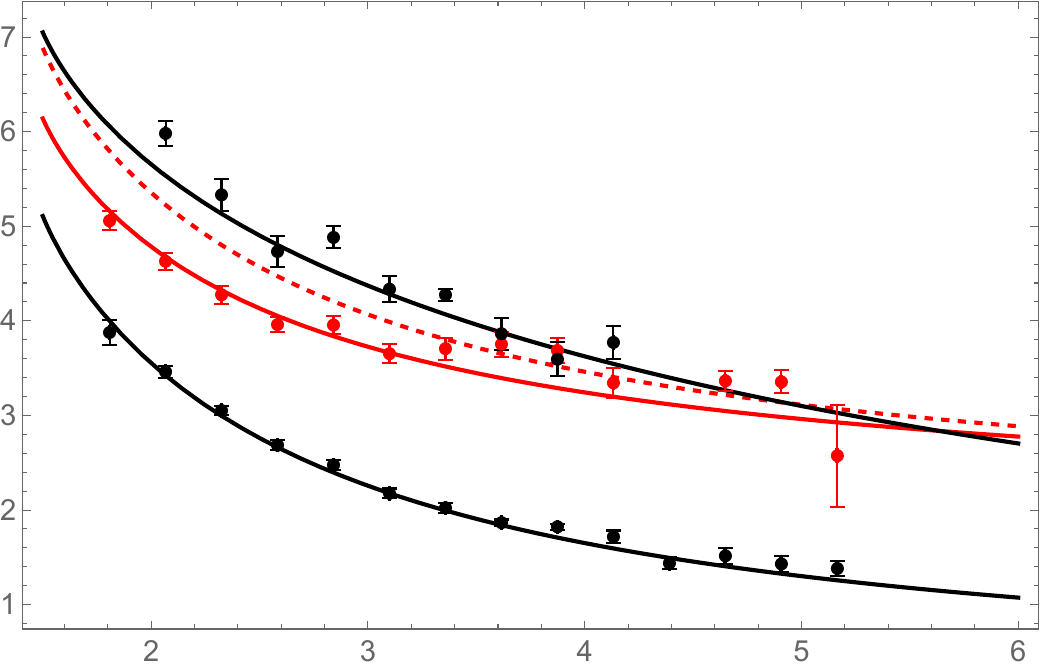}\put(-330,205){ \small $\Delta E\ell_s$}\put(0,0){\small $R/\ell_s$}\put(0,20){\scriptsize $N_L=1,N_R=0$}\put(0,62){\scriptsize $N_L=2,N_R=1$}\put(0,72){\scriptsize\color{red} Axion+1 phonon}}
\centering
\caption{Three lowest lying states in the $1^{\pm}$, $q=1$ sector. Black dots are the ground and second excited states, agreeing well with GGRT spectra at levels $N_L=1,N_R=0$ and $N_L=2,N_R=1$ (black lines). Red dots are the first excited states. The solid and dashed red solid lines are the theoretical predictions of $A_0 \left( a_1^+ \mp a_1^- \right) | 0 \rangle$ with and without $T\bar{T}$ dressing respectively. }
\label{fig:q1j1}
\end{figure}

Let us turn now to spin 1,  $q=1$  doublets. Here the ground state is expected to describe a single GGRT phonon, $a_1^\pm| 0 \rangle$. As the first excited states one expects to find a phonon with an additional axion at rest, 
\[
A_0 \left( a_1^+ \mp a_1^- \right) | 0 \rangle\;.
\] 
In Fig.~\ref{fig:q1j1} we presented the available data in this sector, which provides also the second excited state. As a theory comparison we use the $N_L=1,N_R=0$ and $N_L=2,N_R=1$ GGRT energies and a $T\bar{T}$ dressing result for the axion-phonon state. We find a good agreement for all three states, especially for relatively short strings $R\lesssim  3.5 \ell_s$, indicating that the second excited state is one of the three phonon $N_L=2,N_R=1$ GGRT states. 
In principle, one can obtain a more accurate description of the second excited states including the higher order corrections to the phonon phase shift, but we leave this analysis for the future. For $R \gtrsim 3.5 \ell_s$ the data is somewhat above the dressing prediction for the axion-phonon state.
Probably, this is mostly related to lattice systematics for these heavy states. Also in this regime the energies of the first and second excited states are quite close, which may also introduce additional systematics due to mixing between the two states.

We stress that the axion-phonon interaction in this channel is described very well by the $T\bar{T}$ dressing in this scattering channel. More specifically, it verifies that over a range of center of mass energy, the axion-phonon phase shift is well approximated by the $T\bar{T}$ dressing phase shift. For illustration we plot the phase shift obtained from the data in Fig.~\ref{fig:axion_phonon_phase}.

\subsubsection{  $2^\pm$,  $q=1$ states} 
\begin{figure}[htb]
\scalebox{1}{\includegraphics[width=0.65\textwidth]{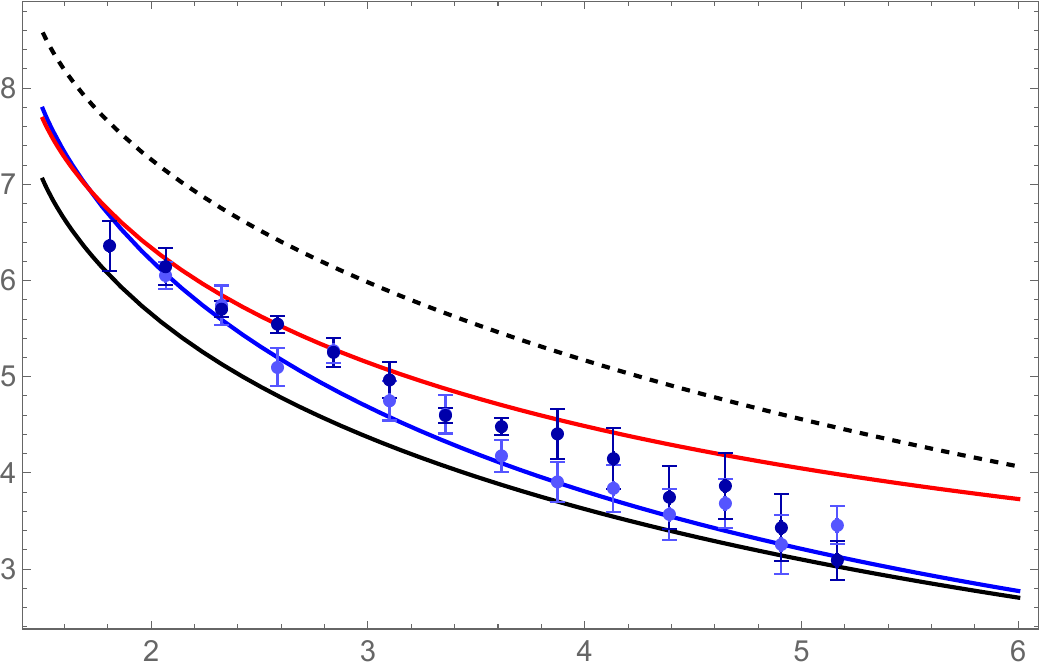}\put(-330,205){ \small $\Delta E\ell_s$}\put(0,0){\small $R/\ell_s$}\put(0,16){\scriptsize $N_L=2,N_R=1$}\put(0,58){\scriptsize $N_L=3,N_R=2$}\put(0,24){\scriptsize\color{blue} 2 phonons}\put(0,48){\scriptsize\color{red} Axion+2 phonons}}
\centering
\caption{Light blue and dark blue dots are the ground states in the $2^+$ and $2^-$ sectors with $q=1$ respectively, agreeing well with the TBA prediction of two phonons after introducing axions, which is shown as a blue curve. We also show the spectrum of two phonons and one axion as a red curve. The dashed and solid black lines are GGRT spectrum at levels $N_L=3, N_R=2$ and $N_L=2,N_R=1$ respectively.  }
\label{fig:q1j2}
\end{figure}

For the  spin 2,    $q=1$ states, we only have  data for the ground state.
This is expected to be a two phonon  $N_L=2, N_R=1$ GGRT state. We computed the corresponding energy using the phase in the spin 2 channel, including the PS correction and the axion exchange. As follows from Fig.~\ref{fig:q1j2}, this calculation provides a good agreement with the data, accounting for the error bars. Note that the first excited state in this sector is expected to have an axionic origin---it is obtained by adding a $q=1$ axion to the $q=0$ spin 2 $N_L=N_R=1$ GGRT state. We presented the dressing estimate  for the energy of this first excited state by a red solid line in Fig.~\ref{fig:q1j2}.  It turns out to be quite close to the ground state and significantly below the $N_L=3, N_R=2$ GGRT state. It will be interesting to test this prediction with future higher quality lattice data.

\subsection{Axionic $q=2$ states}
\label{subsec:q2computation}
Let us conclude our analysis with a discussion of even higher boosted $q=2$ states. These states are quite heavy, so the lattice data is available only for few of them and one expects that the measured energy can be appreciably overestimated. 

\subsubsection{$0^-$, $q=2$ sectors} 
\begin{figure}[htb]
\scalebox{1}{\includegraphics[width=0.65\textwidth]{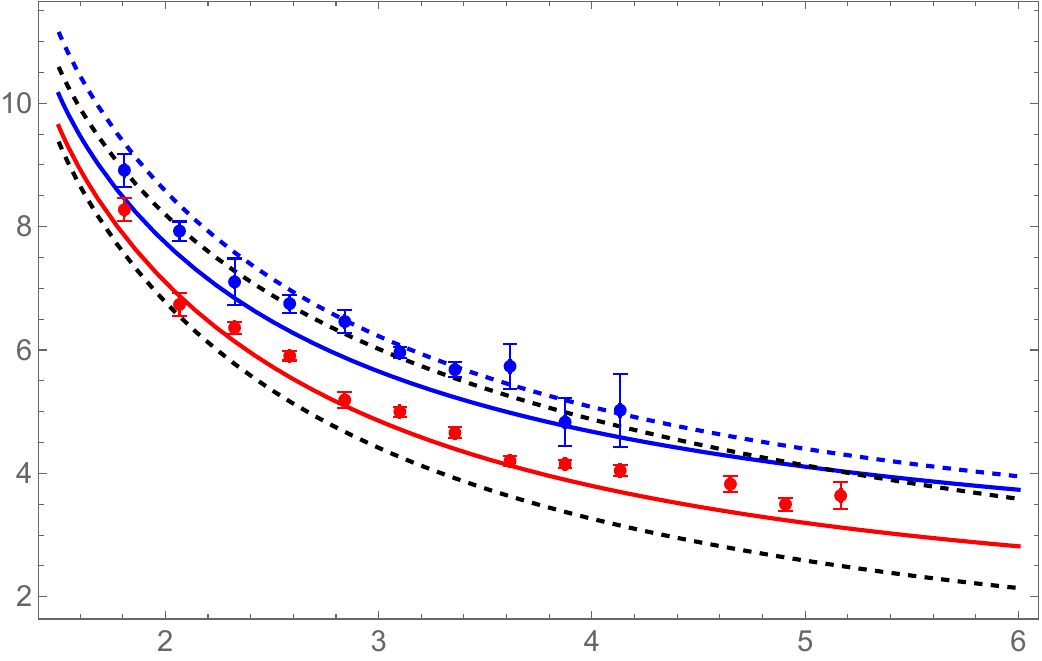}\put(-330,205){ \small $\Delta E\ell_s$}\put(0,0){\small $R/\ell_s$}\put(0,20){\scriptsize $N_L=2,N_R=0$}\put(0,46){\scriptsize $N_L=3,N_R=1$}\put(0,32){\scriptsize\color{red} Axion}\put(0,54){\scriptsize\color{blue} Axion+2 phonons}}
\centering
\caption{The ground state (red dots) and first excited state (blue dots) in the  $0^{-}$, $q=2$ sector. The red solid line is the theoretical prediction of $A_2 | 0 \rangle$. The solid and dashed blue lines are the theoretical predictions of $A_0 a_1^+ a_1^- |0\rangle$ with and without $T\bar{T}$ dressing respectively. The dashed black lines are GGRT spectra at levels $N_L=2,N_R=0$ and $N_L=3,N_R=1$ from bottom to top. }
\label{fig:q2j0}
\end{figure}

We first consider the  $0^-$,  $q=2$ sector. 
Similarly to the $0^-$,  $q=1$ sector the ground state here is expected to be a single boosted axion, $A_2 |0\rangle$. As far as the first excited state is concerned, the situation is different. Namely, in the $q=1$ case the first excited state is a two phonon one, given by (\ref{q1}). An analogous $q=2$ state is expected to be above the $N_L=3$, $N_R=1$ GGRT level for the same reason that (\ref{q1}) is above the $N_L=2$, $N_R=1$ GGRT level. However, in the $q=2$ case there is also another axionic state 
$A_0 a_1^+ a_1^- |0\rangle$. The dressing calculation predicts this state to be a bit below the $N_L=3$, $N_R=1$ GGRT level, as illustrated in Fig.~\ref{fig:q2j0}. In addition, in the $q=2$ case one also expects the four phonon $N_L=3$, $N_R=1$ GGRT states to be close by.

As follows from Fig.~\ref{fig:q2j0}, the lattice data agrees quite well with the predicted ground state, although the measured energies all lie slightly above the theoretical course.
Most likely, this should be attributed to the systematic energy overestimate for this heavy state.
 As far as the first excited state is concerned, it shows a somewhat better agreement with the $N_L=3$, $N_R=1$ GGRT level than with the dressing calculation for the axionic $A_0 a_1^+ a_1^- |0\rangle$ state. However, given that the energy overestimate for this state is expected to be even more pronounced then for the ground one, 
  we do not feel that this should be considered as a serious problem for the axionic model.

\subsubsection{$1^{\pm}$, $q=2$ states} 
\begin{figure}[htb]
\scalebox{1}{\includegraphics[width=0.65\textwidth]{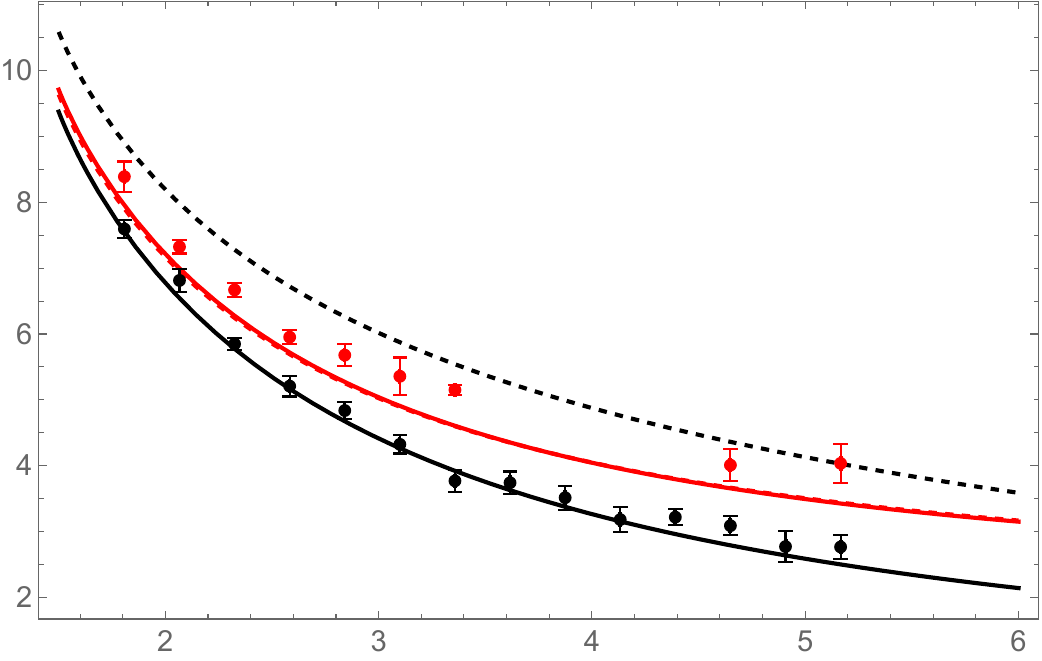}\put(-330,205){ \small $\Delta E\ell_s$}\put(0,0){\small $R/\ell_s$}\put(0,20){\scriptsize $N_L=2,N_R=0$}\put(0,50){\scriptsize $N_L=3,N_R=1$}\put(0,40){\scriptsize\color{red} Axion+1 phonon}}
\centering
\caption{The ground state (black dots) and first excited state (red dots) in the $q=2$; $1^{\pm}$ sector. The solid and dashed red lines are the theoretical predictions of $A_1 \left(a_1^+ \mp a_1^-\right) | 0 \rangle$ with and without $T\bar{T}$ dressing respectively. The solid and dashed black lines are GGRT spectra at levels $N_L=2,N_R=0$ and $N_L=3,N_R=1$ respectively. }
\label{fig:q2j1}
\end{figure}

We now turn to the spin 1, $q=2$  states. Analogously to the spin 1 $q=1$ case the ground state in this sector is the single GGRT phonon, $a_2^\pm|0\rangle$. As for the first excited state, there are two axionic candidates,
\[
A_1 \left(a_1^+ \mp a_1^-\right) | 0 \rangle
\]
and 
\[
A_0 \left(a_2^+ \mp a_2^-\right) | 0 \rangle\;.
\]
The first one of these states has lower energy as a consequence of the inequality\footnote{This argument applies in a free theory. However, dressing does not change the energy ordering of states with the same quantum numbers, so the same is true also in the dressed theory.},
\begin{equation}
    M + p > \sqrt{M^2 + p^2} \,.
\end{equation}
In Fig.~\ref{fig:q2j1} we presented the lattice data for the first excited spin 1, $q=2$ states. We observe that the ground state is well reproduced by the data. For the first excited state, the data are somewhat above the theory curve, which is again likely due to systematic energy overestimation for this heavy state. Still, as expected, the state is definitely below the 
$N_L=3$, $N_R=1$ GGRT level. Note that in this case, the fully dressed and undressed spectra are almost indistinguishable. This is due to the fact that the axion and phonon go in parallel.

\subsubsection{$2^\pm$, $q=2$ states} 
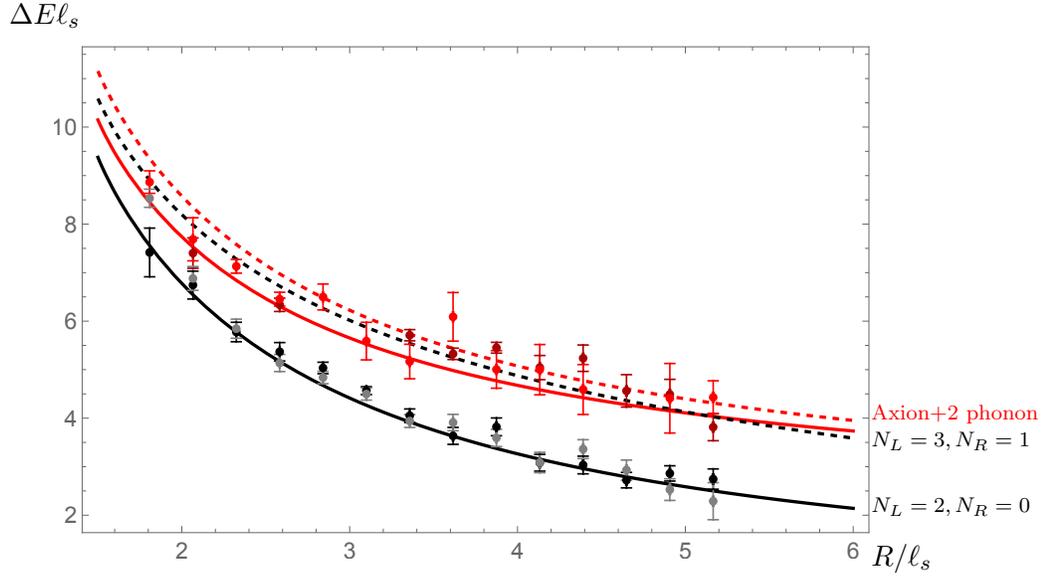
\begin{figure}[htb]
\scalebox{1}{\includegraphics[width=0.65\textwidth]{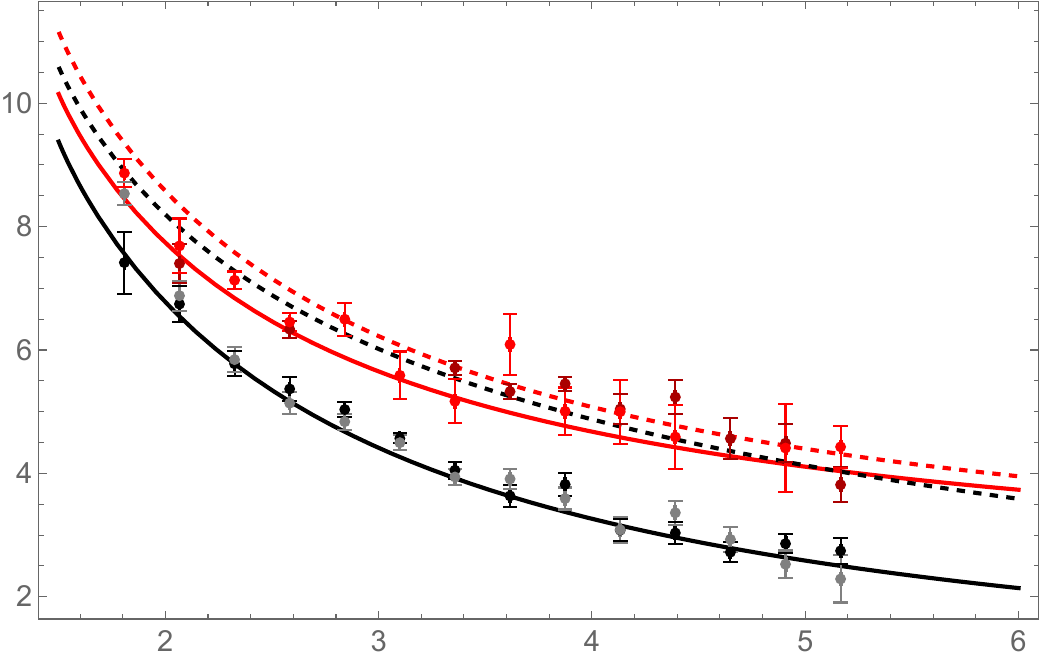}\put(-330,205){ \small $\Delta E\ell_s$}\put(0,0){\small $R/\ell_s$}\put(0,20){\scriptsize $N_L=2,N_R=0$}\put(0,45){\scriptsize $N_L=3,N_R=1$}\put(0,55){\scriptsize\color{red} Axion+2 phonon}}
\centering
\caption{The ground states (black and gray dots) and first excited states (darker red and bright red dots) in the $q=2$; $2^{+}$ and $2^-$ sectors respectively. The solid and dashed red lines are the theoretical predictions of $A_0 \left( a_1^+ a_1^+ \mp a_1^- a_1^- \right) | 0 \rangle$ with and without $T\bar{T}$ dressing respectively. The solid and dashed black lines are GGRT spectra at levels $N_L=2,N_R=0$ and $N_L=3,N_R=1$ respectively. }
\label{fig:q2j2}
\end{figure}

Finally, let us conclude by taking a look at the spin 2, $q=2$ doublet. The ground state is expected to be the two phonon GGRT state,
\[
 \left( a_1^+ a_1^+ \mp a_1^- a_1^- \right) | 0 \rangle\;,
\]
and the first excited one is obtained by adding an axion at rest,
\[
 A_0\left ( a_1^+ a_1^+ \mp a_1^- a_1^- \right) | 0 \rangle\;.
\]
As follows from Fig.~\ref{fig:q2j2}, the first excited states turns out to be quite close to the $N_L=3,N_R=1$ GGRT level. As a result, even though the lattice data agrees within error bars with 
the dressing prediction both for the ground and the first excited states, it is impossible to tell apart the first excited state from the $N_L=3,N_R=1$ GGRT level.

\subsection{Further comments on the ``free" $T\bar{T}$ dressing}
\label{sec:TTcomments}
To summarize, we find that the current data is in a good agreement with a simple model, which can be called a ``free" $T\bar{T}$ dressing. Namely, this is the $T\bar{T}$ dressing of a theory which describes a free massive axion and undressed phonons. To start with we aimed to use this model as a qualitative benchmark, 
which should hopefully allow us to check whether the Nambu-Goto string supplemented with a massive axion is in a broad agreement with the observed spectrum, at least as far the state's quantum numbers are concerned. However, we see that in addition to providing a good qualitative description of the spectrum, in many cases this model
captures very well the measured energies without introducing any new fitting parameters. This raises the question of whether one should be surprised by this, and even whether it is possible that this model provides an exact description of the confining string. 

To address this, let us first point out that the ``free" $T\bar{T}$ dressing definitely cannot be the exact string model. Indeed, interactions between axion and  phonons are constrained by the non-linearly realized Poincar\'e symmetry. This symmetry implies in particular that the leading order terms in the axion Lagrangian take the form (\ref{axion_interaction}). It is straightforward to see that the $T\bar{T}$ dressing is inconsistent with this form already at the level of interactions involving  the first derivative terms only ({\it i.e.}, neglecting the last term in (\ref{axion_interaction})). Indeed, the $T\bar{T}$ dressing of free phonons and a massive axion\footnote{In our ``free" $T\bar{T}$ dressing, the phonons are actually not free, but rather interact via the universal PS phase shift. However, this effect is higher order in the derivative expansion than the one which we discuss at the moment.} at the tree level is described by the  action given by Eq.~(2.23) of \cite{Conti:2018jho}.
The complete form of this action is somewhat cumbersome, so let us present here only its expansion in the axion mass up to the ${\cal O}(m^2)$ order,
\be
\label{Stefano}
S_{T\bar{T}}\simeq\int d^2\sigma\l \ell_s^{-2}\sqrt{-h_\phi}-{m^2\phi^2\over 4} \l 1+{{1+\ell_s^2(\d Y^i)^2+\ell_s^2(\d\phi)^2}\over \sqrt{-h_\phi}} \r \r\;.
\ee
Here, $h_\phi$ is the determinant of the ``axionic" induced metric,
\[
h_\phi=\det (h_{\alpha\beta}+\ell_s^2\d_\alpha\phi\d_\beta\phi)\;.
\]
Recall that $h_{\alpha\beta}$ is the induced metric without axions~\eqref{induced_metric}. In the massless limit the $T\bar{T}$ deformed action (\ref{Stefano}) coincides with the 5d Nambu--Goto theory and, consequently, is compatible with the non-linearly realized Poincar\'e symmetry. However, the ${\cal O}(m^2)$ part of (\ref{Stefano}) does not take a covariant form as in (\ref{axion_interaction}) and thus violates the target space Poincar\'e. In particular, by expanding the ${\cal O}(m^2)$ part of (\ref{Stefano}) up to quartic order in fields one finds that it takes the form
\[
-{m^2\phi^2\over 2}(1+{\cal O}((\d Y)^4)\;.
\]
Consequently, the action  (\ref{Stefano}) does not reproduce the universal quartic vertex
\be
\label{uni}
\ell_s^2 m^2\phi^2(\d Y)^2\;,
\ee
which is present in the Poincare invariant action (\ref{axion_interaction}).

 In fact, this discrepancy can be seen directly from  the $T\bar{T}$ deformed $S$-matrix, without any need to inspect the action (\ref{Stefano}). Indeed, the universal quartic vertex (\ref{uni}) leads to a non-trivial reflection amplitude, where a colliding axion and a left-moving phonon scatter into an axion and a right-moving phonon. $T\bar{T}$ dressing of the non-interacting axions and phonons does not introduce such a process, and thus the corresponding interaction needs to be added in the undressed theory to restore Poincar\'e invariance after dressing. Interestingly, this process would lead to the splitting between $q=0$, $1^+$ and $1^-$ states, which is indeed observed in the data, in the conflict with the free $T\bar{T}$ dressing, as we discussed in section~\ref{sec:q01}.

\begin{figure}[htb]
\scalebox{1}{\includegraphics[width=0.65\textwidth]{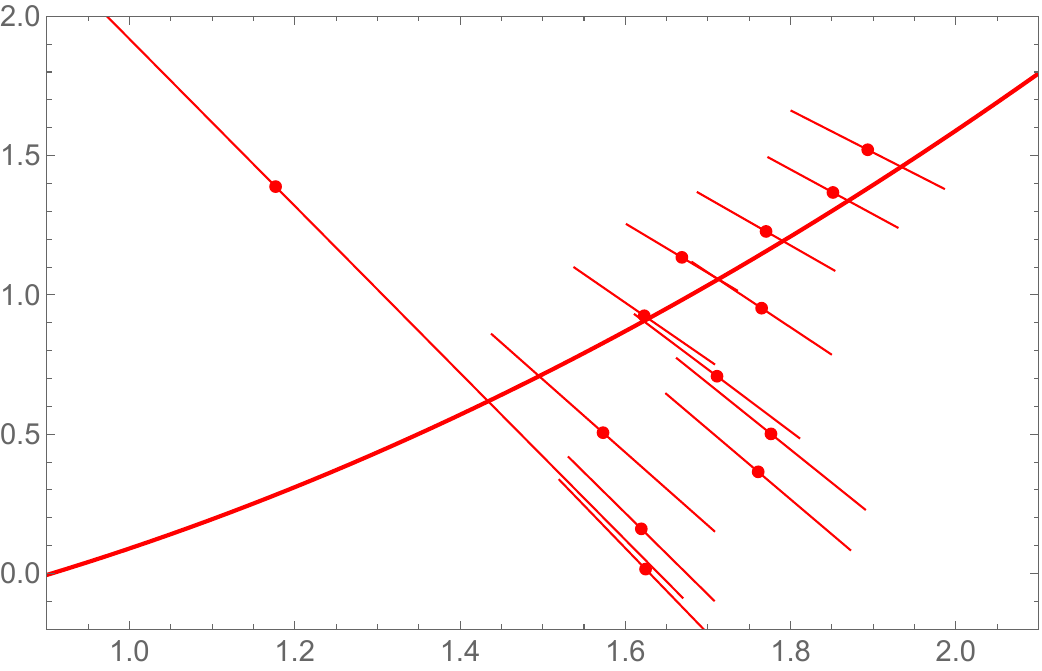}\put(-315,205){ \small $\delta$}\put(4,5){\small $\sqrt{s}\ell_s/2$}}
\centering
\caption{The phase shift data extracted from the spectrum of $A_0 \left( a_1^+ \mp a_1^- \right) | 0 \rangle$ in the $q=1, \;1^{\pm}$ sector are denoted as red dots. The corresponding $T\bar{T}$ dressing phase shift is plotted as a red line for comparison.  }
\label{fig:axion_phonon_phase}
\end{figure}

On the other hand, it is conceivable that some of the success of the free   $T\bar{T}$ dressing in reproducing the correct energies can be explained by the fact that it correctly captures the universal structure of the derivative quartic interactions between two axions and two phonons, $(\d\phi)^2(\d Y)^2$.  In particular, this provides at least a partial explanation for the success of the free $T\bar{T}$ dressing for the $q=1$ colliding axion-phonon state, discussed in section~\ref{sec:anaxph}. However, as illustrated by Fig.~\ref{fig:axion_phonon_phase}, the agreement extends to rather high momenta, where one may expect additional non-universal higher derivative operators to play a role, so it still appears somewhat surprising.

\begin{figure}[htb]
\scalebox{1}{\includegraphics[width=0.65\textwidth]{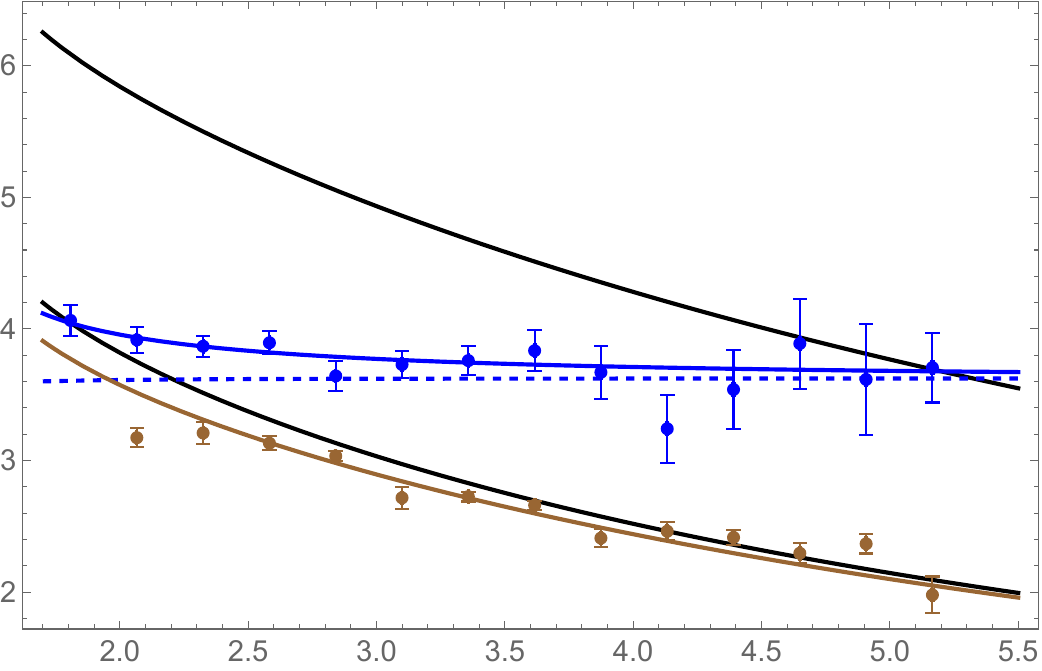}\put(-330,205){ \small $\Delta E\ell_s$}\put(0,0){\small $R/\ell_s$} \put(0,30){\scriptsize $N_L=N_R=1$}\put(0,20){\color{brown} \scriptsize $N_L=N_R=1$}\put(0,12){\color{brown} \scriptsize with corrections}\put(0,80){\scriptsize $N_L=N_R=2$}\put(0,88){\scriptsize\color{blue} 2 axions}}
\centering
\caption{The 1st and 2nd excited states in the $0^{++}$ sector with total longitudinal momentum quanta $q=0$, labeled as brown and blue dots. The solid and dashed blue lines represent the predictions of two massive excitations with and without $T\bar{T}$ dressing. The black lines are GGRT spectra at levels $N_L=N_R=1$ and $N_L=N_R=2$. The brown line is the prediction for the $N_L=N_R=1$ GGRT state with two phonon excitations, after introducing the axionic contribution to the phase shift~\eqref{res_phase}.  }
\label{fig:0pp}
\end{figure}

Even more surprising is the the success of the free   $T\bar{T}$ dressing in describing the two axion state, as discussed in section~(\ref{sec:2ax}). 
We did not provide the analogue of Fig.~\ref{fig:axion_phonon_phase} in this case, because the data represented in Fig.~\ref{fig:0pp} probes only a very narrow range of small spatial momenta of colliding axions. Still, an impressive agreement of the data presented in Fig.~\ref{fig:0pp} with the the free   $T\bar{T}$ dressing indicates that the latter correctly captures the non-universal interactions of two axions at rest.

It will be interesting both to check that the $q=0$ $1^\pm$ splitting can indeed be explained by the universal axion-phonon reflection and to 
quantify how strongly the lattice data prefers the  $T\bar{T}$ description of axion-axion interactions, compared to a generic Poincar\'e invariant Lagrangian. 
At the moment it appears that the observed agreement between the data and the free $T\bar{T}$ dressing provides an additional non-trivial support of the ASA.
We leave further detailed analysis of these questions for the future work.
\section{Conclusions}
\label{sec:conclusion}
In this study, by means of lattice Monte--Carlo simulations we determined energies of about $\sim 35$ excitations of a confining flux tube wound around a spatial circle.
We refer to such a flux tube as a closed confining string or torelon.
We studied these energies as a function of the circle size $R$ by varying  $R$ from a very short size close the critical value $R \ell^{-1}_s \sim 1.4$ where the theory deconfines,  to significantly larger size of order $R \ell^{-1}_s \sim 7$.  To measure the energies  we implemented the  Variational Technique with an extended basis of operators that varies from $\sim 50$ to $\sim 200$ operators depending on the set of quantum numbers of flux tube states. This setup enables us to extract states for all different irreducible representations determined by the quantum numbers of $J$, $P_{\perp}$, $P_{\parallel}$ and momentum $p=2 \pi q / R$.

In order to obtain reliable results that accurately represent the physics in the continuum limit with a large number of colors (large $N$), we followed the following approach. Initially, we set the lattice spacing to approximately $a  \sim 0.19\ell_s$ and conducted a thorough and high-statistics investigation for the $SU(3)$ gauge group by performing our measurements over a substantial number (6-7 million) of gauge configurations.

Subsequently, we replicated a similar investigation for the $SU(5)$ and $SU(6)$ gauge groups, albeit with a slightly reduced statistics (approximately half) compared to the $SU(3)$ case. After this stage, we maintained a constant value of $N$ for our simulations, specifically $N=3$ and $N=5$, while adjusting the lattice spacing to roughly $a \sim 0.13\ell_s$. This allowed us to approach the continuum limit and to further explore the behavior of the system.

In addition to performing the highest precision available computations of closed confining string spectrum, we also provide a systematic theoretical analysis of the results.
 Our analysis is based on the effective theory of a long stings which in addition to massless translational Goldstone modes (``phonons") includes also a massive pesudoscalar worldsheet axion. The axion can decay into two phonons via a cubic vertex originating for the axion coupling to the topological density that counts the self-intersection number of the string worldsheet. The calculation of the finite volume spectrum proceeds in two steps. First one calculates the infinite volume $S$-matrix and then one relates the $S$-matrix to the finite volume spectrum. The latter step is performed either by using the TBA technique or with the help of the $T\bar{T}$ dressing.

 The latter approach allows us to analyze also multiparticle states, which are currently hard to treat using the TBA approach. When applying the $T\bar{T}$ dressing we start with decoupled axion and phonons, so that all axion-phonon and axion-axion interactions are induced by dressing.
 This description is compatible with the target space Poincar\'e symmetry only in strict the limit when the axion is massless. Nevertheless, we find that this approximation not only provides a good qualitative guidance for the quantum numbers of axionic excitations of a confining strings, but in many cases also provides a surprisingly 
accurate prediction for their energies. This may be considered as an evidence in support of the Axionic String Ansatz (ASA), according to which the worldsheet theory is approximated by the $T\bar{T}$ dressing of free axion and phonons in the high energy limit. Definitely, more analysis is required to evaluate to what extent  this agreement follows from the universality and to what extent provides a non-trivial support of the ASA. But in any case our study conclusively demonstrates that no additional low lying sharp resonance are present on the string worldsheet, at least up the energy scale of order $\Delta E \sim 4\div 5 \ell_s^{-1}$, which is probed by our data.

There are still some loose ends yet to be fixed with more details in terms of low-lying spectrum of the closed confining strings in 4D Yang-Mills theory. Firstly, it would be good to conduct a more careful study of the spin 4 flux tube states, which seem to have a bad overlap with our basis at least in symmetry sectors (see sections~\ref{sec:results_0-+_channel}, \ref{sec:TBA0mp}).
Also, we still lack a systematic approach to compute the multiparticle spectrum  beyond the free $T\bar{T}$ approximation. Such formalism would be useful to disentangle the 2-phonon states and 4-phonon states, as well for making predictions on 3-phonon states, and this would be useful to extract the information of non-universal Wilson coefficients of the effective string theory in a theory without low-lying massive resonances.

As a further test for the ASA, it is crucial to improve the precision of the determination of the axion coupling, which is currently limited by the spurious splitting between $2^{++}$ and $2^{-+}$ states. Also it will be interesting to see how it changes in the presence of additional adjoint flavors. According to the ASA it should be 
proportional to the one-loop   $\beta$-function so as to restore the high energy integrability\cite{Dubovsky:2018vde}. It will be interesting to design future Monte-Carlo simulations which would test this in adjoint QCD \footnote{We thank Victor Gorbenko for emphasizing to us this point.}.

\section*{Acknowledgements}

We thank Victor Gorbenko, Andrea Guerrieri, Jo\~ao Penedones for helpful discussions. AA would also like to thank Jacob~Sonnenschein, Elias~Kiritsis and Koji~Hashimoto for interesting discussions. SD and MT acknowledge support by the Simons Collaboration on Confinement and QCD Strings. We are indebted to the participants of Simons collaboration meetings for useful discussions.  SD is also supported in part by the NSF grant PHY-2210349, and by the BSF grant 2018068 and by the IBM Einstein Fellow Fund at the IAS. AA has been financially supported by the European Union's Horizon 2020 research and innovation programme ``Tips in SCQFT'' under the Marie Sklodowska-Curie grant agreement No. 791122 as well as by the EuroCC2 project funded by the Deputy Ministry of Research, Innovation and Digital Policy and the Cyprus Research and Innovation Foundation and the European High-Performance Computing Joint Undertaking (JU) under grant agreement No
101101903. Numerical simulations have been carried out in HYDRA Cluster at Oxford.
\newpage
\section*{Tables}

\addcontentsline{toc}{section}{Tables}

\begin{table}[h]
\begin{center}
{\scriptsize
\begin{tabular}{cc|c|c|c|c}\hline \hline 
$N$ & $\beta$ & $L_x=R/a$ & $L_y\times L_z\times L_t$ & sweeps$\times 10^6$ 
& num ops \\ \hline \hline 
 3  & 6.0625 & 10  & $24 \times 24 \times 36$ & $7.8125$ &  $940$ \\
   &        & 12  & $16 \times 16 \times 24$ & $6.25$ &  $940$ \\
   &        & 12  & $20 \times 20 \times 24$ & $6.25$ &  $940$ \\
   &        & 12  & $24 \times 24 \times 24$ & $6.25$ &  $940$ \\
   &        & 14  & $16 \times 16 \times 24$ & $6.25$ &  $940$ \\
   &        & 14  & $20 \times 20 \times 24$ & $6.25$ &  $940$ \\
   &        & 16  & $16 \times 16 \times 16$ & $6.25$ &  $1175$ \\
\hline \hline   
\end{tabular}
\caption{Parameters of our $SU(3)$, $SU(5)$ and $SU(6)$ calculations: lattice sizes, statistics, and numbers of operators.}
\label{table_param}}
\end{center}
\vspace{-4cm}
\end{table}

\begin{table}[h]
\begin{center}
{\footnotesize
\begin{tabular}{cccc|cccc|ccc}\hline \hline
\multicolumn{11}{c}{ $aE_{\rm gs}(q,R)$ } \\ \hline \hline
$q$ & $|J|$ & $P_{\perp}$ & $P_{||}$& $R/a$ & $N=3$ & $N=5$ & $N=6$ & $R/a$ & $N=3f$ & $N=5f$\\ \hline \hline
 0 & 0 & + & + & 10 &  0.2440(13)  &  0.2527(18) & 0.2718(20) & 14 & 0.1298(15)  &   \\
   &   &   &   & 12 &  0.3544(17)  &  0.3637(21) & 0.3853(28) & 16 & 0.1845(13)  &   \\
   &   &   &   & 14 &  0.4489(18)  &  0.4563(27) & 0.4859(23) & 18 & 0.2324(11)  &   \\
   &   &   &   & 16 &  0.5388(18)  &  0.5519(37) & 0.5765(34) & 20 & 0.2753(16)  &  0.2838(23) \\
   &   &   &   & 18 &  0.6219(23)  &  0.6387(45) & 0.6724(32) & 22 & 0.3187(19)  &  0.3246(24) \\
   &   &   &   & 20 &  0.7036(39)  &  0.7256(47) & 0.7571(60) & 24 & 0.3550(16)  &  0.3614(26) \\ 
   &   &   &   & 22 &  0.7850(36)  &  0.8050(76) & 0.8427(80) & 26 & 0.3893(18)  &  0.3990(34) \\ 
   &   &   &   & 24 &  0.8734(48)  &  0.8777(92) & 0.9305(116) & 28 & 0.4284(31) &  0.4396(29) \\ 
   &   &   &   & 26 &  0.9460(73)  &             &             & 30 & 0.4603(33) &   \\ 
   &   &   &   & 28 &  1.0228(82)  &             &             & 32 & 0.4976(40) &   \\
   &   &   &   & 30 &  1.1075(96)  &             &             & 34 & 0.5382(31) &   \\
   &   &   &   & 32 &  1.1927(98)  &             &             & 36 & 0.5650(38) &   \\ 
   &   &   &   & 36 &  1.3567(131) &             &             & 38 & 0.5968(68) &   \\ 
   &   &   &   &    &              &             &             & 40 & 0.6328(148) &  \\ \hline \hline
\end{tabular}}
\caption{The energy, $E_{\rm gs}(q,R)$, of the lightest flux tube state with 
length $R$ and zero longitudinal momentum. 
Quantum numbers are as shown.
Calculations are: SU(3) at $\beta=6.0625$ ($N=3$), SU(5) at $\beta=17.630$ ($N=5$), SU(6) at $\beta=25.550$ ($N=6$) and SU(3) at $\beta=6.338$ ($N=3f$) as well as  $\beta=18.375$ ($N=5f$).}
\label{tab:table_absolute_ground_state}
\end{center}
\end{table}

\begin{table}[h]
\begin{center}
{\footnotesize
\begin{tabular}{cccc|cccc|ccc}\hline \hline
\multicolumn{11}{c}{ $E_{1^{\rm st} {\rm ex.}}(q,R)$ } \\ \hline \hline
$q$ & $|J|$ & $P_{\perp}$ & $P_{||}$ & $R/a$ & $N=3$ & $N=5$ & $N=6$ & $R/a$ & $N=3f$ & $N=5f$\\ \hline \hline
 0 & 0 & + & + & 10 &  0.9007(133)   &  0.9223(151)  & 0.9787(199) & 14 & 0.6532(319)   &   \\
   &   &   &  & 12 &  1.0078(72)    &  0.9912(99)   & 1.0163(109) & 16 & 0.6003(93)    &   \\
   &   &   &  & 14 &  1.0548(96)    &  1.0607(144)  & 1.0813(122) & 18 & 0.6494(106)   &   \\
   &   &   &  & 16 &  1.0984(111)   &  1.0929(166)  & 1.1403(134) & 20 & 0.6807(70)    &  0.6848(139) \\
   &   &   &  & 18 &  1.1203(170)   &  1.1436(132)  & 1.1842(141) & 22 & 0.7076(50)    &  0.6933(112) \\
   &   &   &  & 20 &  1.2002(140)   &  1.2266(230)  & 1.2131(215) & 24 & 0.7048(108)   &  0.7084(140) \\ 
   &   &   &  & 22 &  1.2566(199)   &  1.2708(258)  & 1.3110(258) & 26 & 0.7433(47)    &  0.7523(217) \\ 
   &   &   &  & 24 &  1.3009(179)   &  1.2824(392)  & 1.3182(416) & 28 & 0.7715(42)    &   0.7910(128) \\ 
   &   &   &  & 26 &  1.3670(235)   &               &             & 30 & 0.7754(85)    &   \\ 
   &   &   &  & 28 &  1.3754(256)   &               &             & 32 & 0.8179(90)    &   \\
   &   &   &  & 30 &  1.4876(308)   &               &             & 34 & 0.8473(70)    &   \\
   &   &   &  & 32 &  1.5917(255)   &               &             & 36 & 0.8670(96)    &   \\ 
   &   &   &  & 36 &  1.7109(276)   &               &             & 38 & 0.9113(95)    &   \\ 
   &   &   &  &    &                &               &             & 40 & 0.8959(181)   &  \\ \hline \hline
\end{tabular}}
\caption{The energy, $E_{1^{\rm st} {\rm ex.}}(q,R)$, of the first excited flux tube state with 
length $R$ and zero longitudinal momentum. Quantum numbers are as shown.
Calculations are: SU(3) at $\beta=6.0625$ ($N=3$), SU(5) at $\beta=17.630$ ($N=5$), SU(6) at $\beta=25.550$ ($N=6$) and SU(3) at $\beta=6.338$ ($N=3f$) as well as  $\beta=18.375$ ($N=5f$).}
\label{tab:0++_first_excited}
\end{center}
\end{table}

\begin{table}[h]
\begin{center}
{\footnotesize
\begin{tabular}{cccc|cccc|ccc}\hline \hline
\multicolumn{10}{c}{ $E_{2^{\rm nd} ex.}(q,R)$ } \\ \hline \hline
$q$ & $|J|$ & $P_{\perp}$ & $P_{||}$ & $R/a$ & $N=3$ & $N=5$ & $N=6$ & $R/a$ & $N=3f$ & $N=5f$\\ \hline \hline
 0 & 0 & + & + & 10 &  1.0437(223)   &  1.0443(219)   &  1.0753(279)  & 14  &  0.6649(154)    &   \\
   &   &   &   & 12 &  1.1160(292)   &  1.0878(306)   &  1.1089(157)  & 16  &  0.6963(125)    &   \\
   &   &   &   & 14 &  1.1999(146)   &  1.1797(163)   &  1.1887(181)  & 18  &  0.7345(105)    &   \\
   &   &   &   & 16 &  1.2573(188)   &  1.2658(227)   &  1.3092(212)  & 20  &  0.7793(114)    &  0.7673(203) \\
   &   &   &   & 18 &  1.3492(106)   &  1.3308(369)   &  1.4145(306)  & 22  &  0.7862(150)    &  0.8093(243) \\
   &   &   &   & 20 &  1.4246(303)   &  1.3720(426)   &  1.4261(502)  & 24  &  0.8356(130)    &  0.8023(192) \\ 
   &   &   &   & 22 &  1.5098(394)   &  1.5125(560)   &  1.4940(706)  & 26  &  0.8768(140)    &  0.8886(423) \\ 
   &   &   &   & 24 &  1.6182(126)   &  1.6511(809)   &  1.7563(238)  & 28  &  0.9232(201)    &  0.9137(428) \\ 
   &   &   &   & 26 &  1.6342(647)   &                &               & 30  &  0.9381(260)    &   \\ 
   &   &   &   & 28 &  1.7361(739)   &                &               & 32  &  0.9185(335)    &   \\
   &   &   &   & 30 &  1.7133(884)   &                &               & 34  &  0.9925(386)    &   \\
   &   &   &   & 32 &  1.9085(363)   &                &               & 36  &  1.0728(442)    &   \\ 
   &   &   &   & 36 &  1.9795(497)   &                &               & 38  &  1.0726(544)    &   \\ 
   &   &   &   &    &                &                &               & 40  &  1.1191(343)    &  \\ \hline \hline
\end{tabular}}
\caption{The energy, $E_{2^{\rm nd} ex.}(q,R)$, of the second excited flux tube state with 
length $R$ and zero longitudinal momentum. 
Quantum numbers are as shown.
Calculations are: SU(3) at $\beta=6.0625$ ($N=3$), SU(5) at $\beta=17.630$ ($N=5$), SU(6) at $\beta=25.550$ ($N=6$) and SU(3) at $\beta=6.338$ ($N=3f$) as well as  $\beta=18.375$ ($N=5f$)}
\label{tab:0++_second_excited}
\end{center}
\end{table}

\begin{table}[h]
\begin{center}
{\footnotesize
\begin{tabular}{cccc|cccc|ccc}\hline \hline
\multicolumn{11}{c}{ $aE_{\rm gs}(q,R)$ } \\ \hline \hline
$q$ & $|J|$ & $P_{\perp}$ & $P_{||}$& $R/a$ & $N=3$ & $N=5$ & $N=6$ & $R/a$ & $N=3f$ & $N=5f$\\ \hline \hline
 0 & 0 & + & - & 10 &  1.9160(744)  &              &             & 14 & 1.2020(236)  &   \\
   &   &   &   & 12 &  1.9926(207)  &  2.0029(257) & 2.0814(341) & 16 & 1.2196(290)  &   \\
   &   &   &   & 14 &  2.0209(221)  &  2.0185(272) & 2.1425(372) & 18 & 1.2297(287)  &   \\
   &   &   &   & 16 &  1.9860(210)  &  1.9540(309) &             & 20 & 1.2711(251)  &  1.2951(414) \\
   &   &   &   & 18 &               &  2.0834(346) & 2.0928(239) & 22 & 1.3134(333)  &  1.2615(486) \\
   &   &   &   & 20 &  2.1081(260)  &  2.0408(326) & 2.1555(452) & 24 & 1.3088(263)  &              \\ 
   &   &   &   & 22 &  2.1385(261)  &  2.1215(429) &             & 26 & 1.3109(262)  &  1.4781(943) \\ 
   &   &   &   & 24 &  2.0995(353)  &  2.1840(478) & 2.2523(694) & 28 & 1.3714(337)  &  1.4147(722) \\ 
   &   &   &   & 26 &  2.1523(324)  &              &             & 30 & 1.3439(380)  &  \\ 
   &   &   &   & 28 &  2.1729(461)  &              &             & 32 & 1.3240(448)  &  \\
   &   &   &   & 30 &               &              &             & 34 & 1.3659(428)  &  \\
   &   &   &   & 32 &  2.2520(956)  &              &             & 36 & 1.4983(610)  &  \\ 
   &   &   &   & 36 &               &              &             & 38 & 1.4133(422)  &  \\ 
   &   &   &   &    &               &              &             & 40 & 1.4227(1052) &  \\ \hline \hline
\end{tabular}}
\caption{The energy, $E_{\rm gs}(q,R)$, of the lightest flux tube state with 
length $R$ and zero longitudinal momentum. 
Quantum numbers are as shown.
Calculations are: SU(3) at $\beta=6.0625$ ($N=3$), SU(5) at $\beta=17.630$ ($N=5$), SU(6) at $\beta=25.550$ ($N=6$) and SU(3) at $\beta=6.338$ ($N=3f$) as well as  $\beta=18.375$ ($N=5f$).}
\label{tab:0+-_ground_state}
\end{center}
\end{table}

\begin{table}[h]
\begin{center}
{\footnotesize
\begin{tabular}{cccc|cccc|ccc}\hline \hline
\multicolumn{11}{c}{ $aE_{\rm gs}(q,R)$ } \\ \hline \hline
$q$ & $|J|$ & $P_{\perp}$ & $P_{||}$& $R/a$ & $N=3$ & $N=5$ & $N=6$ & $R/a$ & $N=3f$ & $N=5f$\\ \hline \hline
 0 & 0 & - & + & 10 &  1.6556(444)  &  1.6379(128) &              & 14 & 1.1030(342)  &   \\
   &   &   &   & 12 &  1.5998(475)  &  1.6743(172) & 1.7352(126)  & 16 & 0.9808(421)  &   \\
   &   &   &   & 14 &  1.6044(634)  &              & 1.7773(184)  & 18 & 1.0599(124)  &   \\
   &   &   &   & 16 &               &  1.7627(228) & 1.8001(186)  & 20 & 1.0573(122)  &  1.0864(343) \\
   &   &   &   & 18 &  1.6211(868)  &  1.8358(290) & 1.8496(205)  & 22 & 1.0616(108)  &  1.0554(272) \\
   &   &   &   & 20 &  1.8214(194)  &  1.8321(289) & 1.9309(260)  & 24 & 1.0910(133)  &  1.0811(191) \\ 
   &   &   &   & 22 &               &              & 1.9403(421)  & 26 & 1.1173(160)  &  1.1404(380) \\
   &   &   &   & 24 &  1.9048(187)  &              &              & 28 & 1.1366(168)  &  1.1659(392) \\ 
   &   &   &   & 26 &  2.0054(216)  &              &              & 30 & 1.1781(185)  &  \\ 
   &   &   &   & 28 &  2.0104(265)  &              &              & 32 & 1.1862(333)  &  \\
   &   &   &   & 30 &               &              &              & 34 & 1.2264(416)  &  \\
   &   &   &   & 32 &               &              &              & 36 &              &  \\ 
   &   &   &   & 36 &               &              &              & 38 &              &  \\ 
   &   &   &   &    &               &              &              & 40 &              &  \\ \hline \hline
\end{tabular}}
\caption{The energy, $E_{\rm gs}(q,R)$, of the lightest flux tube state with 
length $R$ and zero longitudinal momentum. 
Quantum numbers are as shown.
Calculations are: SU(3) at $\beta=6.0625$ ($N=3$), SU(5) at $\beta=17.630$ ($N=5$), SU(6) at $\beta=25.550$ ($N=6$) and SU(3) at $\beta=6.338$ ($N=3f$) as well as  $\beta=18.375$ ($N=5f$).}
\label{tab:0-+_ground_state}
\end{center}
\end{table}

\begin{table}[h]
\begin{center}
{\footnotesize
\begin{tabular}{cccc|cccc|ccc}\hline \hline
\multicolumn{11}{c}{ $E_{{\rm gs}}(q,R)$ } \\ \hline \hline
$q$ & $|J|$ & $P_{\perp}$ & $P_{||}$ & $R/a$ & $N=3$ & $N=5$ & $N=6$ & $R/a$ & $N=3f$ & $N=5f$\\ \hline \hline
 0 & 0 & - & - & 10 &  0.6583(59)    &  0.6353(102)   &  0.6609(104)  & 14  &  0.4163(47)     &   \\
   &   &   &   & 12 &  0.7316(80)    &  0.7207(107)   &  0.7377(99)   & 16  &  0.4443(38)     &   \\
   &   &   &   & 14 &  0.8221(91)    &  0.7888(146)   &  0.8348(138)  & 18  &  0.4832(51)     &   \\
   &   &   &   & 16 &  0.8943(150)   &  0.8866(74)    &  0.9363(78)   & 20  &  0.5368(50)     &  0.5229(61) \\
   &   &   &   & 18 &  0.9783(109)   &  0.9874(122)   &  1.0289(84)   & 22  &  0.5578(77)     &  0.5602(82) \\
   &   &   &   & 20 &  1.0540(200)   &  1.0469(99)    &  1.1094(145)  & 24  &  0.6003(50)     &  0.5800(116) \\ 
   &   &   &   & 22 &  1.1563(137)   &  1.1948(202)   &  1.1733(205)  & 26  &  0.6366(57)     &  0.6374(125) \\ 
   &   &   &   & 24 &  1.2536(189)   &  1.2629(271)   &  1.3221(121)  & 28  &  0.6709(75)     &  0.6469(145) \\ 
   &   &   &   & 26 &  1.3009(216)   &                &               & 30  &  0.7085(80)     &   \\ 
   &   &   &   & 28 &  1.3728(237)   &                &               & 32  &  0.7390(110)    &   \\
   &   &   &   & 30 &  1.4908(403)   &                &               & 34  &  0.7712(113)    &   \\
   &   &   &   & 32 &  1.5648(219)   &                &               & 36  &  0.8011(128)    &   \\ 
   &   &   &   & 36 &  1.7378(322)   &                &               & 38  &  0.8497(208)    &   \\ 
   &   &   &   &    &                &                &               & 40  &  0.9033(213)    &  \\ \hline \hline
\end{tabular}}
\caption{The energy, $E_{gs} (q,R)$, of the lightest flux tube state with 
length $R$ and zero longitudinal momentum. 
Quantum numbers are as shown.
Calculations are: SU(3) at $\beta=6.0625$ ($N=3$), SU(5) at $\beta=17.630$ ($N=5$), SU(6) at $\beta=25.550$ ($N=6$) and SU(3) at $\beta=6.338$ ($N=3f$) as well as  $\beta=18.375$ ($N=5f$)}
\label{tab:0--_ground_state}
\end{center}
\end{table}

\begin{table}[h]
\begin{center}
{\footnotesize
\begin{tabular}{cccc|cccc|ccc}\hline \hline
\multicolumn{11}{c}{ $E_{1^{\rm st} {\rm ex}}(q,R)$ } \\ \hline \hline
$q$ & $|J|$ & $P_{\perp}$ & $P_{||}$ & $R/a$ & $N=3$ & $N=5$ & $N=6$ & $R/a$ & $N=3f$ & $N=5f$\\ \hline \hline
 0 & 0 & - & - & 10 &  1.1569(85)    &  1.1561(527)   &               & 14  &  0.7981(208)    &   \\
   &   &   &   & 12 &  1.2159(139)   &  1.1972(176)   &  1.2633(211)  & 16  &  0.7849(77)     &   \\
   &   &   &   & 14 &  1.2997(193)   &  1.2738(213)   &  1.3636(240)  & 18  &  0.8079(119)    &   \\
   &   &   &   & 16 &  1.2963(235)   &  1.3228(331)   &  1.3969(272)  & 20  &  0.7889(209)    &  0.8386(282) \\
   &   &   &   & 18 &  1.2868(215)   &  1.2981(269)   &  1.3052(259)  & 22  &  0.8518(110)    &  0.8554(257) \\
   &   &   &   & 20 &  1.3660(380)   &  1.3328(395)   &  1.3538(400)  & 24  &  0.8503(182)    &  0.8420(206) \\ 
   &   &   &   & 22 &  1.3504(246)   &  1.3352(343)   &  1.5158(729)  & 26  &  0.8795(76)     &  0.8614(361) \\ 
   &   &   &   & 24 &  1.4346(247)   &  1.4559(487)   &               & 28  &  0.8859(82)     &  0.8781(371) \\ 
   &   &   &   & 26 &  1.4370(307)   &                &               & 30  &  0.8695(195)    &   \\ 
   &   &   &   & 28 &  1.5264(396)   &                &               & 32  &  0.9247(150)    &   \\
   &   &   &   & 30 &  1.5784(445)   &                &               & 34  &  0.9317(189)    &   \\
   &   &   &   & 32 &  1.6330(790)   &                &               & 36  &  0.9612(201)    &   \\ 
   &   &   &   & 36 &                &                &               & 38  &  0.9375(223)    &   \\ 
   &   &   &   &    &                &                &               & 40  &  0.8923(477)    &  \\ \hline \hline
\end{tabular}}
\caption{The energy, $E_{1^{\rm st} {\rm ex}}(q,R)$, of the first excited flux tube state with 
length $R$ and zero longitudinal momentum. 
Quantum numbers are as shown. Calculations are: SU(3) at $\beta=6.0625$ ($N=3$), SU(5) at $\beta=17.630$ ($N=5$), SU(6) at $\beta=25.550$ ($N=6$) and SU(3) at $\beta=6.338$ ($N=3f$) as well as  $\beta=18.375$ ($N=5f$)}
\label{tab:0--_first_excited_state}
\end{center}
\end{table}

\begin{table}[h]
\begin{center}
{\footnotesize
\begin{tabular}{cccc|cccc|ccc}\hline \hline
\multicolumn{11}{c}{ $E_{{\rm g.s}}(q,R)$ } \\ \hline \hline
$q$ & $|J|$ & $P_{\perp}$ & $P_{||}$ & $R/a$ & $N=3$ & $N=5$ & $N=6$ & $R/a$ & $N=3f$ & $N=5f$\\ \hline \hline
 0 & 1 & $\pm$ & + & 10 &  1.1566(307)   &                &               & 14  &  0.8368(182)    &   \\
   &   &   &       & 12 &  1.2290(568)   &  1.2798(289)   &               & 16  &  0.7969(141)     &   \\
   &   &   &       & 14 &  1.2664(552)   &  1.3715(317)   &  1.4325(429)  & 18  &  0.7910(272)    &   \\
   &   &   &       & 16 &  1.3558(289)   &  1.3951(369)   &  1.3292(335)  & 20  &  0.8986(176)    &  0.8738(284) \\
   &   &   &       & 18 &  1.4173(389)   &  1.3710(388)   &  1.4180(378)  & 22  &  0.8878(196)    &  0.8285(281) \\
   &   &   &       & 20 &  1.4297(313)   &  1.4050(465)   &               & 24  &  0.8853(144)    &  0.9106(277) \\ 
   &   &   &       & 22 &  1.5318(492)   &  1.5499(552)   &  1.5541(730)  & 26  &  0.9015(204)    &  0.9419(279) \\ 
   &   &   &       & 24 &  1.5619(512)   &  1.5040(678)   &               & 28  &  0.9431(221)    &  0.9622(217) \\ 
   &   &   &       & 26 &  1.6123(612)   &                &               & 30  &  0.9597(244)    &   \\ 
   &   &   &       & 28 &  1.6470(706)   &                &               & 32  &  1.0141(136)    &   \\
   &   &   &       & 30 &  1.8617(204)   &                &               & 34  &  1.0355(130)    &   \\
   &   &   &       & 32 &                &                &               & 36  &  1.0259(169)    &   \\ 
   &   &   &       & 36 &                &                &               & 38  &  1.0707(215)    &   \\ 
   &   &   &       &    &                &                &               & 40  &  1.0782(202)    &  \\ \hline \hline
\end{tabular}}
\caption{The energy, $E_{{\rm gs}}(q,R)$, of the lightest flux tube state with 
length $R$ and zero longitudinal momentum. 
Quantum numbers are as shown. Calculations are: SU(3) at $\beta=6.0625$ ($N=3$), SU(5) at $\beta=17.630$ ($N=5$), SU(6) at $\beta=25.550$ ($N=6$) and SU(3) at $\beta=6.338$ ($N=3f$) as well as  $\beta=18.375$ ($N=5f$)}
\label{tab:1+_ground_state}
\end{center}
\end{table}

\begin{table}[h]
\begin{center}
{\footnotesize
\begin{tabular}{cccc|cccc|ccc}\hline \hline
\multicolumn{11}{c}{ $E_{{\rm g.s}}(q,R)$ } \\ \hline \hline
$q$ & $|J|$ & $P_{\perp}$ & $P_{||}$ & $R/a$ & $N=3$ & $N=5$ & $N=6$ & $R/a$ & $N=3f$ & $N=5f$\\ \hline \hline
 0 & 1 & $\pm$ & - & 10 &  1.0564(179)   &  1.0594(288)   &  1.0695(310)   & 14  &  0.7045(186)    &   \\
   &   &   &       & 12 &  1.1279(281)   &  1.1372(484)   &  1.1455(485)   & 16  &  0.7188(154)    &   \\
   &   &   &       & 14 &  1.1664(370)   &  1.1591(559)   &  1.2022(440)   & 18  &  0.7486(112)    &   \\
   &   &   &       & 16 &  1.2281(173)   &  1.2422(280)   &  1.2745(213)   & 20  &  0.7712(113)    &  0.7785(194) \\
   &   &   &       & 18 &  1.2958(345)   &  1.3200(326)   &  1.3238(209)   & 22  &  0.7809(150)    &  0.7942(178) \\
   &   &   &       & 20 &  1.3158(268)   &  1.4180(340)   &  1.3938(443)   & 24  &  0.8070(172)    &  0.8152(206) \\ 
   &   &   &       & 22 &  1.4016(336)   &  1.5431(537)   &  1.4913(456)   & 26  &  0.8260(124)    &  0.8701(357) \\ 
   &   &   &       & 24 &  1.5174(445)   &                &                & 28  &  0.8658(173)    &  0.9096(400) \\ 
   &   &   &       & 26 &  1.6184(472)   &                &                & 30  &  0.9083(216)    &   \\ 
   &   &   &       & 28 &  1.5890(576)   &                &                & 32  &  0.9418(292)    &   \\
   &   &   &       & 30 &  1.7779(974)   &                &                & 34  &  1.0106(134)    &   \\
   &   &   &       & 32 &                &                &                & 36  &  1.0278(142)    &   \\ 
   &   &   &       & 36 &                &                &                & 38  &  1.0717(164)    &   \\ 
   &   &   &       &    &                &                &                & 40  &  1.1510(297)    &  \\ \hline \hline
\end{tabular}}
\caption{The energy, $E_{{\rm gs}}(q,R)$, of the first excited flux tube state with 
length $R$ and zero longitudinal momentum. 
Quantum numbers are as shown. Calculations are: SU(3) at $\beta=6.0625$ ($N=3$), SU(5) at $\beta=17.630$ ($N=5$), SU(6) at $\beta=25.550$ ($N=6$) and SU(3) at $\beta=6.338$ ($N=3f$) as well as  $\beta=18.375$ ($N=5f$)}
\label{tab:1-_ground_state}
\end{center}
\end{table}

\begin{table}[h]
\begin{center}
{\footnotesize
\begin{tabular}{cccc|cccc|ccc}\hline \hline
\multicolumn{11}{c}{ $E_{{\rm gs}}(q,R)$ } \\ \hline \hline
$q$ & $|J|$ & $P_{\perp}$ & $P_{||}$ & $R/a$ & $N=3$ & $N=5$ & $N=6$ & $R/a$ & $N=3f$ & $N=5f$\\ \hline \hline
 0 & 2 & + & + & 10 &  1.0238(56)    &  1.0426(263)   &  1.0604(274)  &  14  &   0.7043(239)    &              \\
   &   &   &   & 12 &  1.0444(260)   &  1.0663(132)   &  1.0940(114)  &  16  &   0.7359(164)    &              \\
   &   &   &   & 14 &  1.0830(277)   &  1.1210(123)   &  1.1558(164)  &  18  &   0.7185(70)     &              \\
   &   &   &   & 16 &  1.1526(110)   &  1.1565(154)   &  1.1921(147)  &  20  &   0.7218(101)    &  0.7388(167) \\
   &   &   &   & 18 &  1.1673(149)   &  1.2039(204)   &  1.2486(155)  &  22  &   0.7404(94)     &  0.7336(159) \\
   &   &   &   & 20 &  1.2407(150)   &  1.2076(214)   &  1.2747(223)  &  24  &   0.7559(82)     &  0.7673(113) \\ 
   &   &   &   & 22 &  1.2552(162)   &  1.2629(261)   &  1.3666(293)  &  26  &   0.7612(121)    &  0.7936(128) \\ 
   &   &   &   & 24 &  1.3065(210)   &  1.3555(309)   &  1.3818(503)  &  28  &   0.7863(115)    &  0.7879(235) \\ 
   &   &   &   & 26 &  1.3562(233)   &                &               &  30  &   0.7942(118)    &              \\ 
   &   &   &   & 28 &  1.4423(300)   &                &               &  32  &   0.8146(161)    &              \\
   &   &   &   & 30 &  1.4591(319)   &                &               &  34  &   0.8697(195)    &              \\
   &   &   &   & 32 &  1.5271(184)   &                &               &  36  &   0.8511(176)    &              \\ 
   &   &   &   & 36 &  1.7247(260)   &                &               &  38  &   0.9006(222)    &              \\ 
   &   &   &   &    &                &                &               &  40  &   0.9499(160)    &              \\ \hline \hline
\end{tabular}}
\caption{The energy, $E_{gs}(q,R)$, of the lightest flux tube state with length $R$ and zero longitudinal momentum. 
Quantum numbers are as shown.
Calculations are: SU(3) at $\beta=6.0625$ ($N=3$), SU(5) at $\beta=17.630$ ($N=5$), SU(6) at $\beta=25.550$ ($N=6$) and SU(3) at $\beta=6.338$ ($N=3f$) as well as  $\beta=18.375$ ($N=5f$)}
\label{tab:2++_ground_state}
\end{center}
\end{table}

\begin{table}[h]
\begin{center}
{\footnotesize
\begin{tabular}{cccc|cccc|ccc}\hline \hline
\multicolumn{11}{c}{ $E_{1^{\rm st} {\rm ex.s}}(q,R)$ } \\ \hline \hline
$q$ & $|J|$ & $P_{\perp}$ & $P_{||}$ & $R/a$ & $N=3$ & $N=5$ & $N=6$ & $R/a$ & $N=3f$ & $N=5f$\\ \hline \hline
 0 & 2 & + & + & 10 &  1.3961(223)    &  1.4503(360)  &  1.4693(363)  &  14  &   1.0081(398)    &              \\
   &   &   &   & 12 &  1.4827(301)   &  1.4681(465)   &  1.4850(527)  &  16  &   0.9946(207)    &              \\
   &   &   &   & 14 &  1.4971(333)   &  1.4943(531)   &  1.5731(579)  &  18  &   0.9060(221)     &              \\
   &   &   &   & 16 &  1.5310(379)   &                &  1.5159(454)  &  20  &   0.9513(178)     &  0.9825(159) \\
   &   &   &   & 18 &  1.5220(446)   &  1.5958(608)   &  1.6330(497)  &  22  &   0.9557(177)     &  0.9804(390) \\
   &   &   &   & 20 &  1.5530(366)   &                &  1.6734(785)  &  24  &   0.9794(216)     &  1.0070(387) \\ 
   &   &   &   & 22 &  1.6000(526)   &  1.5937(762)   &  1.7093(911)  &  26  &   1.0100(264)     &  1.0183(243) \\ 
   &   &   &   & 24 &  1.6910(572)   &  1.5869(856)   &  1.8073(284)  &  28  &                   &  1.0120(242) \\ 
   &   &   &   & 26 &  1.5890(534)   &                &               &  30  &   1.0350(349)     &              \\ 
   &   &   &   & 28 &  1.6465(846)   &                &               &  32  &   1.0954(431)     &              \\
   &   &   &   & 30 &  1.8744(1094)  &                &               &  34  &   1.0584(509)     &              \\
   &   &   &   & 32 &  1.9196(474)   &                &               &  36  &   1.0995(554)     &              \\ 
   &   &   &   & 36 &  2.0635(466)   &                &               &  38  &   1.0630(393)     &              \\ 
   &   &   &   &    &                &                &               &  40  &   1.1904(357)     &              \\ \hline \hline
\end{tabular}}
\caption{The energy, $E_{gs}(q,R)$, of the lightest flux tube state with 
length $R$ and zero longitudinal momentum. 
Quantum numbers are as shown.
Calculations are: SU(3) at $\beta=6.0625$ ($N=3$), SU(5) at $\beta=17.630$ ($N=5$), SU(6) at $\beta=25.550$ ($N=6$) and SU(3) at $\beta=6.338$ ($N=3f$) as well as  $\beta=18.375$ ($N=5f$)}
\label{tab:2++_first_excited_state}
\end{center}
\end{table}

\begin{table}[h]
\begin{center}
{\footnotesize
\begin{tabular}{cccc|cccc|ccc}\hline \hline
\multicolumn{11}{c}{ $E_{{\rm gs}}(q,R)$ } \\ \hline \hline
$q$ & $|J|$ & $P_{\perp}$ & $P_{||}$ & $R/a$ & $N=3$ & $N=5$ & $N=6$ & $R/a$ & $N=3f$ & $N=5f$\\ \hline \hline
 0 & 2 & - &  + & 10 &   1.0483(206)  &  1.0574(263)   &  1.0570(287)  &  14  &   0.7164(296)    &                 \\
   &   &   &   & 12 &   1.0864(345)  &  1.1355(146)   &  1.1667(171)  &  16  &    0.7368(91)   &                  \\
   &   &   &   & 14 &   1.1487(415)  &  1.1981(177)   &  1.0871(567)  &  18  &    0.7264(97)   &                  \\
   &   &   &   & 16 &   1.1703(122)  &  1.1127(184)   &  1.1688(148)  &  20  &    0.7419(112)   & 0.7500(81)   \\
   &   &   &   & 18 &   1.1791(141)  &  1.2078(246)   &  1.2275(131)  &  22  &    0.7563(101)   & 0.7765(101)   \\
   &   &   &   & 20 &   1.2319(194)  &  1.2667(248)   &  1.2399(239)  &  24  &    0.7600(119)   & 0.7898(79)   \\ 
   &   &   &   & 22 &   1.2528(211)  &  1.2857(255)   &  1.3291(287)  &  26  &    0.7928(108)   & 0.8148(164)   \\ 
   &   &   &   & 24 &   1.3332(208)  &  1.3279(331)   &  1.4215(550)  &  28  &    0.8148(124)   & 0.8147(133)  \\ 
   &   &   &   & 26 &   1.4058(254)  &                &               &  30  &    0.8397(152)   &               \\ 
   &   &   &   & 28 &   1.4686(281)  &                &               &  32  &    0.8462(157)   &                \\
   &   &   &   & 30 &   1.4963(400)  &                &               &  34  &    0.8348(273)   &               \\
   &   &   &   & 32 &   1.6377(248)  &                &               &  36  &    0.8776(180)   &               \\ 
   &   &   &   & 36 &   1.7521(314)  &                &               &  38  &    0.9367(192)   &               \\ 
   &   &   &   &    &                &                &               &  40  &    0.9453(182)   &               \\ \hline \hline
\end{tabular}}
\caption{The energy, $E_{gs}(q,R)$, of the lightest flux tube state with 
length $R$ and zero longitudinal momentum. 
Quantum numbers are as shown.
Calculations are: SU(3) at $\beta=6.0625$ ($N=3$), SU(5) at $\beta=17.630$ ($N=5$), SU(6) at $\beta=25.550$ ($N=6$) and SU(3) at $\beta=6.338$ ($N=3f$) as well as  $\beta=18.375$ ($N=5f$)}
\label{tab:2-+_ground_state}
\end{center}
\end{table}

\begin{table}[h]
\begin{center}
{\footnotesize
\begin{tabular}{cccc|cccc|ccc}\hline \hline
\multicolumn{11}{c}{ $E_{1^{\rm st} {\rm ex.s}}(q,R)$ } \\ \hline \hline
$q$ & $|J|$ & $P_{\perp}$ & $P_{||}$& $R/a$ & $N=3$ & $N=5$ & $N=6$ & $R/a$ & $N=3f$ & $N=5f$\\ \hline \hline
 0 & 2 & - &  + & 10 &   1.3763(244)  &                &  1.4426(391)  &  14  &        &                 \\
   &   &   &    & 12 &   1.5048(360)  &  1.6395(727)   &  1.5624(615)  &  16  &    0.9135(168)   &                  \\
   &   &   &    & 14 &   1.5568(445)  &  1.6319(834)    &  1.7431(851)  &  18  &   0.9383(204)   &                  \\
   &   &   &    & 16 &   1.5221(406)  &  1.5770(649)   &  1.6271(518)  &  20  &    0.9391(255)   & 0.7500(81)   \\
   &   &   &    & 18 &   1.5250(477)  &  1.6263(664)   &  1.6866(771)  &  22  &    1.0307(308)   & 0.7765(101)   \\
   &   &   &    & 20 &   1.5948(601)  &  1.6743(895)   &  1.5640(707)  &  24  &    0.9959(412)   & 0.7898(79)   \\ 
   &   &   &    & 22 &   1.6922(670)  &  1.7035(984)   &  1.7486(1384)  &  26  &   1.0947(455)   & 0.8148(164)   \\ 
   &   &   &    & 24 &   1.7389(834)  &  1.6827(1012)   &  1.8130(1677)  &  28  &     & 0.8147(133)  \\ 
   &   &   &    & 26 &   1.7767(741)  &                &               &  30  &       &               \\ 
   &   &   &    & 28 &   1.7681(919)  &                &               &  32  &    1.0900(533)    &                \\
   &   &   &    & 30 &   1.8676(939)  &                &               &  34  &    1.1102(302)   &               \\
   &   &   &    & 32 &   1.9389(561)  &                &               &  36  &    1.1084(334)   &               \\ 
   &   &   &    & 36 &   2.0549(612)  &                &               &  38  &    1.1546(327)   &               \\ 
   &   &   &    &    &                &                &               &  40  &                  &               \\ \hline \hline
\end{tabular}}
\caption{The energy, $E_{1^{\rm st} \rm ex.s}(q,R)$, of the first excited flux tube state with 
length $R$ and zero longitudinal momentum. 
Quantum numbers are as shown.
Calculations are: SU(3) at $\beta=6.0625$ ($N=3$), SU(5) at $\beta=17.630$ ($N=5$), SU(6) at $\beta=25.550$ ($N=6$) and SU(3) at $\beta=6.338$ ($N=3f$) as well as  $\beta=18.375$ ($N=5f$)}
\label{tab:2-+_first_excited_state}
\end{center}
\end{table}

\begin{table}[h]
\begin{center}
{\footnotesize
\begin{tabular}{cccc|cccc|ccc}\hline \hline
\multicolumn{11}{c}{ $E_{{\rm gs}}(q,R)$ } \\ \hline \hline
$q$ & $|J|$ & $P_{\perp}$ & $P_{||}$& $R/a$ & $N=3$ & $N=5$ & $N=6$ & $R/a$ & $N=3f$ & $N=5f$\\ \hline \hline
 0 & 2 & + &  - & 10 &   1.3207(117)  &  1.2183(641)    &  1.3074(806)    &  14  &   0.8792,  0.0180    &               \\
   &   &   &    & 12 &   1.3221(224)  &  1.2758(840)    &  1.4010(299)    &  16  &   0.8622,  0.0127    &                 \\
   &   &   &    & 14 &   1.3622(195)  &  1.3622(374)    &  1.4362(403)    &  18  &   0.8942,  0.0158    &                \\
   &   &   &    & 16 &   1.4335(277)  &  1.4849(506)    &  1.4970(432)    &  20  &   0.9110,  0.0193    &    0.9013(329)  \\
   &   &   &    & 18 &   1.5401(373)  &  1.5299(705)    &  1.5987(573)    &  22  &   0.9047,  0.0252    &    0.9091(368)  \\
   &   &   &    & 20 &   1.5455(561)  &  1.5628(698)    &  1.6979(867)    &  24  &   0.9241,  0.0155    &    0.9806(395)  \\ 
   &   &   &    & 22 &   1.6652(565)  &                 &                 &  26  &   1.0113,  0.0260    &    0.9977(299)  \\ 
   &   &   &    & 24 &   1.6548(765)  &                 &                 &  28  &   0.9839,  0.0270    &    1.0065(311)  \\ 
   &   &   &    & 26 &                &                 &                 &  30  &   1.0591,  0.0366    &                 \\ 
   &   &   &    & 28 &                &                 &                 &  32  &   1.0937,  0.0181    &                 \\
   &   &   &    & 30 &                &                 &                 &  34  &   1.0935,  0.0197    &                 \\
   &   &   &    & 32 &                &                 &                 &  36  &   1.0958,  0.0378    &                 \\ 
   &   &   &    & 36 &                &                 &                 &  38  &                      &                \\ 
   &   &   &    &    &                &                 &                 &  40  &                      &                \\ \hline \hline
\end{tabular}}
\caption{The energy, $E_{ \rm gs}(q,R)$, of the lightest flux tube state with 
length $R$ and zero longitudinal momentum. 
Quantum numbers are as shown.
Calculations are: SU(3) at $\beta=6.0625$ ($N=3$), SU(5) at $\beta=17.630$ ($N=5$), SU(6) at $\beta=25.550$ ($N=6$) and SU(3) at $\beta=6.338$ ($N=3f$) as well as  $\beta=18.375$ ($N=5f$)}
\label{tab:2+-_ground_state}
\end{center}
\end{table}

\begin{table}[h]
\begin{center}
{\footnotesize
\begin{tabular}{cccc|cccc|ccc}\hline \hline
\multicolumn{11}{c}{ $E_{{\rm gs}}(q,R)$ } \\ \hline \hline
$q$ & $|J|$ & $P_{\perp}$ & $P_{||}$& $R/a$ & $N=3$ & $N=5$ & $N=6$ & $R/a$ & $N=3f$ & $N=5f$\\ \hline \hline
 0 & 2 & - &  - & 10 &   1.2933(784)  &  1.2607(791)    &  1.3108(1309)   &  14  &    0.8574(209)    &               \\
   &   &   &    & 12 &   1.4419(350)  &  1.4205(479)    &                 &  16  &    0.8963(144)  &                 \\
   &   &   &    & 14 &                &  1.4865(680)    &  1.4366(739)    &  18  &    0.8891(208)   &                \\
   &   &   &    & 16 &   1.4758(493)  &  1.5038(724)    &  1.5253(727)    &  20  &    0.9057(235)   &  0.9144(370)   \\
   &   &   &    & 18 &                &                 &  1.6650(877)    &  22  &    0.9966(285)   &  0.9679(643)   \\
   &   &   &    & 20 &                &  1.6211(1204)   &                 &  24  &    0.9788(300)   &  1.0091(424)   \\ 
   &   &   &    & 22 &                &  1.6823(1473)   &                 &  26  &    1.0476(379)   &  1.0555(853)   \\ 
   &   &   &    & 24 &                &                 &                 &  28  &                  &                \\ 
   &   &   &    & 26 &                &                 &                 &  30  &    1.0842(517)   &                \\ 
   &   &   &    & 28 &                &                 &                 &  32  &                  &                \\
   &   &   &    & 30 &                &                 &                 &  34  &    1.0980(834)   &                \\
   &   &   &    & 32 &                &                 &                 &  36  &                  &                \\ 
   &   &   &    & 36 &                &                 &                 &  38  &                  &                \\ 
   &   &   &    &    &                &                 &                 &  40  &                  &                \\ \hline \hline
\end{tabular}}
\caption{The energy, $E_{ \rm gs}(q,R)$, of the lightest flux tube state with 
length $R$ and zero longitudinal momentum. 
Quantum numbers are as shown.
Calculations are: SU(3) at $\beta=6.0625$ ($N=3$), SU(5) at $\beta=17.630$ ($N=5$), SU(6) at $\beta=25.550$ ($N=6$) and SU(3) at $\beta=6.338$ ($N=3f$) as well as  $\beta=18.375$ ($N=5f$)}
\label{tab:2--_ground_state}
\end{center}
\end{table}

\begin{table}[h]
\begin{center}
{\footnotesize
\begin{tabular}{cccc|cccc|ccc}\hline \hline
\multicolumn{11}{c}{ $E_{{\rm gs}}(q,R)$ } \\ \hline \hline
$q$ & $|J|$ & $P_{\perp}$ & $P_{||}$& $R/a$ & $N=3$ & $N=5$ & $N=6$ & $R/a$ & $N=3f$ & $N=5f$\\ \hline \hline
 1 & 0 & $+$     &  $\pm$ &    10   &   1.3876(237)   &  1.3416(303)    &  1.4040(286)     &  14  &    0.9523(270)  &                 \\
   &   &         &        &    12   &   1.3560(216)   &  1.3770(255)    &  1.3958(303)     &  16  &    0.9378(160)   &                 \\
   &   &         &        &    14   &   1.3757(229)   &  1.4113(317)    &  1.4291(348)     &  18  &    0.8758(194)   &                 \\
   &   &         &        &    16   &   1.3461(286)   &  1.4032(483)    &  1.4360(305)     &  20  &    0.8889(195)   &  0.8767(123)    \\
   &   &         &        &    18   &   1.3414(261)   &  1.4202(346)    &  1.4473(291)     &  22  &    0.8847(126)   &                 \\
   &   &         &        &    20   &   1.4546(257)   &  1.4076(463)    &  1.4093(397)     &  24  &    0.9224(192)   &  0.8982(122)    \\
\hline \hline
\end{tabular}}
\caption{The energy, $E_{ \rm gs}(q,R)$, of the lightest flux tube state with length $R$ and zero longitudinal momentum. 
Quantum numbers are as shown.
Calculations are: SU(3) at $\beta=6.0625$ ($N=3$), SU(5) at $\beta=17.630$ ($N=5$), SU(6) at $\beta=25.550$ ($N=6$) and SU(3) at $\beta=6.338$ ($N=3f$) as well as  $\beta=18.375$ ($N=5f$)}
\label{tab:J0_Pp+_Q1_ground_state}
\end{center}
\end{table}

\newpage

\begin{table}[h]
\begin{center}
{\footnotesize
\begin{tabular}{cccc|cccc|ccc}\hline \hline
\multicolumn{11}{c}{ $E_{{\rm gs}}(q,R)$ } \\ \hline \hline
$q$ & $|J|$ & $P_{\perp}$ & $P_{||}$& $R/a$ & $N=3$ & $N=5$ & $N=6$ & $R/a$ & $N=3f$ & $N=5f$\\ \hline \hline
 1 & 0 & $-$     &  $\pm$ &    10   &   1.0872(64)    &  1.0769(109)    &  1.0895(355)     &  14  &    0.7302(220)  &                 \\
   &   &         &        &    12   &   1.0695(76)    &  1.0697(87)     &  1.0841(353)     &  16  &    0.7028(137)  &                 \\
   &   &         &        &    14   &   1.0800(77)    &  1.0789(105)    &  1.1004(153)     &  18  &    0.6964(84)   &                 \\
   &   &         &        &    16   &   1.1134(111)   &  1.1188(178)    &  1.1345(151)     &  20  &    0.7115(83)   &  0.6819(134)    \\
   &   &         &        &    18   &   1.1652(116)   &  1.1981(196)    &  1.1844(119)     &  22  &    0.7481(138)  &  0.7199(137)               \\
   &   &         &        &    20   &   1.2203(150)   &  1.1954(295)    &  1.2900(243)     &  24  &    0.7389(48)   &  0.7474(127)    \\
\hline \hline
\end{tabular}}
\caption{The energy, $E_{ \rm gs}(q,R)$, of the lightest flux tube state with 
length $R$ and zero longitudinal momentum. 
Quantum numbers are as shown.
Calculations are: SU(3) at $\beta=6.0625$ ($N=3$), SU(5) at $\beta=17.630$ ($N=5$), SU(6) at $\beta=25.550$ ($N=6$) and SU(3) at $\beta=6.338$ ($N=3f$) as well as  $\beta=18.375$ ($N=5f$)}
\label{tab:J0_Pp-_Q1_ground_state}
\end{center}
\end{table}

\begin{table}[h]
\begin{center}
{\footnotesize
\begin{tabular}{cccc|cccc|ccc}\hline \hline
\multicolumn{11}{c}{ $E_{1^{\rm st} {\rm ex.s}}(q,R)$ } \\ \hline \hline
$q$ & $|J|$ & $P_{\perp}$ & $P_{||}$& $R/a$ & $N=3$ & $N=5$ & $N=6$ & $R/a$ & $N=3f$ & $N=5f$\\ \hline \hline
 1 & 0 & $-$     &  $\pm$ &    10   &                 &  1.3896(319)     &  1.4820(350)     &  14  &    0.9575(307)  &                 \\
   &   &         &        &    12   &   1.4229(275)   &  1.3935(294)     &  1.3961(404)     &  16  &    0.9368(406)  &                 \\
   &   &         &        &    14   &   1.4067(221)   &                  &  1.5483(489)     &  18  &    0.9242(235)   &                 \\
   &   &         &        &    16   &   1.4307(330)   &  1.3830(393)     &  1.5018(412)     &  20  &    0.9570(90)   &  0.9635(219)    \\
   &   &         &        &    18   &   1.5187(435)   &  1.5909(675)     &  1.5596(531)     &  22  &    0.9619(179)  &  0.9685(217)    \\
   &   &         &        &    20   &   1.5959(474)   &  1.6051(1055)    &  1.7911(867)     &  24  &    0.9984(96)   &  1.0284(130)    \\
\hline \hline
\end{tabular}}
\caption{The energy, $E_{1^{\rm st} {\rm ex.s}}(q,R)$, of the first excited flux tube state with 
length $R$ and zero longitudinal momentum. 
Quantum numbers are as shown.
Calculations are: SU(3) at $\beta=6.0625$ ($N=3$), SU(5) at $\beta=17.630$ ($N=5$), SU(6) at $\beta=25.550$ ($N=6$) and SU(3) at $\beta=6.338$ ($N=3f$) as well as  $\beta=18.375$ ($N=5f$)}
\label{tab:J0_Pp-_Q1_first_excited_state}
\end{center}
\end{table}

\begin{table}[h]
\begin{center}
{\footnotesize
\begin{tabular}{cccc|cccc|ccc}\hline \hline
\multicolumn{11}{c}{ $E_{{\rm gs}}(q,R)$ } \\ \hline \hline
$q$ & $|J|$ & $P_{\perp}$ & $P_{||}$& $R/a$ & $N=3$ & $N=5$ & $N=6$ & $R/a$ & $N=3f$ & $N=5f$\\ \hline \hline
 1 & 1 & $\pm$   &  $\pm$ &    10   &   0.9669(50)   &  0.9559(193)    &  0.9972(210)    &  14  &    0.6406(176)  &                 \\
   &   &         &        &    12   &   0.9187(139)  &  0.9516(72)     &  0.9383(230)    &  16  &    0.6375(80)   &                 \\
   &   &         &        &    14   &   0.9355(55)   &  0.9566(83)     &  0.9741(78)     &  18  &    0.6290(60)   &                 \\
   &   &         &        &    16   &   0.9759(76)   &  0.9796(39)     &  1.0087(100)    &  20  &    0.6231(67)   &  0.6381(97)    \\
   &   &         &        &    18   &   0.9798(98)   &  1.0221(117)    &  1.0341(71)     &  22  &    0.6355(73)   &  0.6389(89)    \\
   &   &         &        &    20   &   1.0354(78)   &  1.0586(160)    &  1.0945(49)     &  24  &    0.6350(66)   &  0.6468(88)    \\
\hline \hline
\end{tabular}}
\caption{The energy, $E_{ \rm gs}(q,R)$, of the lightest flux tube state with 
length $R$ and zero longitudinal momentum. 
Quantum numbers are as shown.
Calculations are: SU(3) at $\beta=6.0625$ ($N=3$), SU(5) at $\beta=17.630$ ($N=5$), SU(6) at $\beta=25.550$ ($N=6$) and SU(3) at $\beta=6.338$ ($N=3f$) as well as  $\beta=18.375$ ($N=5f$)}
\label{tab:Q1_J1_ground_state}
\end{center}
\end{table}

\begin{table}[h]
\begin{center}
{\footnotesize
\begin{tabular}{cccc|cccc|ccc}\hline \hline
\multicolumn{11}{c}{ $E_{1^{\rm st} {\rm ex.s}}(q,R)$ } \\ \hline \hline
$q$ & $|J|$ & $P_{\perp}$ & $P_{||}$& $R/a$ & $N=3$ & $N=5$ & $N=6$ & $R/a$ & $N=3f$ & $N=5f$\\ \hline \hline
 1 & 1 & $\pm$   &  $\pm$ &    10   &   1.1954(122)   &  1.1940(585)    &  1.2124(169)    &  14  &    0.7931(128)   &                 \\
   &   &         &        &    12   &   1.1675(106)   &  1.1893(182)    &  1.2348(202)    &  16  &    0.7883(113)   &                \\
   &   &         &        &    14   &   1.2294(140)   &  1.2004(222)    &  1.2609(241)    &  18  &    0.7840(137)   &                 \\
   &   &         &        &    16   &   1.2514(151)   &  1.2759(337)    &  1.3138(229)    &  20  &    0.7833(112)   &  0.6381(97)    \\
   &   &         &        &    18   &   1.2930(212)   &  1.2879(278)    &  1.3744(250)    &  22  &    0.8266(126)   &  0.6389(89)    \\
   &   &         &        &    20   &                 &  1.4135(452)    &  1.4391(439)    &  24  &    0.8257(128)   &  0.6468(88)    \\
\hline \hline
\end{tabular}}
\caption{The energy, $E_{1^{\rm st} {\rm ex.s}}(q,R)$, of the first excited flux tube state with 
length $R$ and zero longitudinal momentum. 
Quantum numbers are as shown.
Calculations are: SU(3) at $\beta=6.0625$ ($N=3$), SU(5) at $\beta=17.630$ ($N=5$), SU(6) at $\beta=25.550$ ($N=6$) and SU(3) at $\beta=6.338$ ($N=3f$) as well as  $\beta=18.375$ ($N=5f$)}
\label{tab:Q1_J1_first_excited_state}
\end{center}
\end{table}

\begin{table}[h]
\begin{center}
{\footnotesize
\begin{tabular}{cccc|cccc|ccc}\hline \hline
\multicolumn{11}{c}{ $E_{2^{\rm nd} {\rm ex.s}}(q,R)$ } \\ \hline \hline
$q$ & $|J|$ & $P_{\perp}$ & $P_{||}$& $R/a$ & $N=3$ & $N=5$ & $N=6$ & $R/a$ & $N=3f$ & $N=5f$\\ \hline \hline
 1 & 1 & $\pm$   &  $\pm$ &    10   &                 &  1.4526(375)    &  1.4151(349)    &  14  &                     &                 \\
   &   &         &        &    12   &   1.3999(202)   &  1.4298(412)    &  1.3825(374)    &  16  &    0.9629(168)   &                \\
   &   &         &        &    14   &   1.4187(234)   &  1.4140(388)    &  1.3942(389)    &  18  &    0.9235(220)   &                 \\
   &   &         &        &    16   &   1.3821(255)   &                 &  1.3830(306)    &  20  &    0.8875(212)   &  0.9346(421)    \\
   &   &         &        &    18   &   1.4174(284)   &  1.3658(442)    &  1.4268(281)    &  22  &    0.9462(150)   &  0.9182(344)    \\
   &   &         &        &    20   &   1.4153(317)   &  1.4618(523)    &  1.4356(469)    &  24  &    0.9137(179)   &  0.9512(364)    \\
\hline \hline
\end{tabular}}
\caption{The energy, $E_{2^{\rm nd} {\rm ex.s}}(q,R)$, of the second excited flux tube state with 
length $R$ and zero longitudinal momentum. 
Quantum numbers are as shown.
Calculations are: SU(3) at $\beta=6.0625$ ($N=3$), SU(5) at $\beta=17.630$ ($N=5$), SU(6) at $\beta=25.550$ ($N=6$) and SU(3) at $\beta=6.338$ ($N=3f$) as well as  $\beta=18.375$ ($N=5f$)}
\label{tab:Q1_J1_second_excited_state}
\end{center}
\end{table}

\begin{table}[h]
\begin{center}
{\footnotesize
\begin{tabular}{cccc|cccc|ccc}\hline \hline
\multicolumn{11}{c}{ $E_{{\rm g.s}}(q,R)$ } \\ \hline \hline
$q$ & $|J|$ & $P_{\perp}$ & $P_{||}$& $R/a$ & $N=3$ & $N=5$ & $N=6$ & $R/a$ & $N=3f$ & $N=5f$\\ \hline \hline
 1 & 2 & $+$     &  $\pm$ &    10   &   1.4514(251)   &  1.5014(409) &  1.5429(439)   &  14  &                     &                 \\
   &   &         &        &    12   &   1.4880(334)   &  1.4970(396) &  1.4864(519)   &  16  &  0.9721(182)        &                 \\
   &   &         &        &    14   &   1.4797(340)   &  1.4747(419) &  1.5607(484)   &  18  &  0.9766(265)        &                 \\
   &   &         &        &    16   &   1.4695(376)   &             &  1.5361(490)   &  20  &  0.9345(255)         &  0.9833(164)    \\
   &   &         &        &    18   &   1.4312(356)   &  1.5004(450) &  1.4370(299)   &  22  &  0.9964(169)        &  0.9769(175)    \\
   &   &         &        &    20   &   1.4776(333)   &  1.5106(561) &  1.5694(760)   &  24  &  0.9674(264)        &  0.9910(165)    \\
\hline \hline
\end{tabular}}
\caption{The energy, $E_{{\rm g.s}}(q,R)$, of the lightest flux tube state with 
length $R$ and zero longitudinal momentum. 
Quantum numbers are as shown.
Calculations are: SU(3) at $\beta=6.0625$ ($N=3$), SU(5) at $\beta=17.630$ ($N=5$), SU(6) at $\beta=25.550$ ($N=6$) and SU(3) at $\beta=6.338$ ($N=3f$) as well as  $\beta=18.375$ ($N=5f$)}
\label{tab:J2_Pp+_Q1_ground_state}
\end{center}
\end{table}

\begin{table}[h]
\begin{center}
{\footnotesize
\begin{tabular}{cccc|cccc|ccc}\hline \hline
\multicolumn{11}{c}{ $E_{{\rm g.s}}(q,R)$ } \\ \hline \hline
$q$ & $|J|$ & $P_{\perp}$ & $P_{||}$& $R/a$ & $N=3$ & $N=5$ & $N=6$ & $R/a$ & $N=3f$ & $N=5f$\\ \hline \hline
 1 & 2 & $-$     &  $\pm$ &    10   &                 & 1.5629(508)   &  1.5894(510)   &  14  &  0.9615(333)        &                 \\
   &   &         &        &    12   &   1.5431(441)   & 1.6172(565)   &  1.4894(520)   &  16  &  0.9839(249)        &                 \\
   &   &         &        &    14   &   1.5985(417)   & 1.4518(565)   &  1.5059(659)   &  18  &  0.9713(104)        &                 \\
   &   &         &        &    16   &   1.4991(377)   & 1.6104(745)   &  1.6225(669)   &  20  &  0.9926(114)        &  1.0398(230)  \\
   &   &         &        &    18   &   1.4868(390)   & 1.3965(455)   &  1.5383(477)   &  22  &  0.9942(196)        &  1.0048(486)  \\
   &   &         &        &    20   &   1.4790(315)   & 1.5049(646)   &  1.5744(727)   &  24  &  0.9954(244)        &  1.0154(388) \\
\hline \hline
\end{tabular}}
\caption{The energy, $E_{{\rm g.s}}(q,R)$, of the lightest flux tube state with 
length $R$ and zero longitudinal momentum. 
Quantum numbers are as shown.
Calculations are: SU(3) at $\beta=6.0625$ ($N=3$), SU(5) at $\beta=17.630$ ($N=5$), SU(6) at $\beta=25.550$ ($N=6$) and SU(3) at $\beta=6.338$ ($N=3f$) as well as  $\beta=18.375$ ($N=5f$)}
\label{tab:J2_Pp-_Q1_ground_state}
\end{center}
\end{table}
\newpage

\begin{table}[h]
\begin{center}
{\footnotesize
\begin{tabular}{cccc|cccc|ccc}\hline \hline
\multicolumn{11}{c}{ $E_{{\rm g.s}}(q,R)$ } \\ \hline \hline
$q$ & $|J|$ & $P_{\perp}$ & $P_{||}$& $R/a$ & $N=3$ & $N=5$ & $N=6$ & $R/a$ & $N=3f$ & $N=5f$\\ \hline \hline
 2 & 0 & $+$     &  $\pm$ &    10   &    1.6617(430)  &   1.6065(605)  &   1.6465(572)  &  14  &  1.1817(211)        &                 \\
   &   &         &        &    12   &    1.5029(317)  &                &   1.5897(679)  &  16  &  1.0605(97)       &                 \\
   &   &         &        &    14   &    1.4217(230)  &   1.4798(396)  &   1.4963(436)  &  18  &  0.9898(108)        &                 \\
   &   &         &        &    16   &    1.3991(210)  &   1.3617(407)  &   1.4215(358)  &  20  &  0.9682(100)        &  0.9843(183)  \\
   &   &         &        &    18   &    1.3591(279)  &   1.3997(429)  &   1.4272(278)  &  22  &  0.9263(128)        &  0.9013(359)  \\
   &   &         &        &    20   &    1.3894(207)  &   1.3481(364)  &   1.4324(447)  &  24  &  0.9240(206)        &                  \\
\hline \hline
\end{tabular}}
\caption{The energy, $E_{{\rm g.s}}(q,R)$, of the lightest flux tube state with 
length $R$ and zero longitudinal momentum. 
Quantum numbers are as shown.
Calculations are: SU(3) at $\beta=6.0625$ ($N=3$), SU(5) at $\beta=17.630$ ($N=5$), SU(6) at $\beta=25.550$ ($N=6$) and SU(3) at $\beta=6.338$ ($N=3f$) as well as  $\beta=18.375$ ($N=5f$)}
\label{tab:J0_Pp+_Q2_ground_state}
\end{center}
\end{table}

\begin{table}[h]
\begin{center}
{\footnotesize
\begin{tabular}{cccc|cccc|ccc}\hline \hline
\multicolumn{11}{c}{ $E_{1^{\rm st} {\rm ex.s}}(q,R)$ } \\ \hline \hline
$q$ & $|J|$ & $P_{\perp}$ & $P_{||}$& $R/a$ & $N=3$ & $N=5$ & $N=6$ & $R/a$ & $N=3f$ & $N=5f$\\ \hline \hline
 2 & 0 & $+$     &  $\pm$ &    10   &  1.7867(604)    &                   &                    &  14  &   1.2815(313)      &                 \\
   &   &         &        &    12   &  1.7797(675)    &                   &                    &  16  &   1.2143(201)      &                 \\
   &   &         &        &    14   &  1.6879(586)    &   1.7522(1039)  &   1.6640(901)   &  18  &   1.1978(185)      &                 \\
   &   &         &        &    16   &  1.7153(738)    &   1.7834(998)  &   1.7645(1059)   &  20  &   1.1757(219)      &   1.2066(321)  \\
   &   &         &        &    18   &  1.5669(579)    &   1.7745(1138)  &   1.7874(852)   &  22  &   1.1800(360)      &   1.1089(263)    \\
   &   &         &        &    20   &  1.6915(790)    &   1.6619(1138)  &   1.7537(1023)   &  24  &   1.1488(510)      &                  \\
\hline \hline
\end{tabular}}
\caption{The energy, $E_{1^{\rm st}{\rm ex.s}}(q,R)$, of the first excited flux tube state with 
length $R$ and zero longitudinal momentum. 
Quantum numbers are as shown.
Calculations are: SU(3) at $\beta=6.0625$ ($N=3$), SU(5) at $\beta=17.630$ ($N=5$), SU(6) at $\beta=25.550$ ($N=6$) and SU(3) at $\beta=6.338$ ($N=3f$) as well as  $\beta=18.375$ ($N=5f$)}
\label{tab:J0_Pp+_Q2_first_excited_state}
\end{center}
\end{table}
\newpage

\begin{table}[h]
\begin{center}
{\footnotesize
\begin{tabular}{cccc|cccc|ccc}\hline \hline
\multicolumn{11}{c}{ $E_{{\rm g.s}}(q,R)$ } \\ \hline \hline
$q$ & $|J|$ & $P_{\perp}$ & $P_{||}$& $R/a$ & $N=3$ & $N=5$ & $N=6$ & $R/a$ & $N=3f$ & $N=5f$\\ \hline \hline
 2 & 0 & $-$     &  $\pm$ &    10   &    1.6597(484)  &    1.7516(939)    &   1.7127(824)   &  14  &  1.2084(240)        &                \\
   &   &         &        &    12   &    1.5779(433)  &     1.5576(586)   &   1.5623(564)   &  16  &  1.0605(232)       &                 \\
   &   &         &        &    14   &    1.5121(338)  &    1.5555(624)    &   1.5282(599)   &  18  &   1.0563(121)        &               \\
   &   &         &        &    16   &    1.5186(430)  &     1.4836(490)   &   1.5524(426)   &  20  &   1.0388(96)        &   1.0242(257)  \\
   &   &         &        &    18   &    1.4403(346)  &    1.5157(400)    &   1.5660(485)   &  22  &   0.9861(164)        &   0.9882(409) \\
   &   &         &        &    20   &    1.5365(384)  &     1.5505(700)   &   1.5513(736)   &  24  &  0.9998(103)        &    1.0019(153) \\
\hline \hline
\end{tabular}}
\caption{The energy, $E_{{\rm g.s}}(q,R)$, of the lightest flux tube state with 
length $R$ and zero longitudinal momentum. 
Quantum numbers are as shown.
Calculations are: SU(3) at $\beta=6.0625$ ($N=3$), SU(5) at $\beta=17.630$ ($N=5$), SU(6) at $\beta=25.550$ ($N=6$) and SU(3) at $\beta=6.338$ ($N=3f$) as well as  $\beta=18.375$ ($N=5f$)}
\label{tab:J0_Pp-_Q2_ground_state}
\end{center}
\end{table}

\begin{table}[h]
\begin{center}
{\footnotesize
\begin{tabular}{cccc|cccc|ccc}\hline \hline
\multicolumn{11}{c}{ $E_{1^{\rm st}{\rm ex.s}}(q,R)$  } \\ \hline \hline
$q$ & $|J|$ & $P_{\perp}$ & $P_{||}$& $R/a$ & $N=3$ & $N=5$ & $N=6$ & $R/a$ & $N=3f$ & $N=5f$\\ \hline \hline
 2 & 0 & $-$     &  $\pm$ &    10   &      &        &      &  14  &   1.2910(344)    &                    \\
   &   &         &        &    12   &      &        &      &  16  &   1.2143(200)    &                    \\
   &   &         &        &    14   &      &        &      &  18  &   1.1523(489)    &                    \\
   &   &         &        &    16   &      &        &      &  20  &   1.1482(188)    &    1.1072(276)     \\
   &   &         &        &    18   &      &        &      &  22  &   1.1501(241)    &                    \\
   &   &         &        &    20   &      &        &      &  24  &   1.1238(116)    &    1.1458(223)     \\ 
   &   &         &        &    22   &      &        &      &  26  &   1.1249(156)    &    1.1073(365)      \\ 
   &   &         &        &    24   &      &        &      &  28  &   1.1687(467)    &    1.1961(427)      \\ 
   &   &         &        &    26   &      &        &      &  30  &   1.0881(496)    &                    \\ 
   &   &         &        &    28   &      &        &      &  32  &   1.1487(766)    &                    \\
   &   &         &        &    30   &      &        &      &  34  &                  &                    \\
   &   &         &        &    32   &      &        &      &  36  &                  &                    \\ 
   &   &         &        &    36   &      &        &      &  38  &                  &                    \\ 
   &   &         &        &         &      &        &      &  40  &                  &                    \\ \hline \hline
\end{tabular}}
\caption{The energy, $E_{1^{\rm st}{\rm ex.s}}(q,R)$, of the first excited flux tube state with 
length $R$ and zero longitudinal momentum. 
Quantum numbers are as shown.
Calculations are: SU(3) at $\beta=6.0625$ ($N=3$), SU(5) at $\beta=17.630$ ($N=5$), SU(6) at $\beta=25.550$ ($N=6$) and SU(3) at $\beta=6.338$ ($N=3f$) as well as  $\beta=18.375$ ($N=5f$)}
\label{tab:J0_Pp-_Q2_excited_state}
\end{center}
\end{table}

\newpage

\begin{table}[h]
\begin{center}
{\footnotesize
\begin{tabular}{cccc|cccc|ccc}\hline \hline
\multicolumn{11}{c}{ $E_{{\rm g.s}}(q,R)$ } \\ \hline \hline
$q$ & $|J|$ & $P_{\perp}$ & $P_{||}$& $R/a$ & $N=3$ & $N=5$ & $N=6$ & $R/a$ & $N=3f$ & $N=5f$\\ \hline \hline
 2 & 1 & $\pm$   &  $\pm$ &   10   &  1.5531(347)  &    1.5677(418)  &   1.6354(531)   &  14  &  1.1209(173)      &                \\
   &   &         &        &   12   &  1.4878(304)  &    1.4556(452)  &   1.4689(429)   &  16  &   1.0704(231)      &                 \\
   &   &         &        &   14   &  1.4138(249)  &    1.4397(311)  &   1.4359(410)   &  18  &   0.9902(115)      &               \\
   &   &         &        &   16   &  1.3894(236)  &    1.4114(398)  &   1.4111(325)   &  20  &   0.9489(203)      &   0.9591(162)  \\
   &   &         &        &   18   &  1.3257(224)  &    1.3832(395)  &   1.4116(275)   &  22  &  0.9405(164)      &   0.9362(176) \\
   &   &         &        &   20   &  1.3648(216)  &    1.3500(403)  &   1.4046(392)   &  24  &  0.9405(164)      &   0.9186(287) \\
\hline \hline
\end{tabular}}
\caption{The energy, $E_{{\rm g.s}}(q,R)$, of the lightest flux tube state with 
length $R$ and zero longitudinal momentum. 
Quantum numbers are as shown.
Calculations are: SU(3) at $\beta=6.0625$ ($N=3$), SU(5) at $\beta=17.630$ ($N=5$), SU(6) at $\beta=25.550$ ($N=6$) and SU(3) at $\beta=6.338$ ($N=3f$) as well as  $\beta=18.375$ ($N=5f$)}
\label{tab:J1_Q2_ground_state}
\end{center}
\end{table}

\begin{table}[h]
\begin{center}
{\footnotesize
\begin{tabular}{cccc|cccc|ccc}\hline \hline
\multicolumn{11}{c}{ $E_{1^{\rm st}{\rm ex.s}}(q,R)$ } \\ \hline \hline
$q$ & $|J|$ & $P_{\perp}$ & $P_{||}$& $R/a$ & $N=3$ & $N=5$ & $N=6$ & $R/a$ & $N=3f$ & $N=5f$\\ \hline \hline
 2 & 1 & $\pm$   &  $\pm$ &   10   &  1.7747(633)  &    1.6777(801)  &                &  14  &  1.2234(297)      &                \\
   &   &         &        &   12   &  1.6331(493)  &                 &   1.6338(823)   &  16  &  1.1364(132)     &                 \\
   &   &         &        &   14   &  1.6030(351)  &    1.5171(506)  &   1.6118(665)   &  18  &  1.0963(140)      &               \\
   &   &         &        &   16   &  1.4751(495)  &    1.5396(509)  &   1.6392(576)   &  20  &  1.0450(138)     &   1.0952(196)  \\
   &   &         &        &   18   &              &    1.4858(578)  &                &  22  &  1.0492(210)      &  1.0533(244)    \\
   &   &         &        &   20   &  1.5414(531)  &    1.6167(811)  &   1.6336(895)   &  24  &  1.0459(369)      &  1.0492(417) \\
\hline \hline
\end{tabular}}
\caption{The energy, $E_{1^{\rm st}{\rm ex.s}}(q,R)$, of the first excited flux tube state with 
length $R$ and zero longitudinal momentum. 
Quantum numbers are as shown.
Calculations are: SU(3) at $\beta=6.0625$ ($N=3$), SU(5) at $\beta=17.630$ ($N=5$), SU(6) at $\beta=25.550$ ($N=6$) and SU(3) at $\beta=6.338$ ($N=3f$) as well as  $\beta=18.375$ ($N=5f$)}
\label{tab:J1_Q2_first_excited_state}
\end{center}
\end{table}
 
\begin{table}[h]
\begin{center}
{\footnotesize
\begin{tabular}{cccc|cccc|ccc}\hline \hline
\multicolumn{11}{c}{ $E_{{\rm g.s}}(q,R)$ } \\ \hline \hline
$q$ & $|J|$ & $P_{\perp}$ & $P_{||}$& $R/a$ & $N=3$ & $N=5$ & $N=6$ & $R/a$ & $N=3f$ & $N=5f$\\ \hline \hline
 2 & 2 &   $+$   &  $\pm$ &   10   &  1.6186(335)  &    1.6528(577)  &                 &  14  &  1.0977(648)      &                 \\
   &   &         &        &   12   &  1.4981(338)  &    1.4984(578)  &   1.5807(636)   &  16  &  1.0614(371)      &                 \\
   &   &         &        &   14   &  1.4356(289)  &    1.4988(509)  &   1.5008(479)   &  18  &  0.9809(259)      &                 \\
   &   &         &        &   16   &  1.4231(278)  &    1.4114(398)  &   1.4539(276)   &  20  &  0.9696(245)      &   0.9911(169)   \\
   &   &         &        &   18   &  1.3029(281)  &    1.4141(342)  &   1.4040(270)   &  22  &  0.9660(152)      &   0.9862(145)   \\
   &   &         &        &   20   &  1.3899(323)  &    1.4395(397)  &   1.4579(351)   &  24  &  0.9438(105)      &   0.9582(106)   \\
\hline \hline
\end{tabular}}
\caption{The energy, $E_{{\rm g.s}}(q,R)$, of the lightest flux tube state with 
length $R$ and zero longitudinal momentum. 
Quantum numbers are as shown.
Calculations are: SU(3) at $\beta=6.0625$ ($N=3$), SU(5) at $\beta=17.630$ ($N=5$), SU(6) at $\beta=25.550$ ($N=6$) and SU(3) at $\beta=6.338$ ($N=3f$) as well as  $\beta=18.375$ ($N=5f$)}
\label{tab:J2_Pp+_Q2_ground_state}
\end{center}
\end{table}

\begin{table}[h]
\begin{center}
{\footnotesize
\begin{tabular}{cccc|cccc|ccc}\hline \hline
\multicolumn{11}{c}{ $E_{1^{\rm st} {\rm ex.s}}(q,R)$ } \\ \hline \hline
$q$ & $|J|$ & $P_{\perp}$ & $P_{||}$& $R/a$ & $N=3$ & $N=5$ & $N=6$ & $R/a$ & $N=3f$ & $N=5f$\\ \hline \hline
 2 & 2 &   $+$   &  $\pm$ &   10   &               &                  &                 &  14  &               &                 \\
   &   &         &        &   12   &  1.7051(737)  &    1.7431(981)   &                 &  16  &  1.1464(405)  &                 \\
   &   &         &        &   14   &               &    1.6716(902)   &   1.7391(826)   &  18  &               &                 \\
   &   &         &        &   16   &               &                  &   1.7036(786)   &  20  &  1.0941(177)  &    1.1086(277)   \\
   &   &         &        &   18   &               &    1.6819(972)   &   1.7432(646)   &  22  &               &    1.1521(254)    \\
   &   &         &        &   20   &               &                  &   1.6511(1097)  &  24  &               &    1.1188(208)    \\
\hline \hline
\end{tabular}}
\caption{The energy, $E_{1^{\rm st} {\rm ex.s}}(q,R)$, of the first excited flux tube state with 
length $R$ and zero longitudinal momentum. 
Quantum numbers are as shown.
Calculations are: SU(3) at $\beta=6.0625$ ($N=3$), SU(5) at $\beta=17.630$ ($N=5$), SU(6) at $\beta=25.550$ ($N=6$) and SU(3) at $\beta=6.338$ ($N=3f$) as well as  $\beta=18.375$ ($N=5f$)}
\label{tab:J2_Pp+_Q2_first_excited_state}
\end{center}
\end{table}

\begin{table}[h]
\begin{center}
{\footnotesize
\begin{tabular}{cccc|cccc|ccc}\hline \hline
\multicolumn{11}{c}{ $E_{{\rm g.s}}(q,R)$ } \\ \hline \hline
$q$ & $|J|$ & $P_{\perp}$ & $P_{||}$& $R/a$ & $N=3$ & $N=5$ & $N=6$ & $R/a$ & $N=3f$ & $N=5f$\\ \hline \hline
 2 & 2 &   $-$   &  $\pm$ &   10   &  1.6597(484)  &    1.7516(939)  &   1.7127(824)   &  14  &  1.2084(240)      &                 \\
   &   &         &        &   12   &  1.5779(433)  &    1.5576(586)  &   1.5623(564)   &  16  &  1.0605(232)      &                 \\
   &   &         &        &   14   &  1.5121(338)  &    1.5555(624)  &   1.5282(599)   &  18  &  1.0563(121)      &                 \\
   &   &         &        &   16   &  1.5186(430)  &    1.4836(490)  &   1.5524(426)   &  20  &  1.0388(96)      &  1.0242(257)   \\
   &   &         &        &   18   &  1.4403(346)  &    1.5157(400)  &   1.5660(485)   &  22  &  0.9861(164)      &  0.9882(409)   \\
   &   &         &        &   20   &  1.5365(384)  &    1.5505(700)  &   1.5513(736)   &  24  &  0.9998(103)      &  1.0019(153)   \\
\hline \hline
\end{tabular}}
\caption{The energy, $E_{{\rm g.s}}(q,R)$, of the lightest flux tube state with 
length $R$ and zero longitudinal momentum. 
Quantum numbers are as shown.
Calculations are: SU(3) at $\beta=6.0625$ ($N=3$), SU(5) at $\beta=17.630$ ($N=5$), SU(6) at $\beta=25.550$ ($N=6$) and SU(3) at $\beta=6.338$ ($N=3f$) as well as  $\beta=18.375$ ($N=5f$)}
\label{tab:J2_Pp-_Q2_ground_state}
\end{center}
\end{table}

\begin{table}[h]
\begin{center}
{\footnotesize
\begin{tabular}{cccc|cccc|ccc}\hline \hline
\multicolumn{11}{c}{ $E_{1^{\rm st} {\rm ex.s}}(q,R)$ } \\ \hline \hline
$q$ & $|J|$ & $P_{\perp}$ & $P_{||}$& $R/a$ & $N=3$ & $N=5$ & $N=6$ & $R/a$ & $N=3f$ & $N=5f$\\ \hline \hline
 2 & 2 &   $-$   &  $\pm$ &   10   &               &                  &                 &  14  &  1.2910(344)  &                 \\
   &   &         &        &   12   &               &                  &                 &  16  &  1.2143(200)  &                 \\
   &   &         &        &   14   &               &                  &                 &  18  &  1.1523(489)  &                 \\
   &   &         &        &   16   &               &                  &                 &  20  &  1.1482(188)  &    1.1072(276)   \\
   &   &         &        &   18   &               &                  &                 &  22  &  1.1501(241)  &                      \\
   &   &         &        &   20   &               &                  &                 &  24  &  1.1238(116)  &    1.1458(223)    \\
\hline \hline
\end{tabular}}
\caption{The energy, $E_{1^{\rm st} {\rm ex.s}}(q,R)$, of the first excited flux tube state with 
length $R$ and zero longitudinal momentum. 
Quantum numbers are as shown.
Calculations are: SU(3) at $\beta=6.0625$ ($N=3$), SU(5) at $\beta=17.630$ ($N=5$), SU(6) at $\beta=25.550$ ($N=6$) and SU(3) at $\beta=6.338$ ($N=3f$) as well as  $\beta=18.375$ ($N=5f$)}
\label{tab:J2_Pp-_Q2_first_excited_state}
\end{center}
\end{table}

\clearpage

\section*{Figures}
\addcontentsline{toc}{section}{Figures}

\clearpage

\clearpage

\clearpage

\clearpage
\clearpage

\clearpage

\clearpage
\bibliographystyle{utphys}
\bibliography{references}

\providecommand{\href}[2]{#2}\begingroup\raggedright\begin{thebibliography}{10}

\bibitem{Luscher:2004ib}
M.~Luscher and P.~Weisz, ``{String excitation energies in SU(N) gauge theories beyond the free-string approximation},'' {\em JHEP} {\bf 07} (2004) 014, \href{http://www.arXiv.org/abs/hep-th/0406205}{{\tt hep-th/0406205}}.

\bibitem{Drummond:2006su}
J.~M. Drummond, ``{Reply to hep-th/0606265},'' \href{http://www.arXiv.org/abs/hep-th/0608109}{{\tt hep-th/0608109}}.

\bibitem{Dass:2006ud}
N.~D.~H. Dass and P.~Matlock, ``{Our response to the response hep-th/0608109 by Drummond},'' \href{http://www.arXiv.org/abs/hep-th/0611215}{{\tt hep-th/0611215}}.

\bibitem{Aharony:2009gg}
O.~Aharony and E.~Karzbrun, ``{On the effective action of confining strings},'' {\em JHEP} {\bf 06} (2009) 012, \href{http://www.arXiv.org/abs/0903.1927}{{\tt 0903.1927}}.

\bibitem{Dass:2009xe}
N.~D.~H. Dass, ``{All Conformal Effective String Theories are Isospectral to Nambu-Goto Theory},'' \href{http://www.arXiv.org/abs/0911.3236}{{\tt 0911.3236}}.

\bibitem{HariDass:2009ub}
N.~D. Hari~Dass, P.~Matlock, and Y.~Bharadwaj, ``{Spectrum to all orders of Polchinski-Strominger {Effective} String Theory of Polyakov-Liouville Type},'' \href{http://www.arXiv.org/abs/0910.5615}{{\tt 0910.5615}}.

\bibitem{HariDass:2009ud}
N.~D. Hari~Dass and Y.~Bharadwaj, ``{Spectrum to all orders of Polchinski-Strominger Effective String Theories of the Drummond Type},'' \href{http://www.arXiv.org/abs/0910.5620}{{\tt 0910.5620}}.

\bibitem{Dubovsky:2012sh}
S.~Dubovsky, R.~Flauger, and V.~Gorbenko, ``{Effective String Theory Revisited},'' {\em JHEP} {\bf 09} (2012) 044, \href{http://www.arXiv.org/abs/1203.1054}{{\tt 1203.1054}}.

\bibitem{Aharony:2013ipa}
O.~Aharony and Z.~Komargodski, ``{The Effective Theory of Long Strings},'' {\em JHEP} {\bf 05} (2013) 118, \href{http://www.arXiv.org/abs/1302.6257}{{\tt 1302.6257}}.

\bibitem{Bringoltz:2006zg}
B.~Bringoltz and M.~Teper, ``{A Precise calculation of the fundamental string tension in SU(N) gauge theories in 2+1 dimensions},'' {\em Phys. Lett. B} {\bf 645} (2007) 383--388, \href{http://www.arXiv.org/abs/hep-th/0611286}{{\tt hep-th/0611286}}.

\bibitem{Athenodorou:2011rx}
A.~Athenodorou, B.~Bringoltz, and M.~Teper, ``{Closed flux tubes and their string description in D=2+1 SU(N) gauge theories},'' {\em JHEP} {\bf 05} (2011) 042, \href{http://www.arXiv.org/abs/1103.5854}{{\tt 1103.5854}}.

\bibitem{Athenodorou:2010cs}
A.~Athenodorou, B.~Bringoltz, and M.~Teper, ``{Closed flux tubes and their string description in D=3+1 SU(N) gauge theories},'' {\em JHEP} {\bf 02} (2011) 030, \href{http://www.arXiv.org/abs/1007.4720}{{\tt 1007.4720}}.

\bibitem{goddard1973quantum}
P.~Goddard, J.~Goldstone, C.~Rebbi, and C.~B. Thorn, ``Quantum dynamics of a massless relativistic string,'' {\em Nuclear Physics B} {\bf 56} (1973), no.~1, 109--135.

\bibitem{arvis1983exactqq}
J.~Arvis, ``The exactqq{\={}} potential in nambu string theory,'' {\em Physics Letters B} {\bf 127} (1983), no.~1-2, 106--108.

\bibitem{zamolodchikov1990thermodynamic}
A.~B. Zamolodchikov, ``Thermodynamic bethe ansatz in relativistic models: Scaling 3-state potts and lee-yang models,'' {\em Nuclear Physics B} {\bf 342} (1990), no.~3, 695--720.

\bibitem{Dorey:1996re}
P.~Dorey and R.~Tateo, ``{Excited states by analytic continuation of TBA equations},'' {\em Nucl. Phys. B} {\bf 482} (1996) 639--659, \href{http://www.arXiv.org/abs/hep-th/9607167}{{\tt hep-th/9607167}}.

\bibitem{Dubovsky:2012wk}
S.~Dubovsky, R.~Flauger, and V.~Gorbenko, ``{Solving the Simplest Theory of Quantum Gravity},'' {\em JHEP} {\bf 09} (2012) 133, \href{http://www.arXiv.org/abs/1205.6805}{{\tt 1205.6805}}.

\bibitem{Luscher:1990ux}
M.~Luscher, ``{Two particle states on a torus and their relation to the scattering matrix},'' {\em Nucl. Phys. B} {\bf 354} (1991) 531--578.

\bibitem{Dubovsky:2013gi}
S.~Dubovsky, R.~Flauger, and V.~Gorbenko, ``{Evidence from Lattice Data for a New Particle on the Worldsheet of the QCD Flux Tube},'' {\em Phys. Rev. Lett.} {\bf 111} (2013), no.~6, 062006, \href{http://www.arXiv.org/abs/1301.2325}{{\tt 1301.2325}}.

\bibitem{Dubovsky:2014fma}
S.~Dubovsky, R.~Flauger, and V.~Gorbenko, ``{Flux Tube Spectra from Approximate Integrability at Low Energies},'' {\em J. Exp. Theor. Phys.} {\bf 120} (2015) 399--422, \href{http://www.arXiv.org/abs/1404.0037}{{\tt 1404.0037}}.

\bibitem{Dubovsky:2015zey}
S.~Dubovsky and V.~Gorbenko, ``{Towards a Theory of the QCD String},'' {\em JHEP} {\bf 02} (2016) 022, \href{http://www.arXiv.org/abs/1511.01908}{{\tt 1511.01908}}.

\bibitem{Athenodorou:2017cmw}
A.~Athenodorou and M.~Teper, ``{On the mass of the world-sheet 'axion' in $SU(N)$ gauge theories in 3$+$1 dimensions},'' {\em Phys. Lett. B} {\bf 771} (2017) 408--414, \href{http://www.arXiv.org/abs/1702.03717}{{\tt 1702.03717}}.

\bibitem{Chen:2018keo}
C.~Chen, P.~Conkey, S.~Dubovsky, and G.~Hern\'andez-Chifflet, ``{Undressing Confining Flux Tubes with $T\bar T$},'' {\em Phys. Rev. D} {\bf 98} (2018), no.~11, 114024, \href{http://www.arXiv.org/abs/1808.01339}{{\tt 1808.01339}}.

\bibitem{Dubovsky:2016cog}
S.~Dubovsky and G.~Hernandez-Chifflet, ``{Yang--Mills Glueballs as Closed Bosonic Strings},'' {\em JHEP} {\bf 02} (2017) 022, \href{http://www.arXiv.org/abs/1611.09796}{{\tt 1611.09796}}.

\bibitem{Conkey:2019blu}
P.~Conkey, S.~Dubovsky, and M.~Teper, ``{Glueball spins in $D = 3$ Yang-Mills},'' {\em JHEP} {\bf 10} (2019) 175, \href{http://www.arXiv.org/abs/1909.07430}{{\tt 1909.07430}}.

\bibitem{Dubovsky:2021cor}
S.~Dubovsky, G.~Hern\'andez-Chifflet, and S.~Zare, ``{3D Yang-Mills glueballs vs closed effective strings},'' {\em JHEP} {\bf 07} (2021) 216, \href{http://www.arXiv.org/abs/2104.02154}{{\tt 2104.02154}}.

\bibitem{Perantonis:1988uz}
S.~Perantonis, A.~Huntley, and C.~Michael, ``{Static Potentials From Pure SU(2) Lattice Gauge Theory},'' {\em Nucl. Phys. B} {\bf 326} (1989) 544--556.

\bibitem{Griffiths:1983ah}
L.~A. Griffiths, C.~Michael, and P.~E.~L. Rakow, ``{Mesons With Excited Glue},'' {\em Phys. Lett. B} {\bf 129} (1983) 351--356.

\bibitem{Perantonis:1990dy}
S.~Perantonis and C.~Michael, ``{Static potentials and hybrid mesons from pure SU(3) lattice gauge theory},'' {\em Nucl. Phys. B} {\bf 347} (1990) 854--868.

\bibitem{Caselle:1995fh}
M.~Caselle, F.~Gliozzi, U.~Magnea, and S.~Vinti, ``{Width of long color flux tubes in lattice gauge systems},'' {\em Nucl. Phys. B} {\bf 460} (1996) 397--412, \href{http://www.arXiv.org/abs/hep-lat/9510019}{{\tt hep-lat/9510019}}.

\bibitem{Caselle:1996ii}
M.~Caselle, R.~Fiore, F.~Gliozzi, M.~Hasenbusch, and P.~Provero, ``{String effects in the Wilson loop: A High precision numerical test},'' {\em Nucl. Phys. B} {\bf 486} (1997) 245--260, \href{http://www.arXiv.org/abs/hep-lat/9609041}{{\tt hep-lat/9609041}}.

\bibitem{Michael:1994ej}
{\bf UKQCD} Collaboration, C.~Michael and P.~W. Stephenson, ``{The Nature of the hadronic string},'' {\em Phys. Rev. D} {\bf 50} (1994) 4634--4638, \href{http://www.arXiv.org/abs/hep-lat/9403004}{{\tt hep-lat/9403004}}.

\bibitem{Kuti:2005xg}
J.~Kuti, ``{Lattice QCD and string theory},'' {\em PoS} {\bf LAT2005} (2006) 001, \href{http://www.arXiv.org/abs/hep-lat/0511023}{{\tt hep-lat/0511023}}.

\bibitem{Bicudo:2022lnq}
P.~Bicudo, A.~Sharifian, and N.~Cardoso, ``{Eight spectra of very excited flux tubes in SU(3) gauge theory},'' {\em PoS} {\bf LATTICE2022} (2023) 256, \href{http://www.arXiv.org/abs/2209.00132}{{\tt 2209.00132}}.

\bibitem{Sharifian:2023idc}
A.~Sharifian, N.~Cardoso, and P.~Bicudo, ``{Eight very excited flux tube spectra and possible axions in SU(3) lattice gauge theory},'' {\em Phys. Rev. D} {\bf 107} (2023), no.~11, 114507, \href{http://www.arXiv.org/abs/2303.15152}{{\tt 2303.15152}}.

\bibitem{Chernodub:2018aix}
M.~N. Chernodub, V.~A. Goy, and A.~V. Molochkov, ``{Phase structure of lattice Yang-Mills theory on ${\mathbb T}^2 \times {\mathbb R}^2$},'' {\em Phys. Rev. D} {\bf 99} (2019), no.~7, 074021, \href{http://www.arXiv.org/abs/1811.01550}{{\tt 1811.01550}}.

\bibitem{deForcrand:1984wzs}
P.~de~Forcrand, G.~Schierholz, H.~Schneider, and M.~Teper, ``{The String and Its Tension in SU(3) Lattice Gauge Theory: Towards Definitive Results},'' {\em Phys. Lett. B} {\bf 160} (1985) 137--143.

\bibitem{Juge:2003vw}
K.~J. Juge, J.~Kuti, F.~Maresca, C.~Morningstar, and M.~J. Peardon, ``{Excitations of torelon},'' {\em Nucl. Phys. B Proc. Suppl.} {\bf 129} (2004) 703--705, \href{http://www.arXiv.org/abs/hep-lat/0309180}{{\tt hep-lat/0309180}}.

\bibitem{Meyer:2004hv}
H.~Meyer and M.~Teper, ``{Confinement and the effective string theory in SU(N ---\ensuremath{>} infinity): A Lattice study},'' {\em JHEP} {\bf 12} (2004) 031, \href{http://www.arXiv.org/abs/hep-lat/0411039}{{\tt hep-lat/0411039}}.

\bibitem{Athenodorou:2021vkw}
A.~Athenodorou and M.~Teper, ``{The torelon spectrum and the world-sheet axion},'' {\em PoS} {\bf LATTICE2021} (2022) 103, \href{http://www.arXiv.org/abs/2112.11213}{{\tt 2112.11213}}.

\bibitem{Athenodorou:2022tsj}
A.~Athenodorou, ``{Confining strings, axions and glueballs in the planar limit},'' {\em PoS} {\bf CORFU2021} (2022) 168, \href{http://www.arXiv.org/abs/2205.03642}{{\tt 2205.03642}}.

\bibitem{Dubovsky:2013ira}
S.~Dubovsky, V.~Gorbenko, and M.~Mirbabayi, ``{Natural Tuning: Towards A Proof of Concept},'' {\em JHEP} {\bf 09} (2013) 045, \href{http://www.arXiv.org/abs/1305.6939}{{\tt 1305.6939}}.

\bibitem{Smirnov:2016lqw}
F.~A. Smirnov and A.~B. Zamolodchikov, ``{On space of integrable quantum field theories},'' {\em Nucl. Phys. B} {\bf 915} (2017) 363--383, \href{http://www.arXiv.org/abs/1608.05499}{{\tt 1608.05499}}.

\bibitem{Cavaglia:2016oda}
A.~Cavagli\`a, S.~Negro, I.~M. Sz\'ecs\'enyi, and R.~Tateo, ``{$T \bar{T}$-deformed 2D Quantum Field Theories},'' {\em JHEP} {\bf 10} (2016) 112, \href{http://www.arXiv.org/abs/1608.05534}{{\tt 1608.05534}}.

\bibitem{Athenodorou:2018sab}
A.~Athenodorou and M.~Teper, ``{On the spectrum and string tension of U(1) lattice gauge theory in 2 + 1 dimensions},'' {\em JHEP} {\bf 01} (2019) 063, \href{http://www.arXiv.org/abs/1811.06280}{{\tt 1811.06280}}.

\bibitem{Athenodorou:2022pmz}
A.~Athenodorou, S.~Dubovsky, C.~Luo, and M.~Teper, ``{Excitations of Ising strings on a lattice},'' {\em JHEP} {\bf 05} (2023) 082, \href{http://www.arXiv.org/abs/2301.00034}{{\tt 2301.00034}}.

\bibitem{Luscher:1984is}
M.~Luscher and P.~Weisz, ``{Definition and General Properties of the Transfer Matrix in Continuum Limit Improved Lattice Gauge Theories},'' {\em Nucl. Phys. B} {\bf 240} (1984) 349--361.

\bibitem{Luscher:1990ck}
M.~Luscher and U.~Wolff, ``{How to Calculate the Elastic Scattering Matrix in Two-dimensional Quantum Field Theories by Numerical Simulation},'' {\em Nucl. Phys. B} {\bf 339} (1990) 222--252.

\bibitem{Athenodorou:2021qvs}
A.~Athenodorou and M.~Teper, ``{SU(N) gauge theories in 3+1 dimensions: glueball spectrum, string tensions and topology},'' {\em JHEP} {\bf 12} (2021) 082, \href{http://www.arXiv.org/abs/2106.00364}{{\tt 2106.00364}}.

\bibitem{Meyer:2004gx}
H.~B. Meyer, ``{Glueball regge trajectories},'' \href{http://www.arXiv.org/abs/hep-lat/0508002}{{\tt hep-lat/0508002}}.

\bibitem{Lucini:2004my}
B.~Lucini, M.~Teper, and U.~Wenger, ``{Glueballs and k-strings in SU(N) gauge theories: Calculations with improved operators},'' {\em JHEP} {\bf 06} (2004) 012, \href{http://www.arXiv.org/abs/hep-lat/0404008}{{\tt hep-lat/0404008}}.

\bibitem{Teper:1987wt}
M.~Teper, ``{An Improved Method for Lattice Glueball Calculations},'' {\em Phys. Lett. B} {\bf 183} (1987) 345.

\bibitem{Lucini:2002ku}
B.~Lucini, M.~Teper, and U.~Wenger, ``{The Deconfinement transition in SU(N) gauge theories},'' {\em Phys. Lett. B} {\bf 545} (2002) 197--206, \href{http://www.arXiv.org/abs/hep-lat/0206029}{{\tt hep-lat/0206029}}.

\bibitem{Lucini:2003zr}
B.~Lucini, M.~Teper, and U.~Wenger, ``{The High temperature phase transition in SU(N) gauge theories},'' {\em JHEP} {\bf 01} (2004) 061, \href{http://www.arXiv.org/abs/hep-lat/0307017}{{\tt hep-lat/0307017}}.

\bibitem{Lucini:2005vg}
B.~Lucini, M.~Teper, and U.~Wenger, ``{Properties of the deconfining phase transition in SU(N) gauge theories},'' {\em JHEP} {\bf 02} (2005) 033, \href{http://www.arXiv.org/abs/hep-lat/0502003}{{\tt hep-lat/0502003}}.

\bibitem{EliasMiro:2019kyf}
J.~Elias~Mir\'o, A.~L. Guerrieri, A.~Hebbar, J.~a. Penedones, and P.~Vieira, ``{Flux Tube S-matrix Bootstrap},'' {\em Phys. Rev. Lett.} {\bf 123} (2019), no.~22, 221602, \href{http://www.arXiv.org/abs/1906.08098}{{\tt 1906.08098}}.

\bibitem{Conti:2018jho}
R.~Conti, L.~Iannella, S.~Negro, and R.~Tateo, ``{Generalised Born-Infeld models, Lax operators and the $ \mathrm{T}\overline{\mathrm{T}} $ perturbation},'' {\em JHEP} {\bf 11} (2018) 007, \href{http://www.arXiv.org/abs/1806.11515}{{\tt 1806.11515}}.

\bibitem{Dubovsky:2018vde}
S.~Dubovsky, ``{The QCD $\beta$-function On The String Worldsheet},'' {\em Phys. Rev. D} {\bf 98} (2018), no.~11, 114025, \href{http://www.arXiv.org/abs/1807.00254}{{\tt 1807.00254}}.

\end{thebibliography}\endgroup

\end{document}